\documentclass[superscriptaddress, 10pt, amsmath,amssymb, aps, pra, twocolumn,preprintnumbers]{revtex4-2}

\usepackage[utf8]{inputenc}
\usepackage[T1]{fontenc}
\usepackage{amsmath}
\usepackage{amsfonts}
\usepackage{amssymb}
\usepackage{amsthm}
\usepackage{dsfont}
\usepackage{graphicx}
\usepackage[dvipsnames]{xcolor}
\usepackage{enumitem}
\usepackage{hyperref}
\usepackage{physics}
\usepackage{bbm}
\usepackage{indentfirst}
\usepackage{thmtools}
\usepackage{thm-restate}
\usepackage{soul}
\usepackage{comment}
\setstcolor{red}
\usepackage{breakurl}

\usepackage{algorithm,algorithmic}

\usepackage{setspace}

\newtheorem{theorem}{Theorem}

\newcommand{\ACProb}{\mathsf{AC}[B_{\min}, \Delta \omega]}

\newcommand{\QSPerror}{\varepsilon_{\scriptscriptstyle \text{QSP}}}
\newcommand{\tauconv}{\tau_{\scriptscriptstyle \mathrm{conv}}}
\newcommand{\tauqss}{\tau_{\scriptscriptstyle \mathrm{QSS}}}
\newcommand{\Tmem}{T_{\mathrm{mem}}}

\newcommand{\figtwo}[2]{FIG.~\ref{#1}\,#2}

\begin{document}

\title{Quantum Computing Enhanced Sensing}

\author{Richard R. Allen}
\thanks{These authors contributed equally to this work.}
\affiliation{Department of Physics, Massachusetts Institute of Technology, Cambridge, MA 02139, USA}
\author{Francisco Machado}
\thanks{These authors contributed equally to this work.}
\affiliation{ITAMP, Harvard-Smithsonian Center for Astrophysics, Cambridge, MA 02138, USA}
\affiliation{Department of Physics, Harvard University, Cambridge, MA 02138, USA}
\author{Isaac L. Chuang}
\affiliation{Department of Physics, Massachusetts Institute of Technology, Cambridge, MA 02139, USA}
\author{Hsin-Yuan Huang}
\affiliation{Google Quantum AI, Venice, CA 90291, USA}
\affiliation{Department of Physics, Massachusetts Institute of Technology, Cambridge, MA 02139, USA}
\affiliation{Division of Physics, Mathematics and Astronomy, California Institute of Technology, Pasadena, CA 91125, USA}
\author{Soonwon Choi}
\thanks{Corresponding author: soonwon@mit.edu}
\affiliation{Department of Physics, Massachusetts Institute of Technology, Cambridge, MA 02139, USA}
\preprint{MIT-CTP/5804}

\begin{abstract}
Quantum computing and quantum sensing represent two distinct frontiers of quantum information science. In this work, we harness quantum computing to solve a fundamental and practically important sensing problem: the detection of weak oscillating fields with unknown strength and frequency. We present a quantum computing enhanced sensing protocol that outperforms all existing approaches. Furthermore, we prove our approach is optimal by establishing the \emph{Grover-Heisenberg limit} --- a fundamental lower bound on the minimum sensing time. The key idea is to robustly \emph{digitize} the continuous, analog signal into a discrete operation, which is then integrated into a quantum algorithm. Our metrological gain originates from quantum computation, distinguishing our protocol from conventional sensing approaches. Indeed, we prove that broad classes of protocols based on quantum Fisher information, finite-lifetime quantum memory, or classical signal processing are strictly less powerful. Our protocol is compatible with multiple experimental platforms. We propose and analyze a proof-of-principle experiment using nitrogen-vacancy centers, where meaningful improvements are achievable using current technology. This work establishes quantum computation as a powerful new resource for advancing sensing capabilities.
\end{abstract}

\maketitle

\section{Introduction}

The development of large-scale, coherent quantum devices opens new paradigms for addressing scientific and technological challenges, beyond what is possible using classical approaches~\cite{cerezo:2021, daley:2022,hangleiter:2023}.
A particularly impactful direction is quantum sensing~\cite{degen:2017}.
By controlling individual quantum particles, one can sense at precisions limited only by the quantum fluctuations of each sensor, the so-called standard quantum limit (SQL).
Moreover, by engineering entanglement between sensors, it is possible to surpass the SQL and approach the ultimate sensitivity allowed by quantum mechanics, the Heisenberg limit~\cite{giovannetti:2004, pezze:2018a}.
These insights and techniques now power our most advanced measurements of 
magnetic fields~\cite{zhang:2021},
time~\cite{aeppli:2024},
gravitational redshift~\cite{zheng:2023},
gravitational waves~\cite{ligoo4detectorcollaboration:2023}, and
fundamental constants~\cite{bass:2024}, and they enable sensing under novel conditions, with applications in biology~\cite{aslam:2023} and materials science~\cite{bhattacharyya:2024}.
More recently, ideas from quantum error correction~\cite{dur:2014, demkowicz-dobrzanski:2017,zhou:2018,niroula:2024}, learning theory~\cite{aharonov:2022, huang:2022, huang:2023}, information theory~\cite{shaw:2024,cao:2024a}, and
classical optimization~\cite{marciniak:2022}
have enabled protocols which are more robust, more efficient, and detect signals across a larger range of parameters.

An emerging advancement in these quantum platforms is the ability to perform quantum computation~\cite{acharya:2024a, bluvstein:2024,reichardt:2024}.
Quantum computing is one of the most promising long-term applications of quantum devices, and it is expected to outperform classical alternatives in many settings~\cite{
dalzell:2023}.
This raises a key question: can quantum computation be harnessed to enhance sensing?

\begin{figure*}[t]
    \centering
    \includegraphics[width=\linewidth]{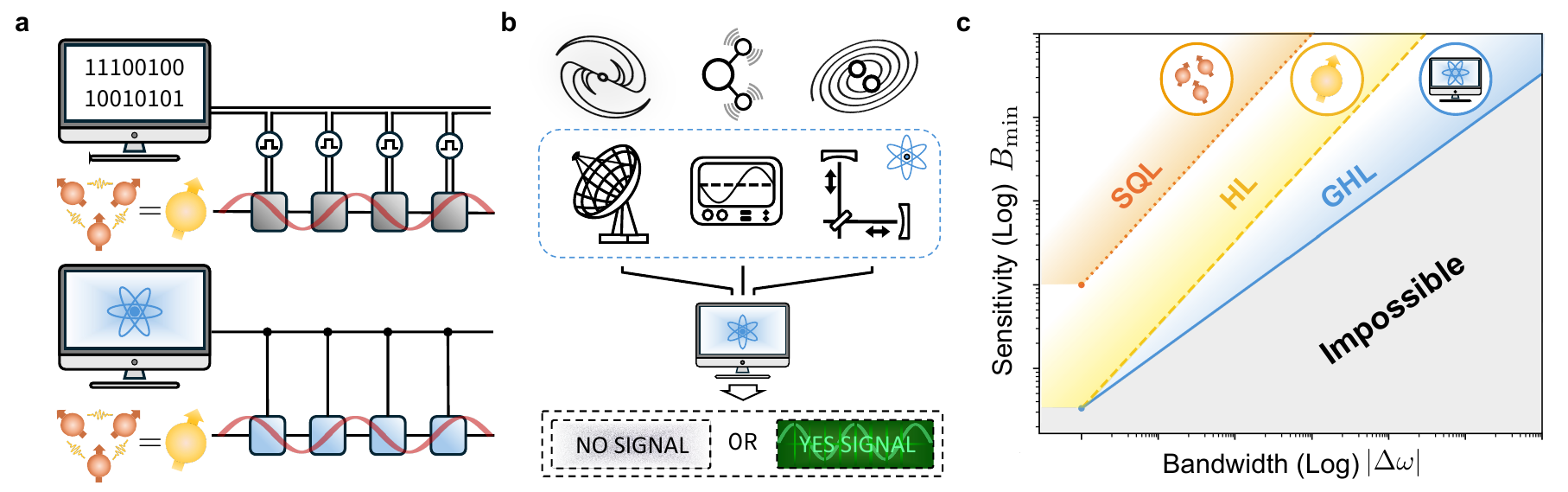}
    \caption{
    {\bf a}  Conventional versus quantum computing enhanced sensing.
    In conventional quantum sensing, 
    entangled sensor particles are controlled by a classical computer (top). 
    We consider a new setting, where the sensor particles are controlled by a quantum computer, which can operate in superposition and run quantum algorithms (bottom). 
    {\bf b} The ubiquity of AC sensing. Many important tasks in science and technology depend on \emph{AC sensing}: the detection of a weak oscillating field of unknown frequency.
    {\bf c} Fundamental limits in AC sensing.
    For a fixed sensing time $\tau$ with $n_S$ quantum sensors, the achievable sensitivity $B_{\min}$ and bandwidth $|\Delta\omega|$ for AC sensing depend on the class of protocols considered, leading to three fundamental limits
    (i) the standard quantum limit (SQL) for conventional protocols using unentangled sensors, $B_{\min} \sim (n_S \tau)^{-1/2} |\Delta \omega|^{1/2}$; (ii) the Heisenberg limit (HL) for conventional protocols using entangled sensors, $B_{\min} \sim n_S^{-1} \tau^{-1/2} |\Delta \omega|^{1/2}$; and (iii) the Grover-Heisenberg limit (GHL) for any protocol based on quantum mechanics, $B_{\min} \sim n_S^{-1} \tau^{-2/3} |\Delta \omega|^{1/3}$.
    Our quantum search sensing algorithm saturates the GHL (up to polylogarithmic corrections in $|\Delta \omega|/(n_S B_{\min})$). See Theorems~\ref{thm: qss_main_text} and~\ref{thm: ghlimit_main_text}.
    }
    \label{fig: fig1}
\end{figure*}
    
In this work, we demonstrate that by augmenting quantum sensors with a quantum processor, computational insights can be leveraged to more efficiently accomplish practically useful sensing tasks.
We focus on the problem of \emph{AC sensing}: the detection of weak oscillating fields with unknown strength and frequency. 
AC sensing is ubiquitous in the physical sciences, with applications as varied as the search for axionic dark matter~\cite{graham:2015}, high resolution nuclear spectroscopy~\cite{reif:2021}, and gravitational wave detection~\cite{tse:2019,ligoscientificcollaborationandvirgocollaboration:2016,ligoo4detectorcollaboration:2023}.
We present a quantum computing enhanced protocol, dubbed \emph{quantum search sensing}, which outperforms all existing strategies for AC sensing (FIG.~\ref{fig: fig1}).
At its core, our protocol transforms time evolution under the analog AC signal into a digital multi-qubit gate, which we integrate as a subroutine in a quantum algorithm. Furthermore, we establish a fundamental limit on the efficiency of AC sensing, which we call the \emph{Grover-Heisenberg limit}. Our algorithm saturates this lower bound on sensing time up to polylogarithmic factors.
Our protocol is compatible with multiple existing experimental platforms. As an example, we present and analyze a detailed experimental proposal based on nitrogen-vacancy centers, where meaningful improvements are within reach using current hardware.

Our approach is fundamentally different from conventional techniques in quantum sensing. Unlike existing quantum sensing protocols, our metrological enhancement originates from a quantum computational speedup. We illustrate this by proving that several large classes of sensing protocols --- based on maximizing quantum Fisher information, using finite-lifetime quantum memory, or using classical signal processing --- are strictly less powerful than our algorithm. Despite the connection between metrological and quantum computational enhancement, their exact relationship is subtle, leading to two important consequences. First, even in the presence of decoherence and without quantum error correction, our approach outperforms conventional strategies in certain regimes. Second, the metrological gain can be dramatically larger if additional information is known about the AC signal, enabling a constant-versus-polynomial sensing speedup. We discuss future directions and opportunities in quantum computing enhanced sensing.

\section{\texorpdfstring{Quantum Computing\\ Enhanced Sensing}{Quantum Computing Enhanced Sensing}}

\subsection{AC Sensing} 

We consider the problem of detecting a weak oscillating field over a broad frequency range: $B(t) = B \cos(\omega t + \phi)$ with unknown $B$, $\omega$, and $\phi$. The signal couples to $n_S$ sensor qubits via the Hamiltonian $H(t) = \hbar g B(t) \sum_{i=1}^{n_S} Z_i$, where $Z_i$ is the Pauli $Z$ operator for the $i$-th sensor. We choose units such that $g = \hbar = 1$, so $B$ has units of inverse time. This task is characterized by two parameters: a minimum detectable signal strength $B_{\min}$ (setting the sensitivity) and a frequency interval $\Delta \omega=[\omega_\textrm{min}, \omega_\textrm{max}]$ (setting the bandwidth $|\Delta \omega| = \omega_\textrm{max} - \omega_\textrm{min}$). We assume $\omega_\textrm{min} \ge B_\textrm{min}$, because otherwise the signal is quasi-static and can be detected using conventional sensing techniques for time-independent fields, as elaborated in the Supplementary Material (SM). Our goal is to determine if any signal exists with $B \ge B_{\min}$, $\omega \in \Delta \omega$, and arbitrary $\phi$ as quickly as possible, minimizing total sensing time $\tau$. We call this problem $\ACProb$. While formulated as a decision problem, a protocol for this task can be used as a subroutine to efficiently estimate $B$ and $\omega$, connecting $\ACProb$ to parameter estimation.
Despite the ubiquity of AC sensing~(\figtwo{fig: fig1}{b}), the optimal protocol and fundamental limit for $\ACProb$ are not known.

\begin{figure}[t]
    \centering\includegraphics[width=\linewidth]{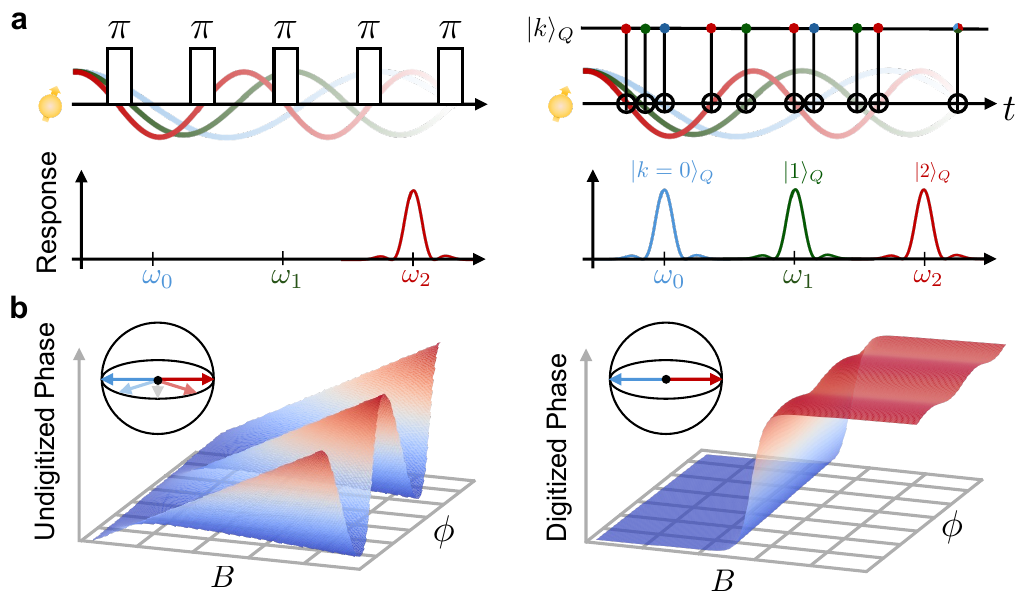}    \caption{\textbf{a} Sensing in superposition. In conventional AC sensing (left), a single target frequency is probed at a time. This frequency is determined, for instance, by the interpulse spacing between periodically applied $\pi$-pulses ($X$ gates). In quantum computing enhanced sensing (right), we probe multiple target frequencies in superposition by conditioning the interpulse spacing, and hence target frequency, on the state of the quantum computer $\ket{k}_Q$.
    \textbf{b} Digitization in quantum sensing. Conventional sensing protocols are sensitive to small perturbations in the unknown signal parameters $B$ and $\phi$, whereas a digital sensing protocol discretizes the response subject to some logical condition, e.g., $B \ge B_{\min}$ and $\omega$ in bin $k_*$. Our sensing oracle accomplishes both these goals. 
    }
    \label{fig: fig2}
\end{figure}

Existing approaches solve $\ACProb$ by probing signals at a particular target frequency $\omega_t$, then scanning the target frequency across $\Delta \omega$~\cite{degen:2017}. Each sensing step can be accomplished, for instance, by periodically applying $X$ gates to the sensor qubits with interpulse spacing $\pi/\omega_t$. The resulting effective signal has a time-independent Fourier component, which can be detected using Ramsey interferometry (\figtwo{fig: fig2}{a}).
This approach requires a total sensing time
\begin{align}\label{eq: conventscaling}
    \tauconv = O
    \left( \frac{1}{n_S B_{\min}} \bigg\lceil \frac{|\Delta \omega|}{n_S B_{\min}} \bigg\rceil \right).
\end{align}
The first factor is the Heisenberg-limited sensing time $T \sim 1/(n_S B_\textrm{min})$ necessary to detect a signal of strength $B_\textrm{min}$.
The second term is the number of sensing steps required to scan across $\Delta \omega$; each step can detect a signal over a Fourier-limited bandwidth $\delta \omega \sim 1/T \sim n_S B_\textrm{min}$ around $\omega_t$. 
The scaling of $\tauconv$ determines the range of detection problems that can be solved in fixed total time, resulting in a trade-off between sensitivity and bandwidth (\figtwo{fig: fig1}{c}). Later, we will show that this scaling is optimal in conventional settings;  any protocol relying on quantum Fisher information or which processes the signal classically requires $\tauconv$ time to solve $\ACProb$.

Here, we introduce a new sensing paradigm, in which the quantum sensors are controlled by a quantum computer~(\figtwo{fig: fig1}{a}). Specifically, we consider protocols which utilize additional computational qubits that are not coupled to the signal, and which can perform arbitrary, instantaneous unitary gates between the sensors and the quantum computer (henceforth denoted $S$ and $Q$). Using these resources, we design the quantum search sensing (QSS) algorithm, which solves $\ACProb$ faster than all existing approaches:
\begin{theorem}\label{thm: qss_main_text} 
The quantum search sensing algorithm solves $\ACProb$ in time
\begin{align}\label{eq: qss_scaling}
    \tauqss = \widetilde{O}\left( \frac{1}{n_S B_{\min}} \sqrt{\bigg\lceil \frac{|\Delta \omega|}{n_S B_{\min}} \bigg\rceil} \right).
\end{align}
\end{theorem}
\noindent The tilde means that this scaling neglects polylogarithmic factors in $|\Delta \omega|/(n_S B_{\min})$. We describe QSS in the following section.

The performance of QSS implies a modified trade-off between $B_{\min}$ and $|\Delta \omega|$ at fixed sensing time $\tau$. 
Consequently, when searching over a given bandwidth, one can detect parametrically weaker signals compared to existing approaches, and vice versa (\figtwo{fig: fig1}{c}).
We demonstrate that QSS is asymptotically optimal (up to polylogarithmic factors) by proving a fundamental limit:
\begin{theorem}\label{thm: ghlimit_main_text} 
Any quantum computing enhanced sensing protocol that solves $\ACProb$ must take time at least
\begin{align} \label{eq:GHL-bound}
    \tau = \Omega \left( \frac{1}{n_S B_{\min}} \sqrt{\bigg\lceil \frac{|\Delta \omega|}{n_S B_{\min}} \bigg\rceil} \right).
\end{align}
\end{theorem}
We call this lower bound the Grover-Heisenberg limit (GHL).
The GHL holds regardless of the number of computational qubits or unitary gates and thus establishes a fundamental limitation on our ability to detect AC signals. The proof of the GHL is presented in the SM.

\subsection{Quantum Search Sensing Algorithm}

The key insight of our quantum search sensing algorithm is that $\ACProb$ can be reduced to an unstructured search problem with a known quantum speedup. While classical search requires $O(N)$ queries to find a marked element $k_*$ in a list of $N$ items, Grover's algorithm achieves this with only $O(\sqrt{N})$ quantum queries. This speedup relies on a \emph{quantum oracle} $\mathcal{O}$ that acts as $\mathcal{O}\ket{k} = (-1)^{\delta_{k,k_*}} \ket{k}$, enabling coherent queries over multiple elements in superposition.

To harness this computational advantage for sensing, we partition $\Delta \omega$ into $N \sim \lceil |\Delta \omega|/(n_S B_{\min}) \rceil$ frequency bins. Our algorithm searches for the bin $k_*$ containing the signal using a quantum \emph{sensing oracle} $\mathcal{O}_\textrm{AC}$ constructed from the analog classical field:
\begin{align} \label{eq:Oracle}
\mathcal{O}_\textrm{AC}\ket{k}_Q\ket{\psi}_S =
        \left\{
    \begin{array}{rl}
    -\ket{k}_Q\ket{\psi}_S & \textrm{if signal in bin $k$}\\
    \ket{k}_Q\ket{\psi}_S & \hspace{1cm}\textrm{else}
    \end{array}
    \right.
\end{align}
where $\ket{k}_Q$ is the quantum computer state associated with the bin $k$, and $\ket{\psi}_S$ is a specific initial state of the sensors. By ensuring that the sensors always return to $\ket{\psi}_S$, we can repeatedly apply the sensing oracle in Grover's algorithm. If the signal exists with $B \ge B_{\min}$ and $\omega \in \Delta \omega$, the algorithm will output the correct bin $k_*$. Otherwise, the algorithm outputs an arbitrary bin, which we verify using conventional AC sensing. With the chosen $N$, the width of each bin is sufficiently small that all frequencies in a bin can be probed simultaneously using conventional techniques.

Our sensing oracle $\mathcal{O}_\textrm{AC}$ has two key features that distinguish it from conventional sensing protocols. First, it probes many target frequencies in superposition, conditioned on the state of the quantum computer (\figtwo{fig: fig2}{a}). Second, it \emph{digitizes} the analog signal into a discrete quantum operation. To robustly evaluate the logical condition ``if signal in bin $k$,'' the sensing oracle must transform a continuous family of signals into the same quantum operation (\figtwo{fig: fig2}{b}). This digitization is particularly challenging due to the time dependence of the oscillating field $B(t)$, which means each oracle application experiences a different signal profile.

\begin{figure}[t]
    \centering
    \includegraphics[width=\linewidth]{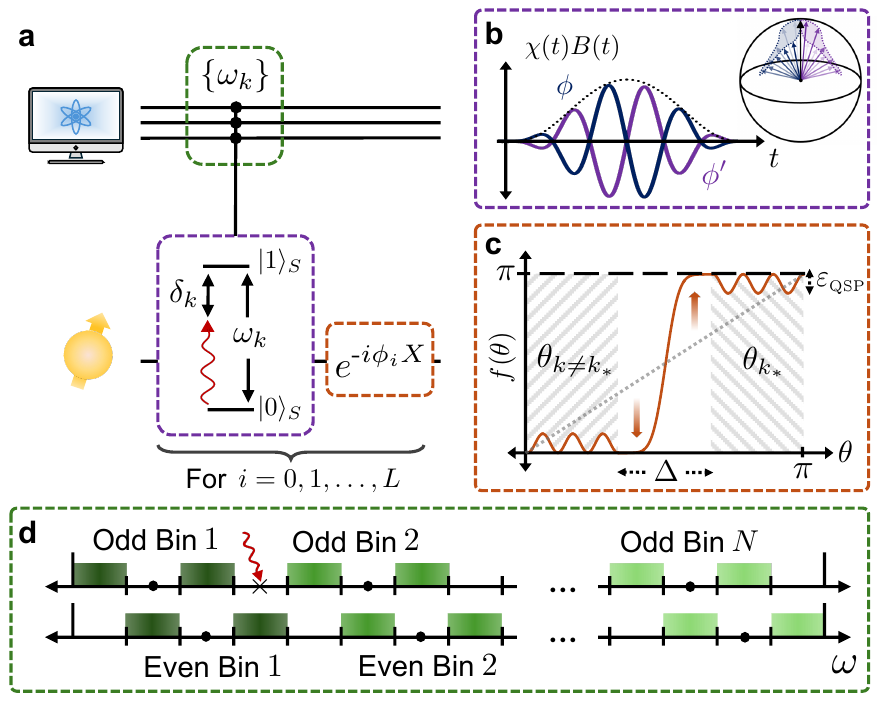}
    \caption{\textbf{a} The sensing oracle. The construction of the sensing oracle has two steps: 
    (i) designing an elementary sensing unit (ESU) (green and purple) and (ii) quantum signal processing (QSP) (orange).
    \textbf{b} ESU via superadiabaticity. By smoothly turning the signal coupling on and off via $\chi(t)$, the system undergoes an approximately adiabatic evolution
    and the sensor accumulates a $\phi$-independent dynamical phase. The inset depicts the evolution in the rotating frame, drawn assuming the signal couples to the sensor in the $X$ basis.
    \textbf{c} Digitization via QSP. By using the ESU as the signal unitary in a QSP sequence, we transform the rotation angles according to a nonlinear function $f(\theta)$. This rectifies the ESU angles, which we ensure are separated by at least $\Delta$, so that $f(\theta_{k_*}) \approx \pi$ and $f(\theta_{k \neq k_*}) \approx 0$ up to an error $\QSPerror$.
    \textbf{d} Binning strategy. We guarantee that all errors in our protocol are well-controlled by a careful choice of frequency bins. We partition $\Delta \omega$ into two sets of bins (odd and even), probed separately. Each bin excludes the frequencies closest to $\omega_k$, so that $B/\delta_k$ is not too large (bounding the ESU error) and $\theta_{k_*}$ is larger in magnitude than all $\theta_{k \neq k_*}$ (bounding the QSP error). See SM for details.
    }
    \label{fig: fig3}
\end{figure}

We implement $\mathcal{O}_\textrm{AC}$ in two steps (\figtwo{fig: fig3}{a}). To simplify our presentation, we assume $n_S = 1$; our construction generalizes to any number of sensors by defining an effective sensor qubit with collectively enhanced coupling $n_S g$, using $\ket{0}^{\otimes n_S}$ and $\ket{1}^{\otimes n_S}$. The first step is to engineer an elementary sensing unit (ESU) --- a conditional rotation of the form $\sum_k \ketbra{k}_Q\otimes \exp[-i \theta_k Z_S]$, realized by evolution under a particular effective Hamiltonian. The key is to ensure that the rotation angles $\theta_k = \theta_k(B,\omega)$ are independent of the signal phase $\phi$, making each ESU application identical, while guaranteeing that $\theta_{k_*}$ has larger magnitude than all other $\theta_{k \neq k_*}$.
The second step transforms these rotations nonlinearly using quantum signal processing (QSP)~\cite{martyn:2021},
effectively discretizing the sensor response to the desired $\pm 1$ phase factor.

To construct the ESU, we combine techniques from quantum optics and adiabatic quantum computing. We first apply partial dynamical decoupling (via a rapid sequence of $X$ gates) to effectively modulate the sensor's coupling to the signal: $B(t) \mapsto \chi(t) B(t)$, where $\chi(t)$ is a smooth control function satisfying $|\chi(t)|\le 1$. We then introduce a control Hamiltonian $H_c = \sum_k \omega_k\ketbra{k}_Q \otimes X_S/2$, where $\omega_k$ is the center frequency of the $k$-th bin. The system evolves under $\chi(t)H(t) + H_c$ for time $T$ as $\chi(t)$ is adiabatically ramped up from $0$ to $1$ and back down to $0$ (\figtwo{fig: fig3}{b}). A final corrective rotation $e^{+iH_{c}T} = Z_S e^{-iH_cT}Z_S$ removes the accumulated dynamical phase in the absence of the signal. Up to a basis change ($Z_S\leftrightarrow X_S$), this sequence implements the ESU.

To understand why this applies a $\phi$-independent conditional rotation, we analyze the dynamics in the rotating frame defined by $H_{\text{rot}} = \omega X_S$. Under the rotating wave approximation, the effective Hamiltonian takes the form $H_\textrm{eff} = \sum_k \ketbra{k}_Q \otimes H_\textrm{eff}^{(k)}$, where:
\begin{align}
    H_\textrm{eff}^{(k)} =  \frac{\delta_k}{2}  X_S + \frac{B}{2}  \chi(t) \left(\cos(\phi) Z_S - \sin(\phi) Y_S \right),
\end{align}
with $\delta_k = \omega_k - \omega$. Crucially, while $H_{\mathrm{eff}}^{(k)}$ depends on $\phi$, its instantaneous eigenvalues do not. Therefore, adiabatic evolution along a closed loop in the Bloch sphere generates a $\phi$-independent dynamical phase, leaving the sensor otherwise unchanged (\figtwo{fig: fig3}{b}). By choosing $\chi(t)$ to be both slowly varying and smooth, we can make the diabatic errors exponentially small in $B/\delta_k$ using superadiabatic techniques~\cite{ge:2016, nenciu:1993} (see SM). After the corrective rotation, we obtain the desired ESU with rotation angles
\begin{align}
    \theta_k = \frac{1}{2} \int_0^T dt~\sqrt{(B \chi(t))^2 + \delta_k^2} - \frac{\delta_k T}{2}= O\left(\frac{B^2 T}{\delta_k}\right)
\end{align}
in the original frame if $B/\delta_k$ is not too large, which we ensure through our frequency binning strategy (\figtwo{fig: fig3}{d} and SM). The angles $\theta_k$ are independent of $\phi$ and decrease in magnitude with $\delta_k$; this ensures that $\theta_{k_*}$ is separated from all $\theta_{k \neq k_*}$ by at least some gap $\Delta$. 

The second step employs QSP~\cite{martyn:2021} to implement a nonlinear transformation of the rotation angles $\theta_k$. By interleaving multiple ESU applications with optimized single-qubit sensor rotations about the $X$ axis, we engineer the operation $\sum_k \ketbra{k}_Q\otimes \exp[-i f(\theta_k) Z_S]$, where $f(\theta)$ approximates a step function of height $\pi$ (\figtwo{fig: fig3}{c}). Given the gap $\Delta$ between $\theta_{k_*}$ and the other rotation angles $\theta_{k \neq k_*}$, we ensure $f(\theta_{k_*}) \approx \pi$ while $f(\theta_{k\neq k_*}) \approx 0$ across all relevant values of $B$ and $\omega$. The approximation error $\QSPerror$ decreases exponentially with both the number of ESU applications $L$ and the gap $\Delta$, enabling efficient digitization of the sensor response.

\setstretch{1.05}
This completes our construction of the sensing oracle $\mathcal{O}_\textrm{AC}$. To perform the quantum search, we apply the sequence $R_s \mathcal{O}_\textrm{AC}$ a total of $O(\sqrt{N})$ times, where $R_s = 2 \ketbra{s}_Q - 1$ is the Grover reflection operator with $\ket{s}_Q = \sum_k \ket{k}_Q / \sqrt{N}$. By choosing $N \sim \lceil |\Delta \omega| / (n_S B_{\min}) \rceil$ and $T \sim (n_S B_{\min})^{-1}$, we ensure the Grover search succeeds with high probability. The total sensing time is thus $\tau_{\scriptscriptstyle \text{QSS}} = \widetilde{O}((n_S B_{\min})^{-1} \sqrt{\lceil |\Delta \omega|/(n_S B_{\min}) \rceil})$, achieving the scaling promised in Theorem~\ref{thm: qss_main_text}.
\setstretch{1}

\subsection{Experimental Realization}

\begin{figure}[t]
    \centering
    \includegraphics[width=0.9\linewidth]{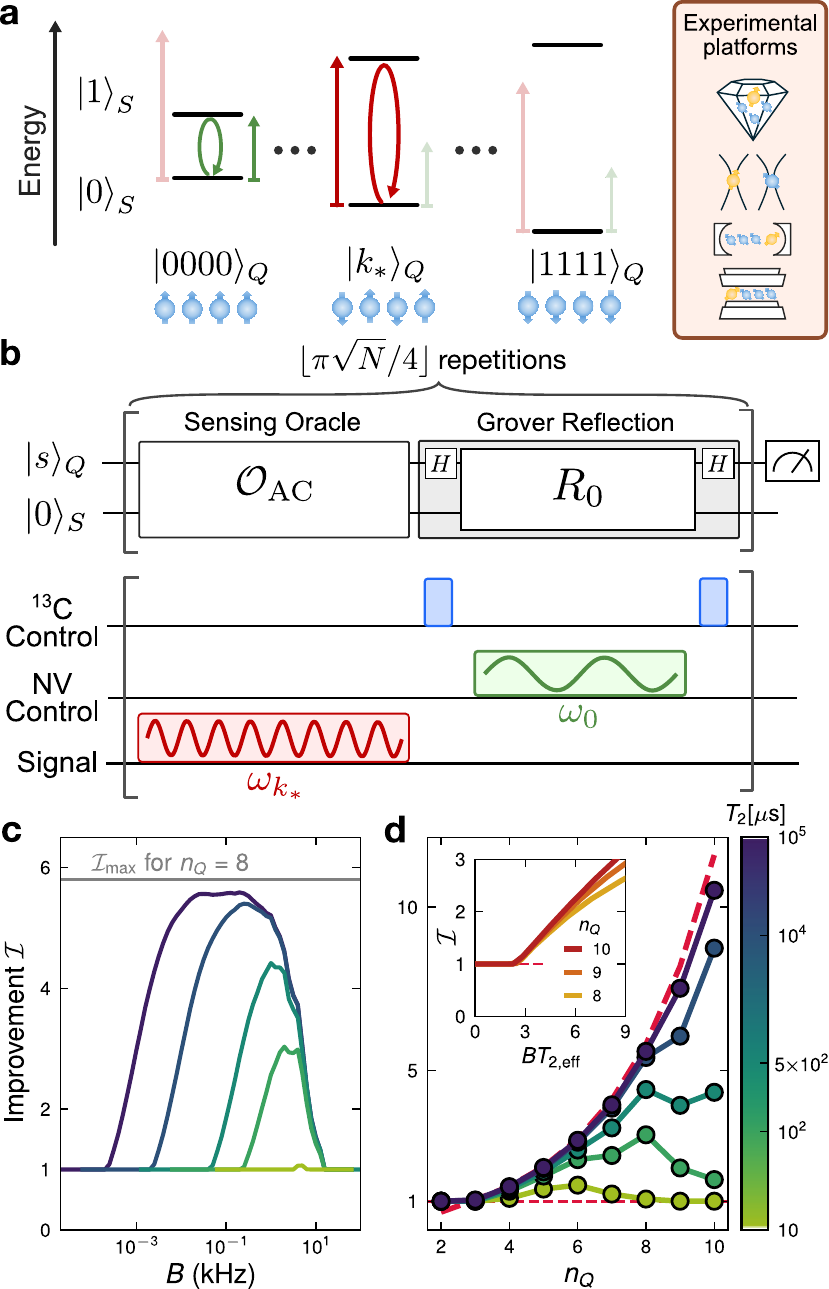}    
    \caption{ {\bf a} 
    Energy level diagram for the sensor qubit conditioned on various computational qubit configurations; the conditional energy shifts naturally arise from Ising interactions, which are present in many quantum sensing and computing platforms.
    In our simplified protocol, both the sensing oracle (red) and the Grover reflection (green) can be implemented using a resonant $2\pi$-rotation of the sensor qubit.
    {\bf b} 
    The control pulse sequence (bottom) for our simplified proposal (top) in an NV-nuclear spin ensemble.
    {\bf c} The improvement factor $\mathcal{I}$ as a function of $B$ for $n_Q=8$ and various $T_2$ (different colors) ranging from $10$~ms~\cite{bradley:2019} (yellow) to $10$~s (dark blue).
    By increasing $T_2$, our quantum computing enhanced sensing protocol exhibits an improvement that approaches the maximum Grover speedup over a wide range of signal strengths $B$.
    At large $B$, the improvement is limited by power broadening.
    {\bf d}
    Estimation of the maximum achievable $\mathcal{I}$ as a function of $n_Q$ for various $T_2$.
    For sufficiently large $T_2$, the improvement grows exponentially in the number of computational qubits. For large $n_Q$ (inset), we observe that the protocol's improvement $\mathcal{I}$ scales linearly with $BT_{2,\text{eff}}$.
    }\vspace{-1.8em}
    \label{fig:Expt}
\end{figure}

Our algorithm is readily implementable in many existing experimental platforms.
Here, we show that both the sensing oracle $\mathcal{O}_{\text{AC}}$ and the Grover reflection $R_s$ can be realized using only three ingredients: (i) arbitrary rotations of the sensor qubit, (ii) global rotations of the computational qubits, and (iii) Ising interactions between the sensor and computational qubits: 
\begin{equation}
    H_\textrm{Ising} = \frac{\omega_S}{2} Z_S + \sum_{i=1}^{n_Q} \frac{J_i}{2} Z_{Q,i} Z_S, 
\end{equation}
where $\omega_S$ is the bare transition frequency of the sensor qubit, $Z_{Q,i}$ is the Pauli $Z$ operator for the $i$-th computational qubit, and $J_i$ is the interaction strength.
A variety of experimental platforms realize this Hamiltonian, including solid state defects \cite{bradley:2019}, Rydberg atoms~\cite{evered:2023a}, qubits coupled to a cavity mode~\cite{blais:2021,pedrozo-penafiel:2020}, and trapped ions~\cite{monroe:2021}.

The key observation is that the spectral structure of the Ising interactions mirrors that of the previously discussed $H_c$:
$H_\textrm{Ising} = \sum_k \omega_k \ketbra{k}_Q \otimes  Z_S/2$.  
Here, the index $k$ enumerates the configurations of the computational qubits $\ket{k}_Q = | k_1  k_2  \dots k_{n_Q} \rangle_Q$ with $k_i \in \{0,1\}$, and the couplings $J_i$ determine the centers of the frequency bins
$\omega_k = \omega_S + \sum_{i=1}^{n_Q}  (-1)^{k_i} J_i $. One can engineer harmonically spaced $\omega_k$ by fine-tuning the $J_i$, but random couplings still provide $2^{n_Q}$ distinct frequency bins. 

Given $H_\textrm{Ising}$, we can implement $R_s$ using the decomposition $R_s = H \,R_0\,H$, where $R_0 = 2 \ketbra{0 \cdots 0}_Q - 1$ is the reflection around the all-zero computational state and $H$ is the $n_Q$-qubit Hadamard gate. 
To implement $R_0$, one applies a time-dependent drive to the sensor, in the presence of the Ising interactions, using a pulse with frequency $\omega_{0}$ and Rabi frequency $B_{R_0}$ for a duration $2\pi/B_{R_0}$.
For sufficiently weak $B_{R_0}$,
the pulse realizes a resonant $2\pi$-rotation of the sensor qubit conditional on the computational qubits being in $\ket{0 \cdots 0}_Q$.
This rotation imparts a phase $-1$ while leaving the sensor otherwise unchanged~(\figtwo{fig:Expt}{a}~green). 
The implementation of the sensing oracle follows our previous discussion by identifying $H_{\text{Ising}}$ with $H_{c}$ and using single qubit rotations for dynamical decoupling and QSP.

The construction of the sensing oracle can be drastically simplified if the value of $B$ is known and $\omega$ belongs to the discrete set $\{\omega_k\}$.
In this case, the oracle can be implemented analogously to $R_0$.
The sensor is exposed to the signal field at $\omega=\omega_{k_*}$ (rotated to the transverse direction) for duration $2\pi/B$ in the presence of the Ising interactions, realizing a resonant $2\pi$-rotation of the sensor qubit conditional on the computational qubits being in $| k_* \rangle_Q$~(\figtwo{fig:Expt}{a}~red).
The resulting, simplified protocol (\figtwo{fig:Expt}{b}) is particularly suitable for proof-of-principle demonstrations of quantum computing enhancement.

As a representative example of a mature quantum sensing platform, we consider an implementation of the simplified protocol using solid state nitrogen-vacancy (NV) centers (other platforms are discussed in the SM). 
The NV center is a spin-1 defect in diamond which can sense a wide variety of signals (magnetic, electric, temperature, pressure, etc.)~\cite{barry:2020}.
It can be easily initialized, controlled, and read out~\cite{doherty:2013} and, crucially, dispersively couples to nearby nuclear spins (both $^{13}$C and the host $^{14}$N), enabling quantum logic gates~\cite{bradley:2019}. 
The electronic spin of the NV and $n_Q$ nuclear spins serve as the sensor and computational qubits, respectively.
We numerically simulate our proposed experiment, incorporating decoherence as well as details of the energy structure of the NV-nuclear spin ensemble based on measured interaction parameters~\cite{abobeih:2019}.
We assume that all spins have the same dephasing rate $1/T_2$, conservatively, informed by the electronic coherence time~(see SM for details).
We quantify the improvement by computing the ratio $\mathcal{I}$ between the times to detect the signal using a conventional approach versus our simplified protocol, averaged over the possible signal frequencies in $\{\omega_k\}$.

We find that $\mathcal{I}$ is limited by two factors: the coherence time $T_2$ and the interaction strength.
For sufficiently weak signals, the protocol's duration exceeds the system's coherence time, preventing the error-free completion of Grover's algorithm.
For sufficiently strong signals, power broadening relative to the separation between nearby frequencies $\omega_k$ prevents the faithful implementation of the sensing oracle.
The interplay of these two effects leads to nontrivial dependence of the improvement factor $\mathcal{I}$ (\figtwo{fig:Expt}{c,d}); the improvement scales as $\mathcal{I}\sim B T_{2,\text{eff}}$, where $T_{2,\text{eff}} = T_2/(n_Q + 1)$ is the system's effective coherence time, until it is limited by power broadening (large $B$) or the total number of computational qubits (large $T_2$). 
For realistic choices of $T_2$,
achievable in experiments using various forms of dynamical decoupling~\cite{abobeih:2018,bradley:2019}, our analysis predicts an improvement in the range of $\mathcal{I} \sim 2-4$, without additional technical innovations (\figtwo{fig:Expt}{c}).
In ~\figtwo{fig:Expt}{d}, we estimate the maximum achievable improvement in NV-nuclear spin ensemble experiments as a function of $n_Q$ for various coherence times.

\section{Improving Sensing via Quantum Computational Enhancement}
\label{sec: advantage}

Our algorithm marks a significant departure from conventional approaches in quantum sensing. Traditional quantum sensing protocols often derive their enhancement by maximizing sensitivity to infinitesimal parameter changes, as quantified by the quantum Fisher information (QFI) $\mathcal{F}_Q$~\cite{degen:2017}. However, these conventional QFI-based approaches cannot achieve the Grover-Heisenberg limit in Eq.~\eqref{eq:GHL-bound}, as demonstrated below (see also~\cite{polloreno:2023}):
\begin{theorem}\label{thm: QFI} Any QFI-based quantum sensing protocol that solves $\ACProb$ must take time at least
\begin{align}
    \tau = \Omega \left( \frac{1}{n_S B_{\min}} \bigg\lceil \frac{|\Delta \omega|}{n_S B_{\min}} \bigg\rceil \right).
\end{align}
\end{theorem}
\noindent The time required by the best known method, Eq.~\eqref{eq: conventscaling}, matches this lower bound. Here, a QFI-based protocol is any protocol with well-defined QFI with respect to perturbatively small $B$ which satisfies $\mathcal{F}_Q(\omega) \ge C/B_\textrm{min}^2$ for all $\omega \in \Delta \omega$, for some constant $C$. From the quantum Cramér-Rao bound, this is the QFI necessary to distinguish between $B = 0$ and $B = B_{\min}$ across the entire bandwidth by performing an unbiased estimate of the signal strength~\cite{degen:2017}. Theorem~\ref{thm: QFI} also implies that any protocol solving $\ACProb$ at or near the GHL must exhibit QFI which vanishes on average over $\Delta \omega$ in the asymptotic limit $n_S B_\textrm{min}/|\Delta \omega| \rightarrow 0$ (see SM for details).

Our enhanced performance instead originates from quantum computing.
Arguing this precisely, however, is challenging, since there is no sharp theoretical distinction between the quantum sensors and the quantum computer. In principle, with arbitrary quantum gates and sufficient coherence time, a subset of the quantum sensors could effectively function as a quantum computer. Despite this, we now identify two key ingredients that differentiate quantum computing enhanced sensing from conventional approaches. To do so, we introduce two sensing models --- one with limited coherent quantum computation capability and one with limited signal processing ability --- and demonstrate that both are strictly less powerful than our paradigm. We then show how a beyond-quadratic speedup over conventional approaches can be achieved in the strong field regime.

\subsection{Sensing with Constant Memory Lifetime}

In the first model, termed \emph{sensing with constant memory lifetime}, we have complete control over both the quantum sensors and a quantum computer of unbounded size. However, after time $\Tmem$, we must measure all qubits and initiate a new experiment (chosen adaptively). We prove the following fundamental limitation:
\begin{theorem}\label{thm: ClQS} Any quantum sensing protocol with constant memory lifetime that solves $\ACProb$ must take time at least
\begin{align} \label{eq:ClQS}
    \tau = \Omega \left(\frac{1}{n_S B_{\min} \Tmem} \cdot \frac{1}{n_S B_{\min}} \bigg\lceil \frac{|\Delta \omega|}{n_S B_{\min}} \bigg\rceil \right).
\end{align}
\end{theorem}
\noindent This lower bound highlights the necessity of long-lived quantum memory for optimal AC sensing, even allowing arbitrary quantum computations in each experiment. 

Notably, a variant of our algorithm saturates this lower bound up to polylogarithmic factors: use QSS to probe the maximum possible frequency range within time $\Tmem$, then linearly iterate across the remainder of $\Delta \omega$. 
This observation reveals a surprising result: our metrological gain exhibits partial robustness to decoherence. We demonstrate in the SM that this iterative QSS outperforms conventional, error-free approaches in the strong field regime $B_{\min} T_2 \gtrsim 1$, where $T_2$ is the single sensor coherence time.
This robustness to error stems from a fundamental distinction between the standard oracle model and quantum sensing. While complexity in the former is measured by oracle query count (each with unit cost), sensing complexity is measured by total physical time. For QSS, this time depends not only on the number of oracle calls, but also on the duration of each individual sensing oracle, which scales as $1/(n_S B_{\min})$. Consequently, with constant coherence time, we can perform more sensing oracle calls as $B_{\min}$ increases, eventually surpassing conventional approaches.

\subsection{Sensing with Classical Signal Processing}

In the second model, termed \emph{sensing with classical signal processing}, we can adaptively conduct experiments of the following form: prepare arbitrary entangled states of the quantum sensors, interrogate the signal, and perform general measurements. During the interrogation phase, our control over the sensors is constrained to applying classical modulation functions that vary the sensors' interactions with the field according to $Z_i \mapsto \chi_i(t) Z_i$, where $|\chi_i(t)| \le 1$ for each sensor $i = 1, \hdots, n_S$.
This framework encompasses existing quantum sensing protocols in which all gates either commute or anticommute with the signal Hamiltonian. In the SM, we establish the following:
\begin{theorem}\label{thm: CpQS} Any quantum sensing protocol with classical signal processing that solves $\ACProb$ must take time at least
\begin{align}
    \tau = \Omega \left( \frac{1}{n_S B_{\min}} \bigg\lceil \frac{|\Delta \omega|}{n_S B_{\min}} \bigg\rceil \right).
\end{align}
\end{theorem}
\noindent This lower bound matches the scaling achieved by conventional protocols in Eq.~\eqref{eq: conventscaling}. Our result demonstrates that any classical signal processing approach based on modulation of the sensor-field interaction (which can realize any linear system~\cite{oppenheim:1996}) is insufficient to achieve the Grover-Heisenberg limit; non-linear signal processing is necessary.

\vspace{-1em}
\subsection{Metrological vs. Computational Enhancement}

While we have attributed our AC sensing improvement to the quadratic speedup of Grover's algorithm, computational enhancement and metrological gain can be fundamentally different. We demonstrate how this distinction can enable beyond-quadratic metrological gain, surpassing the expectation from the Grover speedup alone.

Consider a scenario where the strength of the unknown AC signal increases with frequency: $B_{\min}(\omega) \sim \omega^\mu$ for $\mu > 0$ (or equivalently, the coupling constant scales as $g(\omega) \sim \omega^\mu$ with fixed signal strength).  For sufficiently large $\mu$, the signal strength increases so rapidly that the total sensing time becomes independent of $|\Delta \omega|$.
For $1/3 < \mu < 1/2$, there is a quantum computing enhanced sensing protocol with constant sensing time, while any classical signal processing protocol requires $\Omega(|\Delta \omega|^{1-2\mu})$ time, yielding a \emph{constant-versus-polynomial} separation (see SM). 
Note that this separation only reflects the improved sensing time; in both settings, the gate count remains polynomial in $|\Delta \omega|$, reducing the speedup in realistic scenarios when the gates are not instantaneous.
Given that sensing with classical signal processing (Theorem~\ref{thm: CpQS}) and QFI-based sensing (Theorem~\ref{thm: QFI}) share the same scaling, this demonstrates that our QSS algorithm can provide significant advantages over conventional QFI-based approaches. 
This example illustrates how integrating quantum sensing and computing can achieve substantially larger, beyond-quadratic metrological gain.

\section{Discussion and Outlook}\label{sec: disc_and_outlook}

We have demonstrated that quantum computing enables provably more efficient protocols for practically useful sensing tasks. 
A key insight in our construction is that, in the context of sensing, the unknown signal is naturally queried as a black-box. 
As such, we can treat \emph{nature as an oracle}. 
While the internal structure of the oracle is assumed unknown in conventional query complexity, here we provide an explicit construction of the sensing oracle, which transforms the unknown external signal into an oracular black-box. Our work thus provides a physical instantiation of quantum oracles and their associated speedups.
While we have focused on quantum search, there are many other problems with known (and much larger) quantum-classical query separations~\cite{dalzell:2023}, and it is natural to ask whether we can leverage these speedups for greater metrological gain. In the SM, we show how to implement the quantum oracle associated with \emph{any} Boolean function by sensing a time-dependent, single-qubit field. 
This construction offers a path forward to developing novel quantum computing enhanced sensing protocols; if an oracular problem has a suitable sensing interpretation, we may lift its computational speedup to a sensing speedup. Extending our digitization techniques to this setting could potentially enable significant and robust metrological gain.

This expressivity suggests a connection between query complexity and the complexity of sensing time-dependent Hamiltonians~\cite{pang:2017, declercq:2016, krastanov:2019, siva:2023}. However, in sensing, each ``query'' interacts with a different instantiation of the time-varying signal. This temporal variation challenges core assumptions of oracle models, particularly the ability to repeatedly apply identical operations and implement their inverses. This fundamental distinction was identified nearly thirty years ago by Grover as a potential barrier to realizing quantum search speedups in physical systems~\cite{grover:1998}. While our protocol overcomes this challenge through signal digitization, the broader consequences for sensing and learning time-dependent Hamiltonian dynamics remain largely unexplored. Understanding these implications could reveal new opportunities unique to the sensing paradigm and lead to novel quantum advantages beyond those predicted by standard query complexity.

We have shown that by augmenting quantum sensors with a quantum computer, significant metrological gains are achievable, but fundamental questions about the minimal computational requirements remain open. Our no-go theorems do not rule out the possibility that the Grover-Heisenberg limit can be achieved with only a single qubit; such a protocol would dramatically accelerate on-going AC sensing experiments.
Beyond AC sensing, understanding whether small quantum computers could enable substantial improvements in sensing is a question of considerable practical and theoretical significance.

Finally, it remains an open question whether the Grover-Heisenberg limit is achievable in the presence of decoherence through quantum error correction.
In parameter estimation, achieving the Heisenberg limit is provably impossible when the sensors are subject to generic noise~\cite{zhou:2018, demkowicz-dobrzanski:2017}.
However, these no-go theorems rely crucially on high quantum Fisher information for weak field detection.
Our metrological gain is qualitatively distinct, and our digitization approach may provide a path to bypass these no-go theorems in certain regimes, enabling fault-tolerant quantum computing enhanced sensing.
By co-designing metrological codes with novel sensing protocols, the resulting \emph{logical quantum sensors} could robustly interface with the physical world at the hardware level, opening new possibilities in quantum sensing and algorithms.

\begin{acknowledgments}

We thank J. Borregaard, S. Croke, S. Garratt, A. Harrow, A. Lowe, M. Lukin, J. Preskill, D. Ranard, Q. Zhuang, and P. Zoller for valuable discussions and feedback.
We acknowledge support from the Heising-Simons Foundation (Grant No. 2024-4851), the NSF QCLI (Grant No. 2016245), the NSF QuSeC-TAQS (Grant No. 2326787), and the Center for Ultracold Atoms, an NSF Physics Frontiers
Center (Grant No. 1734011). RRA acknowledges support from the NSF Graduate Research Fellowship Program (Grant No. 2141064). 
FM acknowledges support from the NSF through a grant for ITAMP at Harvard University. 
HH acknowledges the visiting position at the Center for Theoretical Physics, MIT.

\end{acknowledgments}

\bibliography{QuantumSensing}

\begin{thebibliography}{50}%
\makeatletter
\providecommand \@ifxundefined [1]{%
 \@ifx{#1\undefined}
}%
\providecommand \@ifnum [1]{%
 \ifnum #1\expandafter \@firstoftwo
 \else \expandafter \@secondoftwo
 \fi
}%
\providecommand \@ifx [1]{%
 \ifx #1\expandafter \@firstoftwo
 \else \expandafter \@secondoftwo
 \fi
}%
\providecommand \natexlab [1]{#1}%
\providecommand \enquote  [1]{``#1''}%
\providecommand \bibnamefont  [1]{#1}%
\providecommand \bibfnamefont [1]{#1}%
\providecommand \citenamefont [1]{#1}%
\providecommand \href@noop [0]{\@secondoftwo}%
\providecommand \href [0]{\begingroup \@sanitize@url \@href}%
\providecommand \@href[1]{\@@startlink{#1}\@@href}%
\providecommand \@@href[1]{\endgroup#1\@@endlink}%
\providecommand \@sanitize@url [0]{\catcode `\\12\catcode `\$12\catcode
  `\&12\catcode `\#12\catcode `\^12\catcode `\_12\catcode `\%12\relax}%
\providecommand \@@startlink[1]{}%
\providecommand \@@endlink[0]{}%
\providecommand \url  [0]{\begingroup\@sanitize@url \@url }%
\providecommand \@url [1]{\endgroup\@href {#1}{\urlprefix }}%
\providecommand \urlprefix  [0]{URL }%
\providecommand \Eprint [0]{\href }%
\providecommand \doibase [0]{https://doi.org/}%
\providecommand \selectlanguage [0]{\@gobble}%
\providecommand \bibinfo  [0]{\@secondoftwo}%
\providecommand \bibfield  [0]{\@secondoftwo}%
\providecommand \translation [1]{[#1]}%
\providecommand \BibitemOpen [0]{}%
\providecommand \bibitemStop [0]{}%
\providecommand \bibitemNoStop [0]{.\EOS\space}%
\providecommand \EOS [0]{\spacefactor3000\relax}%
\providecommand \BibitemShut  [1]{\csname bibitem#1\endcsname}%
\let\auto@bib@innerbib\@empty
\bibitem [{\citenamefont {Cerezo}\ \emph {and others}(2021)\citenamefont
  {Cerezo}, \citenamefont {Arrasmith}, \citenamefont {Babbush}, \citenamefont
  {Benjamin}, \citenamefont {Endo}, \citenamefont {Fujii}, \citenamefont
  {McClean}, \citenamefont {Mitarai}, \citenamefont {Yuan}, \citenamefont
  {Cincio},\ and\ \citenamefont {Coles}}]{cerezo:2021}%
  \BibitemOpen
  \bibfield  {author} {\bibinfo {author} {\bibfnamefont {M.}~\bibnamefont
  {Cerezo}}, \bibinfo {author} {\bibfnamefont {A.}~\bibnamefont {Arrasmith}},
  \bibinfo {author} {\bibfnamefont {R.}~\bibnamefont {Babbush}}, \bibinfo
  {author} {\bibfnamefont {S.~C.}\ \bibnamefont {Benjamin}}, \bibinfo {author}
  {\bibfnamefont {S.}~\bibnamefont {Endo}}, \bibinfo {author} {\bibfnamefont
  {K.}~\bibnamefont {Fujii}}, \bibinfo {author} {\bibfnamefont {J.~R.}\
  \bibnamefont {McClean}}, \bibinfo {author} {\bibfnamefont {K.}~\bibnamefont
  {Mitarai}}, \bibinfo {author} {\bibfnamefont {X.}~\bibnamefont {Yuan}},
  \bibinfo {author} {\bibfnamefont {L.}~\bibnamefont {Cincio}}, and others,\
  }\href {https://doi.org/10.1038/s42254-021-00348-9} {\bibfield  {journal}
  {\bibinfo  {journal} {Nature Reviews Physics}\ }\textbf {\bibinfo {volume}
  {3}},\ \bibinfo {pages} {625} (\bibinfo {year} {2021})}\BibitemShut {NoStop}%
\bibitem [{\citenamefont {Daley}\ \emph {and others}(2022)\citenamefont
  {Daley}, \citenamefont {Bloch}, \citenamefont {Kokail}, \citenamefont
  {Flannigan}, \citenamefont {Pearson}, \citenamefont {Troyer},\ and\
  \citenamefont {Zoller}}]{daley:2022}%
  \BibitemOpen
  \bibfield  {author} {\bibinfo {author} {\bibfnamefont {A.~J.}\ \bibnamefont
  {Daley}}, \bibinfo {author} {\bibfnamefont {I.}~\bibnamefont {Bloch}},
  \bibinfo {author} {\bibfnamefont {C.}~\bibnamefont {Kokail}}, \bibinfo
  {author} {\bibfnamefont {S.}~\bibnamefont {Flannigan}}, \bibinfo {author}
  {\bibfnamefont {N.}~\bibnamefont {Pearson}}, \bibinfo {author} {\bibfnamefont
  {M.}~\bibnamefont {Troyer}},\ and\ \bibinfo {author} {\bibfnamefont
  {P.}~\bibnamefont {Zoller}},\ }\href
  {https://doi.org/10.1038/s41586-022-04940-6} {\bibfield  {journal} {\bibinfo
  {journal} {Nature}\ }\textbf {\bibinfo {volume} {607}},\ \bibinfo {pages}
  {667} (\bibinfo {year} {2022})}\BibitemShut {NoStop}%
\bibitem [{\citenamefont {Hangleiter}\ and\ \citenamefont
  {Eisert}(2023)}]{hangleiter:2023}%
  \BibitemOpen
  \bibfield  {author} {\bibinfo {author} {\bibfnamefont {D.}~\bibnamefont
  {Hangleiter}}\ and\ \bibinfo {author} {\bibfnamefont {J.}~\bibnamefont
  {Eisert}},\ }\href {https://doi.org/10.1103/RevModPhys.95.035001} {\bibfield
  {journal} {\bibinfo  {journal} {Reviews of Modern Physics}\ }\textbf
  {\bibinfo {volume} {95}},\ \bibinfo {pages} {035001} (\bibinfo {year}
  {2023})}\BibitemShut {NoStop}%
\bibitem [{\citenamefont {Degen}\ \emph {and others}(2017)\citenamefont
  {Degen}, \citenamefont {Reinhard},\ and\ \citenamefont
  {Cappellaro}}]{degen:2017}%
  \BibitemOpen
  \bibfield  {author} {\bibinfo {author} {\bibfnamefont {C.~L.}\ \bibnamefont
  {Degen}}, \bibinfo {author} {\bibfnamefont {F.}~\bibnamefont {Reinhard}},\
  and\ \bibinfo {author} {\bibfnamefont {P.}~\bibnamefont {Cappellaro}},\
  }\href {https://doi.org/10.1103/RevModPhys.89.035002} {\bibfield  {journal}
  {\bibinfo  {journal} {Reviews of Modern Physics}\ }\textbf {\bibinfo {volume}
  {89}},\ \bibinfo {pages} {035002} (\bibinfo {year} {2017})}\BibitemShut
  {NoStop}%
\bibitem [{\citenamefont {Giovannetti}\ \emph {and others}(2004)\citenamefont
  {Giovannetti}, \citenamefont {Lloyd},\ and\ \citenamefont
  {Maccone}}]{giovannetti:2004}%
  \BibitemOpen
  \bibfield  {author} {\bibinfo {author} {\bibfnamefont {V.}~\bibnamefont
  {Giovannetti}}, \bibinfo {author} {\bibfnamefont {S.}~\bibnamefont {Lloyd}},\
  and\ \bibinfo {author} {\bibfnamefont {L.}~\bibnamefont {Maccone}},\ }\href
  {https://doi.org/10.1126/science.1104149} {\bibfield  {journal} {\bibinfo
  {journal} {Science}\ }\textbf {\bibinfo {volume} {306}},\ \bibinfo {pages}
  {1330} (\bibinfo {year} {2004})}\BibitemShut {NoStop}%
\bibitem [{\citenamefont {Pezz{\`e}}\ \emph {and others}(2018)\citenamefont
  {Pezz{\`e}}, \citenamefont {Smerzi}, \citenamefont {Oberthaler},
  \citenamefont {Schmied},\ and\ \citenamefont {Treutlein}}]{pezze:2018a}%
  \BibitemOpen
  \bibfield  {author} {\bibinfo {author} {\bibfnamefont {L.}~\bibnamefont
  {Pezz{\`e}}}, \bibinfo {author} {\bibfnamefont {A.}~\bibnamefont {Smerzi}},
  \bibinfo {author} {\bibfnamefont {M.~K.}\ \bibnamefont {Oberthaler}},
  \bibinfo {author} {\bibfnamefont {R.}~\bibnamefont {Schmied}},\ and\ \bibinfo
  {author} {\bibfnamefont {P.}~\bibnamefont {Treutlein}},\ }\href
  {https://doi.org/10.1103/RevModPhys.90.035005} {\bibfield  {journal}
  {\bibinfo  {journal} {Reviews of Modern Physics}\ }\textbf {\bibinfo {volume}
  {90}},\ \bibinfo {pages} {035005} (\bibinfo {year} {2018})}\BibitemShut
  {NoStop}%
\bibitem [{\citenamefont {Zhang}\ \emph {and others}(2021)\citenamefont
  {Zhang}, \citenamefont {Shagieva}, \citenamefont {Widmann}, \citenamefont
  {K{\"u}bler}, \citenamefont {Vorobyov}, \citenamefont {Kapitanova},
  \citenamefont {Nenasheva}, \citenamefont {Corkill}, \citenamefont {Rhrle},
  \citenamefont {Nakamura}, \citenamefont {Sumiya}, \citenamefont {Onoda},
  \citenamefont {Isoya},\ and\ \citenamefont {Wrachtrup}}]{zhang:2021}%
  \BibitemOpen
  \bibfield  {author} {\bibinfo {author} {\bibfnamefont {C.}~\bibnamefont
  {Zhang}}, \bibinfo {author} {\bibfnamefont {F.}~\bibnamefont {Shagieva}},
  \bibinfo {author} {\bibfnamefont {M.}~\bibnamefont {Widmann}}, \bibinfo
  {author} {\bibfnamefont {M.}~\bibnamefont {K{\"u}bler}}, \bibinfo {author}
  {\bibfnamefont {V.}~\bibnamefont {Vorobyov}}, \bibinfo {author}
  {\bibfnamefont {P.}~\bibnamefont {Kapitanova}}, \bibinfo {author}
  {\bibfnamefont {E.}~\bibnamefont {Nenasheva}}, \bibinfo {author}
  {\bibfnamefont {R.}~\bibnamefont {Corkill}}, \bibinfo {author} {\bibfnamefont
  {O.}~\bibnamefont {Rhrle}}, \bibinfo {author} {\bibfnamefont
  {K.}~\bibnamefont {Nakamura}}, and others,\ }\href
  {https://doi.org/10.1103/PhysRevApplied.15.064075} {\bibfield  {journal}
  {\bibinfo  {journal} {Physical Review Applied}\ }\textbf {\bibinfo {volume}
  {15}},\ \bibinfo {pages} {064075} (\bibinfo {year} {2021})}\BibitemShut
  {NoStop}%
\bibitem [{\citenamefont {Aeppli}\ \emph {and others}(2024)\citenamefont
  {Aeppli}, \citenamefont {Kim}, \citenamefont {Warfield}, \citenamefont
  {Safronova},\ and\ \citenamefont {Ye}}]{aeppli:2024}%
  \BibitemOpen
  \bibfield  {author} {\bibinfo {author} {\bibfnamefont {A.}~\bibnamefont
  {Aeppli}}, \bibinfo {author} {\bibfnamefont {K.}~\bibnamefont {Kim}},
  \bibinfo {author} {\bibfnamefont {W.}~\bibnamefont {Warfield}}, \bibinfo
  {author} {\bibfnamefont {M.~S.}\ \bibnamefont {Safronova}},\ and\ \bibinfo
  {author} {\bibfnamefont {J.}~\bibnamefont {Ye}},\ }\href
  {https://doi.org/10.1103/PhysRevLett.133.023401} {\bibfield  {journal}
  {\bibinfo  {journal} {Physical Review Letters}\ }\textbf {\bibinfo {volume}
  {133}},\ \bibinfo {pages} {023401} (\bibinfo {year} {2024})}\BibitemShut
  {NoStop}%
\bibitem [{\citenamefont {Zheng}\ \emph {and others}(2023)\citenamefont
  {Zheng}, \citenamefont {Dolde}, \citenamefont {Cambria}, \citenamefont
  {Lim},\ and\ \citenamefont {Kolkowitz}}]{zheng:2023}%
  \BibitemOpen
  \bibfield  {author} {\bibinfo {author} {\bibfnamefont {X.}~\bibnamefont
  {Zheng}}, \bibinfo {author} {\bibfnamefont {J.}~\bibnamefont {Dolde}},
  \bibinfo {author} {\bibfnamefont {M.~C.}\ \bibnamefont {Cambria}}, \bibinfo
  {author} {\bibfnamefont {H.~M.}\ \bibnamefont {Lim}},\ and\ \bibinfo {author}
  {\bibfnamefont {S.}~\bibnamefont {Kolkowitz}},\ }\href
  {https://doi.org/10.1038/s41467-023-40629-8} {\bibfield  {journal} {\bibinfo
  {journal} {Nature Communications}\ }\textbf {\bibinfo {volume} {14}},\
  \bibinfo {pages} {4886} (\bibinfo {year} {2023})}\BibitemShut {NoStop}%
\bibitem [{\citenamefont {{LIGO O4 Detector Collaboration}}\ \emph {and
  others}(2023)\citenamefont {{LIGO O4 Detector Collaboration}}, \citenamefont
  {Ganapathy}, \citenamefont {Jia}, \citenamefont {Nakano}, \citenamefont {Xu},
  \citenamefont {Aritomi}, \citenamefont {Cullen}, \citenamefont {Kijbunchoo},
  \citenamefont {Dwyer}, \citenamefont {Mullavey}, \citenamefont {McCuller},
  \citenamefont {Abbott}, \citenamefont {Abouelfettouh}, \citenamefont
  {Adhikari}, \citenamefont {Ananyeva}, \citenamefont {Appert}, \citenamefont
  {Arai}, \citenamefont {Aston}, \citenamefont {Ball}, \citenamefont {Ballmer},
  \citenamefont {Barker}, \citenamefont {Barsotti}, \citenamefont {Berger},
  \citenamefont {Betzwieser}, \citenamefont {Bhattacharjee}, \citenamefont
  {Billingsley}, \citenamefont {Biscans}, \citenamefont {Bode}, \citenamefont
  {Bonilla}, \citenamefont {Bossilkov}, \citenamefont {Branch}, \citenamefont
  {Brooks}, \citenamefont {Brown}, \citenamefont {Bryant}, \citenamefont
  {Cahillane}, \citenamefont {Cao}, \citenamefont {Capote}, \citenamefont
  {Clara}, \citenamefont {Collins}, \citenamefont {Compton}, \citenamefont
  {Cottingham}, \citenamefont {Coyne}, \citenamefont {Crouch}, \citenamefont
  {Csizmazia}, \citenamefont {Dartez}, \citenamefont {Demos}, \citenamefont
  {Dohmen}, \citenamefont {Driggers}, \citenamefont {Effler}, \citenamefont
  {Ejlli}, \citenamefont {Etzel}, \citenamefont {Evans}, \citenamefont
  {Feicht}, \citenamefont {Frey}, \citenamefont {Frischhertz}, \citenamefont
  {Fritschel}, \citenamefont {Frolov}, \citenamefont {Fulda}, \citenamefont
  {Fyffe}, \citenamefont {Gateley}, \citenamefont {Giaime}, \citenamefont
  {Giardina}, \citenamefont {Glanzer}, \citenamefont {Goetz}, \citenamefont
  {Goetz}, \citenamefont {{Goodwin-Jones}}, \citenamefont {Gras}, \citenamefont
  {Gray}, \citenamefont {Griffith}, \citenamefont {Grote}, \citenamefont
  {Guidry}, \citenamefont {Hall}, \citenamefont {Hanks}, \citenamefont
  {Hanson}, \citenamefont {Heintze}, \citenamefont {{Helmling-Cornell}},
  \citenamefont {Holland}, \citenamefont {Hoyland}, \citenamefont {Huang},
  \citenamefont {Inoue}, \citenamefont {James}, \citenamefont {Jennings},
  \citenamefont {Karat}, \citenamefont {Karki}, \citenamefont {Kasprzack},
  \citenamefont {Kawabe}, \citenamefont {King}, \citenamefont {Kissel},
  \citenamefont {Komori}, \citenamefont {Kontos}, \citenamefont {Kumar},
  \citenamefont {Kuns}, \citenamefont {Landry}, \citenamefont {Lantz},
  \citenamefont {Laxen}, \citenamefont {Lee}, \citenamefont {Lesovsky},
  \citenamefont {Llamas}, \citenamefont {Lormand}, \citenamefont {Loughlin},
  \citenamefont {Macas}, \citenamefont {MacInnis}, \citenamefont {Makarem},
  \citenamefont {Mannix}, \citenamefont {Mansell}, \citenamefont {Martin},
  \citenamefont {Mason}, \citenamefont {Matichard}, \citenamefont {Mavalvala},
  \citenamefont {Maxwell}, \citenamefont {McCarrol}, \citenamefont {McCarthy},
  \citenamefont {McClelland}, \citenamefont {McCormick}, \citenamefont {McRae},
  \citenamefont {Mera}, \citenamefont {Merilh}, \citenamefont {Meylahn},
  \citenamefont {Mittleman}, \citenamefont {Moraru}, \citenamefont {Moreno},
  \citenamefont {Nelson}, \citenamefont {Neunzert}, \citenamefont {Notte},
  \citenamefont {Oberling}, \citenamefont {O'Hanlon}, \citenamefont
  {Osthelder}, \citenamefont {Ottaway}, \citenamefont {Overmier}, \citenamefont
  {Parker}, \citenamefont {Pele}, \citenamefont {Pham}, \citenamefont
  {Pirello}, \citenamefont {Quetschke}, \citenamefont {Ramirez}, \citenamefont
  {Reyes}, \citenamefont {Richardson}, \citenamefont {Robinson}, \citenamefont
  {Rollins}, \citenamefont {Romel}, \citenamefont {Romie}, \citenamefont
  {Ross}, \citenamefont {Ryan}, \citenamefont {Sadecki}, \citenamefont
  {Sanchez}, \citenamefont {Sanchez}, \citenamefont {Sanchez}, \citenamefont
  {Savage}, \citenamefont {Schaetzl}, \citenamefont {Schiworski}, \citenamefont
  {Schnabel}, \citenamefont {Schofield}, \citenamefont {Schwartz},
  \citenamefont {Sellers}, \citenamefont {Shaffer}, \citenamefont {Short},
  \citenamefont {Sigg}, \citenamefont {Slagmolen}, \citenamefont {Soike},
  \citenamefont {Soni}, \citenamefont {Srivastava}, \citenamefont {Sun},
  \citenamefont {Tanner}, \citenamefont {Thomas}, \citenamefont {Thomas},
  \citenamefont {Thorne}, \citenamefont {Torrie}, \citenamefont {Traylor},
  \citenamefont {Ubhi}, \citenamefont {Vajente}, \citenamefont {Vanosky},
  \citenamefont {Vecchio}, \citenamefont {Veitch}, \citenamefont {Vibhute},
  \citenamefont {{von Reis}}, \citenamefont {Warner}, \citenamefont {Weaver},
  \citenamefont {Weiss}, \citenamefont {Whittle}, \citenamefont {Willke},
  \citenamefont {Wipf}, \citenamefont {Yamamoto}, \citenamefont {Zhang},\ and\
  \citenamefont {Zucker}}]{ligoo4detectorcollaboration:2023}%
  \BibitemOpen
  \bibfield  {author} {\bibinfo {author} {\bibnamefont {{LIGO O4 Detector
  Collaboration}}}, \bibinfo {author} {\bibfnamefont {D.}~\bibnamefont
  {Ganapathy}}, \bibinfo {author} {\bibfnamefont {W.}~\bibnamefont {Jia}},
  \bibinfo {author} {\bibfnamefont {M.}~\bibnamefont {Nakano}}, \bibinfo
  {author} {\bibfnamefont {V.}~\bibnamefont {Xu}}, \bibinfo {author}
  {\bibfnamefont {N.}~\bibnamefont {Aritomi}}, \bibinfo {author} {\bibfnamefont
  {T.}~\bibnamefont {Cullen}}, \bibinfo {author} {\bibfnamefont
  {N.}~\bibnamefont {Kijbunchoo}}, \bibinfo {author} {\bibfnamefont {S.~E.}\
  \bibnamefont {Dwyer}}, \bibinfo {author} {\bibfnamefont {A.}~\bibnamefont
  {Mullavey}}, and others,\ }\href {https://doi.org/10.1103/PhysRevX.13.041021}
  {\bibfield  {journal} {\bibinfo  {journal} {Physical Review X}\ }\textbf
  {\bibinfo {volume} {13}},\ \bibinfo {pages} {041021} (\bibinfo {year}
  {2023})}\BibitemShut {NoStop}%
\bibitem [{\citenamefont {Bass}\ and\ \citenamefont {Doser}(2024)}]{bass:2024}%
  \BibitemOpen
  \bibfield  {author} {\bibinfo {author} {\bibfnamefont {S.~D.}\ \bibnamefont
  {Bass}}\ and\ \bibinfo {author} {\bibfnamefont {M.}~\bibnamefont {Doser}},\
  }\href {https://doi.org/10.1038/s42254-024-00714-3} {\bibfield  {journal}
  {\bibinfo  {journal} {Nature Reviews Physics}\ }\textbf {\bibinfo {volume}
  {6}},\ \bibinfo {pages} {329} (\bibinfo {year} {2024})}\BibitemShut {NoStop}%
\bibitem [{\citenamefont {Aslam}\ \emph {and others}(2023)\citenamefont
  {Aslam}, \citenamefont {Zhou}, \citenamefont {Urbach}, \citenamefont
  {Turner}, \citenamefont {Walsworth}, \citenamefont {Lukin},\ and\
  \citenamefont {Park}}]{aslam:2023}%
  \BibitemOpen
  \bibfield  {author} {\bibinfo {author} {\bibfnamefont {N.}~\bibnamefont
  {Aslam}}, \bibinfo {author} {\bibfnamefont {H.}~\bibnamefont {Zhou}},
  \bibinfo {author} {\bibfnamefont {E.~K.}\ \bibnamefont {Urbach}}, \bibinfo
  {author} {\bibfnamefont {M.~J.}\ \bibnamefont {Turner}}, \bibinfo {author}
  {\bibfnamefont {R.~L.}\ \bibnamefont {Walsworth}}, \bibinfo {author}
  {\bibfnamefont {M.~D.}\ \bibnamefont {Lukin}},\ and\ \bibinfo {author}
  {\bibfnamefont {H.}~\bibnamefont {Park}},\ }\href
  {https://doi.org/10.1038/s42254-023-00558-3} {\bibfield  {journal} {\bibinfo
  {journal} {Nature Reviews Physics}\ }\textbf {\bibinfo {volume} {5}},\
  \bibinfo {pages} {157} (\bibinfo {year} {2023})}\BibitemShut {NoStop}%
\bibitem [{\citenamefont {Bhattacharyya}\ \emph {and others}(2024)\citenamefont
  {Bhattacharyya}, \citenamefont {Chen}, \citenamefont {Huang}, \citenamefont
  {Chatterjee}, \citenamefont {Huang}, \citenamefont {Kobrin}, \citenamefont
  {Lyu}, \citenamefont {Smart}, \citenamefont {Block}, \citenamefont {Wang},
  \citenamefont {Wang}, \citenamefont {Wu}, \citenamefont {Hsieh},
  \citenamefont {Ma}, \citenamefont {Mandyam}, \citenamefont {Chen},
  \citenamefont {Davis}, \citenamefont {Geballe}, \citenamefont {Zu},
  \citenamefont {Struzhkin}, \citenamefont {Jeanloz}, \citenamefont {Moore},
  \citenamefont {Cui}, \citenamefont {Galli}, \citenamefont {Halperin},
  \citenamefont {Laumann},\ and\ \citenamefont {Yao}}]{bhattacharyya:2024}%
  \BibitemOpen
  \bibfield  {author} {\bibinfo {author} {\bibfnamefont {P.}~\bibnamefont
  {Bhattacharyya}}, \bibinfo {author} {\bibfnamefont {W.}~\bibnamefont {Chen}},
  \bibinfo {author} {\bibfnamefont {X.}~\bibnamefont {Huang}}, \bibinfo
  {author} {\bibfnamefont {S.}~\bibnamefont {Chatterjee}}, \bibinfo {author}
  {\bibfnamefont {B.}~\bibnamefont {Huang}}, \bibinfo {author} {\bibfnamefont
  {B.}~\bibnamefont {Kobrin}}, \bibinfo {author} {\bibfnamefont
  {Y.}~\bibnamefont {Lyu}}, \bibinfo {author} {\bibfnamefont {T.~J.}\
  \bibnamefont {Smart}}, \bibinfo {author} {\bibfnamefont {M.}~\bibnamefont
  {Block}}, \bibinfo {author} {\bibfnamefont {E.}~\bibnamefont {Wang}}, and
  others,\ }\href {https://doi.org/10.1038/s41586-024-07026-7} {\bibfield
  {journal} {\bibinfo  {journal} {Nature}\ }\textbf {\bibinfo {volume} {627}},\
  \bibinfo {pages} {73} (\bibinfo {year} {2024})}\BibitemShut {NoStop}%
\bibitem [{\citenamefont {D{\"u}r}\ \emph {and others}(2014)\citenamefont
  {D{\"u}r}, \citenamefont {Skotiniotis}, \citenamefont {Fr{\"o}wis},\ and\
  \citenamefont {Kraus}}]{dur:2014}%
  \BibitemOpen
  \bibfield  {author} {\bibinfo {author} {\bibfnamefont {W.}~\bibnamefont
  {D{\"u}r}}, \bibinfo {author} {\bibfnamefont {M.}~\bibnamefont
  {Skotiniotis}}, \bibinfo {author} {\bibfnamefont {F.}~\bibnamefont
  {Fr{\"o}wis}},\ and\ \bibinfo {author} {\bibfnamefont {B.}~\bibnamefont
  {Kraus}},\ }\href {https://doi.org/10.1103/PhysRevLett.112.080801} {\bibfield
   {journal} {\bibinfo  {journal} {Physical Review Letters}\ }\textbf {\bibinfo
  {volume} {112}},\ \bibinfo {pages} {080801} (\bibinfo {year}
  {2014})}\BibitemShut {NoStop}%
\bibitem [{\citenamefont {{Demkowicz-Dobrza{\'n}ski}}\ \emph {and
  others}(2017)\citenamefont {{Demkowicz-Dobrza{\'n}ski}}, \citenamefont
  {Czajkowski},\ and\ \citenamefont {Sekatski}}]{demkowicz-dobrzanski:2017}%
  \BibitemOpen
  \bibfield  {author} {\bibinfo {author} {\bibfnamefont {R.}~\bibnamefont
  {{Demkowicz-Dobrza{\'n}ski}}}, \bibinfo {author} {\bibfnamefont
  {J.}~\bibnamefont {Czajkowski}},\ and\ \bibinfo {author} {\bibfnamefont
  {P.}~\bibnamefont {Sekatski}},\ }\href
  {https://doi.org/10.1103/PhysRevX.7.041009} {\bibfield  {journal} {\bibinfo
  {journal} {Physical Review X}\ }\textbf {\bibinfo {volume} {7}},\ \bibinfo
  {pages} {041009} (\bibinfo {year} {2017})}\BibitemShut {NoStop}%
\bibitem [{\citenamefont {Zhou}\ \emph {and others}(2018)\citenamefont {Zhou},
  \citenamefont {Zhang}, \citenamefont {Preskill},\ and\ \citenamefont
  {Jiang}}]{zhou:2018}%
  \BibitemOpen
  \bibfield  {author} {\bibinfo {author} {\bibfnamefont {S.}~\bibnamefont
  {Zhou}}, \bibinfo {author} {\bibfnamefont {M.}~\bibnamefont {Zhang}},
  \bibinfo {author} {\bibfnamefont {J.}~\bibnamefont {Preskill}},\ and\
  \bibinfo {author} {\bibfnamefont {L.}~\bibnamefont {Jiang}},\ }\href
  {https://doi.org/10.1038/s41467-017-02510-3} {\bibfield  {journal} {\bibinfo
  {journal} {Nature Communications}\ }\textbf {\bibinfo {volume} {9}},\
  \bibinfo {pages} {78} (\bibinfo {year} {2018})}\BibitemShut {NoStop}%
\bibitem [{\citenamefont {Niroula}\ \emph {and others}(2024)\citenamefont
  {Niroula}, \citenamefont {Dolde}, \citenamefont {Zheng}, \citenamefont
  {Bringewatt}, \citenamefont {Ehrenberg}, \citenamefont {Cox}, \citenamefont
  {Thompson}, \citenamefont {Gullans}, \citenamefont {Kolkowitz},\ and\
  \citenamefont {Gorshkov}}]{niroula:2024}%
  \BibitemOpen
  \bibfield  {author} {\bibinfo {author} {\bibfnamefont {P.}~\bibnamefont
  {Niroula}}, \bibinfo {author} {\bibfnamefont {J.}~\bibnamefont {Dolde}},
  \bibinfo {author} {\bibfnamefont {X.}~\bibnamefont {Zheng}}, \bibinfo
  {author} {\bibfnamefont {J.}~\bibnamefont {Bringewatt}}, \bibinfo {author}
  {\bibfnamefont {A.}~\bibnamefont {Ehrenberg}}, \bibinfo {author}
  {\bibfnamefont {K.~C.}\ \bibnamefont {Cox}}, \bibinfo {author} {\bibfnamefont
  {J.}~\bibnamefont {Thompson}}, \bibinfo {author} {\bibfnamefont {M.~J.}\
  \bibnamefont {Gullans}}, \bibinfo {author} {\bibfnamefont {S.}~\bibnamefont
  {Kolkowitz}},\ and\ \bibinfo {author} {\bibfnamefont {A.~V.}\ \bibnamefont
  {Gorshkov}},\ }\href {https://doi.org/10.1103/PhysRevLett.133.080801}
  {\bibfield  {journal} {\bibinfo  {journal} {Physical Review Letters}\
  }\textbf {\bibinfo {volume} {133}},\ \bibinfo {pages} {080801} (\bibinfo
  {year} {2024})}\BibitemShut {NoStop}%
\bibitem [{\citenamefont {Aharonov}\ \emph {and others}(2022)\citenamefont
  {Aharonov}, \citenamefont {Cotler},\ and\ \citenamefont
  {Qi}}]{aharonov:2022}%
  \BibitemOpen
  \bibfield  {author} {\bibinfo {author} {\bibfnamefont {D.}~\bibnamefont
  {Aharonov}}, \bibinfo {author} {\bibfnamefont {J.}~\bibnamefont {Cotler}},\
  and\ \bibinfo {author} {\bibfnamefont {X.-L.}\ \bibnamefont {Qi}},\ }\href
  {https://doi.org/10.1038/s41467-021-27922-0} {\bibfield  {journal} {\bibinfo
  {journal} {Nature Communications}\ }\textbf {\bibinfo {volume} {13}},\
  \bibinfo {pages} {887} (\bibinfo {year} {2022})}\BibitemShut {NoStop}%
\bibitem [{\citenamefont {Huang}\ \emph {and others}(2022)\citenamefont
  {Huang}, \citenamefont {Broughton}, \citenamefont {Cotler}, \citenamefont
  {Chen}, \citenamefont {Li}, \citenamefont {Mohseni}, \citenamefont {Neven},
  \citenamefont {Babbush}, \citenamefont {Kueng}, \citenamefont {Preskill},\
  and\ \citenamefont {McClean}}]{huang:2022}%
  \BibitemOpen
  \bibfield  {author} {\bibinfo {author} {\bibfnamefont {H.-Y.}\ \bibnamefont
  {Huang}}, \bibinfo {author} {\bibfnamefont {M.}~\bibnamefont {Broughton}},
  \bibinfo {author} {\bibfnamefont {J.}~\bibnamefont {Cotler}}, \bibinfo
  {author} {\bibfnamefont {S.}~\bibnamefont {Chen}}, \bibinfo {author}
  {\bibfnamefont {J.}~\bibnamefont {Li}}, \bibinfo {author} {\bibfnamefont
  {M.}~\bibnamefont {Mohseni}}, \bibinfo {author} {\bibfnamefont
  {H.}~\bibnamefont {Neven}}, \bibinfo {author} {\bibfnamefont
  {R.}~\bibnamefont {Babbush}}, \bibinfo {author} {\bibfnamefont
  {R.}~\bibnamefont {Kueng}}, \bibinfo {author} {\bibfnamefont
  {J.}~\bibnamefont {Preskill}}, and others,\ }\href
  {https://doi.org/10.1126/science.abn7293} {\bibfield  {journal} {\bibinfo
  {journal} {Science}\ }\textbf {\bibinfo {volume} {376}},\ \bibinfo {pages}
  {1182} (\bibinfo {year} {2022})}\BibitemShut {NoStop}%
\bibitem [{\citenamefont {Huang}\ \emph {and others}(2023)\citenamefont
  {Huang}, \citenamefont {Tong}, \citenamefont {Fang},\ and\ \citenamefont
  {Su}}]{huang:2023}%
  \BibitemOpen
  \bibfield  {author} {\bibinfo {author} {\bibfnamefont {H.-Y.}\ \bibnamefont
  {Huang}}, \bibinfo {author} {\bibfnamefont {Y.}~\bibnamefont {Tong}},
  \bibinfo {author} {\bibfnamefont {D.}~\bibnamefont {Fang}},\ and\ \bibinfo
  {author} {\bibfnamefont {Y.}~\bibnamefont {Su}},\ }\href
  {https://doi.org/10.1103/PhysRevLett.130.200403} {\bibfield  {journal}
  {\bibinfo  {journal} {Physical Review Letters}\ }\textbf {\bibinfo {volume}
  {130}},\ \bibinfo {pages} {200403} (\bibinfo {year} {2023})}\BibitemShut
  {NoStop}%
\bibitem [{\citenamefont {Shaw}\ \emph {and others}(2024)\citenamefont {Shaw},
  \citenamefont {Finkelstein}, \citenamefont {Tsai}, \citenamefont {Scholl},
  \citenamefont {Yoon}, \citenamefont {Choi},\ and\ \citenamefont
  {Endres}}]{shaw:2024}%
  \BibitemOpen
  \bibfield  {author} {\bibinfo {author} {\bibfnamefont {A.~L.}\ \bibnamefont
  {Shaw}}, \bibinfo {author} {\bibfnamefont {R.}~\bibnamefont {Finkelstein}},
  \bibinfo {author} {\bibfnamefont {R.~B.-S.}\ \bibnamefont {Tsai}}, \bibinfo
  {author} {\bibfnamefont {P.}~\bibnamefont {Scholl}}, \bibinfo {author}
  {\bibfnamefont {T.~H.}\ \bibnamefont {Yoon}}, \bibinfo {author}
  {\bibfnamefont {J.}~\bibnamefont {Choi}},\ and\ \bibinfo {author}
  {\bibfnamefont {M.}~\bibnamefont {Endres}},\ }\href
  {https://doi.org/10.1038/s41567-023-02323-w} {\bibfield  {journal} {\bibinfo
  {journal} {Nature Physics}\ }\textbf {\bibinfo {volume} {20}},\ \bibinfo
  {pages} {195} (\bibinfo {year} {2024})}\BibitemShut {NoStop}%
\bibitem [{\citenamefont {Cao}\ \emph {and others}(2024)\citenamefont {Cao},
  \citenamefont {Eckner}, \citenamefont {Yelin}, \citenamefont {Young},
  \citenamefont {Jandura}, \citenamefont {Yan}, \citenamefont {Kim},
  \citenamefont {Pupillo}, \citenamefont {Ye}, \citenamefont {Oppong},\ and\
  \citenamefont {Kaufman}}]{cao:2024a}%
  \BibitemOpen
  \bibfield  {author} {\bibinfo {author} {\bibfnamefont {A.}~\bibnamefont
  {Cao}}, \bibinfo {author} {\bibfnamefont {W.~J.}\ \bibnamefont {Eckner}},
  \bibinfo {author} {\bibfnamefont {T.~L.}\ \bibnamefont {Yelin}}, \bibinfo
  {author} {\bibfnamefont {A.~W.}\ \bibnamefont {Young}}, \bibinfo {author}
  {\bibfnamefont {S.}~\bibnamefont {Jandura}}, \bibinfo {author} {\bibfnamefont
  {L.}~\bibnamefont {Yan}}, \bibinfo {author} {\bibfnamefont {K.}~\bibnamefont
  {Kim}}, \bibinfo {author} {\bibfnamefont {G.}~\bibnamefont {Pupillo}},
  \bibinfo {author} {\bibfnamefont {J.}~\bibnamefont {Ye}}, \bibinfo {author}
  {\bibfnamefont {N.~D.}\ \bibnamefont {Oppong}}, and others,\ }\href
  {https://doi.org/10.1038/s41586-024-07913-z} {\bibfield  {journal} {\bibinfo
  {journal} {Nature}\ }\textbf {\bibinfo {volume} {634}},\ \bibinfo {pages}
  {315} (\bibinfo {year} {2024})}\BibitemShut {NoStop}%
\bibitem [{\citenamefont {Marciniak}\ \emph {and others}(2022)\citenamefont
  {Marciniak}, \citenamefont {Feldker}, \citenamefont {Pogorelov},
  \citenamefont {Kaubruegger}, \citenamefont {Vasilyev}, \citenamefont {{van
  Bijnen}}, \citenamefont {Schindler}, \citenamefont {Zoller}, \citenamefont
  {Blatt},\ and\ \citenamefont {Monz}}]{marciniak:2022}%
  \BibitemOpen
  \bibfield  {author} {\bibinfo {author} {\bibfnamefont {C.~D.}\ \bibnamefont
  {Marciniak}}, \bibinfo {author} {\bibfnamefont {T.}~\bibnamefont {Feldker}},
  \bibinfo {author} {\bibfnamefont {I.}~\bibnamefont {Pogorelov}}, \bibinfo
  {author} {\bibfnamefont {R.}~\bibnamefont {Kaubruegger}}, \bibinfo {author}
  {\bibfnamefont {D.~V.}\ \bibnamefont {Vasilyev}}, \bibinfo {author}
  {\bibfnamefont {R.}~\bibnamefont {{van Bijnen}}}, \bibinfo {author}
  {\bibfnamefont {P.}~\bibnamefont {Schindler}}, \bibinfo {author}
  {\bibfnamefont {P.}~\bibnamefont {Zoller}}, \bibinfo {author} {\bibfnamefont
  {R.}~\bibnamefont {Blatt}},\ and\ \bibinfo {author} {\bibfnamefont
  {T.}~\bibnamefont {Monz}},\ }\href
  {https://doi.org/10.1038/s41586-022-04435-4} {\bibfield  {journal} {\bibinfo
  {journal} {Nature}\ }\textbf {\bibinfo {volume} {603}},\ \bibinfo {pages}
  {604} (\bibinfo {year} {2022})}\BibitemShut {NoStop}%
\bibitem [{\citenamefont {Acharya}\ \emph {and others}(2024)\citenamefont
  {Acharya}, \citenamefont {Abanin}, \citenamefont {{Aghababaie-Beni}},
  \citenamefont {Aleiner}, \citenamefont {Andersen}, \citenamefont {Ansmann},
  \citenamefont {Arute}, \citenamefont {Arya}, \citenamefont {Asfaw},
  \citenamefont {Astrakhantsev}, \citenamefont {Atalaya}, \citenamefont
  {Babbush}, \citenamefont {Bacon}, \citenamefont {Ballard}, \citenamefont
  {Bardin}, \citenamefont {Bausch}, \citenamefont {Bengtsson}, \citenamefont
  {Bilmes}, \citenamefont {Blackwell}, \citenamefont {Boixo}, \citenamefont
  {Bortoli}, \citenamefont {Bourassa}, \citenamefont {Bovaird}, \citenamefont
  {Brill}, \citenamefont {Broughton}, \citenamefont {Browne}, \citenamefont
  {Buchea}, \citenamefont {Buckley}, \citenamefont {Buell}, \citenamefont
  {Burger}, \citenamefont {Burkett}, \citenamefont {Bushnell}, \citenamefont
  {Cabrera}, \citenamefont {Campero}, \citenamefont {Chang}, \citenamefont
  {Chen}, \citenamefont {Chen}, \citenamefont {Chiaro}, \citenamefont {Chik},
  \citenamefont {Chou}, \citenamefont {Claes}, \citenamefont {Cleland},
  \citenamefont {Cogan}, \citenamefont {Collins}, \citenamefont {Conner},
  \citenamefont {Courtney}, \citenamefont {Crook}, \citenamefont {Curtin},
  \citenamefont {Das}, \citenamefont {Davies}, \citenamefont {De~Lorenzo},
  \citenamefont {Debroy}, \citenamefont {Demura}, \citenamefont {Devoret},
  \citenamefont {Di~Paolo}, \citenamefont {Donohoe}, \citenamefont {Drozdov},
  \citenamefont {Dunsworth}, \citenamefont {Earle}, \citenamefont {Edlich},
  \citenamefont {Eickbusch}, \citenamefont {Elbag}, \citenamefont {Elzouka},
  \citenamefont {Erickson}, \citenamefont {Faoro}, \citenamefont {Farhi},
  \citenamefont {Ferreira}, \citenamefont {Burgos}, \citenamefont {Forati},
  \citenamefont {Fowler}, \citenamefont {Foxen}, \citenamefont {Ganjam},
  \citenamefont {Garcia}, \citenamefont {Gasca}, \citenamefont {Genois},
  \citenamefont {Giang}, \citenamefont {Gidney}, \citenamefont {Gilboa},
  \citenamefont {Gosula}, \citenamefont {Dau}, \citenamefont {Graumann},
  \citenamefont {Greene}, \citenamefont {Gross}, \citenamefont {Habegger},
  \citenamefont {Hall}, \citenamefont {Hamilton}, \citenamefont {Hansen},
  \citenamefont {Harrigan}, \citenamefont {Harrington}, \citenamefont {Heras},
  \citenamefont {Heslin}, \citenamefont {Heu}, \citenamefont {Higgott},
  \citenamefont {Hill}, \citenamefont {Hilton}, \citenamefont {Holland},
  \citenamefont {Hong}, \citenamefont {Huang}, \citenamefont {Huff},
  \citenamefont {Huggins}, \citenamefont {Ioffe}, \citenamefont {Isakov},
  \citenamefont {Iveland}, \citenamefont {Jeffrey}, \citenamefont {Jiang},
  \citenamefont {Jones}, \citenamefont {Jordan}, \citenamefont {Joshi},
  \citenamefont {Juhas}, \citenamefont {Kafri}, \citenamefont {Kang},
  \citenamefont {Karamlou}, \citenamefont {Kechedzhi}, \citenamefont {Kelly},
  \citenamefont {Khaire}, \citenamefont {Khattar}, \citenamefont {Khezri},
  \citenamefont {Kim}, \citenamefont {Klimov}, \citenamefont {Klots},
  \citenamefont {Kobrin}, \citenamefont {Kohli}, \citenamefont {Korotkov},
  \citenamefont {Kostritsa}, \citenamefont {Kothari}, \citenamefont
  {Kozlovskii}, \citenamefont {Kreikebaum}, \citenamefont {Kurilovich},
  \citenamefont {Lacroix}, \citenamefont {Landhuis}, \citenamefont
  {{Lange-Dei}}, \citenamefont {Langley}, \citenamefont {Laptev}, \citenamefont
  {Lau}, \citenamefont {Le~Guevel}, \citenamefont {Ledford}, \citenamefont
  {Lee}, \citenamefont {Lee}, \citenamefont {Lensky}, \citenamefont {Leon},
  \citenamefont {Lester}, \citenamefont {Li}, \citenamefont {Li}, \citenamefont
  {Lill}, \citenamefont {Liu}, \citenamefont {Livingston}, \citenamefont
  {Locharla}, \citenamefont {Lucero}, \citenamefont {Lundahl}, \citenamefont
  {Lunt}, \citenamefont {Madhuk}, \citenamefont {Malone}, \citenamefont
  {Maloney}, \citenamefont {Mandr{\`a}}, \citenamefont {Manyika}, \citenamefont
  {Martin}, \citenamefont {Martin}, \citenamefont {Martin}, \citenamefont
  {Maxfield}, \citenamefont {McClean}, \citenamefont {McEwen}, \citenamefont
  {Meeks}, \citenamefont {Megrant}, \citenamefont {Mi}, \citenamefont {Miao},
  \citenamefont {Mieszala}, \citenamefont {Molavi}, \citenamefont {Molina},
  \citenamefont {Montazeri}, \citenamefont {Morvan}, \citenamefont {Movassagh},
  \citenamefont {Mruczkiewicz}, \citenamefont {Naaman}, \citenamefont {Neeley},
  \citenamefont {Neill}, \citenamefont {Nersisyan}, \citenamefont {Neven},
  \citenamefont {Newman}, \citenamefont {Ng}, \citenamefont {Nguyen},
  \citenamefont {Nguyen}, \citenamefont {Ni}, \citenamefont {Niu},
  \citenamefont {O'Brien}, \citenamefont {Oliver}, \citenamefont {Opremcak},
  \citenamefont {Ottosson}, \citenamefont {Petukhov}, \citenamefont {Pizzuto},
  \citenamefont {Platt}, \citenamefont {Potter}, \citenamefont {Pritchard},
  \citenamefont {Pryadko}, \citenamefont {Quintana}, \citenamefont
  {Ramachandran}, \citenamefont {Reagor}, \citenamefont {Redding},
  \citenamefont {Rhodes}, \citenamefont {Roberts}, \citenamefont {Rosenberg},
  \citenamefont {Rosenfeld}, \citenamefont {Roushan}, \citenamefont {Rubin},
  \citenamefont {Saei}, \citenamefont {Sank}, \citenamefont {Sankaragomathi},
  \citenamefont {Satzinger}, \citenamefont {Schurkus}, \citenamefont
  {Schuster}, \citenamefont {Senior}, \citenamefont {Shearn}, \citenamefont
  {Shorter}, \citenamefont {Shutty}, \citenamefont {Shvarts}, \citenamefont
  {Singh}, \citenamefont {Sivak}, \citenamefont {Skruzny}, \citenamefont
  {Small}, \citenamefont {Smelyanskiy}, \citenamefont {Smith}, \citenamefont
  {Somma}, \citenamefont {Springer}, \citenamefont {Sterling}, \citenamefont
  {Strain}, \citenamefont {Suchard}, \citenamefont {Szasz}, \citenamefont
  {Sztein}, \citenamefont {Thor}, \citenamefont {Torres}, \citenamefont
  {Torunbalci}, \citenamefont {Vaishnav}, \citenamefont {Vargas}, \citenamefont
  {Vdovichev}, \citenamefont {Vidal}, \citenamefont {Villalonga}, \citenamefont
  {Heidweiller}, \citenamefont {Waltman}, \citenamefont {Wang}, \citenamefont
  {Ware}, \citenamefont {Weber}, \citenamefont {Weidel}, \citenamefont {White},
  \citenamefont {Wong}, \citenamefont {Woo}, \citenamefont {Xing},
  \citenamefont {Yao}, \citenamefont {Yeh}, \citenamefont {Ying}, \citenamefont
  {Yoo}, \citenamefont {Yosri}, \citenamefont {Young}, \citenamefont {Zalcman},
  \citenamefont {Zhang}, \citenamefont {Zhu}, \citenamefont {Zobrist},\ and\
  \citenamefont {{Google Quantum AI and Collaborators}}}]{acharya:2024a}%
  \BibitemOpen
  \bibfield  {author} {\bibinfo {author} {\bibfnamefont {R.}~\bibnamefont
  {Acharya}}, \bibinfo {author} {\bibfnamefont {D.~A.}\ \bibnamefont {Abanin}},
  \bibinfo {author} {\bibfnamefont {L.}~\bibnamefont {{Aghababaie-Beni}}},
  \bibinfo {author} {\bibfnamefont {I.}~\bibnamefont {Aleiner}}, \bibinfo
  {author} {\bibfnamefont {T.~I.}\ \bibnamefont {Andersen}}, \bibinfo {author}
  {\bibfnamefont {M.}~\bibnamefont {Ansmann}}, \bibinfo {author} {\bibfnamefont
  {F.}~\bibnamefont {Arute}}, \bibinfo {author} {\bibfnamefont
  {K.}~\bibnamefont {Arya}}, \bibinfo {author} {\bibfnamefont {A.}~\bibnamefont
  {Asfaw}}, \bibinfo {author} {\bibfnamefont {N.}~\bibnamefont
  {Astrakhantsev}}, and others,\ }\href
  {https://doi.org/10.1038/s41586-024-08449-y} {\bibfield  {journal} {\bibinfo
  {journal} {Nature}\ ,\ \bibinfo {pages} {1}} (\bibinfo {year}
  {2024})}\BibitemShut {NoStop}%
\bibitem [{\citenamefont {Bluvstein}\ \emph {and others}(2024)\citenamefont
  {Bluvstein}, \citenamefont {Evered}, \citenamefont {Geim}, \citenamefont
  {Li}, \citenamefont {Zhou}, \citenamefont {Manovitz}, \citenamefont {Ebadi},
  \citenamefont {Cain}, \citenamefont {Kalinowski}, \citenamefont {Hangleiter},
  \citenamefont {Bonilla~Ataides}, \citenamefont {Maskara}, \citenamefont
  {Cong}, \citenamefont {Gao}, \citenamefont {Sales~Rodriguez}, \citenamefont
  {Karolyshyn}, \citenamefont {Semeghini}, \citenamefont {Gullans},
  \citenamefont {Greiner}, \citenamefont {Vuleti{\'c}},\ and\ \citenamefont
  {Lukin}}]{bluvstein:2024}%
  \BibitemOpen
  \bibfield  {author} {\bibinfo {author} {\bibfnamefont {D.}~\bibnamefont
  {Bluvstein}}, \bibinfo {author} {\bibfnamefont {S.~J.}\ \bibnamefont
  {Evered}}, \bibinfo {author} {\bibfnamefont {A.~A.}\ \bibnamefont {Geim}},
  \bibinfo {author} {\bibfnamefont {S.~H.}\ \bibnamefont {Li}}, \bibinfo
  {author} {\bibfnamefont {H.}~\bibnamefont {Zhou}}, \bibinfo {author}
  {\bibfnamefont {T.}~\bibnamefont {Manovitz}}, \bibinfo {author}
  {\bibfnamefont {S.}~\bibnamefont {Ebadi}}, \bibinfo {author} {\bibfnamefont
  {M.}~\bibnamefont {Cain}}, \bibinfo {author} {\bibfnamefont {M.}~\bibnamefont
  {Kalinowski}}, \bibinfo {author} {\bibfnamefont {D.}~\bibnamefont
  {Hangleiter}}, and others,\ }\href
  {https://doi.org/10.1038/s41586-023-06927-3} {\bibfield  {journal} {\bibinfo
  {journal} {Nature}\ }\textbf {\bibinfo {volume} {626}},\ \bibinfo {pages}
  {58} (\bibinfo {year} {2024})}\BibitemShut {NoStop}%
\bibitem [{\citenamefont {Reichardt}\ \emph {and others}(2024)\citenamefont
  {Reichardt}, \citenamefont {Aasen}, \citenamefont {Chao}, \citenamefont
  {Chernoguzov}, \citenamefont {van Dam}, \citenamefont {Gaebler},
  \citenamefont {Gresh}, \citenamefont {Lucchetti}, \citenamefont {Mills},
  \citenamefont {Moses}, \citenamefont {Neyenhuis}, \citenamefont {Paetznick},
  \citenamefont {Paz}, \citenamefont {Siegfried}, \citenamefont {da~Silva},
  \citenamefont {Svore}, \citenamefont {Wang},\ and\ \citenamefont
  {Zanner}}]{reichardt:2024}%
  \BibitemOpen
  \bibfield  {author} {\bibinfo {author} {\bibfnamefont {B.~W.}\ \bibnamefont
  {Reichardt}}, \bibinfo {author} {\bibfnamefont {D.}~\bibnamefont {Aasen}},
  \bibinfo {author} {\bibfnamefont {R.}~\bibnamefont {Chao}}, \bibinfo {author}
  {\bibfnamefont {A.}~\bibnamefont {Chernoguzov}}, \bibinfo {author}
  {\bibfnamefont {W.}~\bibnamefont {van Dam}}, \bibinfo {author} {\bibfnamefont
  {J.~P.}\ \bibnamefont {Gaebler}}, \bibinfo {author} {\bibfnamefont
  {D.}~\bibnamefont {Gresh}}, \bibinfo {author} {\bibfnamefont
  {D.}~\bibnamefont {Lucchetti}}, \bibinfo {author} {\bibfnamefont
  {M.}~\bibnamefont {Mills}}, \bibinfo {author} {\bibfnamefont {S.~A.}\
  \bibnamefont {Moses}}, and others,\ }\href
  {https://doi.org/10.48550/arXiv.2409.04628} {\bibinfo {title} {Demonstration
  of quantum computation and error correction with a tesseract code}} (\bibinfo
  {year} {2024}),\ \Eprint {https://arxiv.org/abs/2409.04628} {arXiv:2409.04628
  [quant-ph]} \BibitemShut {NoStop}%
\bibitem [{\citenamefont {Dalzell}\ \emph {and others}(2023)\citenamefont
  {Dalzell}, \citenamefont {McArdle}, \citenamefont {Berta}, \citenamefont
  {Bienias}, \citenamefont {Chen}, \citenamefont {Gily{\'e}n}, \citenamefont
  {Hann}, \citenamefont {Kastoryano}, \citenamefont {Khabiboulline},
  \citenamefont {Kubica}, \citenamefont {Salton}, \citenamefont {Wang},\ and\
  \citenamefont {Brand{\~a}o}}]{dalzell:2023}%
  \BibitemOpen
  \bibfield  {author} {\bibinfo {author} {\bibfnamefont {A.~M.}\ \bibnamefont
  {Dalzell}}, \bibinfo {author} {\bibfnamefont {S.}~\bibnamefont {McArdle}},
  \bibinfo {author} {\bibfnamefont {M.}~\bibnamefont {Berta}}, \bibinfo
  {author} {\bibfnamefont {P.}~\bibnamefont {Bienias}}, \bibinfo {author}
  {\bibfnamefont {C.-F.}\ \bibnamefont {Chen}}, \bibinfo {author}
  {\bibfnamefont {A.}~\bibnamefont {Gily{\'e}n}}, \bibinfo {author}
  {\bibfnamefont {C.~T.}\ \bibnamefont {Hann}}, \bibinfo {author}
  {\bibfnamefont {M.~J.}\ \bibnamefont {Kastoryano}}, \bibinfo {author}
  {\bibfnamefont {E.~T.}\ \bibnamefont {Khabiboulline}}, \bibinfo {author}
  {\bibfnamefont {A.}~\bibnamefont {Kubica}}, and others,\ }\href
  {https://doi.org/10.48550/arXiv.2310.03011} {\bibinfo {title} {Quantum
  algorithms: {{A}} survey of applications and end-to-end complexities}}
  (\bibinfo {year} {2023}),\ \Eprint {https://arxiv.org/abs/2310.03011}
  {arXiv:2310.03011 [quant-ph]} \BibitemShut {NoStop}%
\bibitem [{\citenamefont {Graham}\ \emph {and others}(2015)\citenamefont
  {Graham}, \citenamefont {Irastorza}, \citenamefont {Lamoreaux}, \citenamefont
  {Lindner},\ and\ \citenamefont {van Bibber}}]{graham:2015}%
  \BibitemOpen
  \bibfield  {author} {\bibinfo {author} {\bibfnamefont {P.~W.}\ \bibnamefont
  {Graham}}, \bibinfo {author} {\bibfnamefont {I.~G.}\ \bibnamefont
  {Irastorza}}, \bibinfo {author} {\bibfnamefont {S.~K.}\ \bibnamefont
  {Lamoreaux}}, \bibinfo {author} {\bibfnamefont {A.}~\bibnamefont {Lindner}},\
  and\ \bibinfo {author} {\bibfnamefont {K.~A.}\ \bibnamefont {van Bibber}},\
  }\href {https://doi.org/10.1146/annurev-nucl-102014-022120} {\bibfield
  {journal} {\bibinfo  {journal} {Annual Review of Nuclear and Particle
  Science}\ }\textbf {\bibinfo {volume} {65}},\ \bibinfo {pages} {485}
  (\bibinfo {year} {2015})}\BibitemShut {NoStop}%
\bibitem [{\citenamefont {Reif}\ \emph {and others}(2021)\citenamefont {Reif},
  \citenamefont {Ashbrook}, \citenamefont {Emsley},\ and\ \citenamefont
  {Hong}}]{reif:2021}%
  \BibitemOpen
  \bibfield  {author} {\bibinfo {author} {\bibfnamefont {B.}~\bibnamefont
  {Reif}}, \bibinfo {author} {\bibfnamefont {S.~E.}\ \bibnamefont {Ashbrook}},
  \bibinfo {author} {\bibfnamefont {L.}~\bibnamefont {Emsley}},\ and\ \bibinfo
  {author} {\bibfnamefont {M.}~\bibnamefont {Hong}},\ }\href
  {https://doi.org/10.1038/s43586-020-00002-1} {\bibfield  {journal} {\bibinfo
  {journal} {Nature Reviews Methods Primers}\ }\textbf {\bibinfo {volume}
  {1}},\ \bibinfo {pages} {1} (\bibinfo {year} {2021})}\BibitemShut {NoStop}%
\bibitem [{\citenamefont {Tse}\ \emph {and others}(2019)\citenamefont {Tse},
  \citenamefont {Yu}, \citenamefont {Kijbunchoo}, \citenamefont
  {{Fernandez-Galiana}}, \citenamefont {Dupej}, \citenamefont {Barsotti},
  \citenamefont {Blair}, \citenamefont {Brown}, \citenamefont {Dwyer},
  \citenamefont {Effler}, \citenamefont {Evans}, \citenamefont {Fritschel},
  \citenamefont {Frolov}, \citenamefont {Green}, \citenamefont {Mansell},
  \citenamefont {Matichard}, \citenamefont {Mavalvala}, \citenamefont
  {McClelland}, \citenamefont {McCuller}, \citenamefont {McRae}, \citenamefont
  {Miller}, \citenamefont {Mullavey}, \citenamefont {Oelker}, \citenamefont
  {Phinney}, \citenamefont {Sigg}, \citenamefont {Slagmolen}, \citenamefont
  {Vo}, \citenamefont {Ward}, \citenamefont {Whittle}, \citenamefont {Abbott},
  \citenamefont {Adams}, \citenamefont {Adhikari}, \citenamefont {Ananyeva},
  \citenamefont {Appert}, \citenamefont {Arai}, \citenamefont {Areeda},
  \citenamefont {Asali}, \citenamefont {Aston}, \citenamefont {Austin},
  \citenamefont {Baer}, \citenamefont {Ball}, \citenamefont {Ballmer},
  \citenamefont {Banagiri}, \citenamefont {Barker}, \citenamefont {Bartlett},
  \citenamefont {Berger}, \citenamefont {Betzwieser}, \citenamefont
  {Bhattacharjee}, \citenamefont {Billingsley}, \citenamefont {Biscans},
  \citenamefont {Blair}, \citenamefont {Bode}, \citenamefont {Booker},
  \citenamefont {Bork}, \citenamefont {Bramley}, \citenamefont {Brooks},
  \citenamefont {Buikema}, \citenamefont {Cahillane}, \citenamefont {Cannon},
  \citenamefont {Chen}, \citenamefont {Ciobanu}, \citenamefont {Clara},
  \citenamefont {Cooper}, \citenamefont {Corley}, \citenamefont {Countryman},
  \citenamefont {Covas}, \citenamefont {Coyne}, \citenamefont {Datrier},
  \citenamefont {Davis}, \citenamefont {Di~Fronzo}, \citenamefont {Driggers},
  \citenamefont {Etzel}, \citenamefont {Evans}, \citenamefont {Feicht},
  \citenamefont {Fulda}, \citenamefont {Fyffe}, \citenamefont {Giaime},
  \citenamefont {Giardina}, \citenamefont {Godwin}, \citenamefont {Goetz},
  \citenamefont {Gras}, \citenamefont {Gray}, \citenamefont {Gray},
  \citenamefont {Gupta}, \citenamefont {Gustafson}, \citenamefont {Gustafson},
  \citenamefont {Hanks}, \citenamefont {Hanson}, \citenamefont {Hardwick},
  \citenamefont {Hasskew}, \citenamefont {Heintze}, \citenamefont
  {{Helmling-Cornell}}, \citenamefont {Holland}, \citenamefont {Jones},
  \citenamefont {Kandhasamy}, \citenamefont {Karki}, \citenamefont {Kasprzack},
  \citenamefont {Kawabe}, \citenamefont {King}, \citenamefont {Kissel},
  \citenamefont {Kumar}, \citenamefont {Landry}, \citenamefont {Lane},
  \citenamefont {Lantz}, \citenamefont {Laxen}, \citenamefont {Lecoeuche},
  \citenamefont {Leviton}, \citenamefont {Liu}, \citenamefont {Lormand},
  \citenamefont {Lundgren}, \citenamefont {Macas}, \citenamefont {MacInnis},
  \citenamefont {Macleod}, \citenamefont {M{\'a}rka}, \citenamefont
  {M{\'a}rka}, \citenamefont {Martynov}, \citenamefont {Mason}, \citenamefont
  {Massinger}, \citenamefont {McCarthy}, \citenamefont {McCormick},
  \citenamefont {McIver}, \citenamefont {Mendell}, \citenamefont {Merfeld},
  \citenamefont {Merilh}, \citenamefont {Meylahn}, \citenamefont {Mistry},
  \citenamefont {Mittleman}, \citenamefont {Moreno}, \citenamefont
  {{Mow-Lowry}}, \citenamefont {Mozzon}, \citenamefont {Nelson}, \citenamefont
  {Nguyen}, \citenamefont {Nuttall}, \citenamefont {Oberling}, \citenamefont
  {Oram}, \citenamefont {O'Reilly}, \citenamefont {Osthelder}, \citenamefont
  {Ottaway}, \citenamefont {Overmier}, \citenamefont {Palamos}, \citenamefont
  {Parker}, \citenamefont {Payne}, \citenamefont {Pele}, \citenamefont {Perez},
  \citenamefont {Pirello}, \citenamefont {Radkins}, \citenamefont {Ramirez},
  \citenamefont {Richardson}, \citenamefont {Riles}, \citenamefont {Robertson},
  \citenamefont {Rollins}, \citenamefont {Romel}, \citenamefont {Romie},
  \citenamefont {Ross}, \citenamefont {Ryan}, \citenamefont {Sadecki},
  \citenamefont {Sanchez}, \citenamefont {Sanchez}, \citenamefont {Saravanan},
  \citenamefont {Savage}, \citenamefont {Schaetzl}, \citenamefont {Schnabel},
  \citenamefont {Schofield}, \citenamefont {Schwartz}, \citenamefont {Sellers},
  \citenamefont {Shaffer}, \citenamefont {Smith}, \citenamefont {Soni},
  \citenamefont {Sorazu}, \citenamefont {Spencer}, \citenamefont {Strain},
  \citenamefont {Sun}, \citenamefont {Szczepa{\'n}czyk}, \citenamefont
  {Thomas}, \citenamefont {Thomas}, \citenamefont {Thorne}, \citenamefont
  {Toland}, \citenamefont {Torrie}, \citenamefont {Traylor}, \citenamefont
  {Urban}, \citenamefont {Vajente}, \citenamefont {Valdes}, \citenamefont
  {{Vander-Hyde}}, \citenamefont {Veitch}, \citenamefont {Venkateswara},
  \citenamefont {Venugopalan}, \citenamefont {Viets}, \citenamefont {Vorvick},
  \citenamefont {Wade}, \citenamefont {Warner}, \citenamefont {Weaver},
  \citenamefont {Weiss}, \citenamefont {Willke}, \citenamefont {Wipf},
  \citenamefont {Xiao}, \citenamefont {Yamamoto}, \citenamefont {Yap},
  \citenamefont {Yu}, \citenamefont {Zhang}, \citenamefont {Zucker},\ and\
  \citenamefont {Zweizig}}]{tse:2019}%
  \BibitemOpen
  \bibfield  {author} {\bibinfo {author} {\bibfnamefont {M.}~\bibnamefont
  {Tse}}, \bibinfo {author} {\bibfnamefont {H.}~\bibnamefont {Yu}}, \bibinfo
  {author} {\bibfnamefont {N.}~\bibnamefont {Kijbunchoo}}, \bibinfo {author}
  {\bibfnamefont {A.}~\bibnamefont {{Fernandez-Galiana}}}, \bibinfo {author}
  {\bibfnamefont {P.}~\bibnamefont {Dupej}}, \bibinfo {author} {\bibfnamefont
  {L.}~\bibnamefont {Barsotti}}, \bibinfo {author} {\bibfnamefont {C.~D.}\
  \bibnamefont {Blair}}, \bibinfo {author} {\bibfnamefont {D.~D.}\ \bibnamefont
  {Brown}}, \bibinfo {author} {\bibfnamefont {S.~E.}\ \bibnamefont {Dwyer}},
  \bibinfo {author} {\bibfnamefont {A.}~\bibnamefont {Effler}}, and others,\
  }\href {https://doi.org/10.1103/PhysRevLett.123.231107} {\bibfield  {journal}
  {\bibinfo  {journal} {Physical Review Letters}\ }\textbf {\bibinfo {volume}
  {123}},\ \bibinfo {pages} {231107} (\bibinfo {year} {2019})}\BibitemShut
  {NoStop}%
\bibitem [{\citenamefont {{LIGO Scientific Collaboration and Virgo
  Collaboration}}\ \emph {and others}(2016)\citenamefont {{LIGO Scientific
  Collaboration and Virgo Collaboration}}, \citenamefont {Abbott},
  \citenamefont {Abbott}, \citenamefont {Abbott}, \citenamefont {Abernathy},
  \citenamefont {Acernese}, \citenamefont {Ackley}, \citenamefont {Adams},
  \citenamefont {Adams}, \citenamefont {Addesso}, \citenamefont {Adhikari},
  \citenamefont {Adya}, \citenamefont {Affeldt}, \citenamefont {Agathos},
  \citenamefont {Agatsuma}, \citenamefont {Aggarwal}, \citenamefont {Aguiar},
  \citenamefont {Aiello}, \citenamefont {Ain}, \citenamefont {Ajith},
  \citenamefont {Allen}, \citenamefont {Allocca}, \citenamefont {Altin},
  \citenamefont {Anderson}, \citenamefont {Anderson}, \citenamefont {Arai},
  \citenamefont {Arain}, \citenamefont {Araya}, \citenamefont {Arceneaux},
  \citenamefont {Areeda}, \citenamefont {Arnaud}, \citenamefont {Arun},
  \citenamefont {Ascenzi}, \citenamefont {Ashton}, \citenamefont {Ast},
  \citenamefont {Aston}, \citenamefont {Astone}, \citenamefont {Aufmuth},
  \citenamefont {Aulbert}, \citenamefont {Babak}, \citenamefont {Bacon},
  \citenamefont {Bader}, \citenamefont {Baker}, \citenamefont {Baldaccini},
  \citenamefont {Ballardin}, \citenamefont {Ballmer}, \citenamefont {Barayoga},
  \citenamefont {Barclay}, \citenamefont {Barish}, \citenamefont {Barker},
  \citenamefont {Barone}, \citenamefont {Barr}, \citenamefont {Barsotti},
  \citenamefont {Barsuglia}, \citenamefont {Barta}, \citenamefont {Bartlett},
  \citenamefont {Barton}, \citenamefont {Bartos}, \citenamefont {Bassiri},
  \citenamefont {Basti}, \citenamefont {Batch}, \citenamefont {Baune},
  \citenamefont {Bavigadda}, \citenamefont {Bazzan}, \citenamefont {Behnke},
  \citenamefont {Bejger}, \citenamefont {Belczynski}, \citenamefont {Bell},
  \citenamefont {Bell}, \citenamefont {Berger}, \citenamefont {Bergman},
  \citenamefont {Bergmann}, \citenamefont {Berry}, \citenamefont {Bersanetti},
  \citenamefont {Bertolini}, \citenamefont {Betzwieser}, \citenamefont
  {Bhagwat}, \citenamefont {Bhandare}, \citenamefont {Bilenko}, \citenamefont
  {Billingsley}, \citenamefont {Birch}, \citenamefont {Birney}, \citenamefont
  {Birnholtz}, \citenamefont {Biscans}, \citenamefont {Bisht}, \citenamefont
  {Bitossi}, \citenamefont {Biwer}, \citenamefont {Bizouard}, \citenamefont
  {Blackburn}, \citenamefont {Blair}, \citenamefont {Blair}, \citenamefont
  {Blair}, \citenamefont {Bloemen}, \citenamefont {Bock}, \citenamefont
  {Bodiya}, \citenamefont {Boer}, \citenamefont {Bogaert}, \citenamefont
  {Bogan}, \citenamefont {Bohe}, \citenamefont {Bojtos}, \citenamefont {Bond},
  \citenamefont {Bondu}, \citenamefont {Bonnand}, \citenamefont {Boom},
  \citenamefont {Bork}, \citenamefont {Boschi}, \citenamefont {Bose},
  \citenamefont {Bouffanais}, \citenamefont {Bozzi}, \citenamefont
  {Bradaschia}, \citenamefont {Brady}, \citenamefont {Braginsky}, \citenamefont
  {Branchesi}, \citenamefont {Brau}, \citenamefont {Briant}, \citenamefont
  {Brillet}, \citenamefont {Brinkmann}, \citenamefont {Brisson}, \citenamefont
  {Brockill}, \citenamefont {Brooks}, \citenamefont {Brown}, \citenamefont
  {Brown}, \citenamefont {Brown}, \citenamefont {Buchanan}, \citenamefont
  {Buikema}, \citenamefont {Bulik}, \citenamefont {Bulten}, \citenamefont
  {Buonanno}, \citenamefont {Buskulic}, \citenamefont {Buy}, \citenamefont
  {Byer}, \citenamefont {Cabero}, \citenamefont {Cadonati}, \citenamefont
  {Cagnoli}, \citenamefont {Cahillane}, \citenamefont {Bustillo}, \citenamefont
  {Callister}, \citenamefont {Calloni}, \citenamefont {Camp}, \citenamefont
  {Cannon}, \citenamefont {Cao}, \citenamefont {Capano}, \citenamefont
  {Capocasa}, \citenamefont {Carbognani}, \citenamefont {Caride}, \citenamefont
  {Diaz}, \citenamefont {Casentini}, \citenamefont {Caudill}, \citenamefont
  {Cavagli{\`a}}, \citenamefont {Cavalier}, \citenamefont {Cavalieri},
  \citenamefont {Cella}, \citenamefont {Cepeda}, \citenamefont {Baiardi},
  \citenamefont {Cerretani}, \citenamefont {Cesarini}, \citenamefont
  {Chakraborty}, \citenamefont {Chalermsongsak}, \citenamefont {Chamberlin},
  \citenamefont {Chan}, \citenamefont {Chao}, \citenamefont {Charlton},
  \citenamefont {{Chassande-Mottin}}, \citenamefont {Chen}, \citenamefont
  {Chen}, \citenamefont {Cheng}, \citenamefont {Chincarini}, \citenamefont
  {Chiummo}, \citenamefont {Cho}, \citenamefont {Cho}, \citenamefont {Chow},
  \citenamefont {Christensen}, \citenamefont {Chu}, \citenamefont {Chua},
  \citenamefont {Chung}, \citenamefont {Ciani}, \citenamefont {Clara},
  \citenamefont {Clark}, \citenamefont {Cleva}, \citenamefont {Coccia},
  \citenamefont {Cohadon}, \citenamefont {Colla}, \citenamefont {Collette},
  \citenamefont {Cominsky}, \citenamefont {Constancio}, \citenamefont {Conte},
  \citenamefont {Conti}, \citenamefont {Cook}, \citenamefont {Corbitt},
  \citenamefont {Cornish}, \citenamefont {Corsi}, \citenamefont {Cortese},
  \citenamefont {Costa}, \citenamefont {Coughlin}, \citenamefont {Coughlin},
  \citenamefont {Coulon}, \citenamefont {Countryman}, \citenamefont {Couvares},
  \citenamefont {Cowan}, \citenamefont {Coward}, \citenamefont {Cowart},
  \citenamefont {Coyne}, \citenamefont {Coyne}, \citenamefont {Craig},
  \citenamefont {Creighton}, \citenamefont {Creighton}, \citenamefont {Cripe},
  \citenamefont {Crowder}, \citenamefont {Cruise}, \citenamefont {Cumming},
  \citenamefont {Cunningham}, \citenamefont {Cuoco}, \citenamefont {Canton},
  \citenamefont {Danilishin}, \citenamefont {D'Antonio}, \citenamefont
  {Danzmann}, \citenamefont {Darman}, \citenamefont {Da~Silva~Costa},
  \citenamefont {Dattilo}, \citenamefont {Dave}, \citenamefont {Daveloza},
  \citenamefont {Davier}, \citenamefont {Davies}, \citenamefont {Daw},
  \citenamefont {Day}, \citenamefont {De}, \citenamefont {DeBra}, \citenamefont
  {Debreczeni}, \citenamefont {Degallaix}, \citenamefont {De~Laurentis},
  \citenamefont {Del{\'e}glise}, \citenamefont {Del~Pozzo}, \citenamefont
  {Denker}, \citenamefont {Dent}, \citenamefont {Dereli}, \citenamefont
  {Dergachev}, \citenamefont {DeRosa}, \citenamefont {De~Rosa}, \citenamefont
  {DeSalvo}, \citenamefont {Dhurandhar}, \citenamefont {D{\'i}az},
  \citenamefont {Di~Fiore}, \citenamefont {Di~Giovanni}, \citenamefont
  {Di~Lieto}, \citenamefont {Di~Pace}, \citenamefont {Di~Palma}, \citenamefont
  {Di~Virgilio}, \citenamefont {Dojcinoski}, \citenamefont {Dolique},
  \citenamefont {Donovan}, \citenamefont {Dooley}, \citenamefont {Doravari},
  \citenamefont {Douglas}, \citenamefont {Downes}, \citenamefont {Drago},
  \citenamefont {Drever}, \citenamefont {Driggers}, \citenamefont {Du},
  \citenamefont {Ducrot}, \citenamefont {Dwyer}, \citenamefont {Edo},
  \citenamefont {Edwards}, \citenamefont {Effler}, \citenamefont {Eggenstein},
  \citenamefont {Ehrens}, \citenamefont {Eichholz}, \citenamefont {Eikenberry},
  \citenamefont {Engels}, \citenamefont {Essick}, \citenamefont {Etzel},
  \citenamefont {Evans}, \citenamefont {Evans}, \citenamefont {Everett},
  \citenamefont {Factourovich}, \citenamefont {Fafone}, \citenamefont {Fair},
  \citenamefont {Fairhurst}, \citenamefont {Fan}, \citenamefont {Fang},
  \citenamefont {Farinon}, \citenamefont {Farr}, \citenamefont {Farr},
  \citenamefont {Favata}, \citenamefont {Fays}, \citenamefont {Fehrmann},
  \citenamefont {Fejer}, \citenamefont {Feldbaum}, \citenamefont {Ferrante},
  \citenamefont {Ferreira}, \citenamefont {Ferrini}, \citenamefont {Fidecaro},
  \citenamefont {Finn}, \citenamefont {Fiori}, \citenamefont {Fiorucci},
  \citenamefont {Fisher}, \citenamefont {Flaminio}, \citenamefont {Fletcher},
  \citenamefont {Fong}, \citenamefont {Fournier}, \citenamefont {Franco},
  \citenamefont {Frasca}, \citenamefont {Frasconi}, \citenamefont {Frede},
  \citenamefont {Frei}, \citenamefont {Freise}, \citenamefont {Frey},
  \citenamefont {Frey}, \citenamefont {Fricke}, \citenamefont {Fritschel},
  \citenamefont {Frolov}, \citenamefont {Fulda}, \citenamefont {Fyffe},
  \citenamefont {Gabbard}, \citenamefont {Gair}, \citenamefont {Gammaitoni},
  \citenamefont {Gaonkar}, \citenamefont {Garufi}, \citenamefont {Gatto},
  \citenamefont {Gaur}, \citenamefont {Gehrels}, \citenamefont {Gemme},
  \citenamefont {Gendre}, \citenamefont {Genin}, \citenamefont {Gennai},
  \citenamefont {George}, \citenamefont {Gergely}, \citenamefont {Germain},
  \citenamefont {Ghosh}, \citenamefont {Ghosh}, \citenamefont {Ghosh},
  \citenamefont {Giaime}, \citenamefont {Giardina}, \citenamefont {Giazotto},
  \citenamefont {Gill}, \citenamefont {Glaefke}, \citenamefont {Gleason},
  \citenamefont {Goetz}, \citenamefont {Goetz}, \citenamefont {Gondan},
  \citenamefont {Gonz{\'a}lez}, \citenamefont {Castro}, \citenamefont
  {Gopakumar}, \citenamefont {Gordon}, \citenamefont {Gorodetsky},
  \citenamefont {Gossan}, \citenamefont {Gosselin}, \citenamefont {Gouaty},
  \citenamefont {Graef}, \citenamefont {Graff}, \citenamefont {Granata},
  \citenamefont {Grant}, \citenamefont {Gras}, \citenamefont {Gray},
  \citenamefont {Greco}, \citenamefont {Green}, \citenamefont {Greenhalgh},
  \citenamefont {Groot}, \citenamefont {Grote}, \citenamefont {Grunewald},
  \citenamefont {Guidi}, \citenamefont {Guo}, \citenamefont {Gupta},
  \citenamefont {Gupta}, \citenamefont {Gushwa}, \citenamefont {Gustafson},
  \citenamefont {Gustafson}, \citenamefont {Hacker}, \citenamefont {Hall},
  \citenamefont {Hall}, \citenamefont {Hammond}, \citenamefont {Haney},
  \citenamefont {Hanke}, \citenamefont {Hanks}, \citenamefont {Hanna},
  \citenamefont {Hannam}, \citenamefont {Hanson}, \citenamefont {Hardwick},
  \citenamefont {Harms}, \citenamefont {Harry}, \citenamefont {Harry},
  \citenamefont {Hart}, \citenamefont {Hartman}, \citenamefont {Haster},
  \citenamefont {Haughian}, \citenamefont {Healy}, \citenamefont {Heefner},
  \citenamefont {Heidmann}, \citenamefont {Heintze}, \citenamefont {Heinzel},
  \citenamefont {Heitmann}, \citenamefont {Hello}, \citenamefont {Hemming},
  \citenamefont {Hendry}, \citenamefont {Heng}, \citenamefont {Hennig},
  \citenamefont {Heptonstall}, \citenamefont {Heurs}, \citenamefont {Hild},
  \citenamefont {Hoak}, \citenamefont {Hodge}, \citenamefont {Hofman},
  \citenamefont {Hollitt}, \citenamefont {Holt}, \citenamefont {Holz},
  \citenamefont {Hopkins}, \citenamefont {Hosken}, \citenamefont {Hough},
  \citenamefont {Houston}, \citenamefont {Howell}, \citenamefont {Hu},
  \citenamefont {Huang}, \citenamefont {Huerta}, \citenamefont {Huet},
  \citenamefont {Hughey}, \citenamefont {Husa}, \citenamefont {Huttner},
  \citenamefont {{Huynh-Dinh}}, \citenamefont {Idrisy}, \citenamefont {Indik},
  \citenamefont {Ingram}, \citenamefont {Inta}, \citenamefont {Isa},
  \citenamefont {Isac}, \citenamefont {Isi}, \citenamefont {Islas},
  \citenamefont {Isogai}, \citenamefont {Iyer}, \citenamefont {Izumi},
  \citenamefont {Jacobson}, \citenamefont {Jacqmin}, \citenamefont {Jang},
  \citenamefont {Jani}, \citenamefont {Jaranowski}, \citenamefont {Jawahar},
  \citenamefont {{Jim{\'e}nez-Forteza}}, \citenamefont {Johnson}, \citenamefont
  {{Johnson-McDaniel}}, \citenamefont {Jones}, \citenamefont {Jones},
  \citenamefont {Jonker}, \citenamefont {Ju}, \citenamefont {Haris},
  \citenamefont {Kalaghatgi}, \citenamefont {Kalogera}, \citenamefont
  {Kandhasamy}, \citenamefont {Kang}, \citenamefont {Kanner}, \citenamefont
  {Karki}, \citenamefont {Kasprzack}, \citenamefont {Katsavounidis},
  \citenamefont {Katzman}, \citenamefont {Kaufer}, \citenamefont {Kaur},
  \citenamefont {Kawabe}, \citenamefont {Kawazoe}, \citenamefont
  {K{\'e}f{\'e}lian}, \citenamefont {Kehl}, \citenamefont {Keitel},
  \citenamefont {Kelley}, \citenamefont {Kells}, \citenamefont {Kennedy},
  \citenamefont {Keppel}, \citenamefont {Key}, \citenamefont {Khalaidovski},
  \citenamefont {Khalili}, \citenamefont {Khan}, \citenamefont {Khan},
  \citenamefont {Khan}, \citenamefont {Khazanov}, \citenamefont {Kijbunchoo},
  \citenamefont {Kim}, \citenamefont {Kim}, \citenamefont {Kim}, \citenamefont
  {Kim}, \citenamefont {Kim}, \citenamefont {Kim}, \citenamefont {King},
  \citenamefont {King}, \citenamefont {Kinzel}, \citenamefont {Kissel},
  \citenamefont {Kleybolte}, \citenamefont {Klimenko}, \citenamefont
  {Koehlenbeck}, \citenamefont {Kokeyama}, \citenamefont {Koley}, \citenamefont
  {Kondrashov}, \citenamefont {Kontos}, \citenamefont {Koranda}, \citenamefont
  {Korobko}, \citenamefont {Korth}, \citenamefont {Kowalska}, \citenamefont
  {Kozak}, \citenamefont {Kringel}, \citenamefont {Krishnan}, \citenamefont
  {Kr{\'o}lak}, \citenamefont {Krueger}, \citenamefont {Kuehn}, \citenamefont
  {Kumar}, \citenamefont {Kumar}, \citenamefont {Kuo}, \citenamefont {Kutynia},
  \citenamefont {Kwee}, \citenamefont {Lackey}, \citenamefont {Landry},
  \citenamefont {Lange}, \citenamefont {Lantz}, \citenamefont {Lasky},
  \citenamefont {Lazzarini}, \citenamefont {Lazzaro}, \citenamefont {Leaci},
  \citenamefont {Leavey}, \citenamefont {Lebigot}, \citenamefont {Lee},
  \citenamefont {Lee}, \citenamefont {Lee}, \citenamefont {Lee}, \citenamefont
  {Lenon}, \citenamefont {Leonardi}, \citenamefont {Leong}, \citenamefont
  {Leroy}, \citenamefont {Letendre}, \citenamefont {Levin}, \citenamefont
  {Levine}, \citenamefont {Li}, \citenamefont {Libson}, \citenamefont
  {Littenberg}, \citenamefont {Lockerbie}, \citenamefont {Logue}, \citenamefont
  {Lombardi}, \citenamefont {London}, \citenamefont {Lord}, \citenamefont
  {Lorenzini}, \citenamefont {Loriette}, \citenamefont {Lormand}, \citenamefont
  {Losurdo}, \citenamefont {Lough}, \citenamefont {Lousto}, \citenamefont
  {Lovelace}, \citenamefont {L{\"u}ck}, \citenamefont {Lundgren}, \citenamefont
  {Luo}, \citenamefont {Lynch}, \citenamefont {Ma}, \citenamefont {MacDonald},
  \citenamefont {Machenschalk}, \citenamefont {MacInnis}, \citenamefont
  {Macleod}, \citenamefont {{Maga{\~n}a-Sandoval}}, \citenamefont {Magee},
  \citenamefont {Mageswaran}, \citenamefont {Majorana}, \citenamefont
  {Maksimovic}, \citenamefont {Malvezzi}, \citenamefont {Man}, \citenamefont
  {Mandel}, \citenamefont {Mandic}, \citenamefont {Mangano}, \citenamefont
  {Mansell}, \citenamefont {Manske}, \citenamefont {Mantovani}, \citenamefont
  {Marchesoni}, \citenamefont {Marion}, \citenamefont {M{\'a}rka},
  \citenamefont {M{\'a}rka}, \citenamefont {Markosyan}, \citenamefont {Maros},
  \citenamefont {Martelli}, \citenamefont {Martellini}, \citenamefont {Martin},
  \citenamefont {Martin}, \citenamefont {Martynov}, \citenamefont {Marx},
  \citenamefont {Mason}, \citenamefont {Masserot}, \citenamefont {Massinger},
  \citenamefont {{Masso-Reid}}, \citenamefont {Matichard}, \citenamefont
  {Matone}, \citenamefont {Mavalvala}, \citenamefont {Mazumder}, \citenamefont
  {Mazzolo}, \citenamefont {McCarthy}, \citenamefont {McClelland},
  \citenamefont {McCormick}, \citenamefont {McGuire}, \citenamefont {McIntyre},
  \citenamefont {McIver}, \citenamefont {McManus}, \citenamefont {McWilliams},
  \citenamefont {Meacher}, \citenamefont {Meadors}, \citenamefont {Meidam},
  \citenamefont {Melatos}, \citenamefont {Mendell}, \citenamefont
  {{Mendoza-Gandara}}, \citenamefont {Mercer}, \citenamefont {Merilh},
  \citenamefont {Merzougui}, \citenamefont {Meshkov}, \citenamefont
  {Messenger}, \citenamefont {Messick}, \citenamefont {Meyers}, \citenamefont
  {Mezzani}, \citenamefont {Miao}, \citenamefont {Michel}, \citenamefont
  {Middleton}, \citenamefont {Mikhailov}, \citenamefont {Milano}, \citenamefont
  {Miller}, \citenamefont {Millhouse}, \citenamefont {Minenkov}, \citenamefont
  {Ming}, \citenamefont {Mirshekari}, \citenamefont {Mishra}, \citenamefont
  {Mitra}, \citenamefont {Mitrofanov}, \citenamefont {Mitselmakher},
  \citenamefont {Mittleman}, \citenamefont {Moggi}, \citenamefont {Mohan},
  \citenamefont {Mohapatra}, \citenamefont {Montani}, \citenamefont {Moore},
  \citenamefont {Moore}, \citenamefont {Moraru}, \citenamefont {Moreno},
  \citenamefont {Morriss}, \citenamefont {Mossavi}, \citenamefont {Mours},
  \citenamefont {{Mow-Lowry}}, \citenamefont {Mueller}, \citenamefont
  {Mueller}, \citenamefont {Muir}, \citenamefont {Mukherjee}, \citenamefont
  {Mukherjee}, \citenamefont {Mukherjee}, \citenamefont {Mukund}, \citenamefont
  {Mullavey}, \citenamefont {Munch}, \citenamefont {Murphy}, \citenamefont
  {Murray}, \citenamefont {Mytidis}, \citenamefont {Nardecchia}, \citenamefont
  {Naticchioni}, \citenamefont {Nayak}, \citenamefont {Necula}, \citenamefont
  {Nedkova}, \citenamefont {Nelemans}, \citenamefont {Neri}, \citenamefont
  {Neunzert}, \citenamefont {Newton}, \citenamefont {Nguyen}, \citenamefont
  {Nielsen}, \citenamefont {Nissanke}, \citenamefont {Nitz}, \citenamefont
  {Nocera}, \citenamefont {Nolting}, \citenamefont {Normandin}, \citenamefont
  {Nuttall}, \citenamefont {Oberling}, \citenamefont {Ochsner}, \citenamefont
  {O'Dell}, \citenamefont {Oelker}, \citenamefont {Ogin}, \citenamefont {Oh},
  \citenamefont {Oh}, \citenamefont {Ohme}, \citenamefont {Oliver},
  \citenamefont {Oppermann}, \citenamefont {Oram}, \citenamefont {O'Reilly},
  \citenamefont {O'Shaughnessy}, \citenamefont {Ott}, \citenamefont {Ottaway},
  \citenamefont {Ottens}, \citenamefont {Overmier}, \citenamefont {Owen},
  \citenamefont {Pai}, \citenamefont {Pai}, \citenamefont {Palamos},
  \citenamefont {Palashov}, \citenamefont {Palomba}, \citenamefont
  {{Pal-Singh}}, \citenamefont {Pan}, \citenamefont {Pan}, \citenamefont
  {Pankow}, \citenamefont {Pannarale}, \citenamefont {Pant}, \citenamefont
  {Paoletti}, \citenamefont {Paoli}, \citenamefont {Papa}, \citenamefont
  {Paris}, \citenamefont {Parker}, \citenamefont {Pascucci}, \citenamefont
  {Pasqualetti}, \citenamefont {Passaquieti}, \citenamefont {Passuello},
  \citenamefont {Patricelli}, \citenamefont {Patrick}, \citenamefont
  {Pearlstone}, \citenamefont {Pedraza}, \citenamefont {Pedurand},
  \citenamefont {Pekowsky}, \citenamefont {Pele}, \citenamefont {Penn},
  \citenamefont {Perreca}, \citenamefont {Pfeiffer}, \citenamefont {Phelps},
  \citenamefont {Piccinni}, \citenamefont {Pichot}, \citenamefont {Pickenpack},
  \citenamefont {Piergiovanni}, \citenamefont {Pierro}, \citenamefont
  {Pillant}, \citenamefont {Pinard}, \citenamefont {Pinto}, \citenamefont
  {Pitkin}, \citenamefont {Poeld}, \citenamefont {Poggiani}, \citenamefont
  {Popolizio}, \citenamefont {Post}, \citenamefont {Powell}, \citenamefont
  {Prasad}, \citenamefont {Predoi}, \citenamefont {Premachandra}, \citenamefont
  {Prestegard}, \citenamefont {Price}, \citenamefont {Prijatelj}, \citenamefont
  {Principe}, \citenamefont {Privitera}, \citenamefont {Prix}, \citenamefont
  {Prodi}, \citenamefont {Prokhorov}, \citenamefont {Puncken}, \citenamefont
  {Punturo}, \citenamefont {Puppo}, \citenamefont {P{\"u}rrer}, \citenamefont
  {Qi}, \citenamefont {Qin}, \citenamefont {Quetschke}, \citenamefont
  {Quintero}, \citenamefont {{Quitzow-James}}, \citenamefont {Raab},
  \citenamefont {Rabeling}, \citenamefont {Radkins}, \citenamefont {Raffai},
  \citenamefont {Raja}, \citenamefont {Rakhmanov}, \citenamefont {Ramet},
  \citenamefont {Rapagnani}, \citenamefont {Raymond}, \citenamefont {Razzano},
  \citenamefont {Re}, \citenamefont {Read}, \citenamefont {Reed}, \citenamefont
  {Regimbau}, \citenamefont {Rei}, \citenamefont {Reid}, \citenamefont
  {Reitze}, \citenamefont {Rew}, \citenamefont {Reyes}, \citenamefont {Ricci},
  \citenamefont {Riles}, \citenamefont {Robertson}, \citenamefont {Robie},
  \citenamefont {Robinet}, \citenamefont {Rocchi}, \citenamefont {Rolland},
  \citenamefont {Rollins}, \citenamefont {Roma}, \citenamefont {Romano},
  \citenamefont {Romano}, \citenamefont {Romanov}, \citenamefont {Romie},
  \citenamefont {Rosi{\'n}ska}, \citenamefont {Rowan}, \citenamefont
  {R{\"u}diger}, \citenamefont {Ruggi}, \citenamefont {Ryan}, \citenamefont
  {Sachdev}, \citenamefont {Sadecki}, \citenamefont {Sadeghian}, \citenamefont
  {Salconi}, \citenamefont {Saleem}, \citenamefont {Salemi}, \citenamefont
  {Samajdar}, \citenamefont {Sammut}, \citenamefont {Sampson}, \citenamefont
  {Sanchez}, \citenamefont {Sandberg}, \citenamefont {Sandeen}, \citenamefont
  {Sanders}, \citenamefont {Sanders}, \citenamefont {Sassolas}, \citenamefont
  {Sathyaprakash}, \citenamefont {Saulson}, \citenamefont {Sauter},
  \citenamefont {Savage}, \citenamefont {Sawadsky}, \citenamefont {Schale},
  \citenamefont {Schilling}, \citenamefont {Schmidt}, \citenamefont {Schmidt},
  \citenamefont {Schnabel}, \citenamefont {Schofield}, \citenamefont
  {Sch{\"o}nbeck}, \citenamefont {Schreiber}, \citenamefont {Schuette},
  \citenamefont {Schutz}, \citenamefont {Scott}, \citenamefont {Scott},
  \citenamefont {Sellers}, \citenamefont {Sengupta}, \citenamefont {Sentenac},
  \citenamefont {Sequino}, \citenamefont {Sergeev}, \citenamefont {Serna},
  \citenamefont {Setyawati}, \citenamefont {Sevigny}, \citenamefont {Shaddock},
  \citenamefont {Shaffer}, \citenamefont {Shah}, \citenamefont {Shahriar},
  \citenamefont {Shaltev}, \citenamefont {Shao}, \citenamefont {Shapiro},
  \citenamefont {Shawhan}, \citenamefont {Sheperd}, \citenamefont {Shoemaker},
  \citenamefont {Shoemaker}, \citenamefont {Siellez}, \citenamefont {Siemens},
  \citenamefont {Sigg}, \citenamefont {Silva}, \citenamefont {Simakov},
  \citenamefont {Singer}, \citenamefont {Singer}, \citenamefont {Singh},
  \citenamefont {Singh}, \citenamefont {Singhal}, \citenamefont {Sintes},
  \citenamefont {Slagmolen}, \citenamefont {Smith}, \citenamefont {Smith},
  \citenamefont {Smith}, \citenamefont {Smith}, \citenamefont {Son},
  \citenamefont {Sorazu}, \citenamefont {Sorrentino}, \citenamefont
  {Souradeep}, \citenamefont {Srivastava}, \citenamefont {Staley},
  \citenamefont {Steinke}, \citenamefont {Steinlechner}, \citenamefont
  {Steinlechner}, \citenamefont {Steinmeyer}, \citenamefont {Stephens},
  \citenamefont {Stevenson}, \citenamefont {Stone}, \citenamefont {Strain},
  \citenamefont {Straniero}, \citenamefont {Stratta}, \citenamefont {Strauss},
  \citenamefont {Strigin}, \citenamefont {Sturani}, \citenamefont {Stuver},
  \citenamefont {Summerscales}, \citenamefont {Sun}, \citenamefont {Sutton},
  \citenamefont {Swinkels}, \citenamefont {Szczepa{\'n}czyk}, \citenamefont
  {Tacca}, \citenamefont {Talukder}, \citenamefont {Tanner}, \citenamefont
  {T{\'a}pai}, \citenamefont {Tarabrin}, \citenamefont {Taracchini},
  \citenamefont {Taylor}, \citenamefont {Theeg}, \citenamefont
  {Thirugnanasambandam}, \citenamefont {Thomas}, \citenamefont {Thomas},
  \citenamefont {Thomas}, \citenamefont {Thorne}, \citenamefont {Thorne},
  \citenamefont {Thrane}, \citenamefont {Tiwari}, \citenamefont {Tiwari},
  \citenamefont {Tokmakov}, \citenamefont {Tomlinson}, \citenamefont {Tonelli},
  \citenamefont {Torres}, \citenamefont {Torrie}, \citenamefont
  {T{\"o}yr{\"a}}, \citenamefont {Travasso}, \citenamefont {Traylor},
  \citenamefont {Trifir{\`o}}, \citenamefont {Tringali}, \citenamefont
  {Trozzo}, \citenamefont {Tse}, \citenamefont {Turconi}, \citenamefont
  {Tuyenbayev}, \citenamefont {Ugolini}, \citenamefont {Unnikrishnan},
  \citenamefont {Urban}, \citenamefont {Usman}, \citenamefont {Vahlbruch},
  \citenamefont {Vajente}, \citenamefont {Valdes}, \citenamefont {Vallisneri},
  \citenamefont {{van Bakel}}, \citenamefont {{van Beuzekom}}, \citenamefont
  {{van den Brand}}, \citenamefont {Van Den~Broeck}, \citenamefont
  {{Vander-Hyde}}, \citenamefont {{van der Schaaf}}, \citenamefont {{van
  Heijningen}}, \citenamefont {{van Veggel}}, \citenamefont {Vardaro},
  \citenamefont {Vass}, \citenamefont {Vas{\'u}th}, \citenamefont {Vaulin},
  \citenamefont {Vecchio}, \citenamefont {Vedovato}, \citenamefont {Veitch},
  \citenamefont {Veitch}, \citenamefont {Venkateswara}, \citenamefont
  {Verkindt}, \citenamefont {Vetrano}, \citenamefont {Vicer{\'e}},
  \citenamefont {Vinciguerra}, \citenamefont {Vine}, \citenamefont {Vinet},
  \citenamefont {Vitale}, \citenamefont {Vo}, \citenamefont {Vocca},
  \citenamefont {Vorvick}, \citenamefont {Voss}, \citenamefont {Vousden},
  \citenamefont {Vyatchanin}, \citenamefont {Wade}, \citenamefont {Wade},
  \citenamefont {Wade}, \citenamefont {Waldman}, \citenamefont {Walker},
  \citenamefont {Wallace}, \citenamefont {Walsh}, \citenamefont {Wang},
  \citenamefont {Wang}, \citenamefont {Wang}, \citenamefont {Wang},
  \citenamefont {Wang}, \citenamefont {Ward}, \citenamefont {Ward},
  \citenamefont {Warner}, \citenamefont {Was}, \citenamefont {Weaver},
  \citenamefont {Wei}, \citenamefont {Weinert}, \citenamefont {Weinstein},
  \citenamefont {Weiss}, \citenamefont {Welborn}, \citenamefont {Wen},
  \citenamefont {We{\ss}els}, \citenamefont {Westphal}, \citenamefont {Wette},
  \citenamefont {Whelan}, \citenamefont {Whitcomb}, \citenamefont {White},
  \citenamefont {Whiting}, \citenamefont {Wiesner}, \citenamefont {Wilkinson},
  \citenamefont {Willems}, \citenamefont {Williams}, \citenamefont {Williams},
  \citenamefont {Williamson}, \citenamefont {Willis}, \citenamefont {Willke},
  \citenamefont {Wimmer}, \citenamefont {Winkelmann}, \citenamefont {Winkler},
  \citenamefont {Wipf}, \citenamefont {Wiseman}, \citenamefont {Wittel},
  \citenamefont {Woan}, \citenamefont {Worden}, \citenamefont {Wright},
  \citenamefont {Wu}, \citenamefont {Yablon}, \citenamefont {Yakushin},
  \citenamefont {Yam}, \citenamefont {Yamamoto}, \citenamefont {Yancey},
  \citenamefont {Yap}, \citenamefont {Yu}, \citenamefont {Yvert}, \citenamefont
  {Zadro{\.z}ny}, \citenamefont {Zangrando}, \citenamefont {Zanolin},
  \citenamefont {Zendri}, \citenamefont {Zevin}, \citenamefont {Zhang},
  \citenamefont {Zhang}, \citenamefont {Zhang}, \citenamefont {Zhang},
  \citenamefont {Zhao}, \citenamefont {Zhou}, \citenamefont {Zhou},
  \citenamefont {Zhu}, \citenamefont {Zucker}, \citenamefont {Zuraw},\ and\
  \citenamefont
  {Zweizig}}]{ligoscientificcollaborationandvirgocollaboration:2016}%
  \BibitemOpen
  \bibfield  {author} {\bibinfo {author} {\bibnamefont {{LIGO Scientific
  Collaboration and Virgo Collaboration}}}, \bibinfo {author} {\bibfnamefont
  {B.~P.}\ \bibnamefont {Abbott}}, \bibinfo {author} {\bibfnamefont
  {R.}~\bibnamefont {Abbott}}, \bibinfo {author} {\bibfnamefont {T.~D.}\
  \bibnamefont {Abbott}}, \bibinfo {author} {\bibfnamefont {M.~R.}\
  \bibnamefont {Abernathy}}, \bibinfo {author} {\bibfnamefont {F.}~\bibnamefont
  {Acernese}}, \bibinfo {author} {\bibfnamefont {K.}~\bibnamefont {Ackley}},
  \bibinfo {author} {\bibfnamefont {C.}~\bibnamefont {Adams}}, \bibinfo
  {author} {\bibfnamefont {T.}~\bibnamefont {Adams}}, \bibinfo {author}
  {\bibfnamefont {P.}~\bibnamefont {Addesso}}, and others,\ }\href
  {https://doi.org/10.1103/PhysRevLett.116.061102} {\bibfield  {journal}
  {\bibinfo  {journal} {Physical Review Letters}\ }\textbf {\bibinfo {volume}
  {116}},\ \bibinfo {pages} {061102} (\bibinfo {year} {2016})}\BibitemShut
  {NoStop}%
\bibitem [{\citenamefont {Martyn}\ \emph {and others}(2021)\citenamefont
  {Martyn}, \citenamefont {Rossi}, \citenamefont {Tan},\ and\ \citenamefont
  {Chuang}}]{martyn:2021}%
  \BibitemOpen
  \bibfield  {author} {\bibinfo {author} {\bibfnamefont {J.~M.}\ \bibnamefont
  {Martyn}}, \bibinfo {author} {\bibfnamefont {Z.~M.}\ \bibnamefont {Rossi}},
  \bibinfo {author} {\bibfnamefont {A.~K.}\ \bibnamefont {Tan}},\ and\ \bibinfo
  {author} {\bibfnamefont {I.~L.}\ \bibnamefont {Chuang}},\ }\href
  {https://doi.org/10.1103/PRXQuantum.2.040203} {\bibfield  {journal} {\bibinfo
   {journal} {PRX Quantum}\ }\textbf {\bibinfo {volume} {2}},\ \bibinfo {pages}
  {040203} (\bibinfo {year} {2021})}\BibitemShut {NoStop}%
\bibitem [{\citenamefont {Ge}\ \emph {and others}(2016)\citenamefont {Ge},
  \citenamefont {Moln{\'a}r},\ and\ \citenamefont {Cirac}}]{ge:2016}%
  \BibitemOpen
  \bibfield  {author} {\bibinfo {author} {\bibfnamefont {Y.}~\bibnamefont
  {Ge}}, \bibinfo {author} {\bibfnamefont {A.}~\bibnamefont {Moln{\'a}r}},\
  and\ \bibinfo {author} {\bibfnamefont {J.~I.}\ \bibnamefont {Cirac}},\ }\href
  {https://doi.org/10.1103/PhysRevLett.116.080503} {\bibfield  {journal}
  {\bibinfo  {journal} {Physical Review Letters}\ }\textbf {\bibinfo {volume}
  {116}},\ \bibinfo {pages} {080503} (\bibinfo {year} {2016})}\BibitemShut
  {NoStop}%
\bibitem [{\citenamefont {Nenciu}(1993)}]{nenciu:1993}%
  \BibitemOpen
  \bibfield  {author} {\bibinfo {author} {\bibfnamefont {G.}~\bibnamefont
  {Nenciu}},\ }\href@noop {} {\bibfield  {journal} {\bibinfo  {journal}
  {Communications in Mathematical Physics}\ }\textbf {\bibinfo {volume}
  {152}},\ \bibinfo {pages} {479} (\bibinfo {year} {1993})}\BibitemShut
  {NoStop}%
\bibitem [{\citenamefont {Bradley}\ \emph {and others}(2019)\citenamefont
  {Bradley}, \citenamefont {Randall}, \citenamefont {Abobeih}, \citenamefont
  {Berrevoets}, \citenamefont {Degen}, \citenamefont {Bakker}, \citenamefont
  {Markham}, \citenamefont {Twitchen},\ and\ \citenamefont
  {Taminiau}}]{bradley:2019}%
  \BibitemOpen
  \bibfield  {author} {\bibinfo {author} {\bibfnamefont {C.~E.}\ \bibnamefont
  {Bradley}}, \bibinfo {author} {\bibfnamefont {J.}~\bibnamefont {Randall}},
  \bibinfo {author} {\bibfnamefont {M.~H.}\ \bibnamefont {Abobeih}}, \bibinfo
  {author} {\bibfnamefont {R.~C.}\ \bibnamefont {Berrevoets}}, \bibinfo
  {author} {\bibfnamefont {M.~J.}\ \bibnamefont {Degen}}, \bibinfo {author}
  {\bibfnamefont {M.~A.}\ \bibnamefont {Bakker}}, \bibinfo {author}
  {\bibfnamefont {M.}~\bibnamefont {Markham}}, \bibinfo {author} {\bibfnamefont
  {D.~J.}\ \bibnamefont {Twitchen}},\ and\ \bibinfo {author} {\bibfnamefont
  {T.~H.}\ \bibnamefont {Taminiau}},\ }\href
  {https://doi.org/10.1103/PhysRevX.9.031045} {\bibfield  {journal} {\bibinfo
  {journal} {Physical Review X}\ }\textbf {\bibinfo {volume} {9}},\ \bibinfo
  {pages} {031045} (\bibinfo {year} {2019})}\BibitemShut {NoStop}%
\bibitem [{\citenamefont {Evered}\ \emph {and others}(2023)\citenamefont
  {Evered}, \citenamefont {Bluvstein}, \citenamefont {Kalinowski},
  \citenamefont {Ebadi}, \citenamefont {Manovitz}, \citenamefont {Zhou},
  \citenamefont {Li}, \citenamefont {Geim}, \citenamefont {Wang}, \citenamefont
  {Maskara}, \citenamefont {Levine}, \citenamefont {Semeghini}, \citenamefont
  {Greiner}, \citenamefont {Vuleti{\'c}},\ and\ \citenamefont
  {Lukin}}]{evered:2023a}%
  \BibitemOpen
  \bibfield  {author} {\bibinfo {author} {\bibfnamefont {S.~J.}\ \bibnamefont
  {Evered}}, \bibinfo {author} {\bibfnamefont {D.}~\bibnamefont {Bluvstein}},
  \bibinfo {author} {\bibfnamefont {M.}~\bibnamefont {Kalinowski}}, \bibinfo
  {author} {\bibfnamefont {S.}~\bibnamefont {Ebadi}}, \bibinfo {author}
  {\bibfnamefont {T.}~\bibnamefont {Manovitz}}, \bibinfo {author}
  {\bibfnamefont {H.}~\bibnamefont {Zhou}}, \bibinfo {author} {\bibfnamefont
  {S.~H.}\ \bibnamefont {Li}}, \bibinfo {author} {\bibfnamefont {A.~A.}\
  \bibnamefont {Geim}}, \bibinfo {author} {\bibfnamefont {T.~T.}\ \bibnamefont
  {Wang}}, \bibinfo {author} {\bibfnamefont {N.}~\bibnamefont {Maskara}}, and
  others,\ }\href {https://doi.org/10.1038/s41586-023-06481-y} {\bibfield
  {journal} {\bibinfo  {journal} {Nature}\ }\textbf {\bibinfo {volume} {622}},\
  \bibinfo {pages} {268} (\bibinfo {year} {2023})}\BibitemShut {NoStop}%
\bibitem [{\citenamefont {Blais}\ \emph {and others}(2021)\citenamefont
  {Blais}, \citenamefont {Grimsmo}, \citenamefont {Girvin},\ and\ \citenamefont
  {Wallraff}}]{blais:2021}%
  \BibitemOpen
  \bibfield  {author} {\bibinfo {author} {\bibfnamefont {A.}~\bibnamefont
  {Blais}}, \bibinfo {author} {\bibfnamefont {A.~L.}\ \bibnamefont {Grimsmo}},
  \bibinfo {author} {\bibfnamefont {S.~M.}\ \bibnamefont {Girvin}},\ and\
  \bibinfo {author} {\bibfnamefont {A.}~\bibnamefont {Wallraff}},\ }\href
  {https://doi.org/10.1103/RevModPhys.93.025005} {\bibfield  {journal}
  {\bibinfo  {journal} {Reviews of Modern Physics}\ }\textbf {\bibinfo {volume}
  {93}},\ \bibinfo {pages} {025005} (\bibinfo {year} {2021})}\BibitemShut
  {NoStop}%
\bibitem [{\citenamefont {{Pedrozo-Pe{\~n}afiel}}\ \emph {and
  others}(2020)\citenamefont {{Pedrozo-Pe{\~n}afiel}}, \citenamefont {Colombo},
  \citenamefont {Shu}, \citenamefont {Adiyatullin}, \citenamefont {Li},
  \citenamefont {Mendez}, \citenamefont {Braverman}, \citenamefont {Kawasaki},
  \citenamefont {Akamatsu}, \citenamefont {Xiao},\ and\ \citenamefont
  {Vuleti{\'c}}}]{pedrozo-penafiel:2020}%
  \BibitemOpen
  \bibfield  {author} {\bibinfo {author} {\bibfnamefont {E.}~\bibnamefont
  {{Pedrozo-Pe{\~n}afiel}}}, \bibinfo {author} {\bibfnamefont {S.}~\bibnamefont
  {Colombo}}, \bibinfo {author} {\bibfnamefont {C.}~\bibnamefont {Shu}},
  \bibinfo {author} {\bibfnamefont {A.~F.}\ \bibnamefont {Adiyatullin}},
  \bibinfo {author} {\bibfnamefont {Z.}~\bibnamefont {Li}}, \bibinfo {author}
  {\bibfnamefont {E.}~\bibnamefont {Mendez}}, \bibinfo {author} {\bibfnamefont
  {B.}~\bibnamefont {Braverman}}, \bibinfo {author} {\bibfnamefont
  {A.}~\bibnamefont {Kawasaki}}, \bibinfo {author} {\bibfnamefont
  {D.}~\bibnamefont {Akamatsu}}, \bibinfo {author} {\bibfnamefont
  {Y.}~\bibnamefont {Xiao}}, and others,\ }\href
  {https://doi.org/10.1038/s41586-020-3006-1} {\bibfield  {journal} {\bibinfo
  {journal} {Nature}\ }\textbf {\bibinfo {volume} {588}},\ \bibinfo {pages}
  {414} (\bibinfo {year} {2020})}\BibitemShut {NoStop}%
\bibitem [{\citenamefont {Monroe}\ \emph {and others}(2021)\citenamefont
  {Monroe}, \citenamefont {Campbell}, \citenamefont {Duan}, \citenamefont
  {Gong}, \citenamefont {Gorshkov}, \citenamefont {Hess}, \citenamefont
  {Islam}, \citenamefont {Kim}, \citenamefont {Linke}, \citenamefont {Pagano},
  \citenamefont {Richerme}, \citenamefont {Senko},\ and\ \citenamefont
  {Yao}}]{monroe:2021}%
  \BibitemOpen
  \bibfield  {author} {\bibinfo {author} {\bibfnamefont {C.}~\bibnamefont
  {Monroe}}, \bibinfo {author} {\bibfnamefont {W.~C.}\ \bibnamefont
  {Campbell}}, \bibinfo {author} {\bibfnamefont {L.-M.}\ \bibnamefont {Duan}},
  \bibinfo {author} {\bibfnamefont {Z.-X.}\ \bibnamefont {Gong}}, \bibinfo
  {author} {\bibfnamefont {A.~V.}\ \bibnamefont {Gorshkov}}, \bibinfo {author}
  {\bibfnamefont {P.~W.}\ \bibnamefont {Hess}}, \bibinfo {author}
  {\bibfnamefont {R.}~\bibnamefont {Islam}}, \bibinfo {author} {\bibfnamefont
  {K.}~\bibnamefont {Kim}}, \bibinfo {author} {\bibfnamefont {N.~M.}\
  \bibnamefont {Linke}}, \bibinfo {author} {\bibfnamefont {G.}~\bibnamefont
  {Pagano}}, and others,\ }\href {https://doi.org/10.1103/RevModPhys.93.025001}
  {\bibfield  {journal} {\bibinfo  {journal} {Reviews of Modern Physics}\
  }\textbf {\bibinfo {volume} {93}},\ \bibinfo {pages} {025001} (\bibinfo
  {year} {2021})}\BibitemShut {NoStop}%
\bibitem [{\citenamefont {Barry}\ \emph {and others}(2020)\citenamefont
  {Barry}, \citenamefont {Schloss}, \citenamefont {Bauch}, \citenamefont
  {Turner}, \citenamefont {Hart}, \citenamefont {Pham},\ and\ \citenamefont
  {Walsworth}}]{barry:2020}%
  \BibitemOpen
  \bibfield  {author} {\bibinfo {author} {\bibfnamefont {J.~F.}\ \bibnamefont
  {Barry}}, \bibinfo {author} {\bibfnamefont {J.~M.}\ \bibnamefont {Schloss}},
  \bibinfo {author} {\bibfnamefont {E.}~\bibnamefont {Bauch}}, \bibinfo
  {author} {\bibfnamefont {M.~J.}\ \bibnamefont {Turner}}, \bibinfo {author}
  {\bibfnamefont {C.~A.}\ \bibnamefont {Hart}}, \bibinfo {author}
  {\bibfnamefont {L.~M.}\ \bibnamefont {Pham}},\ and\ \bibinfo {author}
  {\bibfnamefont {R.~L.}\ \bibnamefont {Walsworth}},\ }\href
  {https://doi.org/10.1103/RevModPhys.92.015004} {\bibfield  {journal}
  {\bibinfo  {journal} {Reviews of Modern Physics}\ }\textbf {\bibinfo {volume}
  {92}},\ \bibinfo {pages} {015004} (\bibinfo {year} {2020})}\BibitemShut
  {NoStop}%
\bibitem [{\citenamefont {Doherty}\ \emph {and others}(2013)\citenamefont
  {Doherty}, \citenamefont {Manson}, \citenamefont {Delaney}, \citenamefont
  {Jelezko}, \citenamefont {Wrachtrup},\ and\ \citenamefont
  {Hollenberg}}]{doherty:2013}%
  \BibitemOpen
  \bibfield  {author} {\bibinfo {author} {\bibfnamefont {M.~W.}\ \bibnamefont
  {Doherty}}, \bibinfo {author} {\bibfnamefont {N.~B.}\ \bibnamefont {Manson}},
  \bibinfo {author} {\bibfnamefont {P.}~\bibnamefont {Delaney}}, \bibinfo
  {author} {\bibfnamefont {F.}~\bibnamefont {Jelezko}}, \bibinfo {author}
  {\bibfnamefont {J.}~\bibnamefont {Wrachtrup}},\ and\ \bibinfo {author}
  {\bibfnamefont {L.~C.~L.}\ \bibnamefont {Hollenberg}},\ }\href
  {https://doi.org/10.1016/j.physrep.2013.02.001} {\bibfield  {journal}
  {\bibinfo  {journal} {Physics Reports}\ }\bibinfo {series} {The
  Nitrogen-Vacancy Colour Centre in Diamond},\ \textbf {\bibinfo {volume}
  {528}},\ \bibinfo {pages} {1} (\bibinfo {year} {2013})}\BibitemShut {NoStop}%
\bibitem [{\citenamefont {Abobeih}\ \emph {and others}(2019)\citenamefont
  {Abobeih}, \citenamefont {Randall}, \citenamefont {Bradley}, \citenamefont
  {Bartling}, \citenamefont {Bakker}, \citenamefont {Degen}, \citenamefont
  {Markham}, \citenamefont {Twitchen},\ and\ \citenamefont
  {Taminiau}}]{abobeih:2019}%
  \BibitemOpen
  \bibfield  {author} {\bibinfo {author} {\bibfnamefont {M.~H.}\ \bibnamefont
  {Abobeih}}, \bibinfo {author} {\bibfnamefont {J.}~\bibnamefont {Randall}},
  \bibinfo {author} {\bibfnamefont {C.~E.}\ \bibnamefont {Bradley}}, \bibinfo
  {author} {\bibfnamefont {H.~P.}\ \bibnamefont {Bartling}}, \bibinfo {author}
  {\bibfnamefont {M.~A.}\ \bibnamefont {Bakker}}, \bibinfo {author}
  {\bibfnamefont {M.~J.}\ \bibnamefont {Degen}}, \bibinfo {author}
  {\bibfnamefont {M.}~\bibnamefont {Markham}}, \bibinfo {author} {\bibfnamefont
  {D.~J.}\ \bibnamefont {Twitchen}},\ and\ \bibinfo {author} {\bibfnamefont
  {T.~H.}\ \bibnamefont {Taminiau}},\ }\href
  {https://doi.org/10.1038/s41586-019-1834-7} {\bibfield  {journal} {\bibinfo
  {journal} {Nature}\ }\textbf {\bibinfo {volume} {576}},\ \bibinfo {pages}
  {411} (\bibinfo {year} {2019})}\BibitemShut {NoStop}%
\bibitem [{\citenamefont {Abobeih}\ \emph {and others}(2018)\citenamefont
  {Abobeih}, \citenamefont {Cramer}, \citenamefont {Bakker}, \citenamefont
  {Kalb}, \citenamefont {Markham}, \citenamefont {Twitchen},\ and\
  \citenamefont {Taminiau}}]{abobeih:2018}%
  \BibitemOpen
  \bibfield  {author} {\bibinfo {author} {\bibfnamefont {M.~H.}\ \bibnamefont
  {Abobeih}}, \bibinfo {author} {\bibfnamefont {J.}~\bibnamefont {Cramer}},
  \bibinfo {author} {\bibfnamefont {M.~A.}\ \bibnamefont {Bakker}}, \bibinfo
  {author} {\bibfnamefont {N.}~\bibnamefont {Kalb}}, \bibinfo {author}
  {\bibfnamefont {M.}~\bibnamefont {Markham}}, \bibinfo {author} {\bibfnamefont
  {D.~J.}\ \bibnamefont {Twitchen}},\ and\ \bibinfo {author} {\bibfnamefont
  {T.~H.}\ \bibnamefont {Taminiau}},\ }\href
  {https://doi.org/10.1038/s41467-018-04916-z} {\bibfield  {journal} {\bibinfo
  {journal} {Nature Communications}\ }\textbf {\bibinfo {volume} {9}},\
  \bibinfo {pages} {2552} (\bibinfo {year} {2018})}\BibitemShut {NoStop}%
\bibitem [{\citenamefont {Polloreno}\ \emph {and others}(2023)\citenamefont
  {Polloreno}, \citenamefont {Beckey}, \citenamefont {Levin}, \citenamefont
  {Shlosberg}, \citenamefont {Thompson}, \citenamefont {{Foss-Feig}},
  \citenamefont {Hayes},\ and\ \citenamefont {Smith}}]{polloreno:2023}%
  \BibitemOpen
  \bibfield  {author} {\bibinfo {author} {\bibfnamefont {A.~M.}\ \bibnamefont
  {Polloreno}}, \bibinfo {author} {\bibfnamefont {J.~L.}\ \bibnamefont
  {Beckey}}, \bibinfo {author} {\bibfnamefont {J.}~\bibnamefont {Levin}},
  \bibinfo {author} {\bibfnamefont {A.}~\bibnamefont {Shlosberg}}, \bibinfo
  {author} {\bibfnamefont {J.~K.}\ \bibnamefont {Thompson}}, \bibinfo {author}
  {\bibfnamefont {M.}~\bibnamefont {{Foss-Feig}}}, \bibinfo {author}
  {\bibfnamefont {D.}~\bibnamefont {Hayes}},\ and\ \bibinfo {author}
  {\bibfnamefont {G.}~\bibnamefont {Smith}},\ }\href
  {https://doi.org/10.1103/PhysRevApplied.19.014029} {\bibfield  {journal}
  {\bibinfo  {journal} {Physical Review Applied}\ }\textbf {\bibinfo {volume}
  {19}},\ \bibinfo {pages} {014029} (\bibinfo {year} {2023})}\BibitemShut
  {NoStop}%
\bibitem [{\citenamefont {Oppenheim}\ \emph {and others}(1996)\citenamefont
  {Oppenheim}, \citenamefont {Willsky},\ and\ \citenamefont
  {Nawab}}]{oppenheim:1996}%
  \BibitemOpen
  \bibfield  {author} {\bibinfo {author} {\bibfnamefont {A.~V.}\ \bibnamefont
  {Oppenheim}}, \bibinfo {author} {\bibfnamefont {A.~S.}\ \bibnamefont
  {Willsky}},\ and\ \bibinfo {author} {\bibfnamefont {S.~H.}\ \bibnamefont
  {Nawab}},\ }\href@noop {} {\emph {\bibinfo {title} {Signals and Systems}}},\
  \bibinfo {edition} {2nd}\ ed.,\ Prentice-{{Hall}} Signal Processing Series\
  (\bibinfo  {publisher} {Prentice Hall},\ \bibinfo {address} {Upper Saddle
  River, N.J.},\ \bibinfo {year} {1996})\BibitemShut {NoStop}%
\bibitem [{\citenamefont {Pang}\ and\ \citenamefont
  {Jordan}(2017)}]{pang:2017}%
  \BibitemOpen
  \bibfield  {author} {\bibinfo {author} {\bibfnamefont {S.}~\bibnamefont
  {Pang}}\ and\ \bibinfo {author} {\bibfnamefont {A.~N.}\ \bibnamefont
  {Jordan}},\ }\href {https://doi.org/10.1038/ncomms14695} {\bibfield
  {journal} {\bibinfo  {journal} {Nature Communications}\ }\textbf {\bibinfo
  {volume} {8}},\ \bibinfo {pages} {14695} (\bibinfo {year}
  {2017})}\BibitemShut {NoStop}%
\bibitem [{\citenamefont {{de Clercq}}\ \emph {and others}(2016)\citenamefont
  {{de Clercq}}, \citenamefont {Oswald}, \citenamefont {Fl{\"u}hmann},
  \citenamefont {Keitch}, \citenamefont {Kienzler}, \citenamefont {Lo},
  \citenamefont {Marinelli}, \citenamefont {Nadlinger}, \citenamefont
  {Negnevitsky},\ and\ \citenamefont {Home}}]{declercq:2016}%
  \BibitemOpen
  \bibfield  {author} {\bibinfo {author} {\bibfnamefont {L.~E.}\ \bibnamefont
  {{de Clercq}}}, \bibinfo {author} {\bibfnamefont {R.}~\bibnamefont {Oswald}},
  \bibinfo {author} {\bibfnamefont {C.}~\bibnamefont {Fl{\"u}hmann}}, \bibinfo
  {author} {\bibfnamefont {B.}~\bibnamefont {Keitch}}, \bibinfo {author}
  {\bibfnamefont {D.}~\bibnamefont {Kienzler}}, \bibinfo {author}
  {\bibfnamefont {H.-Y.}\ \bibnamefont {Lo}}, \bibinfo {author} {\bibfnamefont
  {M.}~\bibnamefont {Marinelli}}, \bibinfo {author} {\bibfnamefont
  {D.}~\bibnamefont {Nadlinger}}, \bibinfo {author} {\bibfnamefont
  {V.}~\bibnamefont {Negnevitsky}},\ and\ \bibinfo {author} {\bibfnamefont
  {J.~P.}\ \bibnamefont {Home}},\ }\href {https://doi.org/10.1038/ncomms11218}
  {\bibfield  {journal} {\bibinfo  {journal} {Nature Communications}\ }\textbf
  {\bibinfo {volume} {7}},\ \bibinfo {pages} {11218} (\bibinfo {year}
  {2016})}\BibitemShut {NoStop}%
\bibitem [{\citenamefont {Krastanov}\ \emph {and others}(2019)\citenamefont
  {Krastanov}, \citenamefont {Zhou}, \citenamefont {Flammia},\ and\
  \citenamefont {Jiang}}]{krastanov:2019}%
  \BibitemOpen
  \bibfield  {author} {\bibinfo {author} {\bibfnamefont {S.}~\bibnamefont
  {Krastanov}}, \bibinfo {author} {\bibfnamefont {S.}~\bibnamefont {Zhou}},
  \bibinfo {author} {\bibfnamefont {S.~T.}\ \bibnamefont {Flammia}},\ and\
  \bibinfo {author} {\bibfnamefont {L.}~\bibnamefont {Jiang}},\ }\href
  {https://doi.org/10.1088/2058-9565/ab18d5} {\bibfield  {journal} {\bibinfo
  {journal} {Quantum Science and Technology}\ }\textbf {\bibinfo {volume}
  {4}},\ \bibinfo {pages} {035003} (\bibinfo {year} {2019})}\BibitemShut
  {NoStop}%
\bibitem [{\citenamefont {Siva}\ \emph {and others}(2023)\citenamefont {Siva},
  \citenamefont {Koolstra}, \citenamefont {Steinmetz}, \citenamefont
  {Livingston}, \citenamefont {Das}, \citenamefont {Chen}, \citenamefont
  {Kreikebaum}, \citenamefont {Stevenson}, \citenamefont {J{\"u}nger},
  \citenamefont {Santiago}, \citenamefont {Siddiqi},\ and\ \citenamefont
  {Jordan}}]{siva:2023}%
  \BibitemOpen
  \bibfield  {author} {\bibinfo {author} {\bibfnamefont {K.}~\bibnamefont
  {Siva}}, \bibinfo {author} {\bibfnamefont {G.}~\bibnamefont {Koolstra}},
  \bibinfo {author} {\bibfnamefont {J.}~\bibnamefont {Steinmetz}}, \bibinfo
  {author} {\bibfnamefont {W.~P.}\ \bibnamefont {Livingston}}, \bibinfo
  {author} {\bibfnamefont {D.}~\bibnamefont {Das}}, \bibinfo {author}
  {\bibfnamefont {L.}~\bibnamefont {Chen}}, \bibinfo {author} {\bibfnamefont
  {J.}~\bibnamefont {Kreikebaum}}, \bibinfo {author} {\bibfnamefont
  {N.}~\bibnamefont {Stevenson}}, \bibinfo {author} {\bibfnamefont
  {C.}~\bibnamefont {J{\"u}nger}}, \bibinfo {author} {\bibfnamefont
  {D.}~\bibnamefont {Santiago}}, and others,\ }\href
  {https://doi.org/10.1103/PRXQuantum.4.040324} {\bibfield  {journal} {\bibinfo
   {journal} {PRX Quantum}\ }\textbf {\bibinfo {volume} {4}},\ \bibinfo {pages}
  {040324} (\bibinfo {year} {2023})}\BibitemShut {NoStop}%
\bibitem [{\citenamefont {Grover}(1998)}]{grover:1998}%
  \BibitemOpen
  \bibfield  {author} {\bibinfo {author} {\bibfnamefont {L.~K.}\ \bibnamefont
  {Grover}},\ }\href {https://doi.org/10.1103/PhysRevLett.80.4329} {\bibfield
  {journal} {\bibinfo  {journal} {Physical Review Letters}\ }\textbf {\bibinfo
  {volume} {80}},\ \bibinfo {pages} {4329} (\bibinfo {year}
  {1998})}\BibitemShut {NoStop}%
\end{thebibliography}%


\begin{thebibliography}{68}%
\makeatletter
\providecommand \@ifxundefined [1]{%
 \@ifx{#1\undefined}
}%
\providecommand \@ifnum [1]{%
 \ifnum #1\expandafter \@firstoftwo
 \else \expandafter \@secondoftwo
 \fi
}%
\providecommand \@ifx [1]{%
 \ifx #1\expandafter \@firstoftwo
 \else \expandafter \@secondoftwo
 \fi
}%
\providecommand \natexlab [1]{#1}%
\providecommand \enquote  [1]{``#1''}%
\providecommand \bibnamefont  [1]{#1}%
\providecommand \bibfnamefont [1]{#1}%
\providecommand \citenamefont [1]{#1}%
\providecommand \href@noop [0]{\@secondoftwo}%
\providecommand \href [0]{\begingroup \@sanitize@url \@href}%
\providecommand \@href[1]{\@@startlink{#1}\@@href}%
\providecommand \@@href[1]{\endgroup#1\@@endlink}%
\providecommand \@sanitize@url [0]{\catcode `\\12\catcode `\$12\catcode
  `\&12\catcode `\#12\catcode `\^12\catcode `\_12\catcode `\%12\relax}%
\providecommand \@@startlink[1]{}%
\providecommand \@@endlink[0]{}%
\providecommand \url  [0]{\begingroup\@sanitize@url \@url }%
\providecommand \@url [1]{\endgroup\@href {#1}{\urlprefix }}%
\providecommand \urlprefix  [0]{URL }%
\providecommand \Eprint [0]{\href }%
\providecommand \doibase [0]{https://doi.org/}%
\providecommand \selectlanguage [0]{\@gobble}%
\providecommand \bibinfo  [0]{\@secondoftwo}%
\providecommand \bibfield  [0]{\@secondoftwo}%
\providecommand \translation [1]{[#1]}%
\providecommand \BibitemOpen [0]{}%
\providecommand \bibitemStop [0]{}%
\providecommand \bibitemNoStop [0]{.\EOS\space}%
\providecommand \EOS [0]{\spacefactor3000\relax}%
\providecommand \BibitemShut  [1]{\csname bibitem#1\endcsname}%
\let\auto@bib@innerbib\@empty
\bibitem [{\citenamefont {Degen}\ \emph {and others}(2017)\citenamefont
  {Degen}, \citenamefont {Reinhard},\ and\ \citenamefont
  {Cappellaro}}]{degen:2017}%
  \BibitemOpen
  \bibfield  {author} {\bibinfo {author} {\bibfnamefont {C.~L.}\ \bibnamefont
  {Degen}}, \bibinfo {author} {\bibfnamefont {F.}~\bibnamefont {Reinhard}},\
  and\ \bibinfo {author} {\bibfnamefont {P.}~\bibnamefont {Cappellaro}},\
  }\href {https://doi.org/10.1103/RevModPhys.89.035002} {\bibfield  {journal}
  {\bibinfo  {journal} {Reviews of Modern Physics}\ }\textbf {\bibinfo {volume}
  {89}},\ \bibinfo {pages} {035002} (\bibinfo {year} {2017})}\BibitemShut
  {NoStop}%
\bibitem [{\citenamefont {Watson}(1922)}]{watson:1922}%
  \BibitemOpen
  \bibfield  {author} {\bibinfo {author} {\bibfnamefont {G.~N.}\ \bibnamefont
  {Watson}},\ }\href@noop {} {\emph {\bibinfo {title} {A {{Treatise}} on the
  {{Theory}} of {{Bessel Functions}}}}},\ Vol.~\bibinfo {volume} {2}\ (\bibinfo
   {publisher} {The Cambridge University Press},\ \bibinfo {year}
  {1922})\BibitemShut {NoStop}%
\bibitem [{\citenamefont {Low}(2017)}]{low:2017}%
  \BibitemOpen
  \bibfield  {author} {\bibinfo {author} {\bibfnamefont {G.~H.}\ \bibnamefont
  {Low}},\ }\emph {\bibinfo {title} {Quantum Signal Processing by Single-Qubit
  Dynamics}},\ \href@noop {} {\bibinfo {type} {Thesis}},\ \bibinfo  {school}
  {Massachusetts Institute of Technology} (\bibinfo {year} {2017})\BibitemShut
  {NoStop}%
\bibitem [{\citenamefont {Martyn}\ \emph {and others}(2021)\citenamefont
  {Martyn}, \citenamefont {Rossi}, \citenamefont {Tan},\ and\ \citenamefont
  {Chuang}}]{martyn:2021}%
  \BibitemOpen
  \bibfield  {author} {\bibinfo {author} {\bibfnamefont {J.~M.}\ \bibnamefont
  {Martyn}}, \bibinfo {author} {\bibfnamefont {Z.~M.}\ \bibnamefont {Rossi}},
  \bibinfo {author} {\bibfnamefont {A.~K.}\ \bibnamefont {Tan}},\ and\ \bibinfo
  {author} {\bibfnamefont {I.~L.}\ \bibnamefont {Chuang}},\ }\href
  {https://doi.org/10.1103/PRXQuantum.2.040203} {\bibfield  {journal} {\bibinfo
   {journal} {PRX Quantum}\ }\textbf {\bibinfo {volume} {2}},\ \bibinfo {pages}
  {040203} (\bibinfo {year} {2021})}\BibitemShut {NoStop}%
\bibitem [{\citenamefont {Scully}\ and\ \citenamefont
  {Zubairy}(1997)}]{scully:1997}%
  \BibitemOpen
  \bibfield  {author} {\bibinfo {author} {\bibfnamefont {M.~O.}\ \bibnamefont
  {Scully}}\ and\ \bibinfo {author} {\bibfnamefont {M.~S.}\ \bibnamefont
  {Zubairy}},\ }\href {https://doi.org/10.1017/CBO9780511813993} {\emph
  {\bibinfo {title} {Quantum {{Optics}}}}}\ (\bibinfo  {publisher} {Cambridge
  University Press},\ \bibinfo {address} {Cambridge},\ \bibinfo {year}
  {1997})\BibitemShut {NoStop}%
\bibitem [{\citenamefont {Burgarth}\ \emph {and others}(2022)\citenamefont
  {Burgarth}, \citenamefont {Facchi}, \citenamefont {Gramegna},\ and\
  \citenamefont {Yuasa}}]{burgarth:2022}%
  \BibitemOpen
  \bibfield  {author} {\bibinfo {author} {\bibfnamefont {D.}~\bibnamefont
  {Burgarth}}, \bibinfo {author} {\bibfnamefont {P.}~\bibnamefont {Facchi}},
  \bibinfo {author} {\bibfnamefont {G.}~\bibnamefont {Gramegna}},\ and\
  \bibinfo {author} {\bibfnamefont {K.}~\bibnamefont {Yuasa}},\ }\href
  {https://doi.org/10.22331/q-2022-06-14-737} {\bibfield  {journal} {\bibinfo
  {journal} {Quantum}\ }\textbf {\bibinfo {volume} {6}},\ \bibinfo {pages}
  {737} (\bibinfo {year} {2022})}\BibitemShut {NoStop}%
\bibitem [{\citenamefont {Stein}\ and\ \citenamefont
  {Shakarchi}(2011)}]{stein:2011}%
  \BibitemOpen
  \bibfield  {author} {\bibinfo {author} {\bibfnamefont {E.~M.}\ \bibnamefont
  {Stein}}\ and\ \bibinfo {author} {\bibfnamefont {R.}~\bibnamefont
  {Shakarchi}},\ }\href@noop {} {\emph {\bibinfo {title} {Fourier
  {{Analysis}}}}},\ Vol.~\bibinfo {volume} {1}\ (\bibinfo  {publisher}
  {Princeton University Press},\ \bibinfo {year} {2011})\BibitemShut {NoStop}%
\bibitem [{\citenamefont {Nenciu}(1993)}]{nenciu:1993}%
  \BibitemOpen
  \bibfield  {author} {\bibinfo {author} {\bibfnamefont {G.}~\bibnamefont
  {Nenciu}},\ }\href@noop {} {\bibfield  {journal} {\bibinfo  {journal}
  {Communications in Mathematical Physics}\ }\textbf {\bibinfo {volume}
  {152}},\ \bibinfo {pages} {479} (\bibinfo {year} {1993})}\BibitemShut
  {NoStop}%
\bibitem [{\citenamefont {Hagedorn}\ and\ \citenamefont
  {Joye}(2002)}]{hagedorn:2002}%
  \BibitemOpen
  \bibfield  {author} {\bibinfo {author} {\bibfnamefont {G.~A.}\ \bibnamefont
  {Hagedorn}}\ and\ \bibinfo {author} {\bibfnamefont {A.}~\bibnamefont
  {Joye}},\ }\href {https://doi.org/10.1006/jmaa.2001.7765} {\bibfield
  {journal} {\bibinfo  {journal} {Journal of Mathematical Analysis and
  Applications}\ }\textbf {\bibinfo {volume} {267}},\ \bibinfo {pages} {235}
  (\bibinfo {year} {2002})}\BibitemShut {NoStop}%
\bibitem [{\citenamefont {Ge}\ \emph {and others}(2016)\citenamefont {Ge},
  \citenamefont {Moln{\'a}r},\ and\ \citenamefont {Cirac}}]{ge:2016}%
  \BibitemOpen
  \bibfield  {author} {\bibinfo {author} {\bibfnamefont {Y.}~\bibnamefont
  {Ge}}, \bibinfo {author} {\bibfnamefont {A.}~\bibnamefont {Moln{\'a}r}},\
  and\ \bibinfo {author} {\bibfnamefont {J.~I.}\ \bibnamefont {Cirac}},\ }\href
  {https://doi.org/10.1103/PhysRevLett.116.080503} {\bibfield  {journal}
  {\bibinfo  {journal} {Physical Review Letters}\ }\textbf {\bibinfo {volume}
  {116}},\ \bibinfo {pages} {080503} (\bibinfo {year} {2016})}\BibitemShut
  {NoStop}%
\bibitem [{\citenamefont {Albash}\ and\ \citenamefont
  {Lidar}(2018)}]{albash:2018}%
  \BibitemOpen
  \bibfield  {author} {\bibinfo {author} {\bibfnamefont {T.}~\bibnamefont
  {Albash}}\ and\ \bibinfo {author} {\bibfnamefont {D.~A.}\ \bibnamefont
  {Lidar}},\ }\href {https://doi.org/10.1103/RevModPhys.90.015002} {\bibfield
  {journal} {\bibinfo  {journal} {Reviews of Modern Physics}\ }\textbf
  {\bibinfo {volume} {90}},\ \bibinfo {pages} {015002} (\bibinfo {year}
  {2018})}\BibitemShut {NoStop}%
\bibitem [{\citenamefont {Avrachenkov}\ \emph {and others}(2013)\citenamefont
  {Avrachenkov}, \citenamefont {Filar},\ and\ \citenamefont
  {Howlett}}]{avrachenkov:2013}%
  \BibitemOpen
  \bibfield  {author} {\bibinfo {author} {\bibfnamefont {K.~E.}\ \bibnamefont
  {Avrachenkov}}, \bibinfo {author} {\bibfnamefont {J.~A.}\ \bibnamefont
  {Filar}},\ and\ \bibinfo {author} {\bibfnamefont {P.~G.}\ \bibnamefont
  {Howlett}},\ }\href {https://doi.org/10.1137/1.9781611973143} {\emph
  {\bibinfo {title} {Analytic {{Perturbation Theory}} and {{Its
  Applications}}}}}\ (\bibinfo  {publisher} {{Society for Industrial and
  Applied Mathematics}},\ \bibinfo {address} {Philadelphia, PA},\ \bibinfo
  {year} {2013})\BibitemShut {NoStop}%
\bibitem [{\citenamefont {Huelga}\ \emph {and others}(1997)\citenamefont
  {Huelga}, \citenamefont {Macchiavello}, \citenamefont {Pellizzari},
  \citenamefont {Ekert}, \citenamefont {Plenio},\ and\ \citenamefont
  {Cirac}}]{huelga:1997}%
  \BibitemOpen
  \bibfield  {author} {\bibinfo {author} {\bibfnamefont {S.~F.}\ \bibnamefont
  {Huelga}}, \bibinfo {author} {\bibfnamefont {C.}~\bibnamefont
  {Macchiavello}}, \bibinfo {author} {\bibfnamefont {T.}~\bibnamefont
  {Pellizzari}}, \bibinfo {author} {\bibfnamefont {A.~K.}\ \bibnamefont
  {Ekert}}, \bibinfo {author} {\bibfnamefont {M.~B.}\ \bibnamefont {Plenio}},\
  and\ \bibinfo {author} {\bibfnamefont {J.~I.}\ \bibnamefont {Cirac}},\ }\href
  {https://doi.org/10.1103/PhysRevLett.79.3865} {\bibfield  {journal} {\bibinfo
   {journal} {Physical Review Letters}\ }\textbf {\bibinfo {volume} {79}},\
  \bibinfo {pages} {3865} (\bibinfo {year} {1997})}\BibitemShut {NoStop}%
\bibitem [{\citenamefont {Fujiwara}\ and\ \citenamefont
  {Imai}(2008)}]{fujiwara:2008}%
  \BibitemOpen
  \bibfield  {author} {\bibinfo {author} {\bibfnamefont {A.}~\bibnamefont
  {Fujiwara}}\ and\ \bibinfo {author} {\bibfnamefont {H.}~\bibnamefont
  {Imai}},\ }\href {https://doi.org/10.1088/1751-8113/41/25/255304} {\bibfield
  {journal} {\bibinfo  {journal} {Journal of Physics A: Mathematical and
  Theoretical}\ }\textbf {\bibinfo {volume} {41}},\ \bibinfo {pages} {255304}
  (\bibinfo {year} {2008})}\BibitemShut {NoStop}%
\bibitem [{\citenamefont {{Demkowicz-Dobrzanski}}\ \emph {and
  others}(2009)\citenamefont {{Demkowicz-Dobrzanski}}, \citenamefont {Dorner},
  \citenamefont {Smith}, \citenamefont {Lundeen}, \citenamefont {Wasilewski},
  \citenamefont {Banaszek},\ and\ \citenamefont
  {Walmsley}}]{demkowicz-dobrzanski:2009}%
  \BibitemOpen
  \bibfield  {author} {\bibinfo {author} {\bibfnamefont {R.}~\bibnamefont
  {{Demkowicz-Dobrzanski}}}, \bibinfo {author} {\bibfnamefont {U.}~\bibnamefont
  {Dorner}}, \bibinfo {author} {\bibfnamefont {B.~J.}\ \bibnamefont {Smith}},
  \bibinfo {author} {\bibfnamefont {J.~S.}\ \bibnamefont {Lundeen}}, \bibinfo
  {author} {\bibfnamefont {W.}~\bibnamefont {Wasilewski}}, \bibinfo {author}
  {\bibfnamefont {K.}~\bibnamefont {Banaszek}},\ and\ \bibinfo {author}
  {\bibfnamefont {I.~A.}\ \bibnamefont {Walmsley}},\ }\href
  {https://doi.org/10.1103/PhysRevA.80.013825} {\bibfield  {journal} {\bibinfo
  {journal} {Physical Review A}\ }\textbf {\bibinfo {volume} {80}},\ \bibinfo
  {pages} {013825} (\bibinfo {year} {2009})}\BibitemShut {NoStop}%
\bibitem [{\citenamefont {{Demkowicz-Dobrza{\'n}ski}}\ \emph {and
  others}(2012)\citenamefont {{Demkowicz-Dobrza{\'n}ski}}, \citenamefont
  {Ko{\l}ody{\'n}ski},\ and\ \citenamefont {Gu{\c t}{\u
  a}}}]{demkowicz-dobrzanski:2012}%
  \BibitemOpen
  \bibfield  {author} {\bibinfo {author} {\bibfnamefont {R.}~\bibnamefont
  {{Demkowicz-Dobrza{\'n}ski}}}, \bibinfo {author} {\bibfnamefont
  {J.}~\bibnamefont {Ko{\l}ody{\'n}ski}},\ and\ \bibinfo {author}
  {\bibfnamefont {M.}~\bibnamefont {Gu{\c t}{\u a}}},\ }\href
  {https://doi.org/10.1038/ncomms2067} {\bibfield  {journal} {\bibinfo
  {journal} {Nature Communications}\ }\textbf {\bibinfo {volume} {3}},\
  \bibinfo {pages} {1063} (\bibinfo {year} {2012})}\BibitemShut {NoStop}%
\bibitem [{\citenamefont {Escher}\ \emph {and others}(2011)\citenamefont
  {Escher}, \citenamefont {{de Matos Filho}},\ and\ \citenamefont
  {Davidovich}}]{escher:2011}%
  \BibitemOpen
  \bibfield  {author} {\bibinfo {author} {\bibfnamefont {B.~M.}\ \bibnamefont
  {Escher}}, \bibinfo {author} {\bibfnamefont {R.~L.}\ \bibnamefont {{de Matos
  Filho}}},\ and\ \bibinfo {author} {\bibfnamefont {L.}~\bibnamefont
  {Davidovich}},\ }\href {https://doi.org/10.1038/nphys1958} {\bibfield
  {journal} {\bibinfo  {journal} {Nature Physics}\ }\textbf {\bibinfo {volume}
  {7}},\ \bibinfo {pages} {406} (\bibinfo {year} {2011})}\BibitemShut {NoStop}%
\bibitem [{\citenamefont {Knysh}\ \emph {and others}(2011)\citenamefont
  {Knysh}, \citenamefont {Smelyanskiy},\ and\ \citenamefont
  {Durkin}}]{knysh:2011}%
  \BibitemOpen
  \bibfield  {author} {\bibinfo {author} {\bibfnamefont {S.}~\bibnamefont
  {Knysh}}, \bibinfo {author} {\bibfnamefont {V.~N.}\ \bibnamefont
  {Smelyanskiy}},\ and\ \bibinfo {author} {\bibfnamefont {G.~A.}\ \bibnamefont
  {Durkin}},\ }\href {https://doi.org/10.1103/PhysRevA.83.021804} {\bibfield
  {journal} {\bibinfo  {journal} {Physical Review A}\ }\textbf {\bibinfo
  {volume} {83}},\ \bibinfo {pages} {021804} (\bibinfo {year}
  {2011})}\BibitemShut {NoStop}%
\bibitem [{\citenamefont {Kessler}\ \emph {and others}(2014)\citenamefont
  {Kessler}, \citenamefont {Lovchinsky}, \citenamefont {Sushkov},\ and\
  \citenamefont {Lukin}}]{kessler:2014}%
  \BibitemOpen
  \bibfield  {author} {\bibinfo {author} {\bibfnamefont {E.~M.}\ \bibnamefont
  {Kessler}}, \bibinfo {author} {\bibfnamefont {I.}~\bibnamefont {Lovchinsky}},
  \bibinfo {author} {\bibfnamefont {A.~O.}\ \bibnamefont {Sushkov}},\ and\
  \bibinfo {author} {\bibfnamefont {M.~D.}\ \bibnamefont {Lukin}},\ }\href
  {https://doi.org/10.1103/PhysRevLett.112.150802} {\bibfield  {journal}
  {\bibinfo  {journal} {Physical Review Letters}\ }\textbf {\bibinfo {volume}
  {112}},\ \bibinfo {pages} {150802} (\bibinfo {year} {2014})}\BibitemShut
  {NoStop}%
\bibitem [{\citenamefont {Arrad}\ \emph {and others}(2014)\citenamefont
  {Arrad}, \citenamefont {Vinkler}, \citenamefont {Aharonov},\ and\
  \citenamefont {Retzker}}]{arrad:2014}%
  \BibitemOpen
  \bibfield  {author} {\bibinfo {author} {\bibfnamefont {G.}~\bibnamefont
  {Arrad}}, \bibinfo {author} {\bibfnamefont {Y.}~\bibnamefont {Vinkler}},
  \bibinfo {author} {\bibfnamefont {D.}~\bibnamefont {Aharonov}},\ and\
  \bibinfo {author} {\bibfnamefont {A.}~\bibnamefont {Retzker}},\ }\href
  {https://doi.org/10.1103/PhysRevLett.112.150801} {\bibfield  {journal}
  {\bibinfo  {journal} {Physical Review Letters}\ }\textbf {\bibinfo {volume}
  {112}},\ \bibinfo {pages} {150801} (\bibinfo {year} {2014})}\BibitemShut
  {NoStop}%
\bibitem [{\citenamefont {D{\"u}r}\ \emph {and others}(2014)\citenamefont
  {D{\"u}r}, \citenamefont {Skotiniotis}, \citenamefont {Fr{\"o}wis},\ and\
  \citenamefont {Kraus}}]{dur:2014}%
  \BibitemOpen
  \bibfield  {author} {\bibinfo {author} {\bibfnamefont {W.}~\bibnamefont
  {D{\"u}r}}, \bibinfo {author} {\bibfnamefont {M.}~\bibnamefont
  {Skotiniotis}}, \bibinfo {author} {\bibfnamefont {F.}~\bibnamefont
  {Fr{\"o}wis}},\ and\ \bibinfo {author} {\bibfnamefont {B.}~\bibnamefont
  {Kraus}},\ }\href {https://doi.org/10.1103/PhysRevLett.112.080801} {\bibfield
   {journal} {\bibinfo  {journal} {Physical Review Letters}\ }\textbf {\bibinfo
  {volume} {112}},\ \bibinfo {pages} {080801} (\bibinfo {year}
  {2014})}\BibitemShut {NoStop}%
\bibitem [{\citenamefont {Sekatski}\ \emph {and others}(2017)\citenamefont
  {Sekatski}, \citenamefont {Skotiniotis}, \citenamefont {Ko{\l}ody{\'n}ski},\
  and\ \citenamefont {D{\"u}r}}]{sekatski:2017}%
  \BibitemOpen
  \bibfield  {author} {\bibinfo {author} {\bibfnamefont {P.}~\bibnamefont
  {Sekatski}}, \bibinfo {author} {\bibfnamefont {M.}~\bibnamefont
  {Skotiniotis}}, \bibinfo {author} {\bibfnamefont {J.}~\bibnamefont
  {Ko{\l}ody{\'n}ski}},\ and\ \bibinfo {author} {\bibfnamefont
  {W.}~\bibnamefont {D{\"u}r}},\ }\href
  {https://doi.org/10.22331/q-2017-09-06-27} {\bibfield  {journal} {\bibinfo
  {journal} {Quantum}\ }\textbf {\bibinfo {volume} {1}},\ \bibinfo {pages} {27}
  (\bibinfo {year} {2017})}\BibitemShut {NoStop}%
\bibitem [{\citenamefont {{Demkowicz-Dobrza{\'n}ski}}\ \emph {and
  others}(2017)\citenamefont {{Demkowicz-Dobrza{\'n}ski}}, \citenamefont
  {Czajkowski},\ and\ \citenamefont {Sekatski}}]{demkowicz-dobrzanski:2017}%
  \BibitemOpen
  \bibfield  {author} {\bibinfo {author} {\bibfnamefont {R.}~\bibnamefont
  {{Demkowicz-Dobrza{\'n}ski}}}, \bibinfo {author} {\bibfnamefont
  {J.}~\bibnamefont {Czajkowski}},\ and\ \bibinfo {author} {\bibfnamefont
  {P.}~\bibnamefont {Sekatski}},\ }\href
  {https://doi.org/10.1103/PhysRevX.7.041009} {\bibfield  {journal} {\bibinfo
  {journal} {Physical Review X}\ }\textbf {\bibinfo {volume} {7}},\ \bibinfo
  {pages} {041009} (\bibinfo {year} {2017})}\BibitemShut {NoStop}%
\bibitem [{\citenamefont {Zhou}\ \emph {and others}(2018)\citenamefont {Zhou},
  \citenamefont {Zhang}, \citenamefont {Preskill},\ and\ \citenamefont
  {Jiang}}]{zhou:2018}%
  \BibitemOpen
  \bibfield  {author} {\bibinfo {author} {\bibfnamefont {S.}~\bibnamefont
  {Zhou}}, \bibinfo {author} {\bibfnamefont {M.}~\bibnamefont {Zhang}},
  \bibinfo {author} {\bibfnamefont {J.}~\bibnamefont {Preskill}},\ and\
  \bibinfo {author} {\bibfnamefont {L.}~\bibnamefont {Jiang}},\ }\href
  {https://doi.org/10.1038/s41467-017-02510-3} {\bibfield  {journal} {\bibinfo
  {journal} {Nature Communications}\ }\textbf {\bibinfo {volume} {9}},\
  \bibinfo {pages} {78} (\bibinfo {year} {2018})}\BibitemShut {NoStop}%
\bibitem [{\citenamefont {Zhou}\ and\ \citenamefont {Jiang}(2021)}]{zhou:2021}%
  \BibitemOpen
  \bibfield  {author} {\bibinfo {author} {\bibfnamefont {S.}~\bibnamefont
  {Zhou}}\ and\ \bibinfo {author} {\bibfnamefont {L.}~\bibnamefont {Jiang}},\
  }\href {https://doi.org/10.1103/PRXQuantum.2.010343} {\bibfield  {journal}
  {\bibinfo  {journal} {PRX Quantum}\ }\textbf {\bibinfo {volume} {2}},\
  \bibinfo {pages} {010343} (\bibinfo {year} {2021})}\BibitemShut {NoStop}%
\bibitem [{\citenamefont {Zhou}\ and\ \citenamefont {Jiang}(2020)}]{zhou:2020}%
  \BibitemOpen
  \bibfield  {author} {\bibinfo {author} {\bibfnamefont {S.}~\bibnamefont
  {Zhou}}\ and\ \bibinfo {author} {\bibfnamefont {L.}~\bibnamefont {Jiang}},\
  }\href {https://doi.org/10.1103/PhysRevResearch.2.013235} {\bibfield
  {journal} {\bibinfo  {journal} {Physical Review Research}\ }\textbf {\bibinfo
  {volume} {2}},\ \bibinfo {pages} {013235} (\bibinfo {year}
  {2020})}\BibitemShut {NoStop}%
\bibitem [{\citenamefont {Layden}\ \emph {and others}(2019)\citenamefont
  {Layden}, \citenamefont {Zhou}, \citenamefont {Cappellaro},\ and\
  \citenamefont {Jiang}}]{layden:2019}%
  \BibitemOpen
  \bibfield  {author} {\bibinfo {author} {\bibfnamefont {D.}~\bibnamefont
  {Layden}}, \bibinfo {author} {\bibfnamefont {S.}~\bibnamefont {Zhou}},
  \bibinfo {author} {\bibfnamefont {P.}~\bibnamefont {Cappellaro}},\ and\
  \bibinfo {author} {\bibfnamefont {L.}~\bibnamefont {Jiang}},\ }\href
  {https://doi.org/10.1103/PhysRevLett.122.040502} {\bibfield  {journal}
  {\bibinfo  {journal} {Physical Review Letters}\ }\textbf {\bibinfo {volume}
  {122}},\ \bibinfo {pages} {040502} (\bibinfo {year} {2019})}\BibitemShut
  {NoStop}%
\bibitem [{\citenamefont {Zhou}\ \emph {and others}(2024)\citenamefont {Zhou},
  \citenamefont {Manes},\ and\ \citenamefont {Jiang}}]{zhou:2024}%
  \BibitemOpen
  \bibfield  {author} {\bibinfo {author} {\bibfnamefont {S.}~\bibnamefont
  {Zhou}}, \bibinfo {author} {\bibfnamefont {A.~G.}\ \bibnamefont {Manes}},\
  and\ \bibinfo {author} {\bibfnamefont {L.}~\bibnamefont {Jiang}},\ }\href
  {https://doi.org/10.1103/PhysRevA.109.042406} {\bibfield  {journal} {\bibinfo
   {journal} {Physical Review A}\ }\textbf {\bibinfo {volume} {109}},\ \bibinfo
  {pages} {042406} (\bibinfo {year} {2024})}\BibitemShut {NoStop}%
\bibitem [{\citenamefont {Wolfowicz}\ \emph {and others}(2021)\citenamefont
  {Wolfowicz}, \citenamefont {Heremans}, \citenamefont {Anderson},
  \citenamefont {Kanai}, \citenamefont {Seo}, \citenamefont {Gali},
  \citenamefont {Galli},\ and\ \citenamefont {Awschalom}}]{wolfowicz:2021}%
  \BibitemOpen
  \bibfield  {author} {\bibinfo {author} {\bibfnamefont {G.}~\bibnamefont
  {Wolfowicz}}, \bibinfo {author} {\bibfnamefont {F.~J.}\ \bibnamefont
  {Heremans}}, \bibinfo {author} {\bibfnamefont {C.~P.}\ \bibnamefont
  {Anderson}}, \bibinfo {author} {\bibfnamefont {S.}~\bibnamefont {Kanai}},
  \bibinfo {author} {\bibfnamefont {H.}~\bibnamefont {Seo}}, \bibinfo {author}
  {\bibfnamefont {A.}~\bibnamefont {Gali}}, \bibinfo {author} {\bibfnamefont
  {G.}~\bibnamefont {Galli}},\ and\ \bibinfo {author} {\bibfnamefont {D.~D.}\
  \bibnamefont {Awschalom}},\ }\href
  {https://doi.org/10.1038/s41578-021-00306-y} {\bibfield  {journal} {\bibinfo
  {journal} {Nature Reviews Materials}\ }\textbf {\bibinfo {volume} {6}},\
  \bibinfo {pages} {906} (\bibinfo {year} {2021})}\BibitemShut {NoStop}%
\bibitem [{\citenamefont {Bradac}\ \emph {and others}(2019)\citenamefont
  {Bradac}, \citenamefont {Gao}, \citenamefont {Forneris}, \citenamefont
  {Trusheim},\ and\ \citenamefont {Aharonovich}}]{bradac:2019}%
  \BibitemOpen
  \bibfield  {author} {\bibinfo {author} {\bibfnamefont {C.}~\bibnamefont
  {Bradac}}, \bibinfo {author} {\bibfnamefont {W.}~\bibnamefont {Gao}},
  \bibinfo {author} {\bibfnamefont {J.}~\bibnamefont {Forneris}}, \bibinfo
  {author} {\bibfnamefont {M.~E.}\ \bibnamefont {Trusheim}},\ and\ \bibinfo
  {author} {\bibfnamefont {I.}~\bibnamefont {Aharonovich}},\ }\href
  {https://doi.org/10.1038/s41467-019-13332-w} {\bibfield  {journal} {\bibinfo
  {journal} {Nature Communications}\ }\textbf {\bibinfo {volume} {10}},\
  \bibinfo {pages} {5625} (\bibinfo {year} {2019})}\BibitemShut {NoStop}%
\bibitem [{\citenamefont {Son}\ \emph {and others}(2020)\citenamefont {Son},
  \citenamefont {Anderson}, \citenamefont {Bourassa}, \citenamefont {Miao},
  \citenamefont {Babin}, \citenamefont {Widmann}, \citenamefont {Niethammer},
  \citenamefont {Ul~Hassan}, \citenamefont {Morioka}, \citenamefont {Ivanov},
  \citenamefont {Kaiser}, \citenamefont {Wrachtrup},\ and\ \citenamefont
  {Awschalom}}]{son:2020}%
  \BibitemOpen
  \bibfield  {author} {\bibinfo {author} {\bibfnamefont {N.~T.}\ \bibnamefont
  {Son}}, \bibinfo {author} {\bibfnamefont {C.~P.}\ \bibnamefont {Anderson}},
  \bibinfo {author} {\bibfnamefont {A.}~\bibnamefont {Bourassa}}, \bibinfo
  {author} {\bibfnamefont {K.~C.}\ \bibnamefont {Miao}}, \bibinfo {author}
  {\bibfnamefont {C.}~\bibnamefont {Babin}}, \bibinfo {author} {\bibfnamefont
  {M.}~\bibnamefont {Widmann}}, \bibinfo {author} {\bibfnamefont
  {M.}~\bibnamefont {Niethammer}}, \bibinfo {author} {\bibfnamefont
  {J.}~\bibnamefont {Ul~Hassan}}, \bibinfo {author} {\bibfnamefont
  {N.}~\bibnamefont {Morioka}}, \bibinfo {author} {\bibfnamefont {I.~G.}\
  \bibnamefont {Ivanov}}, and others,\ }\href
  {https://doi.org/10.1063/5.0004454} {\bibfield  {journal} {\bibinfo
  {journal} {Applied Physics Letters}\ }\textbf {\bibinfo {volume} {116}},\
  \bibinfo {pages} {190501} (\bibinfo {year} {2020})}\BibitemShut {NoStop}%
\bibitem [{\citenamefont {Gottscholl}\ \emph {and others}(2020)\citenamefont
  {Gottscholl}, \citenamefont {Kianinia}, \citenamefont {Soltamov},
  \citenamefont {Orlinskii}, \citenamefont {Mamin}, \citenamefont {Bradac},
  \citenamefont {Kasper}, \citenamefont {Krambrock}, \citenamefont {Sperlich},
  \citenamefont {Toth}, \citenamefont {Aharonovich},\ and\ \citenamefont
  {Dyakonov}}]{gottscholl:2020}%
  \BibitemOpen
  \bibfield  {author} {\bibinfo {author} {\bibfnamefont {A.}~\bibnamefont
  {Gottscholl}}, \bibinfo {author} {\bibfnamefont {M.}~\bibnamefont
  {Kianinia}}, \bibinfo {author} {\bibfnamefont {V.}~\bibnamefont {Soltamov}},
  \bibinfo {author} {\bibfnamefont {S.}~\bibnamefont {Orlinskii}}, \bibinfo
  {author} {\bibfnamefont {G.}~\bibnamefont {Mamin}}, \bibinfo {author}
  {\bibfnamefont {C.}~\bibnamefont {Bradac}}, \bibinfo {author} {\bibfnamefont
  {C.}~\bibnamefont {Kasper}}, \bibinfo {author} {\bibfnamefont
  {K.}~\bibnamefont {Krambrock}}, \bibinfo {author} {\bibfnamefont
  {A.}~\bibnamefont {Sperlich}}, \bibinfo {author} {\bibfnamefont
  {M.}~\bibnamefont {Toth}}, and others,\ }\href
  {https://doi.org/10.1038/s41563-020-0619-6} {\bibfield  {journal} {\bibinfo
  {journal} {Nature Materials}\ }\textbf {\bibinfo {volume} {19}},\ \bibinfo
  {pages} {540} (\bibinfo {year} {2020})}\BibitemShut {NoStop}%
\bibitem [{\citenamefont {Zhong}\ and\ \citenamefont
  {Goldner}(2019)}]{zhong:2019}%
  \BibitemOpen
  \bibfield  {author} {\bibinfo {author} {\bibfnamefont {T.}~\bibnamefont
  {Zhong}}\ and\ \bibinfo {author} {\bibfnamefont {P.}~\bibnamefont
  {Goldner}},\ }\href {https://doi.org/10.1515/nanoph-2019-0185} {\bibfield
  {journal} {\bibinfo  {journal} {Nanophotonics}\ }\textbf {\bibinfo {volume}
  {8}},\ \bibinfo {pages} {2003} (\bibinfo {year} {2019})}\BibitemShut
  {NoStop}%
\bibitem [{\citenamefont {Bradley}\ \emph {and others}(2019)\citenamefont
  {Bradley}, \citenamefont {Randall}, \citenamefont {Abobeih}, \citenamefont
  {Berrevoets}, \citenamefont {Degen}, \citenamefont {Bakker}, \citenamefont
  {Markham}, \citenamefont {Twitchen},\ and\ \citenamefont
  {Taminiau}}]{bradley:2019}%
  \BibitemOpen
  \bibfield  {author} {\bibinfo {author} {\bibfnamefont {C.~E.}\ \bibnamefont
  {Bradley}}, \bibinfo {author} {\bibfnamefont {J.}~\bibnamefont {Randall}},
  \bibinfo {author} {\bibfnamefont {M.~H.}\ \bibnamefont {Abobeih}}, \bibinfo
  {author} {\bibfnamefont {R.~C.}\ \bibnamefont {Berrevoets}}, \bibinfo
  {author} {\bibfnamefont {M.~J.}\ \bibnamefont {Degen}}, \bibinfo {author}
  {\bibfnamefont {M.~A.}\ \bibnamefont {Bakker}}, \bibinfo {author}
  {\bibfnamefont {M.}~\bibnamefont {Markham}}, \bibinfo {author} {\bibfnamefont
  {D.~J.}\ \bibnamefont {Twitchen}},\ and\ \bibinfo {author} {\bibfnamefont
  {T.~H.}\ \bibnamefont {Taminiau}},\ }\href
  {https://doi.org/10.1103/PhysRevX.9.031045} {\bibfield  {journal} {\bibinfo
  {journal} {Physical Review X}\ }\textbf {\bibinfo {volume} {9}},\ \bibinfo
  {pages} {031045} (\bibinfo {year} {2019})}\BibitemShut {NoStop}%
\bibitem [{\citenamefont {Madjarov}\ \emph {and others}(2019)\citenamefont
  {Madjarov}, \citenamefont {Cooper}, \citenamefont {Shaw}, \citenamefont
  {Covey}, \citenamefont {Schkolnik}, \citenamefont {Yoon}, \citenamefont
  {Williams},\ and\ \citenamefont {Endres}}]{madjarov:2019}%
  \BibitemOpen
  \bibfield  {author} {\bibinfo {author} {\bibfnamefont {I.~S.}\ \bibnamefont
  {Madjarov}}, \bibinfo {author} {\bibfnamefont {A.}~\bibnamefont {Cooper}},
  \bibinfo {author} {\bibfnamefont {A.~L.}\ \bibnamefont {Shaw}}, \bibinfo
  {author} {\bibfnamefont {J.~P.}\ \bibnamefont {Covey}}, \bibinfo {author}
  {\bibfnamefont {V.}~\bibnamefont {Schkolnik}}, \bibinfo {author}
  {\bibfnamefont {T.~H.}\ \bibnamefont {Yoon}}, \bibinfo {author}
  {\bibfnamefont {J.~R.}\ \bibnamefont {Williams}},\ and\ \bibinfo {author}
  {\bibfnamefont {M.}~\bibnamefont {Endres}},\ }\href
  {https://doi.org/10.1103/PhysRevX.9.041052} {\bibfield  {journal} {\bibinfo
  {journal} {Physical Review X}\ }\textbf {\bibinfo {volume} {9}},\ \bibinfo
  {pages} {041052} (\bibinfo {year} {2019})}\BibitemShut {NoStop}%
\bibitem [{\citenamefont {Cao}\ \emph {and others}(2024)\citenamefont {Cao},
  \citenamefont {Eckner}, \citenamefont {Yelin}, \citenamefont {Young},
  \citenamefont {Jandura}, \citenamefont {Yan}, \citenamefont {Kim},
  \citenamefont {Pupillo}, \citenamefont {Ye}, \citenamefont {Oppong},\ and\
  \citenamefont {Kaufman}}]{cao:2024a}%
  \BibitemOpen
  \bibfield  {author} {\bibinfo {author} {\bibfnamefont {A.}~\bibnamefont
  {Cao}}, \bibinfo {author} {\bibfnamefont {W.~J.}\ \bibnamefont {Eckner}},
  \bibinfo {author} {\bibfnamefont {T.~L.}\ \bibnamefont {Yelin}}, \bibinfo
  {author} {\bibfnamefont {A.~W.}\ \bibnamefont {Young}}, \bibinfo {author}
  {\bibfnamefont {S.}~\bibnamefont {Jandura}}, \bibinfo {author} {\bibfnamefont
  {L.}~\bibnamefont {Yan}}, \bibinfo {author} {\bibfnamefont {K.}~\bibnamefont
  {Kim}}, \bibinfo {author} {\bibfnamefont {G.}~\bibnamefont {Pupillo}},
  \bibinfo {author} {\bibfnamefont {J.}~\bibnamefont {Ye}}, \bibinfo {author}
  {\bibfnamefont {N.~D.}\ \bibnamefont {Oppong}}, and others,\ }\href
  {https://doi.org/10.1038/s41586-024-07913-z} {\bibfield  {journal} {\bibinfo
  {journal} {Nature}\ }\textbf {\bibinfo {volume} {634}},\ \bibinfo {pages}
  {315} (\bibinfo {year} {2024})}\BibitemShut {NoStop}%
\bibitem [{\citenamefont {Evered}\ \emph {and others}(2023)\citenamefont
  {Evered}, \citenamefont {Bluvstein}, \citenamefont {Kalinowski},
  \citenamefont {Ebadi}, \citenamefont {Manovitz}, \citenamefont {Zhou},
  \citenamefont {Li}, \citenamefont {Geim}, \citenamefont {Wang}, \citenamefont
  {Maskara}, \citenamefont {Levine}, \citenamefont {Semeghini}, \citenamefont
  {Greiner}, \citenamefont {Vuleti{\'c}},\ and\ \citenamefont
  {Lukin}}]{evered:2023}%
  \BibitemOpen
  \bibfield  {author} {\bibinfo {author} {\bibfnamefont {S.~J.}\ \bibnamefont
  {Evered}}, \bibinfo {author} {\bibfnamefont {D.}~\bibnamefont {Bluvstein}},
  \bibinfo {author} {\bibfnamefont {M.}~\bibnamefont {Kalinowski}}, \bibinfo
  {author} {\bibfnamefont {S.}~\bibnamefont {Ebadi}}, \bibinfo {author}
  {\bibfnamefont {T.}~\bibnamefont {Manovitz}}, \bibinfo {author}
  {\bibfnamefont {H.}~\bibnamefont {Zhou}}, \bibinfo {author} {\bibfnamefont
  {S.~H.}\ \bibnamefont {Li}}, \bibinfo {author} {\bibfnamefont {A.~A.}\
  \bibnamefont {Geim}}, \bibinfo {author} {\bibfnamefont {T.~T.}\ \bibnamefont
  {Wang}}, \bibinfo {author} {\bibfnamefont {N.}~\bibnamefont {Maskara}}, and
  others,\ }\href {https://doi.org/10.1038/s41586-023-06481-y} {\bibfield
  {journal} {\bibinfo  {journal} {Nature}\ }\textbf {\bibinfo {volume} {622}},\
  \bibinfo {pages} {268} (\bibinfo {year} {2023})}\BibitemShut {NoStop}%
\bibitem [{\citenamefont {Bluvstein}\ \emph {and others}(2024)\citenamefont
  {Bluvstein}, \citenamefont {Evered}, \citenamefont {Geim}, \citenamefont
  {Li}, \citenamefont {Zhou}, \citenamefont {Manovitz}, \citenamefont {Ebadi},
  \citenamefont {Cain}, \citenamefont {Kalinowski}, \citenamefont {Hangleiter},
  \citenamefont {Bonilla~Ataides}, \citenamefont {Maskara}, \citenamefont
  {Cong}, \citenamefont {Gao}, \citenamefont {Sales~Rodriguez}, \citenamefont
  {Karolyshyn}, \citenamefont {Semeghini}, \citenamefont {Gullans},
  \citenamefont {Greiner}, \citenamefont {Vuleti{\'c}},\ and\ \citenamefont
  {Lukin}}]{bluvstein:2024}%
  \BibitemOpen
  \bibfield  {author} {\bibinfo {author} {\bibfnamefont {D.}~\bibnamefont
  {Bluvstein}}, \bibinfo {author} {\bibfnamefont {S.~J.}\ \bibnamefont
  {Evered}}, \bibinfo {author} {\bibfnamefont {A.~A.}\ \bibnamefont {Geim}},
  \bibinfo {author} {\bibfnamefont {S.~H.}\ \bibnamefont {Li}}, \bibinfo
  {author} {\bibfnamefont {H.}~\bibnamefont {Zhou}}, \bibinfo {author}
  {\bibfnamefont {T.}~\bibnamefont {Manovitz}}, \bibinfo {author}
  {\bibfnamefont {S.}~\bibnamefont {Ebadi}}, \bibinfo {author} {\bibfnamefont
  {M.}~\bibnamefont {Cain}}, \bibinfo {author} {\bibfnamefont {M.}~\bibnamefont
  {Kalinowski}}, \bibinfo {author} {\bibfnamefont {D.}~\bibnamefont
  {Hangleiter}}, and others,\ }\href
  {https://doi.org/10.1038/s41586-023-06927-3} {\bibfield  {journal} {\bibinfo
  {journal} {Nature}\ }\textbf {\bibinfo {volume} {626}},\ \bibinfo {pages}
  {58} (\bibinfo {year} {2024})}\BibitemShut {NoStop}%
\bibitem [{\citenamefont {Levine}\ \emph {and others}(2018)\citenamefont
  {Levine}, \citenamefont {Keesling}, \citenamefont {Omran}, \citenamefont
  {Bernien}, \citenamefont {Schwartz}, \citenamefont {Zibrov}, \citenamefont
  {Endres}, \citenamefont {Greiner}, \citenamefont {Vuleti{\'c}},\ and\
  \citenamefont {Lukin}}]{levine:2018}%
  \BibitemOpen
  \bibfield  {author} {\bibinfo {author} {\bibfnamefont {H.}~\bibnamefont
  {Levine}}, \bibinfo {author} {\bibfnamefont {A.}~\bibnamefont {Keesling}},
  \bibinfo {author} {\bibfnamefont {A.}~\bibnamefont {Omran}}, \bibinfo
  {author} {\bibfnamefont {H.}~\bibnamefont {Bernien}}, \bibinfo {author}
  {\bibfnamefont {S.}~\bibnamefont {Schwartz}}, \bibinfo {author}
  {\bibfnamefont {A.~S.}\ \bibnamefont {Zibrov}}, \bibinfo {author}
  {\bibfnamefont {M.}~\bibnamefont {Endres}}, \bibinfo {author} {\bibfnamefont
  {M.}~\bibnamefont {Greiner}}, \bibinfo {author} {\bibfnamefont
  {V.}~\bibnamefont {Vuleti{\'c}}},\ and\ \bibinfo {author} {\bibfnamefont
  {M.~D.}\ \bibnamefont {Lukin}},\ }\href
  {https://doi.org/10.1103/PhysRevLett.121.123603} {\bibfield  {journal}
  {\bibinfo  {journal} {Physical Review Letters}\ }\textbf {\bibinfo {volume}
  {121}},\ \bibinfo {pages} {123603} (\bibinfo {year} {2018})}\BibitemShut
  {NoStop}%
\bibitem [{\citenamefont {Yuan}\ \emph {and others}(2023)\citenamefont {Yuan},
  \citenamefont {Yang}, \citenamefont {Jing}, \citenamefont {Zhang},
  \citenamefont {Jiao}, \citenamefont {Li}, \citenamefont {Zhang},
  \citenamefont {Xiao},\ and\ \citenamefont {Jia}}]{yuan:2023}%
  \BibitemOpen
  \bibfield  {author} {\bibinfo {author} {\bibfnamefont {J.}~\bibnamefont
  {Yuan}}, \bibinfo {author} {\bibfnamefont {W.}~\bibnamefont {Yang}}, \bibinfo
  {author} {\bibfnamefont {M.}~\bibnamefont {Jing}}, \bibinfo {author}
  {\bibfnamefont {H.}~\bibnamefont {Zhang}}, \bibinfo {author} {\bibfnamefont
  {Y.}~\bibnamefont {Jiao}}, \bibinfo {author} {\bibfnamefont {W.}~\bibnamefont
  {Li}}, \bibinfo {author} {\bibfnamefont {L.}~\bibnamefont {Zhang}}, \bibinfo
  {author} {\bibfnamefont {L.}~\bibnamefont {Xiao}},\ and\ \bibinfo {author}
  {\bibfnamefont {S.}~\bibnamefont {Jia}},\ }\href
  {https://doi.org/10.1088/1361-6633/acf22f} {\bibfield  {journal} {\bibinfo
  {journal} {Reports on Progress in Physics}\ }\textbf {\bibinfo {volume}
  {86}},\ \bibinfo {pages} {106001} (\bibinfo {year} {2023})}\BibitemShut
  {NoStop}%
\bibitem [{\citenamefont {Bornet}\ \emph {and others}(2023)\citenamefont
  {Bornet}, \citenamefont {Emperauger}, \citenamefont {Chen}, \citenamefont
  {Ye}, \citenamefont {Block}, \citenamefont {Bintz}, \citenamefont {Boyd},
  \citenamefont {Barredo}, \citenamefont {Comparin}, \citenamefont {Mezzacapo},
  \citenamefont {Roscilde}, \citenamefont {Lahaye}, \citenamefont {Yao},\ and\
  \citenamefont {Browaeys}}]{bornet:2023}%
  \BibitemOpen
  \bibfield  {author} {\bibinfo {author} {\bibfnamefont {G.}~\bibnamefont
  {Bornet}}, \bibinfo {author} {\bibfnamefont {G.}~\bibnamefont {Emperauger}},
  \bibinfo {author} {\bibfnamefont {C.}~\bibnamefont {Chen}}, \bibinfo {author}
  {\bibfnamefont {B.}~\bibnamefont {Ye}}, \bibinfo {author} {\bibfnamefont
  {M.}~\bibnamefont {Block}}, \bibinfo {author} {\bibfnamefont
  {M.}~\bibnamefont {Bintz}}, \bibinfo {author} {\bibfnamefont {J.~A.}\
  \bibnamefont {Boyd}}, \bibinfo {author} {\bibfnamefont {D.}~\bibnamefont
  {Barredo}}, \bibinfo {author} {\bibfnamefont {T.}~\bibnamefont {Comparin}},
  \bibinfo {author} {\bibfnamefont {F.}~\bibnamefont {Mezzacapo}}, and others,\
  }\href {https://doi.org/10.1038/s41586-023-06414-9} {\bibfield  {journal}
  {\bibinfo  {journal} {Nature}\ }\textbf {\bibinfo {volume} {621}},\ \bibinfo
  {pages} {728} (\bibinfo {year} {2023})}\BibitemShut {NoStop}%
\bibitem [{\citenamefont {Bluvstein}\ \emph {and others}(2023)\citenamefont
  {Bluvstein}, \citenamefont {Evered}, \citenamefont {Geim}, \citenamefont
  {Li}, \citenamefont {Zhou}, \citenamefont {Manovitz}, \citenamefont {Ebadi},
  \citenamefont {Cain}, \citenamefont {Kalinowski}, \citenamefont {Hangleiter},
  \citenamefont {Ataides}, \citenamefont {Maskara}, \citenamefont {Cong},
  \citenamefont {Gao}, \citenamefont {Rodriguez}, \citenamefont {Karolyshyn},
  \citenamefont {Semeghini}, \citenamefont {Gullans}, \citenamefont {Greiner},
  \citenamefont {Vuletic},\ and\ \citenamefont {Lukin}}]{bluvstein:2023}%
  \BibitemOpen
  \bibfield  {author} {\bibinfo {author} {\bibfnamefont {D.}~\bibnamefont
  {Bluvstein}}, \bibinfo {author} {\bibfnamefont {S.~J.}\ \bibnamefont
  {Evered}}, \bibinfo {author} {\bibfnamefont {A.~A.}\ \bibnamefont {Geim}},
  \bibinfo {author} {\bibfnamefont {S.~H.}\ \bibnamefont {Li}}, \bibinfo
  {author} {\bibfnamefont {H.}~\bibnamefont {Zhou}}, \bibinfo {author}
  {\bibfnamefont {T.}~\bibnamefont {Manovitz}}, \bibinfo {author}
  {\bibfnamefont {S.}~\bibnamefont {Ebadi}}, \bibinfo {author} {\bibfnamefont
  {M.}~\bibnamefont {Cain}}, \bibinfo {author} {\bibfnamefont {M.}~\bibnamefont
  {Kalinowski}}, \bibinfo {author} {\bibfnamefont {D.}~\bibnamefont
  {Hangleiter}}, and others,\ }\bibfield  {journal} {\bibinfo  {journal}
  {Nature}\ }\href {https://doi.org/10.1038/s41586-023-06927-3}
  {10.1038/s41586-023-06927-3} (\bibinfo {year} {2023})\BibitemShut {NoStop}%
\bibitem [{\citenamefont {B{\'e}guin}\ \emph {and others}(2013)\citenamefont
  {B{\'e}guin}, \citenamefont {Vernier}, \citenamefont {Chicireanu},
  \citenamefont {Lahaye},\ and\ \citenamefont {Browaeys}}]{beguin:2013}%
  \BibitemOpen
  \bibfield  {author} {\bibinfo {author} {\bibfnamefont {L.}~\bibnamefont
  {B{\'e}guin}}, \bibinfo {author} {\bibfnamefont {A.}~\bibnamefont {Vernier}},
  \bibinfo {author} {\bibfnamefont {R.}~\bibnamefont {Chicireanu}}, \bibinfo
  {author} {\bibfnamefont {T.}~\bibnamefont {Lahaye}},\ and\ \bibinfo {author}
  {\bibfnamefont {A.}~\bibnamefont {Browaeys}},\ }\href
  {https://doi.org/10.48550/arXiv.1302.4262} {\bibinfo {title} {Direct
  measurement of the van der {{Waals}} interaction between two {{Rydberg}}
  atoms}} (\bibinfo {year} {2013}),\ \Eprint {https://arxiv.org/abs/1302.4262}
  {arXiv:1302.4262} \BibitemShut {NoStop}%
\bibitem [{\citenamefont {Zeiher}\ \emph {and others}(2017)\citenamefont
  {Zeiher}, \citenamefont {Choi}, \citenamefont {{Rubio-Abadal}}, \citenamefont
  {Pohl}, \citenamefont {{van Bijnen}}, \citenamefont {Bloch},\ and\
  \citenamefont {Gross}}]{zeiher:2017}%
  \BibitemOpen
  \bibfield  {author} {\bibinfo {author} {\bibfnamefont {J.}~\bibnamefont
  {Zeiher}}, \bibinfo {author} {\bibfnamefont {J.-y.}\ \bibnamefont {Choi}},
  \bibinfo {author} {\bibfnamefont {A.}~\bibnamefont {{Rubio-Abadal}}},
  \bibinfo {author} {\bibfnamefont {T.}~\bibnamefont {Pohl}}, \bibinfo {author}
  {\bibfnamefont {R.}~\bibnamefont {{van Bijnen}}}, \bibinfo {author}
  {\bibfnamefont {I.}~\bibnamefont {Bloch}},\ and\ \bibinfo {author}
  {\bibfnamefont {C.}~\bibnamefont {Gross}},\ }\href
  {https://doi.org/10.1103/PhysRevX.7.041063} {\bibfield  {journal} {\bibinfo
  {journal} {Physical Review X}\ }\textbf {\bibinfo {volume} {7}},\ \bibinfo
  {pages} {041063} (\bibinfo {year} {2017})}\BibitemShut {NoStop}%
\bibitem [{\citenamefont {Anand}\ \emph {and others}(2024)\citenamefont
  {Anand}, \citenamefont {Bradley}, \citenamefont {White}, \citenamefont
  {Ramesh}, \citenamefont {Singh},\ and\ \citenamefont {Bernien}}]{anand:2024}%
  \BibitemOpen
  \bibfield  {author} {\bibinfo {author} {\bibfnamefont {S.}~\bibnamefont
  {Anand}}, \bibinfo {author} {\bibfnamefont {C.~E.}\ \bibnamefont {Bradley}},
  \bibinfo {author} {\bibfnamefont {R.}~\bibnamefont {White}}, \bibinfo
  {author} {\bibfnamefont {V.}~\bibnamefont {Ramesh}}, \bibinfo {author}
  {\bibfnamefont {K.}~\bibnamefont {Singh}},\ and\ \bibinfo {author}
  {\bibfnamefont {H.}~\bibnamefont {Bernien}},\ }\href
  {https://doi.org/10.1038/s41567-024-02638-2} {\bibfield  {journal} {\bibinfo
  {journal} {Nature Physics}\ ,\ \bibinfo {pages} {1}} (\bibinfo {year}
  {2024})}\BibitemShut {NoStop}%
\bibitem [{\citenamefont {Monroe}\ \emph {and others}(1995)\citenamefont
  {Monroe}, \citenamefont {Meekhof}, \citenamefont {King}, \citenamefont
  {Itano},\ and\ \citenamefont {Wineland}}]{monroe:1995}%
  \BibitemOpen
  \bibfield  {author} {\bibinfo {author} {\bibfnamefont {C.}~\bibnamefont
  {Monroe}}, \bibinfo {author} {\bibfnamefont {D.~M.}\ \bibnamefont {Meekhof}},
  \bibinfo {author} {\bibfnamefont {B.~E.}\ \bibnamefont {King}}, \bibinfo
  {author} {\bibfnamefont {W.~M.}\ \bibnamefont {Itano}},\ and\ \bibinfo
  {author} {\bibfnamefont {D.~J.}\ \bibnamefont {Wineland}},\ }\href
  {https://doi.org/10.1103/PhysRevLett.75.4714} {\bibfield  {journal} {\bibinfo
   {journal} {Physical Review Letters}\ }\textbf {\bibinfo {volume} {75}},\
  \bibinfo {pages} {4714} (\bibinfo {year} {1995})}\BibitemShut {NoStop}%
\bibitem [{\citenamefont {Bruzewicz}\ \emph {and others}(2019)\citenamefont
  {Bruzewicz}, \citenamefont {Chiaverini}, \citenamefont {McConnell},\ and\
  \citenamefont {Sage}}]{bruzewicz:2019}%
  \BibitemOpen
  \bibfield  {author} {\bibinfo {author} {\bibfnamefont {C.~D.}\ \bibnamefont
  {Bruzewicz}}, \bibinfo {author} {\bibfnamefont {J.}~\bibnamefont
  {Chiaverini}}, \bibinfo {author} {\bibfnamefont {R.}~\bibnamefont
  {McConnell}},\ and\ \bibinfo {author} {\bibfnamefont {J.~M.}\ \bibnamefont
  {Sage}},\ }\href {https://doi.org/10.1063/1.5088164} {\bibfield  {journal}
  {\bibinfo  {journal} {Applied Physics Reviews}\ }\textbf {\bibinfo {volume}
  {6}},\ \bibinfo {pages} {021314} (\bibinfo {year} {2019})}\BibitemShut
  {NoStop}%
\bibitem [{\citenamefont {Moses}\ \emph {and others}(2023)\citenamefont
  {Moses}, \citenamefont {Baldwin}, \citenamefont {Allman}, \citenamefont
  {Ancona}, \citenamefont {Ascarrunz}, \citenamefont {Barnes}, \citenamefont
  {Bartolotta}, \citenamefont {Bjork}, \citenamefont {Blanchard}, \citenamefont
  {Bohn}, \citenamefont {Bohnet}, \citenamefont {Brown}, \citenamefont
  {Burdick}, \citenamefont {Burton}, \citenamefont {Campbell}, \citenamefont
  {Campora~III}, \citenamefont {Carron}, \citenamefont {Chambers},
  \citenamefont {Chan}, \citenamefont {Chen}, \citenamefont {Chernoguzov},
  \citenamefont {Chertkov}, \citenamefont {Colina}, \citenamefont {Curtis},
  \citenamefont {Daniel}, \citenamefont {DeCross}, \citenamefont {Deen},
  \citenamefont {Delaney}, \citenamefont {Dreiling}, \citenamefont {Ertsgaard},
  \citenamefont {Esposito}, \citenamefont {Estey}, \citenamefont {Fabrikant},
  \citenamefont {Figgatt}, \citenamefont {Foltz}, \citenamefont {{Foss-Feig}},
  \citenamefont {Francois}, \citenamefont {Gaebler}, \citenamefont {Gatterman},
  \citenamefont {Gilbreth}, \citenamefont {Giles}, \citenamefont {Glynn},
  \citenamefont {Hall}, \citenamefont {Hankin}, \citenamefont {Hansen},
  \citenamefont {Hayes}, \citenamefont {Higashi}, \citenamefont {Hoffman},
  \citenamefont {Horning}, \citenamefont {Hout}, \citenamefont {Jacobs},
  \citenamefont {Johansen}, \citenamefont {Jones}, \citenamefont {Karcz},
  \citenamefont {Klein}, \citenamefont {Lauria}, \citenamefont {Lee},
  \citenamefont {Liefer}, \citenamefont {Lytle}, \citenamefont {Lu},
  \citenamefont {Lucchetti}, \citenamefont {Malm}, \citenamefont {Matheny},
  \citenamefont {Mathewson}, \citenamefont {Mayer}, \citenamefont {Miller},
  \citenamefont {Mills}, \citenamefont {Neyenhuis}, \citenamefont {Nugent},
  \citenamefont {Olson}, \citenamefont {Parks}, \citenamefont {Price},
  \citenamefont {Price}, \citenamefont {Pugh}, \citenamefont {Ransford},
  \citenamefont {Reed}, \citenamefont {Roman}, \citenamefont {Rowe},
  \citenamefont {{Ryan-Anderson}}, \citenamefont {Sanders}, \citenamefont
  {Sedlacek}, \citenamefont {Shevchuk}, \citenamefont {Siegfried},
  \citenamefont {Skripka}, \citenamefont {Spaun}, \citenamefont {Sprenkle},
  \citenamefont {Stutz}, \citenamefont {Swallows}, \citenamefont {Tobey},
  \citenamefont {Tran}, \citenamefont {Tran}, \citenamefont {Vogt},
  \citenamefont {Volin}, \citenamefont {Walker}, \citenamefont {Zolot},\ and\
  \citenamefont {Pino}}]{moses:2023}%
  \BibitemOpen
  \bibfield  {author} {\bibinfo {author} {\bibfnamefont {S.~A.}\ \bibnamefont
  {Moses}}, \bibinfo {author} {\bibfnamefont {C.~H.}\ \bibnamefont {Baldwin}},
  \bibinfo {author} {\bibfnamefont {M.~S.}\ \bibnamefont {Allman}}, \bibinfo
  {author} {\bibfnamefont {R.}~\bibnamefont {Ancona}}, \bibinfo {author}
  {\bibfnamefont {L.}~\bibnamefont {Ascarrunz}}, \bibinfo {author}
  {\bibfnamefont {C.}~\bibnamefont {Barnes}}, \bibinfo {author} {\bibfnamefont
  {J.}~\bibnamefont {Bartolotta}}, \bibinfo {author} {\bibfnamefont
  {B.}~\bibnamefont {Bjork}}, \bibinfo {author} {\bibfnamefont
  {P.}~\bibnamefont {Blanchard}}, \bibinfo {author} {\bibfnamefont
  {M.}~\bibnamefont {Bohn}}, and others,\ }\href
  {https://doi.org/10.1103/PhysRevX.13.041052} {\bibfield  {journal} {\bibinfo
  {journal} {Physical Review X}\ }\textbf {\bibinfo {volume} {13}},\ \bibinfo
  {pages} {041052} (\bibinfo {year} {2023})}\BibitemShut {NoStop}%
\bibitem [{\citenamefont {Sorensen}\ and\ \citenamefont
  {Molmer}(1999)}]{sorensen:1999}%
  \BibitemOpen
  \bibfield  {author} {\bibinfo {author} {\bibfnamefont {A.}~\bibnamefont
  {Sorensen}}\ and\ \bibinfo {author} {\bibfnamefont {K.}~\bibnamefont
  {Molmer}},\ }\href {https://doi.org/10.48550/arXiv.quant-ph/9810039}
  {\bibinfo {title} {Quantum computation with ions in thermal motion}}
  (\bibinfo {year} {1999}),\ \Eprint {https://arxiv.org/abs/quant-ph/9810039}
  {arXiv:quant-ph/9810039} \BibitemShut {NoStop}%
\bibitem [{\citenamefont {Monroe}\ \emph {and others}(2020)\citenamefont
  {Monroe}, \citenamefont {Campbell}, \citenamefont {Duan}, \citenamefont
  {Gong}, \citenamefont {Gorshkov}, \citenamefont {Hess}, \citenamefont
  {Islam}, \citenamefont {Kim}, \citenamefont {Linke}, \citenamefont {Pagano},
  \citenamefont {Richerme}, \citenamefont {Senko},\ and\ \citenamefont
  {Yao}}]{monroe:2020}%
  \BibitemOpen
  \bibfield  {author} {\bibinfo {author} {\bibfnamefont {C.}~\bibnamefont
  {Monroe}}, \bibinfo {author} {\bibfnamefont {W.~C.}\ \bibnamefont
  {Campbell}}, \bibinfo {author} {\bibfnamefont {L.-M.}\ \bibnamefont {Duan}},
  \bibinfo {author} {\bibfnamefont {Z.-X.}\ \bibnamefont {Gong}}, \bibinfo
  {author} {\bibfnamefont {A.~V.}\ \bibnamefont {Gorshkov}}, \bibinfo {author}
  {\bibfnamefont {P.}~\bibnamefont {Hess}}, \bibinfo {author} {\bibfnamefont
  {R.}~\bibnamefont {Islam}}, \bibinfo {author} {\bibfnamefont
  {K.}~\bibnamefont {Kim}}, \bibinfo {author} {\bibfnamefont {N.}~\bibnamefont
  {Linke}}, \bibinfo {author} {\bibfnamefont {G.}~\bibnamefont {Pagano}}, and
  others,\ }\href {https://doi.org/10.48550/arXiv.1912.07845} {\bibinfo {title}
  {Programmable {{Quantum Simulations}} of {{Spin Systems}} with {{Trapped
  Ions}}}} (\bibinfo {year} {2020}),\ \Eprint
  {https://arxiv.org/abs/1912.07845} {arXiv:1912.07845} \BibitemShut {NoStop}%
\bibitem [{\citenamefont {{Pedrozo-Pe{\~n}afiel}}\ \emph {and
  others}(2020)\citenamefont {{Pedrozo-Pe{\~n}afiel}}, \citenamefont {Colombo},
  \citenamefont {Shu}, \citenamefont {Adiyatullin}, \citenamefont {Li},
  \citenamefont {Mendez}, \citenamefont {Braverman}, \citenamefont {Kawasaki},
  \citenamefont {Akamatsu}, \citenamefont {Xiao},\ and\ \citenamefont
  {Vuleti{\'c}}}]{pedrozo-penafiel:2020}%
  \BibitemOpen
  \bibfield  {author} {\bibinfo {author} {\bibfnamefont {E.}~\bibnamefont
  {{Pedrozo-Pe{\~n}afiel}}}, \bibinfo {author} {\bibfnamefont {S.}~\bibnamefont
  {Colombo}}, \bibinfo {author} {\bibfnamefont {C.}~\bibnamefont {Shu}},
  \bibinfo {author} {\bibfnamefont {A.~F.}\ \bibnamefont {Adiyatullin}},
  \bibinfo {author} {\bibfnamefont {Z.}~\bibnamefont {Li}}, \bibinfo {author}
  {\bibfnamefont {E.}~\bibnamefont {Mendez}}, \bibinfo {author} {\bibfnamefont
  {B.}~\bibnamefont {Braverman}}, \bibinfo {author} {\bibfnamefont
  {A.}~\bibnamefont {Kawasaki}}, \bibinfo {author} {\bibfnamefont
  {D.}~\bibnamefont {Akamatsu}}, \bibinfo {author} {\bibfnamefont
  {Y.}~\bibnamefont {Xiao}}, and others,\ }\href
  {https://doi.org/10.1038/s41586-020-3006-1} {\bibfield  {journal} {\bibinfo
  {journal} {Nature}\ }\textbf {\bibinfo {volume} {588}},\ \bibinfo {pages}
  {414} (\bibinfo {year} {2020})}\BibitemShut {NoStop}%
\bibitem [{\citenamefont {Wallraff}\ \emph {and others}(2004)\citenamefont
  {Wallraff}, \citenamefont {Schuster}, \citenamefont {Blais}, \citenamefont
  {Frunzio}, \citenamefont {Huang}, \citenamefont {Majer}, \citenamefont
  {Kumar}, \citenamefont {Girvin},\ and\ \citenamefont
  {Schoelkopf}}]{wallraff:2004}%
  \BibitemOpen
  \bibfield  {author} {\bibinfo {author} {\bibfnamefont {A.}~\bibnamefont
  {Wallraff}}, \bibinfo {author} {\bibfnamefont {D.~I.}\ \bibnamefont
  {Schuster}}, \bibinfo {author} {\bibfnamefont {A.}~\bibnamefont {Blais}},
  \bibinfo {author} {\bibfnamefont {L.}~\bibnamefont {Frunzio}}, \bibinfo
  {author} {\bibfnamefont {R.-S.}\ \bibnamefont {Huang}}, \bibinfo {author}
  {\bibfnamefont {J.}~\bibnamefont {Majer}}, \bibinfo {author} {\bibfnamefont
  {S.}~\bibnamefont {Kumar}}, \bibinfo {author} {\bibfnamefont {S.~M.}\
  \bibnamefont {Girvin}},\ and\ \bibinfo {author} {\bibfnamefont {R.~J.}\
  \bibnamefont {Schoelkopf}},\ }\href {https://doi.org/10.1038/nature02851}
  {\bibfield  {journal} {\bibinfo  {journal} {Nature}\ }\textbf {\bibinfo
  {volume} {431}},\ \bibinfo {pages} {162} (\bibinfo {year}
  {2004})}\BibitemShut {NoStop}%
\bibitem [{\citenamefont {Reiserer}\ and\ \citenamefont
  {Rempe}(2015)}]{reiserer:2015}%
  \BibitemOpen
  \bibfield  {author} {\bibinfo {author} {\bibfnamefont {A.}~\bibnamefont
  {Reiserer}}\ and\ \bibinfo {author} {\bibfnamefont {G.}~\bibnamefont
  {Rempe}},\ }\href {https://doi.org/10.1103/RevModPhys.87.1379} {\bibfield
  {journal} {\bibinfo  {journal} {Reviews of Modern Physics}\ }\textbf
  {\bibinfo {volume} {87}},\ \bibinfo {pages} {1379} (\bibinfo {year}
  {2015})}\BibitemShut {NoStop}%
\bibitem [{\citenamefont {Viehmann}\ \emph {and others}(2013)\citenamefont
  {Viehmann}, \citenamefont {{von Delft}},\ and\ \citenamefont
  {Marquardt}}]{viehmann:2013}%
  \BibitemOpen
  \bibfield  {author} {\bibinfo {author} {\bibfnamefont {O.}~\bibnamefont
  {Viehmann}}, \bibinfo {author} {\bibfnamefont {J.}~\bibnamefont {{von
  Delft}}},\ and\ \bibinfo {author} {\bibfnamefont {F.}~\bibnamefont
  {Marquardt}},\ }\href {https://doi.org/10.1103/PhysRevLett.110.030601}
  {\bibfield  {journal} {\bibinfo  {journal} {Physical Review Letters}\
  }\textbf {\bibinfo {volume} {110}},\ \bibinfo {pages} {030601} (\bibinfo
  {year} {2013})}\BibitemShut {NoStop}%
\bibitem [{\citenamefont {Periwal}\ \emph {and others}(2021)\citenamefont
  {Periwal}, \citenamefont {Cooper}, \citenamefont {Kunkel}, \citenamefont
  {Wienand}, \citenamefont {Davis},\ and\ \citenamefont
  {{Schleier-Smith}}}]{periwal:2021}%
  \BibitemOpen
  \bibfield  {author} {\bibinfo {author} {\bibfnamefont {A.}~\bibnamefont
  {Periwal}}, \bibinfo {author} {\bibfnamefont {E.~S.}\ \bibnamefont {Cooper}},
  \bibinfo {author} {\bibfnamefont {P.}~\bibnamefont {Kunkel}}, \bibinfo
  {author} {\bibfnamefont {J.~F.}\ \bibnamefont {Wienand}}, \bibinfo {author}
  {\bibfnamefont {E.~J.}\ \bibnamefont {Davis}},\ and\ \bibinfo {author}
  {\bibfnamefont {M.}~\bibnamefont {{Schleier-Smith}}},\ }\href
  {https://doi.org/10.48550/arXiv.2106.04070} {\bibinfo {title} {Programmable
  {{Interactions}} and {{Emergent Geometry}} in an {{Atomic Array}}}} (\bibinfo
  {year} {2021}),\ \Eprint {https://arxiv.org/abs/2106.04070}
  {arXiv:2106.04070} \BibitemShut {NoStop}%
\bibitem [{\citenamefont {Ho}\ \emph {and others}(2024)\citenamefont {Ho},
  \citenamefont {Lu}, \citenamefont {Xiang}, \citenamefont {Rusconi},
  \citenamefont {Masson}, \citenamefont {{Asenjo-Garcia}}, \citenamefont
  {Yan},\ and\ \citenamefont {{Stamper-Kurn}}}]{ho:2024}%
  \BibitemOpen
  \bibfield  {author} {\bibinfo {author} {\bibfnamefont {J.}~\bibnamefont
  {Ho}}, \bibinfo {author} {\bibfnamefont {Y.-H.}\ \bibnamefont {Lu}}, \bibinfo
  {author} {\bibfnamefont {T.}~\bibnamefont {Xiang}}, \bibinfo {author}
  {\bibfnamefont {C.~C.}\ \bibnamefont {Rusconi}}, \bibinfo {author}
  {\bibfnamefont {S.~J.}\ \bibnamefont {Masson}}, \bibinfo {author}
  {\bibfnamefont {A.}~\bibnamefont {{Asenjo-Garcia}}}, \bibinfo {author}
  {\bibfnamefont {Z.}~\bibnamefont {Yan}},\ and\ \bibinfo {author}
  {\bibfnamefont {D.~M.}\ \bibnamefont {{Stamper-Kurn}}},\ }\href
  {https://doi.org/10.48550/arXiv.2410.12754} {\bibinfo {title} {Optomechanical
  self-organization in a mesoscopic atom array}} (\bibinfo {year} {2024}),\
  \Eprint {https://arxiv.org/abs/2410.12754} {arXiv:2410.12754} \BibitemShut
  {NoStop}%
\bibitem [{\citenamefont {Abobeih}\ \emph {and others}(2022)\citenamefont
  {Abobeih}, \citenamefont {Wang}, \citenamefont {Randall}, \citenamefont
  {Loenen}, \citenamefont {Bradley}, \citenamefont {Markham}, \citenamefont
  {Twitchen}, \citenamefont {Terhal},\ and\ \citenamefont
  {Taminiau}}]{abobeih:2022}%
  \BibitemOpen
  \bibfield  {author} {\bibinfo {author} {\bibfnamefont {M.~H.}\ \bibnamefont
  {Abobeih}}, \bibinfo {author} {\bibfnamefont {Y.}~\bibnamefont {Wang}},
  \bibinfo {author} {\bibfnamefont {J.}~\bibnamefont {Randall}}, \bibinfo
  {author} {\bibfnamefont {S.~J.~H.}\ \bibnamefont {Loenen}}, \bibinfo {author}
  {\bibfnamefont {C.~E.}\ \bibnamefont {Bradley}}, \bibinfo {author}
  {\bibfnamefont {M.}~\bibnamefont {Markham}}, \bibinfo {author} {\bibfnamefont
  {D.~J.}\ \bibnamefont {Twitchen}}, \bibinfo {author} {\bibfnamefont {B.~M.}\
  \bibnamefont {Terhal}},\ and\ \bibinfo {author} {\bibfnamefont {T.~H.}\
  \bibnamefont {Taminiau}},\ }\href
  {https://doi.org/10.1038/s41586-022-04819-6} {\bibfield  {journal} {\bibinfo
  {journal} {Nature}\ }\textbf {\bibinfo {volume} {606}},\ \bibinfo {pages}
  {884} (\bibinfo {year} {2022})}\BibitemShut {NoStop}%
\bibitem [{\citenamefont {Abobeih}\ \emph {and others}(2019)\citenamefont
  {Abobeih}, \citenamefont {Randall}, \citenamefont {Bradley}, \citenamefont
  {Bartling}, \citenamefont {Bakker}, \citenamefont {Degen}, \citenamefont
  {Markham}, \citenamefont {Twitchen},\ and\ \citenamefont
  {Taminiau}}]{abobeih:2019}%
  \BibitemOpen
  \bibfield  {author} {\bibinfo {author} {\bibfnamefont {M.~H.}\ \bibnamefont
  {Abobeih}}, \bibinfo {author} {\bibfnamefont {J.}~\bibnamefont {Randall}},
  \bibinfo {author} {\bibfnamefont {C.~E.}\ \bibnamefont {Bradley}}, \bibinfo
  {author} {\bibfnamefont {H.~P.}\ \bibnamefont {Bartling}}, \bibinfo {author}
  {\bibfnamefont {M.~A.}\ \bibnamefont {Bakker}}, \bibinfo {author}
  {\bibfnamefont {M.~J.}\ \bibnamefont {Degen}}, \bibinfo {author}
  {\bibfnamefont {M.}~\bibnamefont {Markham}}, \bibinfo {author} {\bibfnamefont
  {D.~J.}\ \bibnamefont {Twitchen}},\ and\ \bibinfo {author} {\bibfnamefont
  {T.~H.}\ \bibnamefont {Taminiau}},\ }\href
  {https://doi.org/10.1038/s41586-019-1834-7} {\bibfield  {journal} {\bibinfo
  {journal} {Nature}\ }\textbf {\bibinfo {volume} {576}},\ \bibinfo {pages}
  {411} (\bibinfo {year} {2019})}\BibitemShut {NoStop}%
\bibitem [{\citenamefont {Abobeih}\ \emph {and others}(2018)\citenamefont
  {Abobeih}, \citenamefont {Cramer}, \citenamefont {Bakker}, \citenamefont
  {Kalb}, \citenamefont {Markham}, \citenamefont {Twitchen},\ and\
  \citenamefont {Taminiau}}]{abobeih:2018}%
  \BibitemOpen
  \bibfield  {author} {\bibinfo {author} {\bibfnamefont {M.~H.}\ \bibnamefont
  {Abobeih}}, \bibinfo {author} {\bibfnamefont {J.}~\bibnamefont {Cramer}},
  \bibinfo {author} {\bibfnamefont {M.~A.}\ \bibnamefont {Bakker}}, \bibinfo
  {author} {\bibfnamefont {N.}~\bibnamefont {Kalb}}, \bibinfo {author}
  {\bibfnamefont {M.}~\bibnamefont {Markham}}, \bibinfo {author} {\bibfnamefont
  {D.~J.}\ \bibnamefont {Twitchen}},\ and\ \bibinfo {author} {\bibfnamefont
  {T.~H.}\ \bibnamefont {Taminiau}},\ }\href
  {https://doi.org/10.1038/s41467-018-04916-z} {\bibfield  {journal} {\bibinfo
  {journal} {Nature Communications}\ }\textbf {\bibinfo {volume} {9}},\
  \bibinfo {pages} {2552} (\bibinfo {year} {2018})}\BibitemShut {NoStop}%
\bibitem [{\citenamefont {Mansfield}(1971)}]{mansfield:1971}%
  \BibitemOpen
  \bibfield  {author} {\bibinfo {author} {\bibfnamefont {P.}~\bibnamefont
  {Mansfield}},\ }\href {https://doi.org/10.1088/0022-3719/4/11/020} {\bibfield
   {journal} {\bibinfo  {journal} {Journal of Physics C: Solid State Physics}\
  }\textbf {\bibinfo {volume} {4}},\ \bibinfo {pages} {1444} (\bibinfo {year}
  {1971})}\BibitemShut {NoStop}%
\bibitem [{\citenamefont {Bennett}\ \emph {and others}(1997)\citenamefont
  {Bennett}, \citenamefont {Bernstein}, \citenamefont {Brassard},\ and\
  \citenamefont {Vazirani}}]{bennett:1997}%
  \BibitemOpen
  \bibfield  {author} {\bibinfo {author} {\bibfnamefont {C.~H.}\ \bibnamefont
  {Bennett}}, \bibinfo {author} {\bibfnamefont {E.}~\bibnamefont {Bernstein}},
  \bibinfo {author} {\bibfnamefont {G.}~\bibnamefont {Brassard}},\ and\
  \bibinfo {author} {\bibfnamefont {U.}~\bibnamefont {Vazirani}},\ }\href
  {https://doi.org/10.1137/S0097539796300933} {\bibfield  {journal} {\bibinfo
  {journal} {SIAM Journal on Computing}\ }\textbf {\bibinfo {volume} {26}},\
  \bibinfo {pages} {1510} (\bibinfo {year} {1997})}\BibitemShut {NoStop}%
\bibitem [{\citenamefont {Zalka}(1999)}]{zalka:1999}%
  \BibitemOpen
  \bibfield  {author} {\bibinfo {author} {\bibfnamefont {C.}~\bibnamefont
  {Zalka}},\ }\href {https://doi.org/10.1103/PhysRevA.60.2746} {\bibfield
  {journal} {\bibinfo  {journal} {Physical Review A}\ }\textbf {\bibinfo
  {volume} {60}},\ \bibinfo {pages} {2746} (\bibinfo {year}
  {1999})}\BibitemShut {NoStop}%
\bibitem [{\citenamefont {Polloreno}\ \emph {and others}(2023)\citenamefont
  {Polloreno}, \citenamefont {Beckey}, \citenamefont {Levin}, \citenamefont
  {Shlosberg}, \citenamefont {Thompson}, \citenamefont {{Foss-Feig}},
  \citenamefont {Hayes},\ and\ \citenamefont {Smith}}]{polloreno:2023}%
  \BibitemOpen
  \bibfield  {author} {\bibinfo {author} {\bibfnamefont {A.~M.}\ \bibnamefont
  {Polloreno}}, \bibinfo {author} {\bibfnamefont {J.~L.}\ \bibnamefont
  {Beckey}}, \bibinfo {author} {\bibfnamefont {J.}~\bibnamefont {Levin}},
  \bibinfo {author} {\bibfnamefont {A.}~\bibnamefont {Shlosberg}}, \bibinfo
  {author} {\bibfnamefont {J.~K.}\ \bibnamefont {Thompson}}, \bibinfo {author}
  {\bibfnamefont {M.}~\bibnamefont {{Foss-Feig}}}, \bibinfo {author}
  {\bibfnamefont {D.}~\bibnamefont {Hayes}},\ and\ \bibinfo {author}
  {\bibfnamefont {G.}~\bibnamefont {Smith}},\ }\href
  {https://doi.org/10.1103/PhysRevApplied.19.014029} {\bibfield  {journal}
  {\bibinfo  {journal} {Physical Review Applied}\ }\textbf {\bibinfo {volume}
  {19}},\ \bibinfo {pages} {014029} (\bibinfo {year} {2023})}\BibitemShut
  {NoStop}%
\bibitem [{\citenamefont {Pang}\ and\ \citenamefont
  {Jordan}(2017)}]{pang:2017}%
  \BibitemOpen
  \bibfield  {author} {\bibinfo {author} {\bibfnamefont {S.}~\bibnamefont
  {Pang}}\ and\ \bibinfo {author} {\bibfnamefont {A.~N.}\ \bibnamefont
  {Jordan}},\ }\href {https://doi.org/10.1038/ncomms14695} {\bibfield
  {journal} {\bibinfo  {journal} {Nature Communications}\ }\textbf {\bibinfo
  {volume} {8}},\ \bibinfo {pages} {14695} (\bibinfo {year}
  {2017})}\BibitemShut {NoStop}%
\bibitem [{\citenamefont {Chen}\ \emph {and others}(2021)\citenamefont {Chen},
  \citenamefont {Cotler}, \citenamefont {Huang},\ and\ \citenamefont
  {Li}}]{chen:2021}%
  \BibitemOpen
  \bibfield  {author} {\bibinfo {author} {\bibfnamefont {S.}~\bibnamefont
  {Chen}}, \bibinfo {author} {\bibfnamefont {J.}~\bibnamefont {Cotler}},
  \bibinfo {author} {\bibfnamefont {H.-Y.}\ \bibnamefont {Huang}},\ and\
  \bibinfo {author} {\bibfnamefont {J.}~\bibnamefont {Li}},\ }\href
  {https://doi.org/10.48550/arXiv.2111.05881} {\bibinfo {title} {Exponential
  separations between learning with and without quantum memory}} (\bibinfo
  {year} {2021}),\ \Eprint {https://arxiv.org/abs/2111.05881} {arXiv:2111.05881
  [quant-ph]} \BibitemShut {NoStop}%
\bibitem [{\citenamefont {Chen}\ \emph {and others}(2023)\citenamefont {Chen},
  \citenamefont {Cotler}, \citenamefont {Huang},\ and\ \citenamefont
  {Li}}]{chen:2023}%
  \BibitemOpen
  \bibfield  {author} {\bibinfo {author} {\bibfnamefont {S.}~\bibnamefont
  {Chen}}, \bibinfo {author} {\bibfnamefont {J.}~\bibnamefont {Cotler}},
  \bibinfo {author} {\bibfnamefont {H.-Y.}\ \bibnamefont {Huang}},\ and\
  \bibinfo {author} {\bibfnamefont {J.}~\bibnamefont {Li}},\ }\href
  {https://doi.org/10.1038/s41467-023-41217-6} {\bibfield  {journal} {\bibinfo
  {journal} {Nature Communications}\ }\textbf {\bibinfo {volume} {14}},\
  \bibinfo {pages} {6001} (\bibinfo {year} {2023})}\BibitemShut {NoStop}%
\bibitem [{\citenamefont {Oppenheim}\ \emph {and others}(1996)\citenamefont
  {Oppenheim}, \citenamefont {Willsky},\ and\ \citenamefont
  {Nawab}}]{oppenheim:1996}%
  \BibitemOpen
  \bibfield  {author} {\bibinfo {author} {\bibfnamefont {A.~V.}\ \bibnamefont
  {Oppenheim}}, \bibinfo {author} {\bibfnamefont {A.~S.}\ \bibnamefont
  {Willsky}},\ and\ \bibinfo {author} {\bibfnamefont {S.~H.}\ \bibnamefont
  {Nawab}},\ }\href@noop {} {\emph {\bibinfo {title} {Signals and Systems}}},\
  \bibinfo {edition} {2nd}\ ed.,\ Prentice-{{Hall}} Signal Processing Series\
  (\bibinfo  {publisher} {Prentice Hall},\ \bibinfo {address} {Upper Saddle
  River, N.J.},\ \bibinfo {year} {1996})\BibitemShut {NoStop}%
\bibitem [{\citenamefont {Chattopadhyay}\ \emph {and others}(2021)\citenamefont
  {Chattopadhyay}, \citenamefont {Filmus}, \citenamefont {Koroth},
  \citenamefont {Meir},\ and\ \citenamefont {Pitassi}}]{chattopadhyay:2021}%
  \BibitemOpen
  \bibfield  {author} {\bibinfo {author} {\bibfnamefont {A.}~\bibnamefont
  {Chattopadhyay}}, \bibinfo {author} {\bibfnamefont {Y.}~\bibnamefont
  {Filmus}}, \bibinfo {author} {\bibfnamefont {S.}~\bibnamefont {Koroth}},
  \bibinfo {author} {\bibfnamefont {O.}~\bibnamefont {Meir}},\ and\ \bibinfo
  {author} {\bibfnamefont {T.}~\bibnamefont {Pitassi}},\ }\href
  {https://doi.org/10.1137/19M1310153} {\bibfield  {journal} {\bibinfo
  {journal} {SIAM Journal on Computing}\ }\textbf {\bibinfo {volume} {50}},\
  \bibinfo {pages} {171} (\bibinfo {year} {2021})}\BibitemShut {NoStop}%
\end{thebibliography}%

\end{document}


\title{Supplemental Material: Quantum Computing Enhanced Sensing}
\author{Richard R. Allen}
\thanks{These authors contributed equally to this work.}
\affiliation{Department of Physics, Massachusetts Institute of Technology, Cambridge, MA 02139, USA}
\author{Francisco Machado}
\thanks{These authors contributed equally to this work.}
\affiliation{ITAMP, Harvard-Smithsonian Center for Astrophysics, Cambridge, MA 02138, USA}
\affiliation{Department of Physics, Harvard University, Cambridge, MA 02138, USA}
\author{Isaac L. Chuang}
\affiliation{Department of Physics, Massachusetts Institute of Technology, Cambridge, MA 02139, USA}
\author{Hsin-Yuan Huang}
\affiliation{Google Quantum AI, Venice, CA 90291, USA}
\affiliation{Department of Physics, Massachusetts Institute of Technology, Cambridge, MA 02139, USA}
\affiliation{Division of Physics, Mathematics and Astronomy, California Institute of Technology, Pasadena, CA 91125, USA}
\author{Soonwon Choi}
\thanks{Corresponding author: soonwon@mit.edu}
\affiliation{Department of Physics, Massachusetts Institute of Technology, Cambridge, MA 02139, USA}
\maketitle

\begingroup
  \hypersetup{hidelinks}
  \vspace{-12mm}
  \tableofcontents
\endgroup

\section{Preliminaries}\label{sec: prelims}

\subsection{Definition of AC Sensing}\label{sec: ac_sensing_def}

We begin by formally defining AC sensing. 
The difficulty of detecting an unknown AC signal is quantified by two parameters: the minimum signal strength $B_{\min}$ and the range of possible frequencies $\Delta \omega$. For a given $B_{\min}$ and $\Delta \omega$, AC sensing can be naturally formulated as a decision problem:

\begin{definition}[AC Sensing]\label{def: ac_disc} Let $B_{\min} > 0$, and let $\Delta \omega = [\omega_{\min}, \omega_{\max}] \subset \mathbb{R}_{> 0}$ with $\omega_{\min} \ge g B_{\min}$, where $g > 0$ is the coupling constant of the signal to the sensing qubits. Let $H(t)$ be a Hamiltonian acting on $n_S$ sensing qubits. We define the \textbf{AC sensing problem} $\ACProb$ as the task of deciding between the two cases:
\begin{enumerate}
    \item $H(t) = \hbar g B \cos(\omega t + \phi) \sum_{i=1}^{n_s} Z_i$ with the promise $B \ge B_{\min}$ and $\omega \in \Delta \omega$ (``YES'')
    
    \item $H(t) = 0$ (``NO'')
\end{enumerate}
with success probability at least $2/3$, where $Z_i$ is the Pauli-$Z$ operator on the $i$-th sensing qubit. 
\end{definition}
\noindent As in the main text, from now on we work in units such that $g = 1$, so $B$ and $\omega$ both have units of inverse time. We also set $\hbar = 1$ to simplify our notation.

We impose the additional condition $\omega_{\min} \ge g B_{\min}$ because otherwise the AC signal is quasi-static at low frequencies. In other words, we can treat the signal as time-independent in the regime where $\omega \ll g B_{\min}$. Indeed, in Sec.~\ref{sec: quasi_static}, we describe how to detect quasi-static signals with arbitrarily low frequencies using conventional DC sensing techniques.

\subsection{Quantum Protocols for AC Sensing}\label{sec: sensing_protocol}
We now describe a general framework which encompasses all possible quantum sensing protocols for detecting AC signals. We consider protocols which use a fixed number of sensors $n_S$ but have access to an unbounded number of ancilla qubits $n_Q$, which represent a quantum computer that controls the quantum sensors. A general protocol is specified by a sequence of times $0 = t_0 < t_1 < \cdots < t_P = \tau$ and unitaries $V_0, V_1, \hdots, V_P$ acting on all $n_S + n_Q$ qubits, where $P$ is arbitrarily large. Starting from the state $\ket{0}^{\otimes n_S + n_Q}$ at time $t_0 = 0$, the system is evolved under $H(t)$ for a total sensing time $\tau$ while (instantaneously) applying each unitary $V_j$ at time $t_j$. The final state of the qubits is
\begin{align}
    V_P \left( \prod_{j=0}^{P-1} \text{exp}\left(-i \int_{t_j}^{t_{j+1}} dt~H(t) \otimes \mathbbm{1}_Q \right) V_j \right) \ket{0}^{\otimes n_S + n_Q},
\end{align}
where $\mathbbm{1}_Q$ denotes the identity matrix on all $n_Q$ ancilla qubits, and the product is time-ordered (i.e., the indices increase from right to left). Finally, we perform a two-outcome measurement at time $t_P = \tau$ to solve the decision problem of whether or not the signal exists. 

We note that this framework includes adaptive quantum sensing protocols which perform intermediate measurements and apply different gates depending on the measurement outcomes.
Such a protocol can always be expressed in the form above by increasing $n_Q$ and applying the principle of deferred measurement. We also note that this gate-based framework is equivalent to a Hamiltonian-based framework, in which a protocol is given by time evolution under the joint Hamiltonian,
\begin{align}
    H_{\text{total}}(t) = H(t) + H_c(t)
\end{align}
over a total sensing time $\tau$, where $H_c(t)$ is an arbitrary control Hamiltonian acting on all $n_S + n_Q$ qubits. In this setting, the choice of $H_c(t)$ specifies a quantum sensing protocol.

Finally, when designing quantum sensing protocols for $\ACProb$ in the absence of decoherence, one may focus on the single sensor case $n_S = 1$. To generalize to arbitrary $n_S$, note that $\left( \sum_{i=1}^{n_S} Z_i  \right) \ket{b}^{\otimes n_S} = (-1)^b n_S \ket{b}^{\otimes n_S}$ for $b \in \{0,1\}$. Therefore, the states $\ket{0}^{\otimes n_S}$ and $\ket{1}^{\otimes n_S}$ form an effective sensing qubit whose coupling to the signal is collectively enhanced by a factor of $n_S$. By replacing single qubit states and operators with their counterparts for this effective sensing qubit, we can convert a single sensor protocol to an $n_S$ sensor protocol. In our lower bound proofs, we will return to explicitly considering arbitrary $n_S$ to place restrictions on the most general quantum sensing protocols.

\subsection{Conventional AC Sensing}\label{sec: conventproof}

In this section, we consider conventional approaches to solving $\ACProb$. 
Our goal is twofold. First, we will justify the claim in the main text that there exists a conventional sensing protocol which solves $\ACProb$ using a total sensing time of
\begin{align}\label{eq: conventscaling_sm}
    \tauconv = O
    \left( \frac{1}{n_S B_{\min}} \bigg\lceil \frac{|\Delta \omega|}{n_SB_{\min}} \bigg\rceil \right).
\end{align}
Second, we will use the protocol presented here as a subroutine in our eventual quantum search sensing (QSS) algorithm, for the verification step after the Grover search over frequency bins. 

We present a thorough analysis of a conceptually simple conventional AC sensing protocol for $\ACProb$. As noted in Sec.~\ref{sec: sensing_protocol}, it suffices to consider $n_S = 1$. Our overall strategy is as follows. We first partition $\Delta \omega$ into $N = \lceil |\Delta \omega| / \beta \rceil$ sequential frequency intervals (bins), each of width $\beta$; we will search for the AC signal over each bin separately and sequentially. Suppose we could perform a sensing subroutine of duration $\Tsub$ with the following behavior: if the signal does not exist, the subroutine always outputs ``NO,'' and if the signal does exist and satisfies $\omega \in [\omega_t - \beta/2, \omega_t + \beta/2]$, the subroutine outputs ``YES'' with probability at least $2/3$. Here, $\omega_t$ is a tunable parameter, the target frequency of the subroutine. By applying the sensing subroutine $N$ times, with $\omega_t$ varying over the centers of each frequency bin,
we solve $\ACProb$ in total time $\tauconv = O(N \Tsub) = O(\lceil |\Delta \omega| / \beta \rceil \Tsub)$. Therefore, it suffices to construct the sensing subroutine and show that it achieves the desired behavior with $\beta \sim B_{\min}$ and $\Tsub \sim 1/B_{\min}$, which we now do explicitly.

Our sensing subroutine is inspired by the quantum sensing technique of \emph{variance detection} (or quadratic detection), used for detecting AC signals with unknown phase $\phi$~\cite{degen:2017}. 
Concretely, our subroutine is as follows:
\begin{algorithm}[H]
\caption{Conventional AC sensing subroutine based on Carr-Purcell-Meiboom-Gill (CPMG) pulse sequence }
\setstretch{1.15}
\label{alg:cpmg}
\begin{algorithmic}[1]
\REQUIRE Target sensitivity $B_{\min} > 0$ and frequency interval $\Delta \omega = [\omega_t-\beta/2, \omega_t+\beta/2]$ such that $\omega_t-\beta/2 > B_\textrm{min}$.
\FOR{$i= 1, \hdots, M$}
\STATE Prepare the sensor in initial state $\ket{+} = \frac{1}{\sqrt{2}} (\ket{0} + \ket{1})$. Let $t_0$ denote the time.

\STATE Wait for a time duration $t_w$ chosen uniformly at random from $[0, \frac{2 \pi}{B_{\min}}]$. \label{eq: random_wait}

\STATE Evolve under $H(t)$ for duration $T$, while applying $\pi$-pulses ($X$ gates) at times $t_j = t_0 + t_w+\frac{2j-1}{2} \frac{\pi}{\omega_t}$ for $j = 1, \hdots, P$. Here, $P$ is an even number and $T = \pi \frac{P}{\omega_t}$. \label{eq: pulse_seq_step}

\STATE  Apply a Hadamard gate at time $t_{P+1} = t_0 +t_w +T$ and measure in the $Z$-basis, obtaining outcome $z_i \in \{0,1\}$. 

\ENDFOR

\ENSURE If any $z_i = 1$, output ``YES.'' Otherwise, output ``NO.'' 
\end{algorithmic}
\end{algorithm}

This subroutine performs a well-known AC sensing protocol called the Carr-Purcell-Meiboom-Gill (CPMG) pulse sequence, with minor modifications.
In practical settings, the duration $T$ of each iteration is optimized according to the effective coherence time of the sensors. Here, we assume that the coherence time is unlimited.

We now analyze the success probability of the subroutine. We first suppose the signal does not exist, so the sensor is still in state $\ket{+}$ after all $X$ gates are applied, and the measurement returns $z_i = 0$ for all $i = 1, \hdots, M$. Hence, the subroutine always outputs ``NO'' in this case, as desired. Now, suppose that the signal exists and has frequency $\omega$ within $\beta/2$ of $\omega_t$. We will lower bound the probability that at least one measurement yields $z_i = 1$. 
Consider a particular iteration of the for loop in Algorithm~\ref{alg:cpmg}.
The state immediately prior to the Hadamard gate is of the form $(e^{-i \theta} \ket{0} + e^{i \theta} \ket{1})/\sqrt{2}$, where the metrological information is encoded in the phase angle
\begin{align}
    \theta = B \int_{0}^{T} dt~\chi(t) \cos(\omega t + \phi_0 + \omega t_w) \quad \text{with} \quad \chi(t) = (-1)^j \text{ for } t_j \le t < t_{j+1},
\end{align}
where $\phi_0 = \omega t_0+\phi$ is the (unknown) phase of the signal field at time $t_0$.
Here, $\chi(t)$ is the \emph{modulation function} characterizing the pulse sequence: a square wave with period $2 \pi / \omega_t$ which changes sign every time a $\pi$ pulse is applied (due to the anticommutation of $X$ and $Z$). The final $Z$-basis measurement yields $z_i = 1$ with probability $p=\sin^2(\theta)$; therefore, we focus on analyzing $\theta$.

As is well-known within the sensing community~\cite{degen:2017}, it is convenient to rewrite $\theta$ in the frequency domain:
\begin{align}
    \theta  = BT \cdot \widetilde{\chi}(\omega) \cdot  \cos(\phi_0 + \omega t_w + \frac{\omega T}{2})
\end{align}
as a product of three factors.
The first factor, $BT$, describes the coherent, linear accumulation of the sensing signal over the duration $T$.
We will choose a specific $T \sim 1/B_\textrm{min}$ such that $BT \geq 2\pi$ for any $B\geq B_\textrm{min}$.
The second factor is known as the \emph{filter function},
\begin{align}
    \widetilde{\chi}(\omega)
    \equiv \int_0^T dt~e^{i\omega (t - T/2)} \chi(t) = \frac{\sin\left( \frac{P\pi}{2} \frac{\omega}{\omega_t}\right)}{\frac{P \pi}{2} \frac{\omega}{\omega_t}} \left(1 - \sec(\frac{\pi}{2} \frac{\omega}{\omega_t}) \right),
\end{align}
and characterizes the response of the sensor to signals at frequency $\omega$.
We will choose a specific $\beta \sim B_\textrm{min}$ such that $|\widetilde{\chi}(\omega)| \geq 1/5$ for any $\omega$ within $\beta/2$ of $\omega_t$.
Finally, the third factor describes the phase synchronization of the signal field and the CPMG pulse sequence. For certain values of $\phi_0$, the accumulated sensing phase $\theta$ can vanish; we introduce the random wait time $t_w$ in order to avoid this unlucky situation with high probability. We now show that $\theta$ has a large magnitude with high probability, which will translate into a lower bound on the success probability $p$ after averaging over $t_w$.

We set $P = 2 \lceil \frac{\omega_t}{B_\textrm{min}} \rceil$, and correspondingly $T = \frac{2\pi}{\omega_t} \lceil \frac{\omega_t}{B_\textrm{min}} \rceil$. This choice guarantees that $P$ is a positive, even number.
Clearly, $B_\textrm{min} T \geq 2\pi$, as promised. In addition, we set $\beta = \omega_t/P = \frac{\omega_t}{2}/ \lceil \frac{\omega_t}{B_\textrm{min}} \rceil $.
This choice ensures that $\beta \leq B_\textrm{min}/2$, while saturating this inequality in the limit $B_\textrm{min}/\omega_t \rightarrow 0$. Now, we analyze the magnitude of the filter function evaluated at $\omega = \omega_t + \delta$ for $|\delta|\leq \beta/2$:
\begin{align}
|\widetilde{\chi}(\omega)| &= 
\frac{
\left|
\sin\left( \frac{P\pi }{2} \frac{\omega_t + \delta }{\omega_t}\right)
\right|}{
\left| 
\frac{P\pi}{2}\frac{\omega_t + \delta }{\omega_t}
\right| }
\cdot 
\left|1 - \sec(\frac{\pi}{2} \frac{\omega_t + \delta}{\omega_t}) \right|
=
\frac{
\left|
\sin\left( \frac{P\pi }{2} \frac{\delta }{\omega_t}\right)
\right|}{
\left| 
\frac{P\pi}{2}\frac{\omega_t + \delta }{\omega_t}
\right| } 
\cdot 
\left|1 + \csc(\frac{\pi}{2} \frac{\delta}{\omega_t}) \right|
\\
&=
\frac{
\left|
\sin\left( \frac{P\pi }{2} \frac{\delta }{\omega_t}\right)
\right|}
{
\left| 
\sin(\frac{\pi}{2} \frac{\delta}{\omega_t})
\right|
}
\cdot 
\frac{\left|1 + \sin(\frac{\pi}{2} \frac{\delta}{\omega_t}) \right|}
{
\left| 
\frac{P\pi}{2}\frac{\omega_t + \delta }{\omega_t}
\right| } 
\geq \frac{\left| \frac{P \delta}{\omega_t} \right| }{\frac{\pi}{2} \left| \frac{\delta}{\omega_t}\right|}
\cdot \frac{1-\sin(\pi/8)}{\frac{P \pi}{2} \frac{5}{4} }> \frac{1}{5}.
\end{align}
In the first line, we use that $P$ is an even integer.
In the second line, we recognize $| P \delta/\omega_t| \le P \beta /(2 \omega) = 1/2 < 1$
and use the inequalities $\left| \frac{\pi}{2} x \right| \geq \left| \sin(\frac{\pi}{2} x)\right| \geq |x|$ for any $|x|\leq 1$. We also use that $|\delta / \omega_t| \leq 1/(2P) \leq 1/4 $ since $P$ is a positive even number.
Combined, this shows that $ A \equiv \left|    BT \cdot \widetilde{\chi}(\omega) \right| > 2 \pi/5$ 
for any $B>B_\textrm{min}$ and $\omega \in [\omega_t-\beta/2,\omega+\beta/2]$.

Now we consider the averaging over wait time $t_w$. The probability of measuring $z = 1$ in each round is $p = \sin^2(\theta)$; after averaging we obtain
\begin{align}
    \mathbb{E}_{t_w}[ p] &= \frac{B_{\min}}{2 \pi} \int_0^{\frac{2 \pi}{B_{\min}}} dt_w \sin^2\left( A  \cos(\phi_0 + \omega t_w + \frac{\omega T}{2}) \right) \\
    &\ge \frac{B_{\min}}{2 \pi \omega} \int_0^{2 \pi \lfloor \frac{\omega}{B_{\min}} \rfloor} d \phi_w  \sin^2\left( A  \cos(\phi_0 + \phi_w + \frac{\omega T}{2}) \right) 
\end{align}
where we introduce $\phi_w = \omega t_w$. 
Note that the upper limit of the integral is nonzero since $\omega \geq \omega-\beta/2 \geq B_\textrm{min}$. This integral is now taken over an integer number of periods of the integrand and can be evaluated exactly, leading to
\begin{align}
    \mathbb{E}_{t_w}[ p]  \ge \frac{1}{2} \frac{\lfloor \omega / B_{\min} \rfloor}{\omega / B_{\min}} \left( 1 - J_0(2A) \right) \ge \frac{1}{4} \left( 1 - J_0(2A) \right) > \frac{1}{6}.
\end{align}
where $J_0(\cdot)$ denotes the Bessel function of the first kind, and the second inequality follows from $\omega / B_{\min} \le 1 + \lfloor \omega / B_{\min} \rfloor \le 2 \lfloor \omega / B_{\min} \rfloor $.
The last inequality is obtained by explicitly checking $\frac{1}{4} \left( 1- J_0(2A)\right) >\frac{1}{6}$ for all $A>2\pi/5$ (see also~\cite{watson:1922} for bounds which imply this inequality).

This proves that if the signal exists, it is detected by the measurement in any given iteration with probability at least $1/6$. This bound is independent of the measurement outcomes in different iterations. Therefore, after $M$ repetitions, we detect the signal with probability at least $1-(5/6)^M$. For $M\geq 7$, this success probability exceeds the desired $2/3$.
The total sensing time of this approach is bounded as desired:
\begin{align}
    \Tsub \leq  M \left( \frac{2\pi}{B_\textrm{min}} + \frac{2\pi}{\omega_t} \bigg\lceil \frac{\omega_t}{B_\textrm{min}} \bigg\rceil \right) = O(1/B_\textrm{min}).
\end{align}

\subsection{Quasi-static Sensing}\label{sec: quasi_static}
In this section, we describe a conventional sensing protocol which detects all low frequency signals satisfying $\omega_{\min} < \omega \le B_{\min}$, for an arbitrary $\omega_{\min} > 0$.
The sensing task in this regime is qualitatively different from the case $\omega_\textrm{min} > B_\textrm{min}$ for two reasons.
First, since the signal varies slowly, we can detect it using a simple sensing protocol designed for time-independent signals: Ramsey interferometry~\cite{degen:2017}.
Second, the minimum necessary sensing time depends explicitly on the frequency of the signal. 
To illustrate this point, consider $B(t) = B\cos(\omega t + \phi)$ with $\phi = \pi/2$ and $\omega$ very small. Then, $B(t)$ is indistinguishable from the case of no signal until $\omega t + \phi$ is sufficiently away from $\pi/2$.
This suggests that in this quasi-static regime, we should have a $1 / \omega_{\min}$ scaling in addition to the $1 / B_{\min}$ scaling. We show this in what follows.

The sensing protocol for detecting quasi-static fields is very similar to  Algorithm~\ref{alg:cpmg}, but with two modifications. First, in Line~\ref{eq: random_wait}, we randomly select a wait time $t_w$ from $[0, \frac{2 \pi}{\omega_{\min}}]$. Second, we do not apply $X$ gates, so Line~\ref{eq: pulse_seq_step} is replaced simply by time evolution under $H(t)$ for duration $T$. Again, if the signal does not exist, we always output ``NO.'' Suppose the signal does exist: the state immediately prior to the Hadamard gate is $(e^{-i \theta} \ket{0} + e^{i \theta} \ket{1})/\sqrt{2}$, where $\theta$ is simply
\begin{align}
    \theta = B T \cdot \widetilde{\chi}_R(\omega) \cdot \cos(\phi_0 + \omega t_w + \frac{\omega T}{2}) \quad \text{with} \quad \widetilde{\chi}_R(\omega) = \frac{\sin(\frac{\omega T}{2})}{\frac{\omega T}{2}}.
\end{align}
We choose $T = \pi/B_{\min}$ and bound the magnitude of $\widetilde{\chi}_R(\omega)$:
\begin{align}
    |\widetilde{\chi}_R(\omega)| = \frac{\left|\sin(\frac{\pi}{2} \frac{\omega T}{\pi}) \right|}{\frac{\omega T}{2}} \ge \frac{2}{\pi}.
\end{align}
Here, we use the assumption $|\omega T /\pi| \le B_{\min} T / \pi = 1$ and the inequality $|\sin(\frac{\pi}{2} x)| \ge |x|$ for $|x| \le 1$. We conclude that $A_R \equiv |B T \cdot\widetilde{\chi}_R(\omega)| \ge 2$. 

The probability of measuring $z=1$ in a given round is $p = \sin^2(\theta)$. Averaged over random $t_w$, we obtain by an identical analysis to the previous section
\begin{align}
    \mathbb{E}_{t_w}[p] \ge \frac{1}{2} \frac{\lfloor \omega / \omega_{\min} \rfloor}{\omega / \omega_{\min}} (1 - 2 J_0(2 A_R)) \ge \frac{1}{4} (1 - 2 J_0(2 A_R)) > \frac{1}{6}.
\end{align}
Again, this shows that the signal is detected in a given iteration with probability at least $1/6$, independent of the measurement outcomes of different iterations, so for $M \ge 7$ the total success probability exceeds $2/3$. The total sensing time of this sensing protocol is given by
\begin{align}
    \tau_R \le M \left(\frac{2 \pi}{\omega_{\min}} + \frac{\pi}{B_{\min}}\right) = O(1/\omega_{\min}+1/B_{\min}).
\end{align}
The sensing time increases with decreasing $\omega_{\min}$ because we need to wait until $\omega t + \phi$ is sufficiently away from $\pi / 2$.

\newpage

\section{Quantum Search Sensing Algorithm}

\subsection{Overview of the Algorithm}\label{sec: overview_qss}

In this section, we describe our quantum search sensing (QSS) algorithm in detail and provide a thorough error and sensing time analysis of the protocol. Concretely, we prove the following result, a formal version of Theorem 1 in the main text:

\begin{theorem}[Quantum Search Sensing]\label{thm: qss_formal}
There exists a quantum sensing protocol which solves $\ACProb$ with sensing time $\tau$ and $n_Q$ ancilla qubits, where
\begin{align}
    \tau = O\left(\frac{1}{n_S B_{\min}} \sqrt{\bigg\lceil \frac{|\Delta \omega|}{n_S B_{\min}} \bigg\rceil} \log(1 + \bigg\lceil \frac{|\Delta \omega|}{n_S B_{\min}} \bigg\rceil)^{\frac{5}{4}} \right) \ \ \text{and} \ \ n_Q = O\left( \log(\bigg\lceil \frac{|\Delta \omega|}{n_S B_{\min}} \bigg\rceil) \right).
\end{align}
\end{theorem}
\noindent 
We observe that in the edge case $|\Delta \omega| \le n_S B_{\min}$, the conventional sensing protocol described in Sec.~\ref{sec: conventproof} solves $\ACProb$ in time $\tau = O((n_S B_{\min})^{-1})$ with $n_Q = 0$ ancilla, matching the scaling in Theorem~\ref{thm: qss_formal}. Therefore, we assume $|\Delta \omega| > n_S B_{\min}$ in what follows. Further, as noted in Sec.~\ref{sec: sensing_protocol}, it suffices to consider the case $n_S = 1$.

The main idea behind QSS, as described in the main text, is to divide the bandwidth $\Delta \omega$ into frequency bins, then engineer a sensing oracle and perform a Grover search over the bins. Our sensing oracle will have some error, and we must ensure that this error accumulates negligibly over the duration of the Grover search. To guarantee this is the case, it is helpful to have additional constraints on the signal strength $B$ and frequency $\omega$, which we can enforce by dividing $\ACProb$ into a number of subproblems. Specifically, consider the following variant of $\ACProb$:

\begin{definition}[AC Sensing with High Frequency, Finite Dynamic Range] Let $B_{\min}, \Delta \omega,$ and $H(t)$ be as in Definition~\ref{def: ac_disc}. Let $\gamma > 1$ be a constant. We define $\ACProbDR$ as $\ACProb$ subject to the additional promise that, if the signal exists with strength $B$ and frequency $\omega$, then $B \le \gamma B_{\min}$ and $\omega \ge |\Delta \omega|$.
\end{definition}
\noindent
We comment briefly on why the two additional conditions are helpful in designing the sensing oracle, which will be explained in detail in subsequent sections. First, the condition $B_{\min} \le B \le \gamma B_{\min}$ means we only need to detect signals over a finite dynamic range. This allows us to bound the phase accumulation in our elementary sensing unit, which ensures the success of quantum signal processing. Second, the condition $\omega \ge |\Delta \omega|$ means we only need to detect high frequency fields, so we can apply the rotating wave approximation.

The main technical step in our proof of Theorem~\ref{thm: qss_formal} is captured by the following lemma:
\begin{lemma}\label{lemma: qss_finite_dr} For some constant $\gamma$, there exists a quantum sensing protocol which solves $\ACProbDR$ with sensing time $\tau$ and $n_Q$ ancilla qubits, both scaling as in Theorem~\ref{thm: qss_formal}. Further, the protocol never falsely concludes that the signal exists.
\end{lemma}
\noindent
We prove Lemma~\ref{lemma: qss_finite_dr} in the following subsection, where we will see that the protocol described in the main text is sufficient to solve $\ACProbDR$. 

We now prove Theorem~\ref{thm: qss_formal} assuming Lemma~\ref{lemma: qss_finite_dr}. 
We simply partition the possible signal strengths and frequencies into subintervals 
which satisfy the additional conditions in $\ACProbDR$. First, note that we can detect all signals with strength $B \ge |\Delta \omega|$ in time $O(1/B_{\min})$ using conventional sensing; let $\Delta B \equiv [B_{\min}, |\Delta \omega|]$ denote the remaining signal strengths. Observe that $\Delta B \subseteq \cup_{i=1}^{i_{\max}} \Delta B_i$ where $\Delta B_i = [\gamma^{i-1} B_{\min}, \gamma^i B_{\min}]$ and $i_{\max} = \lceil \log_\gamma(|\Delta \omega|/B_{\min}) \rceil$. Similarly, $\Delta \omega \subseteq \cup_{j=1}^{j_{\max}} \Delta \omega_j$ where $\Delta \omega_j = [2^{j-1} \omega_{\min}, 2^j \omega_{\min}]$ and $j_{\max} = \lceil \log(\omega_{\max}/\omega_{\min}) \rceil$. (It is unnecessary to partition the frequencies in this way if $\omega_{\min} \ge |\Delta \omega|$ already, so we assume $\omega_{\min} < |\Delta \omega|$ without loss of generality.) 
To solve $\ACProb$, we simply solve $\mathsf{AC}_{\gamma}[\gamma^{i-1} B_{\min}, \Delta \omega_j]$ for each $i$ and $j$. Output ``YES'' if any round outputs ``YES,'' and ``NO'' otherwise. We never falsely conclude that the signal exists, so this protocol succeeds with probability at least $2/3$. The total sensing time is at most
\begin{align}
    \tau \le C_\gamma \sum_{i,j} \frac{1}{\gamma^{i-1} B_{\min}} \sqrt{\frac{2^{j-1} \omega_{\min}}{\gamma^{i-1} B_{\min}}} \log(\frac{2^{j-1} \omega_{\min}}{\gamma^{i-1} B_{\min}})^{5/4} \le  C_\gamma \frac{\gamma^{3/2}}{\gamma^{3/2} - 1} \frac{1}{B_{\min}} \sum_j \sqrt{\frac{2^{j-1} \omega_{\min}}{B_{\min}}} \log(\frac{2^{j-1} \omega_{\min}}{B_{\min}})^{5/4}.
\end{align}
The first inequality is true by Lemma~\ref{lemma: qss_finite_dr} for some constant $C_\gamma$, and the second inequality follows by summing over all $i \ge 1$. Summing the partial geometric series over $j$, we obtain
\begin{align}
    \tau \le  C_\gamma \frac{\gamma^{3/2}}{\gamma^{3/2} - 1} \frac{1}{\sqrt{2} - 1}  \frac{1}{B_{\min}} \left( \frac{2 \omega_{\max}}{B_{\min}} \right)^{1/2} \log(\frac{\omega_{\max}}{B_{\min}})^{5/4}.
\end{align}
Since $\omega_{\min} < |\Delta \omega|$ by assumption, we have $\omega_{\max} < 2 |\Delta \omega|$, completing the sensing time analysis. Finally, the $n_Q$ scaling follows immediately by noting that no subproblem requires more than $O(\log(\lceil |\Delta \omega|/B_{\min} \rceil))$ ancilla qubits.

\subsection{AC Sensing with High Frequency, Finite Dynamic Range}\label{sec: high_freq_qss}

In this subsection, we prove Lemma~\ref{lemma: qss_finite_dr} by explicitly constructing and analyzing a quantum sensing protocol to solve $\ACProbDR$. This protocol will be identical (up to minor technical differences) to the algorithm described in the main text, which we now briefly recall. The first step is to partition the bandwidth $\Delta \omega$ into $N$ frequency bins indexed by $k = 1, \hdots, N$ and each associated to a state $\ket{k}_Q$ of the $n_Q$ ancilla qubits. We use Grover's algorithm to search for the bin which contains the signal frequency. 
The Grover search is enabled by constructing a sensing oracle 
$\orac$ which acts as
\begin{align} \label{eq:Oracle_sm}
\mathcal{O}_\textrm{AC}\ket{k}_Q\ket{0}_S \approx
        \left\{
    \begin{array}{rl}
    -\ket{k}_Q\ket{0}_S & \textrm{if signal in bin $k$}\\
    \ket{k}_Q\ket{0}_S & \hspace{1cm}\textrm{else}
    \end{array} \hspace{.2cm}.
    \right.
\end{align}
The sensor begins and ends in a fixed initial state, which we take as $\ket{0}_S$ without loss of generality. We implement $\orac$ in two steps. First, we design an elementary sensing unit (ESU)
which approximates $\sum_k \ketbra{k}_Q \otimes \text{exp}[-i \theta_k Z_S]$ for $\theta_k = \theta_k(B, \omega)$ a sequence of $\phi$-independent rotation angles. Then, we use the ESU as the signal unitary in a quantum signal processing (QSP) sequence applied to the sensing qubit to discretize the conditional phase to $\pm 1$, as desired.

Before describing our protocol in detail, we present three lemmas which encapsulate the main technical results necessary to complete our construction and error analysis of the sensing oracle. We defer proofs of these lemmas to subsections Sec.~\ref{sec: ham_trans_proof} and~\ref{sec: esu_error_proof}. The first lemma describes the Hamiltonian transformation at the core of our ESU:
\begin{lemma}[ESU Hamiltonian Transformation]\label{lemma: ham_trans} Let $\omega_k \ge 0$ and let $\chi(t) \in [0,1]$ be a differentiable function defined on $[0,T]$. There is a sequence of single qubit unitaries which transforms time evolution under $H(t) = B \cos(\omega t + \phi) Z$ into time evolution under 
\begin{align}
    H_{\text{ESU}}^{(k)}(t) = B \chi(t) \cos(\omega t + \phi)\left(\cos(\omega_k t) X - \sin(\omega_k t) Y \right)
\end{align}
for $0 \le t \le T$, with arbitrarily small error in the limit of instantaneous control gates.
\end{lemma}
Let $U_{\text{ESU}}^{(k)}(T)$ denote the unitary corresponding to time evolution under $H_{\text{ESU}}^{(k)}(t)$ for $0 \le t \le T$. $U_{\text{ESU}}^{(k)}(T)$ approximately implements a $\phi$-independent rotation of the sensor, as the following lemma shows ($\| \cdot \|$ denotes operator norm):
\begin{lemma}[ESU Error Bound]\label{lemma: esu_error_new} Let $\delta_k = \omega_k - \omega$ for some $\omega_k \ge 0$ and suppose $0 < |\delta_k| \le \omega$. There exists a choice of $\chi(t) \in [0,1]$ and constants $\mu, \nu > 0$ such that, as long as $B/|\delta_k| \le \mu$ and $\omega T \ge \nu$, we have
\begin{align}
    \| U_{\text{ESU}}^{(k)}(T) - e^{-i \theta_k Z} \| \le \SAerror^{(k)} + \RWAerror
\end{align}
with errors bounded as
\begin{align}
    \SAerror^{(k)} \le C \frac{B}{|\delta_k|} \mathrm{exp}\left(-\left(D \frac{|\delta_k|^3}{B^2}T \right)^{1/2}\right) \quad \text{and} \quad \RWAerror \le C_{\scriptscriptstyle \text{RWA}} \frac{B}{\omega}(1 + B T)
\end{align}
and $\theta_k$ satisfying
\begin{align}\label{eq: esu_angle_errors}
    \left| \theta_k -  \left(\frac{\delta_k}{2} \int_0^T dt \sqrt{1 + (B \chi(t) / \delta_k)^2} - \frac{\delta_kT}{2} \right)\right| \le E \frac{B}{|\delta_k|}
\end{align}
for constants $C, C_{\scriptscriptstyle \text{RWA}},D, E > 0$.
\end{lemma}

Finally, the following lemma describes a polynomial approximation to the shifted sign function, which we will use to choose our QSP sequence:

\begin{lemma}[Corollary 6.27 in~\cite{low:2017}] \label{lemma: qsp_constr_new}
    Let $\mathrm{sgn}(x - x_*)$ denote the shifted sign function centered at $x_*$, defined as
    \begin{align}
        \mathrm{sgn}(x - x_*) = 
        \begin{cases}
            1 & x > x_* \\
            0 & x = x_* \\
            -1 & x < x_*
        \end{cases}.
    \end{align}
    For all $x_* \in [-1,1]$, $\Delta > 0$, and $\varepsilon$ sufficiently small, there is a polynomial $Q_{x_*, \Delta, \varepsilon}(x)$ of degree $O(\frac{1}{\Delta} \log(1/\varepsilon))$ with real-valued coefficients such that
    \begin{align}
        \max_{x \in [-1,1]} |Q_{x_*, \Delta, \varepsilon}(x)| \le 1 \quad \text{and} \quad \max_{x \in [-1, x_* - \Delta/2] \cup [x_* + \Delta/2, 1]} |Q_{x_*, \Delta, \varepsilon}(x) - \mathrm{sgn}(x - x_*)| \le  \varepsilon~.
    \end{align}
\end{lemma}
\noindent
In other words, outside of a failure region of width $\Delta$ centered at $x_*$, $Q_{x_*, \Delta, \varepsilon}(x)$ is an $\varepsilon$-approximation to the shifted sign function. We refer the reader to~\cite{low:2017} for a proof of Lemma~\ref{lemma: qsp_constr_new}.

\subsubsection{Detailed Protocol}

Assuming Lemmas~\ref{lemma: ham_trans},~\ref{lemma: esu_error_new}, and~\ref{lemma: qsp_constr_new}, we now prove Lemma~\ref{lemma: qss_finite_dr}. We begin by describing our quantum sensing protocol for $\ACProbDR$ in detail. First, we handle the edge case that the signal has very high strength. In particular, if $B_{\min} > |\Delta \omega| / x_0$ for some constant $x_0$ (to be determined), we solve $\ACProbDR$ simply by performing the conventional sensing protocol in Sec.~\ref{sec: conventproof}. This takes time $O((B_{\min})^{-1})$ and uses $n_Q = 0$ ancilla qubits. Therefore, we assume without loss of generality that $|\Delta \omega|/ B_{\min} \ge x_0$. 

Next, we specify our choice of frequency bins. Our binning strategy differs from the naive approach (dividing $\Delta \omega$ into $N$ adjacent intervals, as in conventional sensing) in two key ways. First, we divide $\Delta \omega$ into two separate collections of frequency bins, which we call odd and even, defined so that the bins in either collection are non-adjacent. Second, each frequency bin is not a contiguous interval, but rather a pair of disjoint, separated intervals. Explicitly, our bins are defined as follows (see FIG. 3d in the main text for an illustration). First, partition $\Delta \omega$ into $4N$ adjacent intervals for some $N$ (to be determined), and let $\beta$ denote the width of each interval. 
That is, $\Delta \omega = \bigcup_{k,a} \Delta \omega_{k,a}$ with $\Delta \omega_{k,a} = [\omega_\textrm{min} +(4(k-1)+(a-1))\beta, \omega_\textrm{min} + (4(k-1)+(a-1)+1)\beta]$ for $k= 1, \dots N$ and $a=1, 2, 3, 4$.
We define the $k$-th odd frequency bin as $\Delta\omega_{k,1} \cup \Delta\omega_{k,3}$ and the $k$-th even frequency bin as $\Delta\omega_{k,2} \cup \Delta\omega_{k,4}$. 
The odd bins have center frequencies $\omega_k \equiv \omega_\textrm{min} + (4(k-1)+\frac{3}{2}) \beta$, and the even bins have center frequencies $\omega_k \equiv \omega_\textrm{min} + (4(k-1)+\frac{5}{2}) \beta$. We note that no frequency bin contains its own center frequency.
We will perform separate Grover searches over the odd and even bins, so we must construct a sensing oracle for each collection. Since the constructions are identical, we refer to both as $\orac$ for ease of notation.

We now describe our implementation of $\orac$, beginning with the ESU. Our ESU is $U_{\text{ESU}} = \sum_k \ketbra{k}_Q \otimes U_{\text{ESU}}^{(k)}(T)$, which we obtain by evolving under $H(t)$ for time $0 \le t \le T$ while applying the sensor gates in Lemma~\ref{lemma: ham_trans} conditional on the state of the ancilla qubits. (We suppress the $\phi$-dependence of $U_{\text{ESU}}$, since the bounds in Lemma~\ref{lemma: esu_error_new} are $\phi$-independent.) We choose $\omega_k$ as the center frequency of the $k$-th bin (odd or even depending on which Grover search we are performing). Finally, we choose $\chi(t)$ so that the error bound in Lemma~\ref{lemma: esu_error_new} holds. 
Note that this construction differs slightly from the presentation in the main text, which was based on the use of a control Hamiltonian. The two constructions are equivalent, but this approach is simpler to analyze (since it avoids moving into different reference frames).

To complete the construction of $\orac$, we must choose a QSP sequence. Let $Q_{x_*, \Delta, \varepsilon}(x)$ be the polynomial in Lemma~\ref{lemma: qsp_constr_new}, and define the even polynomial $P(x) = \frac{1}{1 + \varepsilon}(Q_{x_*, \Delta, \varepsilon}(x) + Q_{x_*, \Delta, \varepsilon}(-x) + 1)$. Let $\varepsilon = \QSPerror/5$. It follows that, outside of failure regions of width $\Delta$ centered at $\pm x_*$, $P(x)$ is an $\QSPerror$-approximation to $\text{sgn}(|x| - x_*)$. 
Moreover, as long as these failure regions do not overlap (i.e., $|x_*| \ge \Delta / 2$), then we have $\max_{x \in [-1,1]} |P(x)| \le 1$. Consequently, we can apply the main theorem of QSP~\cite{martyn:2021} to find rotation angles $\Phi = (\phi_0, \cdots, \phi_L)$ such that
\begin{align}
    P(\cos(\theta)) = \bra{0} e^{-i \phi_L X} e^{- i \theta Z} \cdots e^{-i \phi_1 X} e^{-i \theta Z} e^{-i \phi_0 X}  \ket{0}
\end{align}
with $L \le \frac{C_{\text{QSP}}}{\Delta} \ln(1/\QSPerror)$ for some constant $C_{\text{QSP}}$. These angles define our QSP sequence: let $\orac$ be given by 
\begin{align}
    \orac = e^{-i \phi_L X_S} U_{\text{ESU}} \cdots e^{-i \phi_1 X_S} U_{\text{ESU}} e^{-i \phi_0 X} .
\end{align}

Finally, we can perform a Grover search over both the odd and even bins. 
Prepare the $n_Q = \log(N)$ ancilla qubits in the uniform superposition state $|s \rangle_Q = \sum_{k=1}^N | k \rangle_Q/\sqrt{N}$ and the sensor in state $\ket{0}_S$ (we may assume $N$ is a power of 2 by increasing the number of frequency bins, at most doubling $|\Delta \omega|$).
Apply $R_s \orac$ a total of $\lfloor \pi \sqrt{N}/4 \rfloor$ times, where $R_s = 2 \ketbra{s}_Q - \mathbbm{1}$.
Measure the ancilla qubits in the computational basis, obtaining a prediction for the marked bin, and check the prediction using the conventional sensing protocol in Sec.~\ref{sec: conventproof}. 
Repeat both the even and odd Grover searches $M$ times to amplify the success probability. 
Output ``YES'' if any conventional check outputs ``YES,'' and ``NO'' otherwise. This completes our sensing protocol for $\ACProbDR$.

\subsubsection{Error and Resource Analysis}\label{sec: qss_error_analysis}

Having described our protocol for $\ACProbDR$ in detail, we now perform a detailed error analysis and quantify the necessary resources, i.e., sensing time and number of ancilla qubits, used by the protocol. We first note that the protocol never falsely concludes that the signal exists, because of the conventional check. It remains to show that if the signals exists, and therefore has frequency in either the even or odd collection of bins, then the corresponding Grover search succeeds with at least constant probability. We can amplify this success probability to $2/3$ with $M = O(1)$, changing the overall sensing time by a constant factor.

We assume without loss of generality that the signal exists and belongs to odd bin $k_*$. 
To prove that the Grover search succeeds with constant probability, it suffices to bound the Euclidean distance (denoted $\| \cdot \|_2$) between the state of the ancilla qubits and sensor immediately prior to measurement in our algorithm, and the \emph{ideal} state, namely, $(R_s \mathcal{O})^{\lfloor \pi \sqrt{N}/4 \rfloor} \ket{s}_Q \ket{0}_S$, where $\mathcal{O} = \mathbbm{1} - 2 \ketbra{k_*}_Q$ is the ideal Grover oracle. 

We now bound this overall error in terms of the errors $\SAerror^{(k)}, \RWAerror,$ and $\QSPerror$. We suppress the subscripts $S$ and $Q$ for simplicity in what follows. First, by repeatedly applying the triangle inequality, we have
\begin{align}
     \| (R_s \orac)^{\lfloor \pi \sqrt{N} / 4 \rfloor} \ket{s} \ket{0} - (R_s \mathcal{O})^{\lfloor \pi \sqrt{N} / 4 \rfloor} \ket{s} \ket{0}\|_2 &\le \sum_{j = 1}^{\lfloor \pi \sqrt{N} / 4 \rfloor} \| \orac (R_s \mathcal{O})^{j-1} \ket{s} \ket{0} 
 - \mathcal{O} (R_s \mathcal{O})^{j-1} \ket{s} \ket{0} \|_2.
     \\ \label{eq: hybrid_eq}
     &\le \sqrt{N} \max_{\ket{\psi} \in \mathcal{H}_Q} \| \orac \ket{\psi} \ket{0} - \mathcal{O} \ket{\psi} \ket{0} \|_2
\end{align}
where $\mathcal{H}_Q$ denotes the Hilbert space of the ancilla qubits. Using triangle inequality, we can separate this error into two terms: one arising from the error in the ESU, and one arising from the error in the QSP polynomial approximation. To do so, we define the following operator:
\begin{align}
    \text{CQSP}_{\Phi}(\Theta) = \sum_{k=1}^N \ketbra{k} \otimes e^{-i \phi_L X} e^{- i \theta_k Z} \cdots e^{- i \theta_k Z} e^{- i \phi_0 X}.
\end{align}
Here, $\Theta = (\theta_1, \hdots, \theta_N)$ are the conditional rotation angles approximately realized by $U_{\text{ESU}}$. Therefore, $\text{CQSP}_{\Phi}(\Theta)$ is the operator which would be realized by $\orac$ if $U_{\text{ESU}}$ were a perfect conditional $Z$ rotation. Applying triangle inequality to~\eqref{eq: hybrid_eq}, we obtain the following upper bound:
\begin{align}\label{eq: two_terms}
    \sqrt{N} \max_{\ket{\psi} \in \mathcal{H}_Q} \| \orac \ket{\psi} \ket{0} - \text{CQSP}_{\Phi}(\Theta) \ket{\psi} \ket{0}  \|_2 +  \sqrt{N} \max_{\ket{\psi} \in \mathcal{H}_Q} \| \mathcal{O} \ket{\psi} \ket{0} - \text{CQSP}_{\Phi}(\Theta) \ket{\psi} \ket{0}  \|_2 .
\end{align}
The first term captures the error in the ESU. Using our expressions for $\orac$ and $\text{CQSP}_{\Phi}(\Theta)$ and writing $\ket{\psi}$ out in a basis, we see that the first term is upper bounded by the following expression:
\begin{align}
    \sqrt{N} \max_k \| e^{-i \phi_L X} U_{\text{ESU}}^{(k)}(T) \cdots U_{\text{ESU}}^{(k)}(T) e^{-i \phi_0 X} \ket{0} - e^{-i \phi_L X} e^{-i \theta_k Z} \cdots e^{-i \theta_k Z} e^{-i \phi_0 X} \ket{0} \|_2 .
\end{align}
Using the definition of operator norm and repeatedly applying triangle inequality again, this expression can be further upper bounded as:
\begin{align}
    \sqrt{N} L \max_k \|U_{\text{ESU}}^{(k)}(T) - e^{-i \theta_k Z}\| \le \sqrt{N} L (\max_k \SAerror^{(k)} + \RWAerror)
\end{align}
where the last line follows from Lemma~\ref{lemma: esu_error_new}. 

This handles the error from the ESU; we now upper bound the error which arises from the QSP polynomial approximation. To do so, we will need to be more specific about our choice of QSP polynomial. Suppose that there exists some positive constant $\Delta$ such that $\cos(\theta_k) - \cos(\theta_{k_*}) \ge \Delta$ for every $k \neq k_*$, and furthermore that $\cos(\theta_{k_*}) \ge 0$. For this choice of $\Delta$, by letting $x_*$ be the midpoint of the promised gap region, we are guaranteed that $|P(\cos(\theta_{k_*})) - (-1)| \le \QSPerror$ and $|P(\cos(\theta_{k})) - 1| \le \QSPerror$ for all $k \neq k_*$. If this is the case, we can easily bound the second term in~\eqref{eq: two_terms}. Writing $\ket{\psi} = \sum_k \psi_k \ket{k}$ and using our expressions for $\mathcal{O}$ and $\text{CQSP}_{\Phi}(\Theta)$, this term can be rewritten as
\begin{align}\label{eq: qsp_second_term}
    \sqrt{N} \max_{\ket{\psi} \in \mathcal{H}_Q} \sqrt{2} \left( 1 - \sum_k |\psi_k|^2 (-1)^{\delta_{k, k_*}} \bra{0} e^{-i \phi_L X} e^{-i \theta_k Z} \cdots e^{-i \theta_k Z}  e^{-i \phi_0 X} \ket{0}\right)^{1/2}
\end{align}
where we have written out the Euclidean distance explicitly and noted that $\bra{0} e^{-i \phi_L X} e^{-i \theta_k Z} \cdots e^{-i \theta_k Z}  e^{-i \phi_0 X} \ket{0} = P(\cos(\theta_k))$ is real. Finally, we have $(-1)^{\delta_{k, k_*}} P(\cos(\theta_k))  \ge 1 - \QSPerror$ by assumption. Therefore,~\eqref{eq: qsp_second_term} is at most $\sqrt{2 N \QSPerror}$. We conclude that the total error between the ideal output of Grover's algorithm and the state output by our protocol is at most
\begin{align}\label{eq: total_error}
    \sqrt{N} \left( \sqrt{2 \QSPerror} + L \left(\max_k \SAerror^{(k)} + \RWAerror \right)\right).
\end{align}

To complete our error analysis, we show that the three terms in~\eqref{eq: total_error} are bounded by arbitrarily small constants $\varepsilon_1, \varepsilon_2, \varepsilon_3 > 0$ for sufficiently large $|\Delta \omega|/B_{\min}$. Therefore, by an appropriate choice of the constant $x_0$ (such that $|\Delta \omega|/B_{\min} \ge x_0$ whenever we perform Grover's algorithm), we can guarantee that whenever we perform Grover's algorithm, it succeeds with high probability. After bounding the error, we will verify the conditions in Lemma~\ref{lemma: esu_error_new}, as well as the conditions which ensure our desired QSP approximation exists.

When bounding the errors in~\eqref{eq: total_error}, we will repeatedly use the conditions $B_{\min} \le B \le \gamma B_{\min}$ and $\omega > |\Delta \omega|$, which hold by assumption, and the condition $\delta_k \ge \beta / 2$ for all $k$, due to our choice of frequency bins. We begin by specifying our choices of $L, T,$ and $\beta$ (which implicitly fixes $N = |\Delta \omega|/(4 \beta)$ as the number of bins). We let
\begin{align}
    \beta = C_\beta B_{\min} \sqrt{\ln(|\Delta \omega|/B_{\min})} \quad , \quad T = C_T \beta/B_{\min}^2 \quad , \quad L = \frac{C_{\text{QSP}}}{\Delta} \ln(2 N/\varepsilon_1^2)
\end{align}
for constants $C_\beta, C_T$ (to be determined). Note that $\beta$ and $T$, the bin width and sensing duration of the ESU, are only polylogarithmically different (in $|\Delta \omega|/B_{\min}$) from the analogous parameters in conventional sensing.

To bound the first error term, note that $\QSPerror \le e^{-\Delta L/C_{\text{QSP}}}$, so $\sqrt{2 N \QSPerror} \le \varepsilon_1$ by construction. To bound the second error term, we use Lemma~\ref{lemma: esu_error_new} and the above conditions to conclude
\begin{align}
    \sqrt{N} L \max_k \SAerror^{(k)} \le C' \frac{x^{1/2 - D'}}{\ln(x)^{3/4}} \ln(\frac{1}{\varepsilon_1'} \frac{x}{\sqrt{\ln(x)}}).
\end{align}
Here, $C' = \gamma C_{\text{QSP}}C/( C_\beta^{3/2} \Delta)$, $D' = (D C_TC_\beta^4/(8 \gamma^2))^{1/2}$, and $\varepsilon_1' = 2 \varepsilon_1^2 C_\beta$, and we define $x = |\Delta \omega|/B_{\min}$. The exact functional form of this error is not important; what matters is that as $x \to \infty$, this expression goes to zero (as long as $D' > 1/2$, which we can ensure by sufficiently large choice of $C_\beta$.) Therefore, for any $\varepsilon_2 > 0$, there exists some $x_2$ such that $\sqrt{N} L \max_k \SAerror^{(k)} \le \varepsilon_2$ for $x > x_2.$ Finally, to bound the third error term, we again use Lemma~\ref{lemma: esu_error_new} and the above conditions to conclude
\begin{align}
    \sqrt{N} L \RWAerror \le C_{\text{RWA}}' \frac{1}{\sqrt{x} \ln(x)^{1/4}} \ln(\frac{1}{\varepsilon_1'} \frac{x}{\sqrt{\ln(x)}})\left(1 + D_{\text{RWA}}' \sqrt{\ln(x)} \right).
\end{align}
Here, $C_{\text{RWA}}' = \gamma C_{\text{QSP}} C_{\text{RWA}}/(2 C_\beta^{1/2} \Delta)$ and $D_{\text{RWA}}' = \gamma C_T C_\beta$. Again, this expression goes to zero as $x \to \infty$, so for any $\varepsilon_3 > 0$, there exists some $x_3$ such that $\sqrt{N} L \RWAerror < \varepsilon_3$ for $x > x_3$. Therefore, as long as $x_0 > x_2, x_3$ the overall error~\eqref{eq: total_error} is at most $\varepsilon_1 + \varepsilon_2 + \varepsilon_3$, as desired.

We conclude the error analysis by showing that all desired conditions hold for sufficiently large $|\Delta \omega|/B_{\min}$. We start with the conditions in Lemma~\ref{lemma: esu_error_new}. We have the following two inequalities:
\begin{align}
    \frac{B}{|\delta_k|} &\le \frac{2 \gamma B_{\min}}{\beta} = \frac{2 \gamma}{C_\beta \sqrt{\ln(x)}}
    \\
    \omega T &\ge |\Delta \omega| T = C_T C_\beta x \sqrt{\ln(x)}.
\end{align}
Therefore, by choosing $x_0$ sufficiently large, the conditions in Lemma~\ref{lemma: esu_error_new} hold for any choice of $\mu, \nu$. 

Finally, we prove that $\cos(\theta_{k_*}) \ge 0$ and $\cos(\theta_k) - \cos(\theta_{k_*}) \ge \Delta$ for some constant $\Delta >0$ and all $k \neq k_*$, so our desired QSP approximation exists. We first note that it suffices to show $|\theta_{k_*}| \le \pi/2$ and $|\theta_{k_*}| > |\theta_k|$. To see why, observe that
\begin{align}
    \cos(\theta_k) - \cos(\theta_{k_*}) = 2 \sin(\frac{\theta_{k_*} + \theta_k}{2})\sin(\frac{\theta_{k_*} - \theta_k}{2})
\end{align}
using the angle sum and difference identities. This expression is nonnegative as long as $|\theta_{k_*}| > |\theta_k|$, so we may replace it by its absolute value. Then, using $|\sin(z)| \ge 2 |z|/\pi$ for $|z| \le \pi/2$, we conclude
\begin{align}
    \left| 2 \sin(\frac{\theta_{k_*} + \theta_k}{2})\sin(\frac{\theta_{k_*} - \theta_k}{2}) \right| \ge \frac{2}{\pi^2}| \theta_{k_*} + \theta_k| \cdot | \theta_{k_*} - \theta_k| \ge \frac{2}{\pi^2} (|\theta_{k_*}| - |\theta_k|)^2
\end{align}
where the last step follows from the triangle inequality. Then, we can simply choose $\Delta = (2/\pi^2) (|\theta_{k_*}| - |\theta_k|)^2$. To prove the desired bounds on the marked and unmarked phase angles, we use the error bound on the ESU angles in Lemma~\ref{lemma: esu_error_new}, as well as the structure of the frequency bins. We first note, using $z^2/3 \le \sqrt{1+z^2} \le z^2/2$ for $|z| \le 1$, that as long as $B/|\delta_k| \le 1$ we can simplify the ESU angle bound as
\begin{align}\label{eq: simpler_angle_bound}
    \frac{I}{6} \frac{B^2 T}{|\delta_k|} - E \frac{B}{|\delta_k|} \le |\theta_k| \le \frac{I}{4} \frac{B^2 T}{|\delta_k|} + E \frac{B}{|\delta_k|}
\end{align}
where $I = \frac{1}{T} \int_0^T dt~\chi(t)$ (which is constant for our choice of $\chi(t)$). Next, observe that our binning strategy implies
\begin{align}\label{eq: bin_condition}
    \frac{1}{2} \beta \le |\delta_{k_*}| \le \frac{3}{2} \beta \quad \text{and} \quad |\delta_{k}| \ge \frac{5}{2} \beta \text{ for all } k \neq k_*.
\end{align}
Combining~\eqref{eq: simpler_angle_bound} and~\eqref{eq: bin_condition}, we deduce the following two inequalities:
\begin{align}\label{eq: k_star_upper}
    |\theta_{k_*}| &\le \frac{I}{2} \gamma^2 \frac{B_{\min}^2 T}{\beta} + 2 E \gamma \frac{B_{\min}}{\beta}
    \\ \label{eq: angle_gap}
    |\theta_{k_*}| - |\theta_k| &\ge \left(\frac{1}{9} - \frac{\gamma^2}{10} \right) I \frac{B_{\min}^2 T}{\beta} - \frac{12}{5} E \gamma \frac{B_{\min}}{\beta} \text{ for all } k \neq k_* .
\end{align}
Since $B_{\min}/\beta = 1/(C_\beta \sqrt{\ln(x)})$, the second term in each inequality is arbitrarily small for sufficiently large $x_0$. By choosing $C_T = \pi/(2 I \gamma^2)$, so that the first term in~\eqref{eq: k_star_upper} is $\pi/4$, and $\gamma = \sqrt{19/18}$, so that the first term in~\eqref{eq: angle_gap} is positive, we can ensure that $|\theta_{k_*}| \le \pi/2$ and $|\theta_{k*}| > |\theta_k|$ for sufficiently large $x_0$, as desired.

This completes the error analysis and proves that our protocol solves $\ACProbDR$. Finally, we consider the number of ancilla qubits and sensing time used by the protocol. The number of ancilla qubits is simply $n_Q = \log(N)$ (again, we may assume $N$ is a power of 2):
\begin{align}
    n_Q = \log(\frac{1}{4 C_\beta} \frac{x}{\sqrt{\ln(x)}}) = O\left(\log(\frac{|\Delta \omega|}{B_{\min}}) \right)
\end{align}
as desired. To bound the total sensing time, note that it suffices to consider the sensing time $\tau_G$ of a single Grover search, since the number of repetitions is $M = O(1)$ and each conventional verification takes subdominant time $O(B_{\min}^{-1} \log(|\Delta \omega|/B_{\min}))$. We have $ \tau_G = \lfloor \frac{\pi}{4} \sqrt{N} \rfloor L T$, which is
\begin{align}
    \tau_G \le C_G \frac{1}{B_{\min}} \sqrt{x} \ln(x)^{1/4} \ln(\frac{1}{\varepsilon_1'} \frac{x}{\sqrt{\ln(x)}}) = O\left( \frac{1}{B_{\min}} \sqrt{\frac{|\Delta \omega|}{B_{\min}}} \log(\frac{|\Delta \omega|}{B_{\min}})^{5/4} \right)
\end{align}
as desired (where $C_G = C_{\text{QSP}} C_T \sqrt{C_\beta}/(2 \Delta)$). This concludes the proof of Lemma~\ref{lemma: qss_finite_dr}.

\subsection{Proof of Lemma \texorpdfstring{\ref{lemma: ham_trans}}{S2}, ESU Hamiltonian Transformation}\label{sec: ham_trans_proof}

\setstretch{1.1}
In this subsection, we provide an explicit sequence of gates which transforms time evolution under $H(t)$ into time evolution under $H^{(k)}_{\text{ESU}}(t)$, proving Lemma~\ref{lemma: ham_trans}. This analysis is for theoretical completeness of our argument in the gate-based model described in Sec.~\ref{sec: sensing_protocol}. In many practical situations, it is possible to engineer the desired Hamiltonian by alternative techniques, using capabilities which depend on the experimental platform of interest, see Sec.~\ref{SM:Exp_Plat}.

Our proof has two steps. First, we use $H(t)$ to simulate $\chi(t) H(t)$, where $\chi(t)$ is an arbitrary differentiable function taking values in $[0,1]$. Then, we use $\chi(t) H(t)$ to simulate $H_{\text{ESU}}^{(k)}(t)$. Both constructions assume we can apply arbitrarily many instantaneous gates (though we will bound the simulation error for any finite number of gates).

\setstretch{1}
We first simulate $\chi(t) H(t)$ for duration $T$. 
The key idea is to do sufficiently fast dynamical decoupling; 
we present a quantitative analysis for completeness.
Concretely, divide $[0,T]$ into $P$ intervals of width $\Delta t$, and let $t_j = j \Delta t$ for $j = 0, \hdots, P$. Evolve under $H(t)$ while applying $X$ gates at times $t_j + (1 + \chi(t_j)) \Delta t /2$ for $j = 0, \hdots, P - 1$, as well as times $t_j$ for $j = 1, \hdots, P$. The effect of this pulse sequence (as can be seen in the toggling frame) is that in the $j$-th interval, we evolve forward under $H(t)$ for duration $(1 + \chi(t_j)) \Delta t/2$ and backward for duration $(1 - \chi(t_j)) \Delta t/2$. Consider the distance in operator norm between the resulting unitary and the unitary describing time evolution under $\chi(t) H(t)$. Using triangle inequality, this distance is bounded by the following expression:

\begin{align}\label{eq: dd_trotter}
    B \sum_{j = 0}^{P - 1} \left| \int_{t_j}^{t_{j+1}} dt~\chi(t) \cos(\omega t + \phi) - \left( \int_{t_j}^{t_j + (1 + \chi(t_j)) \Delta t/2} dt \cos(\omega t + \phi) -  \int_{t_j + (1 + \chi(t_j)) \Delta t/2}^{t_{j+1}} dt \cos(\omega t + \phi) \right) \right|
\end{align}
Consider the $j$-th term in this sum. We will prove a $\phi$-independent upper bound on this term, so we may assume it is non-negative and drop the absolute value (otherwise consider $\phi \to \phi + \pi$). Let $\chi_{\max}$ and $\chi_{\min}$ denote the maximum and minimum of $\chi(t)$ over $[t_j, t_{j+1}]$. Define $c_{\max}$ and $c_{\min}$ equivalently for  $\cos(\omega t + \phi)$. Suppose that $c_{\max} \ge 0$. By individually bounding the three integrals in the $j$-th term, we obtain an overall upper bound of
\begin{align}
    \chi_{\max} c_{\max} \Delta t - \frac{1 + \chi(t_j)}{2} c_{\min} \Delta t + c_{\max} \frac{1 - \chi(t_j)}{2} \Delta t \le (\chi_{\max} - \chi_{\min}) c_{\max} \Delta t + \frac{1 + \chi_{\max}}{2} (c_{\max} - c_{\min}) \Delta t.
\end{align}
A similar analysis shows that if $c_{\max} < 0$, then the $j$-th term is bounded by
\begin{align}
    \chi_{\min} c_{\max} \Delta t - \frac{1 + \chi(t_j)}{2} c_{\min} \Delta t + c_{\max} \frac{1 - \chi(t_j)}{2} \Delta t \le (\chi_{\max} - \chi_{\min}) (-c_{\max}) \Delta t + \frac{1 + \chi_{\max}}{2} (c_{\max} - c_{\min}) \Delta t.
\end{align}
By the mean value theorem, $\chi_{\max} - \chi_{\min} \le \lambda \Delta t$, where $\lambda = \sup_{t \in [0,1]} |\chi'(t)|$ (which is constant by assumption). Similarly, $c_{\max} - c_{\min} \le \omega \Delta t$. Therefore, the $j$-th term is bounded by $(\lambda + \omega) \Delta t^2$, and the overall error is at most
\begin{align}
    B P (\lambda + \omega) \Delta t^2 = \frac{B T^2 (\lambda + \omega)}{P}
\end{align}
which goes to zero as $P \to \infty$.

Next, we simulate $H_{\text{ESU}}^{(k)}(t)$ using $\chi(t) H(t)$ (for duration $T$). Note that $H_{\text{ESU}}^{(k)}(t) = R(t) \chi(t) H(t) R(t)^\dagger$, where $R(t) = e^{i \omega_k t Z/2} H$ and $H$ denotes the Hadamard gate. Again, divide $T$ into $P$ intervals of width $\Delta t$ and let $t_j = j \Delta t$ for $j = 0, \hdots, P$. Evolve under $\chi(t) H(t)$ and apply $R^\dagger(t_0)$ at time $t_0 = 0$, $R^\dagger(t_j) R(t_{j-1})$ at times $t_j$ for $j = 1, \hdots, P-1$, and $R(t_{P-1})$ at time $t_P = T$. The distance in operator norm between the resulting unitary and the unitary describing time evolution under $H_{\text{ESU}}^{(k)}(t)$ is 
\begin{align}
    \left\| \mathcal{T} \text{exp}\left(-i \int_0^T dt~H_{\text{ESU}}^{(k)}(t) \right) - \prod_{j = 0}^{P-1} \text{exp}\left(-i \int_{t_j}^{t_{j+1}} dt~R(t_j) \chi(t) H(t) R^\dagger(t_j) \right) \right\|.
\end{align}
Using triangle inequality and submultiplicativity of the operator norm, this is at most
\begin{align}
    \sum_{j = 0}^{P-1} \int_{t_j}^{t_{j+1}} dt~\| \chi(t) \left(R(t) H(t) R^\dagger(t) - R(t_j) H(t) R^\dagger(t_j) \right) \| &= B \sum_{j = 0}^{P-1} \int_{t_j}^{t_{j+1}} dt~\chi(t) |\cos(\omega t + \phi)| |2 \sin(\omega_k (t-t_j)/2)|
    \\
    &\le\frac{B \omega_c}{2} \frac{T^2}{P}
\end{align}
which again goes to zero as $P \to \infty$. By combining the two steps, this completes the construction.

\subsection{Proof of Lemma \texorpdfstring{\ref{lemma: esu_error_new}}{S3}, ESU Error Bound}\label{sec: esu_error_proof}

In this subsection, we prove Lemma~\ref{lemma: esu_error_new}. For simplicity, we write $\delta_k = \delta$, $H_{\text{ESU}}^{(k)}(t) = H_{\text{ESU}}(t)$, $U_{\text{ESU}}^{(k)}(T) = U_{\text{ESU}}(T)$, etc. We will show that $U_{\text{ESU}}(T)$ approximately implements a $\phi$-independent rotation as long as $\chi(t)$ is a \emph{bump function}, a function which is smooth yet compactly supported on $[0,T]$. Our proof has two main steps. The first step is to apply the rotating wave approximation to $H_{\text{ESU}}(t)$, which gives rise to the error $\RWAerror$. The second step use a strong variant of the adiabatic theorem to bound the error between the resulting unitary and the closest $Z$ rotation (in operator norm). When $\chi(t)$ is a bump function, we can show that this error decays stretched-exponentially in $|\delta|/B$, giving rise to the error $\SAerror$. Moreover, the rotation angle is given by the dynamical phase of the evolution, up to corrections of size $B / |\delta|$.

We begin by bounding the operator distance between  $U_{\text{ESU}}(T)$ and $U_{\text{RWA}}(T)$, the unitary which implements time evolution for $0 \le t \le T$ under the following Hamiltonian:
\begin{align}
    H_{\text{RWA}}(t) = \frac{B}{2} \chi(t) \left( \cos(\delta t - \phi) X - \sin(\delta t - \phi) Y \right).
\end{align}
In quantum optics, replacing $H_{\text{ESU}}(t)$ by $H_{\text{RWA}}(t)$ is known as the \emph{rotating wave approximation} (RWA), since it is equivalent to approximating a linearly-polarized field by a circularly-polarized field~\cite{scully:1997}. We prove the following bound:

\begin{prop}[RWA Error]\label{prop: rwa_error} Suppose that $\chi(t) \in [0,1]$ is a bump function, smooth and compactly supported on $[0,T]$. Further, suppose that $|\delta| \le \omega$. There exists a constant $\nu$ such that, as long as $\omega T \ge \nu$, we have
\begin{align}
    \|U_{\text{ESU}}(T) - U_{\text{RWA}}(T) \| \le C_{\text{RWA}} \frac{B}{\omega}(1 + B T)
\end{align}
for some constant $C_{\text{RWA}} \ge 0$.
\end{prop}

\begin{proof}    
We will use the following result of~\cite{burgarth:2022}:
\begin{lemma}[Lemma 1 in~\cite{burgarth:2022}]\label{lemma:onebound_new} Let $H_1(t)$ and $H_2(t)$ be two time-dependent Hamiltonians, and let $U_1(T)$ and $U_2(T)$ denote the corresponding unitary operators describing time evolution for $0 \le t \le T$. Define
\begin{align}
    S(t) = \int_0^t dt' \ H_2(t') - H_1(t').
\end{align}
Then the following holds:
\begin{align}\label{eq: oneboundstatement_new}
    \|U_2(T) - U_1(T)\| \le \left( \max_{0 \le t \le T} \|S(t)\| \right) \left( 1 + \int_0^T dt \|H_1(t)\| + \int_0^T dt \|H_2(t)\| \right).
\end{align}
\end{lemma}
\noindent
Lemma~\ref{lemma:onebound_new} is especially useful when $H_1(t)$ and $H_2(t)$ are close on average over time.
In particular, it provides a meaningful bound if $H_1(t) - H_2(t)$ oscillates rapidly within the duration $T$, which is the case when the RWA is applicable. To apply the lemma, we let $H_2(t) = H_{\text{ESU}}(t)$ and $H_1(t) = H_{\text{RWA}}(t)$. $\|S(t)\|$ is given explicitly by
\begin{align}
    \| S(t) \| = \frac{B}{2} \left| \int_0^t dt'~\chi(t') e^{i (\delta + 2\omega) t'} \right|
\end{align}
for any $0 \le t \le T$. We can relate this expression to the Fourier transform $\hat{\chi}(k) = \int_{-\infty}^{\infty} dt'~\chi(t') e^{i k t'}$ of $\chi$ as follows. First, define the rectangle function $r_t(t')$ as
\begin{align}
    r_t(t') = 
    \begin{cases}
        1 & -t \le t' \le t \\
        0 & \text{else}
    \end{cases}.
\end{align}
Then, by applying the Fourier convolution theorem, we can re-express $\|S(t)\|$ in the following form:
\begin{align}
    \| S(t) \| &= \frac{B}{2} \left| \int_{-\infty}^{\infty} dt'~\chi(t') r_t(t') e^{i (\delta + 2\omega) t'} \right| =  \frac{B}{4\pi} \left| \int_{-\infty}^{\infty} dk ~\hat{\chi}(k) \hat{r}_t(\delta + 2\omega - k) \right|.
\end{align}
Applying triangle inequality to this integral and substituting an explicit expression for $\hat{r}_t$, the Fourier transform of the rectangle function, we have
\begin{align}\label{eq:fourier_before_split_new}
    \|S(t)\| \le \frac{B}{2 \pi} \int_{-\infty}^{\infty} dk |\hat{\chi}(k)| \frac{|\sin((\delta + 2 \omega - k) t)|}{|\delta + 2 \omega - k|}.
\end{align}
Next, we use a basic fact from Fourier analysis: if a function $f$ has an integrable $p$-th derivative, then $\hat{f}(k) = O(|k|^{-p})$~\cite{stein:2011}. Therefore, as $\chi$ is both smooth and compactly supported, the Fourier transform $\hat{\chi}$ must decay faster than any polynomial. Concretely, for all $p \in \mathbb{N}$, there exist constants $\nu_p$ and $F_p$ such that
\begin{align}
    |\hat{\chi}(k)| \le \frac{F_p T}{(kT)^p}
\end{align}
as long as $|k| \ge \nu_p / T$.~\footnote{The $T$ dependence is obtained by changing to unitless time $t/T$ and applying the decay result.} 
Let $\nu = 2 \nu_2$, so that $\omega T \ge 2 \nu_2$, and define $F = 2 F_2$. Recalling that $|\delta| \le \omega$ by assumption, we have $(\delta + 2\omega)T/2 \ge \omega T/2 \ge \nu_2$. Split the integral~\eqref{eq:fourier_before_split_new} as follows:
\begin{align}
    \|S(t)\| \le \frac{B}{2 \pi} \int_{|k| > (\delta  + 2\omega)/2} dk |\hat{\chi}(k)| \frac{|\sin((\delta + 2 \omega - k) t)|}{|\delta + 2 \omega - k|} + \frac{B}{2 \pi} \int_{|k| \le (\delta  + 2\omega)/2} dk |\hat{\chi}(k)| \frac{|\sin((\delta + 2 \omega - k) t)|}{|\delta + 2 \omega - k|}.
\end{align}
The first integral can be bounded using $\sinc(x) \le 1$ as well as the decay of $|\hat{\chi}(k)|$ as follows:
\begin{align}
    \frac{B}{2 \pi} \int_{|k| > (\delta  + 2\omega)/2} dk \frac{F_2 T}{(kT)^2} t = \frac{2F_2}{\pi} \frac{t}{T} \frac{B}{\delta + 2 \omega} \le \frac{F}{\pi} \frac{B}{\omega}.
\end{align}
The second integral can be bounded using $|\sin(x)| \le 1$ as well as the assumption on the integration range as follows:
\begin{align}
    \frac{B}{2 \pi} \int_{|k| \le (\delta  + 2\omega)/2} dk |\hat{\chi}(k)| \frac{2}{\delta + 2 \omega} \le \frac{1}{\pi} \frac{B}{\omega} \int_{-\infty}^{\infty} |\hat{\chi}(k)|.
\end{align}
Finally, we use another fact from Fourier analysis: %
if a function $f$ and its derivative $f'$ are both in 
$L^2(\mathbb{R})$, then the Fourier transform of $f$ is in $L^1(\mathbb{R})$. Since $\chi$ is a bump function, the conditions hold, and the absolute integral of $\hat{\chi}(k)$ is some constant $F'$. Therefore, we have
\begin{align}
    \|S(t)\| \le \frac{(F + F')}{\pi} \frac{B}{\omega}.
\end{align}
Returning to~\eqref{eq: oneboundstatement_new}, it is easy to see that both $\int_0^T dt \|H_1(t)\|$ and $\int_0^T dt \|H_2(t)\|$ are at most $BT$. Choosing $C_{\scriptscriptstyle \text{RWA}} = 2 (F + F') / \pi$ completes the proof. 
\end{proof}

Proposition~\ref{prop: rwa_error} explains the origin of the error term $\RWAerror$; next we must bound the operator distance between $U_{\text{RWA}}(T)$ and the closest $Z$ rotation $e^{-i \theta Z}$. We begin with some simple observations which make this task easier. First, we can work in the frame rotating at frequency $\delta$; in this frame, the Hamiltonian $H_{\text{RWA}}(t)$ transforms to 
\begin{align}
    \widetilde{H}_{\text{RWA}}(t) = \frac{\delta}{2} Z + \frac{B}{2} \chi(t) (\cos(\phi) X + \sin(Y))
\end{align}
so that all the time dependence is in $\chi(t)$. Let $\widetilde{U}_{\text{RWA}}(T)$ denote $U_{\text{RWA}}(T)$ in this rotating frame; then
\begin{align}
    \|U_{\text{RWA}}(T) - e^{-i \theta Z} \| = \| \widetilde{U}_{\text{RWA}}(T) - e^{-i (\theta + \delta T/2) Z} \| \le \sqrt{2} \max_{i \in \{0,1\}} \| \widetilde{U}_{\text{RWA}}(T) \ket{i} - e^{-i (\theta + \delta T/2) Z} \ket{i} \|_2
\end{align}
where $\| \cdot \|_2$ denotes Euclidean norm, and the inequality holds for any two-by-two matrices. In fact, as $\widetilde{U}_{\text{RWA}}(T) \in \mathrm{SU}(2)$ (since $\widetilde{H}_{\text{RWA}}(t)$ is traceless), the two terms over which the maximization is taken are in fact equal, so it suffices to take $\ket{i} = \ket{0}$. As a final simplification, we shift $\widetilde{H}_{\text{RWA}}(t)$ so that it always has a zero eigenvalue, defining
\begin{align}
    H_0(t) = \widetilde{H}_{\text{RWA}}(t) - \frac{\delta}{2} \sqrt{1 + (B \chi(t) / \delta)^2}.
\end{align}
In conclusion, letting $U_0(T)$ denote the unitary corresponding to time evolution under $H_0(t)$ for $0 \le t \le T$, we have
\begin{align}\label{eq: error_to_bound}
    \| U_{\text{RWA}}(T) - e^{-i \theta Z} \| \le \sqrt{2} \| U_0(T) \ket{0} - e^{- i \eta} \ket{0} \|_2 \quad \text{with} \quad \eta = \theta - \left(\frac{\delta}{2} \int_0^T dt \sqrt{1 + (B \chi(t) / \delta)^2} - \frac{\delta T}{2} \right).
\end{align}

Note that, since $\chi(t)$ is compactly supported on $[0,T]$, it must satisfy $\chi(0) = \chi(T) = 0$. This means that $\ket{0}$ is a zero-energy eigenstate of $H_0(t)$ at the initial and final times. Moreover, $H_0(t)$ is gapped for all $0 \le t \le T$, with minimum gap $|\delta|$. These two observations suggest that we can use the adiabatic theorem to bound the error~\eqref{eq: error_to_bound}. In particular, we will use a variant of the adiabatic theorem, which was originally proven by~\cite{nenciu:1993} using the adiabatic expansion method of~\cite{hagedorn:2002}, then tightened by~\cite{ge:2016}. This variant, which is sometimes referred to as a \emph{superadiabatic} theorem, proves that under certain conditions, the diabatic error between the time-evolved state and the instantaneous ground state decays stretched-exponentially in the minimum gap $|\delta|$ and total evolution time $T$. 

We begin by stating the superadiabatic theorem of~\cite{ge:2016}. Consider a smooth, finite-dimensional time-dependent Hamiltonian $H_T(t)$ defined for $0 \le t \le T$. As is typically the case when proving any form of adiabatic theorem, we assume $H_T(t)$ can be written in unitless time, i.e., $H_T(t) = H(s)$ for $s = t/T$, where $H(\cdot)$ does not depend on $T$~\cite{albash:2018}. Let $\ket{\phi(s)}$ denote an instantaneous zero energy eigenstate of $H(s)$, and let $\ket{\psi(s)}$ denote the solution of the time-dependent Schr\"odinger equation starting from the zero energy state:
\begin{align}
    i T^{-1} |\dot{\psi}(s) \rangle = H(s) \ket{\psi(s)} \quad \text{with} \quad \ket{\psi(0)} = \ket{\phi(0)}.
\end{align}
where $|\dot{\psi}(s) \rangle = \dv{s} \ket{\psi(s)}$. Without loss of generality (by freedom of global phase choice), we can assume $\langle \phi(s) | \dot{\phi}(s) \rangle = 0$ for all $0 \le s \le 1$. Suppose that $H(s)$ has the following properties:

\begin{condition}[Conditions on $H(s)$]\leavevmode\label{cond: ham_conds}
\begin{enumerate}
    \item $\ket{\phi(s)}$ is non-degenerate and has energy gap at least $|\delta| > 0$ for all $0 \le s \le 1$.

    \item $H^{(k)}(0) = H^{(k)}(1) = 0$ for all $k \ge 1$, where $H^{(k)}(s)$ denotes the $k$-th derivative of $H(s)$.

    \item There exist constants $K, c, \alpha \ge 0$ such that $\|H^{(k)}(s)\| \le K c^k \frac{[k!]^{1 + \alpha}}{(k+1)^2}$ for all $0 \le s \le 1$ and all $k \ge 1$.
\end{enumerate}
\end{condition} 
\noindent
Supposing these conditions to be true, the following theorem holds:
\begin{theorem}[Theorem 1 in~\cite{ge:2016}]\label{thm: cirac_superad} Assuming $H(s)$ satisfies Condition~\ref{cond: ham_conds}, then
\begin{align}
    \min_{\eta} \| \ket{\psi(1)} - e^{- i \eta} \ket{\phi(1)} \|_2 \le 8 c e \frac{K}{|\delta|} \left(\frac{4 \pi^2}{3} \right)^3 \mathrm{exp}\left\{- \left(\frac{1}{4 e c^2} \left(\frac{3}{4 \pi^2}\right)^5 T \frac{|\delta|^3}{K^2}\right)^{\frac{1}{1 + \alpha}}\right\}.
\end{align}
\end{theorem}
\noindent
In addition to bounding the error in~\eqref{eq: error_to_bound} for the closest possible $Z$ rotation, minimized over values of the rotation angle, we need to know the approximate value of this angle. We expect that this angle is approximately the dynamical phase of the evolution. In other words, we expect that the value $\eta_*$ which minimizes~\eqref{eq: error_to_bound} over all choices of $\eta$ is close to zero. We make this intuition precise by proving the following proposition extending Theorem~\ref{thm: cirac_superad}:

\begin{prop}[Superadiabatic Phase] \label{lemma: superad_phase} Let $\eta_* = \mathrm{argmin}_{\eta}\left(\| \ket{\psi(1)} - e^{- i \eta} \ket{\phi(1)} \|_2\right)$, where $\eta_* \in (-\pi, \pi]$. Then
\begin{align}
        |\eta_*| \le \frac{c \pi}{4} \frac{K}{|\delta|}.
\end{align}
\end{prop}
\begin{proof}
First, if $\langle \psi(1)|\phi(1)\rangle = 0$, then the Euclidean distance is the same for all choices of $\eta$, so we may choose $\eta_* = 0$. Therefore, assume $\langle \psi(1)|\phi(1)\rangle \neq 0$. $\eta_*$ is the angle such that $\langle \psi(1) | \phi(1) \rangle = e^{i \eta_*} | \langle \psi(1) | \phi(1) \rangle |$. Therefore, we have
\begin{align}
    \| \ket{\psi(1)} - \ket{\phi(1)} \|_2^2 &= 2 - 2 \cos(\eta_*) | \langle \psi(1) | \phi(1) \rangle | .
\end{align}
After rearranging and observing that $| \langle \psi(1) | \phi(1) \rangle | \le 1$ and $\| \ket{\psi(1)} - \ket{\phi(1)} \|_2^2 \le 2$, we have
\begin{align}
    \cos(\eta_*) = \frac{1}{| \langle \psi(1) | \phi(1) \rangle |} \left( 1 - \frac{1}{2} \| \ket{\psi(1)} - \ket{\phi(1)} \|_2^2 \right) \ge 1 - \frac{1}{2} \| \ket{\psi(1)} - \ket{\phi(1)} \|_2^2 .
\end{align}
Combining this inequality with $\cos(\eta_*) \le 1 - 2 \eta_*^2/ \pi^2$, which holds for any $\eta_* \in (-\pi, \pi]$, we conclude that 
\begin{align}
    |\eta_*| \le \frac{\pi}{2} \| \ket{\psi(1)} - \ket{\phi(1)} \|_2.
\end{align}
Now let $U(s)$ denote the unitary corresponding to time evolution under $H(s')$ for $0 \le s' \le s$; then
\begin{align}
    \| \ket{\psi(1)} - \ket{\phi(1)} \|_2 &= \| \ket{\phi(0)} - U^\dagger(1) \ket{\phi(1)} \|_2
    \\
    &= \big\| \int_0^1 \textrm{d}s \dv{s} \left(U^\dagger(s) \ket{\phi(s)} \right) \big\|_2
    \\
    &= \big\| \int_0^1 \textrm{d}s \left( i T U^\dagger(s) H(s) \ket{\phi(s)} + U^\dagger(s) | \dot{\phi}(s) \rangle \right) \big\|_2.
\end{align}
Noting that $H(s) \ket{\phi(s)} = 0$ and applying triangle inequality, we conclude that
\begin{align}
    \| \ket{\psi(1)} - \ket{\phi(1)} \|_2 \le \int_0^1 \textrm{d}s~\| | \dot{\phi}(s) \rangle \|_2.
\end{align}
Next, since $H(s) \ket{\phi(s)} = 0$ and $\langle \dot{\phi}(s) | \phi(s) \rangle = 0$, it follows that $| \dot{\phi}(s) \rangle = - H^{-1}(s) H^{(1)}(s)\ket{\phi(s)}$, where $H^{-1}(s)$ denotes the Moore-Penrose pseudoinverse of $H(s)$. This implies the simple bound
\begin{align}
    \| |\dot{\phi}(s) \rangle \|_2 \le \| H^{-1}(s) \| \| H^{(1)}(s) \| .
\end{align}
By the third condition in Condition~\ref{cond: ham_conds}, we have $\| H^{(1)}(s) \| \le K c / 4$. To bound the operator norm of $H^{-1}(s)$, we use the Cauchy integral representation of the Moore-Penrose pseudoinverse~\cite{avrachenkov:2013}:
\begin{align}
    H^{-1}(s) = \frac{1}{2 \pi i} \oint_{\Gamma} dz  \frac{(H(s) - z)^{-1}}{z}
\end{align}
where $\Gamma = \{z \in \mathbb{C} : |z| = |\delta|/2\}$. We therefore obtain
\begin{align}
     |\eta_*| \le \frac{\pi}{2} \int_0^1 \textrm{d}s~\| H^{-1}(s) \| \| H^{(1)}(s) \| \le \frac{c \pi}{4} \frac{K}{|\delta|}
\end{align}
as desired.
\end{proof}

Using these general results, we can bound the error in~\eqref{eq: error_to_bound}, as long $H_0(t)$ satisfies the necessary conditions to apply Theorem~\ref{thm: cirac_superad}. We choose $\chi(t) = e^{-1/(\frac{t}{T}(1 - \frac{t}{T}))}$, a well-known example of a bump function on $[0,T]$. We abuse notation slightly by writing $\chi(s) = e^{-1/(s(1-s))}$ for $\chi(\cdot)$ in unitless time and do the same for $H_0(s)$. We first bound the derivatives of $\chi(s)$, then we show how these bounds imply that $H_0(s)$ satisfies Condition~\ref{cond: ham_conds}.

\begin{prop}[Derivatives of $\chi(s)$]\label{lemma: chi_gevrey} Let $\chi(s) = e^{-1/(s(1-s))}$. For all $0 \le s \le 1$ and all $k \ge 1$, we have
\begin{align}
    |\chi^{(k)}(s)| \le A c^k \frac{[k!]^{1+\alpha}}{(k+1)^2}
\end{align}
with $A = e^{-1}$, $c = 12$, and $\alpha = 1$.
\end{prop}
\begin{proof}
    Since $\chi(s)$ is an even function about $s = 1/2$, all its derivatives are also even or odd, so it suffices to consider $s \le \frac{1}{2}$. Since $\chi(s)$ is a bump function, all its derivatives must vanish at $s = 0$, so we may also assume $s \neq 0$. Let $\Gamma_s = \{z \in \mathbb{C}: |z - s| = s/2\}$.  $\chi(s)$ admits a Taylor series at $s$ with radius of convergence equal to $s$, so we can analytically continue $\chi(s)$ to $\Gamma_s$ and its interior and compute the derivatives of $\chi(s)$ using Cauchy's integral theorem:
    \begin{align}
        |\chi^{(k)}(s)| &= \left| \frac{k !}{2 \pi i} \oint_{\Gamma_s} dz \frac{e^{-1/(z(1-z))}}{(z-s)^{k+1}} \right| 
        \\
        &\le k! \left(\frac{2}{s}\right)^k \max_{z \in \Gamma_s} \left| e^{-1/(z(1-z))} \right|
        \\
        &= k! \left(\frac{2}{s}\right)^k \max_{z \in \Gamma_s} e^{-\operatorname{Re}(1/(z(1-z)))}.
    \end{align}
    We can write any $z \in \Gamma_s$ as $z = s + \frac{s}{2} e^{i \alpha}$ for $\alpha \in [0,2\pi)$, so that
    \begin{align}
        \operatorname{Re}\left(\frac{1}{z(1-z)}\right) = \frac{4(4 - 4s + (4s-2) \cos(\alpha) + s \cos(2\alpha))}{s(5 + 4\cos(\alpha))(4 + s(5s - 8) + 4(s-1)s\cos(\alpha))} \ge \frac{2}{3s}
    \end{align}
    where the lower bound is obtained by minimizing over $\alpha$. This implies the following upper bound on the derivatives of $\chi(s)$:
    \begin{align}
        |\chi^{(k)}(s)| &\le k! \left(\frac{2}{s}\right)^k e^{-2/(3s)} \le k! \left(\frac{3}{e}\right)^k k^k
    \end{align}
    where we have used $e^{-x} \le \left(\frac{k}{ex} \right)^k$ for $k, x > 0$. Using the inequality $k^k \le e^k  k!/ e$, we obtain 
    \begin{align}
        |\chi^{(k)}(s)| \le \frac{1}{e} 3^k [k!]^2
        \le \frac{1}{e} 12^k \frac{[k!]^2}{(k+1)^2}
    \end{align}
    where in the last step we simply observe that $4^k/(k+1)^2 \ge 1$ for all $k \ge 1$.
\end{proof}

Finally, we show that $H_0(s)$ satisfies the conditions in Condition~\ref{cond: ham_conds}.

\begin{prop}[Conditions on $H_0(s)$] Suppose that $B / |\delta| \le \mu = 3 e / (4 \pi^2)$. Then $H_0(s)$ satisfies Condition~\ref{cond: ham_conds} with $K = B / e$, $c = 12$, and $\alpha = 1$.
\end{prop}
\begin{proof}

$H_0(s)$ has a gap of at least $|\delta| > 0$ for all $0 \le s \le 1$, so the first property in Condition~\ref{cond: ham_conds} holds. Applying Fa\`a di Bruno's formula to $H_0(s)$ and noting that $\chi^{(k)}(0) = \chi^{(k)}(1) = 0$ for all $k \ge 1$, we see that the second property in  Condition~\ref{cond: ham_conds} holds, as well. We now consider the third property. We first observe that 
\begin{align}\label{eq: third_condition}
    \| H_0^{(k)}(s) \| \le \frac{B}{2} |\chi^{(k)}(s)| +  \frac{|\delta|}{2} \left| \dv[k]{s} \sqrt{1 + (B \chi(s)/ \delta)^2} \right|.
\end{align}
The growth of the first term is bounded by Proposition~\ref{lemma: chi_gevrey}. To bound the growth of the second term, we use the Taylor series expansion of $\sqrt{1+x^2}$, which converges as long as $|x| \le 1$. We drop the constant term from the expansion, since we are taking a derivative:
\begin{align}\label{eq: square_root_gevrey}
    \left| \dv[k]{s} \sqrt{1 + (B \chi(s)/ \delta))^2} \right| &= \left| \dv[k]{s} \sum_{n=1}^\infty \binom{1/2}{n} \left(\frac{B \chi(s)}{\delta} \right)^{2n} \right|
    \le -\sum_{n=1}^\infty (-1)^n \binom{1/2}{n} \left( \frac{B}{\delta} \right)^{2n} \left|\dv[k]{s} \chi(s)^{2n} \right|
\end{align}
where we use $| \binom{1/2}{n} | = (-1)^{n+1}  \binom{1/2}{n}$.

We next prove a bound on the growth of derivatives of $\chi(s)$, showing that for any integer $N \ge 1$ we have
\begin{align}\label{eq: inductive_claim}
    \left| \dv[k]{s} \chi(s)^N \right| \le A^N \left(\frac{4 \pi^2}{3}\right)^{N-1} c^k \frac{[k!]^{1 + \alpha}}{(k+1)^2}.
\end{align}
We proceed by induction. The case $N = 1$ follows from Proposition~\ref{lemma: chi_gevrey}. Therefore, assume $N > 1$ and observe that
\begin{align}
    \left| \dv[k]{s} \chi(s)^N \right| &= \left| \sum_{l = 0}^k \binom{k}{l} \dv[l]{s} \chi(s) \dv[k-l]{s} \chi^{N-1}(s) \right| \le A^N \left(\frac{4 \pi^2}{3}\right)^{N-2} c^k \sum_{l = 0}^k \binom{k}{l} \frac{[l!]^{1+\alpha} [(k-l)!]^{1+\alpha}}{(l+1)^2 (k-l+1)^2}
\end{align}
by the inductive hypothesis. Now, observe that $\binom{k}{l} [l!]^{1+\alpha} [(k-l)!]^{1+\alpha} = \binom{k}{l}^{-\alpha}  [k!]^{1+\alpha} \le [k!]^{1+\alpha}$, so we have
\begin{align}\label{eq: sum_before_split}
    \sum_{l = 0}^k \binom{k}{l} \frac{[l!]^{1+\alpha} [(k-l)!]^{1+\alpha}}{(l+1)^2 (k-l+1)^2} \le [k!]^{1 + \alpha} \sum_{l=0}^k \frac{1}{{(l+1)^2 (k-l+1)^2}}.
\end{align}
Next, for each choice of $l$, at least one of $l + 1$ or $k - l + 1$ is greater than or equal to $\lfloor k/2 \rfloor + 1 \ge (k+1)/2$. By considering these two cases individually, we can upper bound the sum in~\eqref{eq: sum_before_split} as
\begin{align}
    \sum_{l=0}^k \frac{1}{{(l+1)^2 (k-l+1)^2}} \le \frac{2}{((k+1)/2)^2} \sum_{l = 1}^{\lfloor k/2 \rfloor} \frac{1}{l^2} \le \frac{4 \pi^2}{3} \frac{1}{(k+1)^2}
\end{align}
where we have used $\sum_{l=1}^\infty l^{-2} = \pi^2/6$. This completes the proof of~\eqref{eq: inductive_claim}.

Combining the bound~\eqref{eq: inductive_claim} with~\eqref{eq: square_root_gevrey}, we obtain
\begin{align}
    \left| \dv[k]{s} \sqrt{1 + (B \chi(s)/\delta)^2} \right| &\le \frac{3}{4 \pi^2} \left[ -\sum_{n = 1}^\infty \binom{1/2}{n} \left(- \left( \frac{4 \pi^2}{3} A \frac{B}{\delta} \right)^2 \right)^{n} \right] c^k \frac{[k!]^{1 + \alpha}}{(k+1)^2}
    \\
    &= \frac{3}{4 \pi^2} \left[1 - \sqrt{1 - \left( \frac{4 \pi^2}{3} A \frac{B}{\delta} \right)^2 } \right] c^k \frac{[k!]^{1 + \alpha}}{(k+1)^2}
\end{align}
where the last step is true because $\frac{4 \pi^2}{3} A \frac{B}{|\delta|} \le 1$ by assumption. Finally, using $1 - \sqrt{1 - x} \le x$ for $x \le 1$, we obtain
\begin{align}
    \left| \dv[k]{s} \sqrt{1 + (B \chi(s)/ \delta)^2} \right| &\le \frac{4 \pi^2}{3} A^2 \left(\frac{B}{\delta}\right)^2 c^k \frac{[k!]^{1 + \alpha}}{(k+1)^2} \le A \frac{B}{|\delta|} c^k \frac{[k!]^{1 + \alpha}}{(k+1)^2}
\end{align}
Plugging this back into~\eqref{eq: third_condition}, we find that
\begin{align}
     \| H_0^{(k)}(s) \| \le  A B c^k \frac{[k!]^{1 + \alpha}}{(k+1)^2}
\end{align}
which concludes the proof.
\end{proof}

Since $H_0(s)$ satisfies the conditions in Condition~\ref{cond: ham_conds}, we can apply Theorem~\ref{thm: cirac_superad} and Proposition~\ref{lemma: superad_phase} to complete the proof, with constants $C, D, E$ given by 
\begin{align}
    C = 96 \sqrt{2} \left(\frac{4 \pi^2}{3} \right)^3 \quad \quad D = \frac{e}{576} \left(\frac{3}{4 \pi^2} \right)^5 \quad \quad  E = \frac{3 \pi}{e}.
\end{align}

\newpage

\section{Compatibility of QSS with Quantum Error Correction}\label{sec: qec}

In this section, we show that QSS is compatible with existing efforts to leverage quantum error correction (QEC) for quantum sensing in the presence of decoherence. In particular, we demonstrate that in the presence of Markovian noise on the sensing qubits which is ``distinguishable'' from the signal Hamiltonian in a precise sense, one can use QEC to make the QSS algorithm robust to noise. The resulting protocol solves $\ACProb$ with the same sensing time (up to constant factors) as the noiseless case.

It is well understood that errors play a detrimental role in quantum metrology, the simplest example being the impossibility of achieving the Heisenberg limit and the degradation to the standard quantum limit in the presence of decoherence~\cite{huelga:1997, fujiwara:2008, demkowicz-dobrzanski:2009, demkowicz-dobrzanski:2012, escher:2011, knysh:2011}. Many works have explored the use of QEC for metrology in the presence of restricted noise~\cite{kessler:2014, arrad:2014, dur:2014, sekatski:2017}, and it has been shown that optimal parameter estimation remains possible if the signal Hamiltonian is not within the span of the noise operators~\cite{demkowicz-dobrzanski:2017, zhou:2018, zhou:2021, zhou:2020, layden:2019, zhou:2024}. 

In particular,~\cite{zhou:2018} considers quantum sensing protocols for estimating the coupling strength $\lambda$ of a time-independent Hamiltonian $H = \lambda G$ in the presence of Markovian noise with jump operators $\{L_j\}_{j=1}^r$. 
They prove that it is possible to perform QEC and achieve the Heisenberg limit, i.e, to estimate $\lambda$ with a precision which scales linearly in the sensing time $\tau$, as long as the Hamiltonian satisfies the Hamiltonian-not-in-Lindblad-span (HNLS) condition: 
\begin{align}
    G \notin \mathcal{S} \equiv \text{span}\{\mathbbm{1}, L_j, L_j^\dagger, L_j^\dagger L_k\}_{j,k = 1}^r.
\end{align}
The HNLS condition can be understood as the formal condition that the signal Hamiltonian is distinguishable from the noise itself.
The key idea behind the construction is to decompose the Hamiltonian as 
$\lambda G = \lambda(G_{\parallel} + G_{\perp})$,
where $G_{\parallel} \in \mathcal{S}$ and $G_{\perp} \perp \mathcal{S}$ under the trace inner product, then define a two-dimensional QEC codespace $\mathcal{C}$ which is preserved by evolution under $\lambda G_{\perp}$, but can detect errors generated by the jump operators. If $H$ is a single-qubit Hamiltonian, for example, $\mathcal{C}$ is simply a repetition code with codewords $\ket{+_{\perp}}_S \otimes \ket{0}_A$ and  $\ket{-_{\perp}}_S \otimes \ket{1}_A$, where $\ket{\pm_{\perp}}_S$ denote the eigenstates of $G_{\perp}$, and $A$ is a noiseless ancilla qubit. Quantum jumps manifest as bit flip errors in this code, which are perfectly correctable since $A$ is noiseless. We refer the reader to~\cite{zhou:2018} for details of the general construction.

We now show that QSS is compatible with quantum error correction. This is not immediate, because QSS must sense in superposition: the response of the sensor to the signal is conditional on the state of the quantum computer. This means that the error syndrome can in principle reveal information about the quantum computer state, making our computation incoherent. This is true even if a quantum jump does not occur; evolution under $\lambda G_{\parallel}$ also rotates outside the codespace, so an error detection event indicates that the sensor is susceptible to the signal. 

The resolution to this potential impediment (at least at a theoretical level) is to increase the rate at which quantum error correction is performed. By performing rapid QEC, projection onto $\mathcal{C}$ succeeds with high probability, and the error syndromes do not reveal any information. Concretely, the analysis in~\cite{zhou:2018} shows that by performing a round of error correction every time step $dt$, the resulting dynamics are equivalent to evolution under $\Pi_{\mathcal{C}} H \Pi_{\mathcal{C}}$, up to $O(dt^2)$ corrections, where $\Pi_{\mathcal{C}}$ denotes the projection operator onto $\mathcal{C}$. Therefore, over a total time $\tau$, the error is at most $O(\tau \cdot dt)$, which can be made arbitrarily small by applying QEC more frequently. This analysis still holds for our signal Hamiltonian $H(t) = B \cos(\omega t + \phi) \sum_{i = 1}^{n_S} Z_i$, since the eigenspace of $H(t)$ (and hence the codespace $\mathcal{C}$) are still time-independent. Therefore, by using rapid QEC as the ``innermost'' layer of QSS and redefining all states and operators with respect to the effective qubit defined by $\mathcal{C}$, we can solve $\ACProb$ with the same sensing time scaling as the noiseless case when the HNLS condition is met.

\newpage

\section{Partial Robustness of QSS Speedup with Decoherence}\label{sec: decohere}

\begin{figure}[h!]
    \centering
    \includegraphics[width=0.8\linewidth]{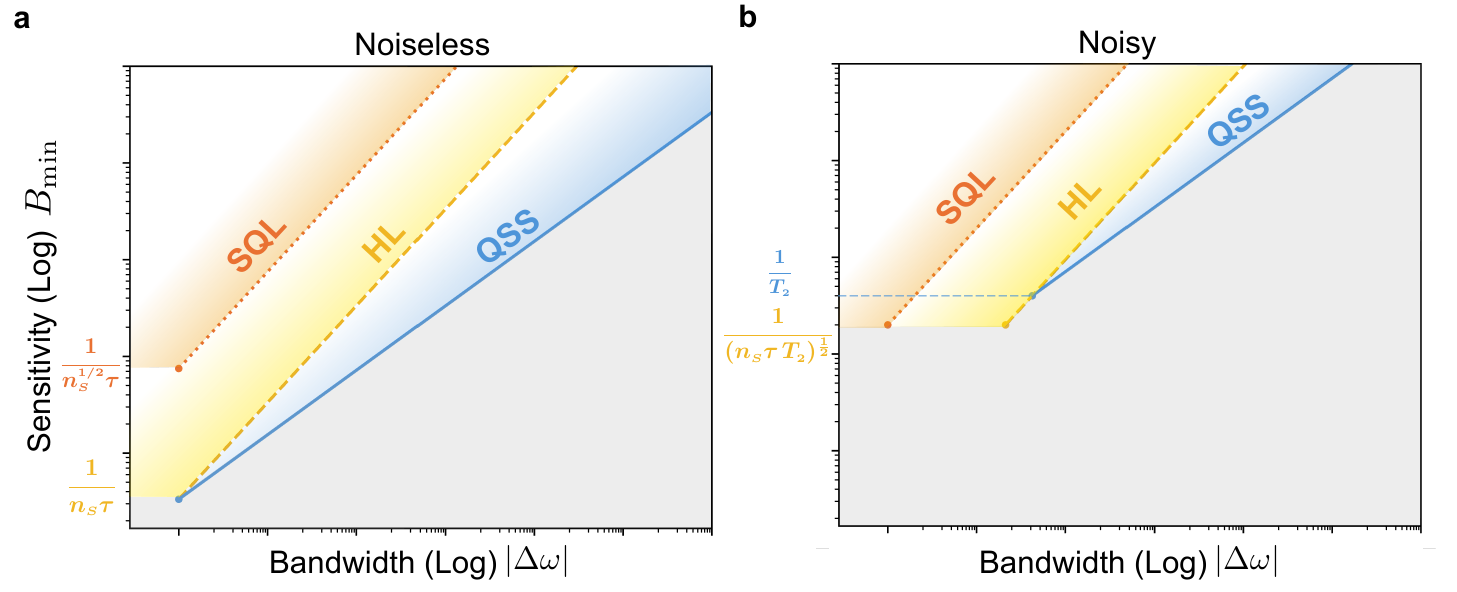}
    \caption{Comparing the achievable sensitivity $B_{\min}$ and bandwidth $|\Delta \omega|$ without decoherence (\textbf{a}, see also FIG. 1c in main text) versus with decoherence (\textbf{b}). We fix the sensing time $\tau$, number of sensors $n_S$, and sensor coherence time $T_2 \sim 1/\Gamma$. Regardless of the presence of decoherence, conventional protocols with unentangled sensors achieve $B_{\min} \sim (n_S \tau)^{-1/2} |\Delta \omega|^{1/2}$ (the standard quantum limit, SQL) and conventional protocols with entangled sensors achieve $B_{\min} \sim n_S^{-1} \tau^{-1/2} |\Delta \omega|^{1/2}$ (the Heisenberg limit, HL). However, in the presence of decoherence, the minimal achievable sensitivity of conventional protocols with entangled or unentangled sensors is the same, scaling as $1/(n_S \tau T_2)^{1/2}$ (due to the reduced effective $T_2^{\text{eff}} = T_2/n_S$ of the GHZ state). Without decoherence, our quantum search sensing (QSS) algorithm achieves $B_{\min} \sim n_S^{-1} \tau^{-2/3} |\Delta \omega|^{1/3}$ (up to polylogarithmic corrections). With decoherence, for $B_{\min} \gtrsim 1/T_2$, the QSS-based protocol in this section achieves $B_{\min} \sim (n_S^2 \tau T_2)^{-1/3}  |\Delta \omega|^{1/3}$ (again, up to polylogarithmic corrections), outperforming conventional approaches.}
    \label{fig:noisy_qss}
\end{figure}

In this section, we consider protocols for AC sensing in the presence of decoherence. We will show that in the presence of local, Markovian noise occurring at a rate $\Gamma$ on each sensor, there is a simple variant of the QSS algorithm which still outperforms conventional sensing protocols in some regimes. In particular, we demonstrate a quantum sensing protocol which solves $\ACProb$ with sensing time
\begin{align}
    \tau = \widetilde{O}\left(\frac{\Gamma}{B_{\min}} \frac{|\Delta \omega|}{(n_S B_{\min})^2} \right).
\end{align}
For $\Gamma/B_{\min}$ sufficiently small, this protocol provides a speedup over conventional sensing (FIG.~\ref{fig:noisy_qss}). Here, we only consider broadband sensing $|\Delta \omega| \ge n_S B_{\min}$, because otherwise conventional sensing is optimal for $\ACProb$, even without noise. 

Our protocol is based on the fact that, for short times, quantum dynamics are not substantially modified by the addition of local noise. Therefore, we can successfully run QSS for a short time $\Tnoise$, detecting signals over a small interval $\Delta \omega_{\scriptscriptstyle \mathrm{noise}}$, and classically repeat to cover the entire bandwidth $\Delta \omega$. 

We model local noise by a Markovian master equation with Lindblad jump operators $\{L_j = \sqrt{\Gamma} M_j \}_{j = 1}^{n_S}$ such that $M_j$ acts non-trivially on the $j$-th sensor only and satisfies $\| M_j \| = 1$. We prove the following lemma bounding the difference between Hamiltonian evolution with and without noise:
\begin{lemma}[Effect of Lindbladian Noise]\label{lemma: effect_of_noise} Let $\mathcal{N}_T(\cdot)$ denote the quantum channel corresponding to Hamiltonian evolution under $H(t)$ for time $0 \le t \le T$. Let $\widetilde{\mathcal{N}}_T(\cdot)$ denote the channel describing evolution under a Markovian master equation with the same $H(t)$, in addition to Lindblad jump operators $\{L_j\}_{j = 1}^r$. Then, for any initial state $\rho_0$, the following bound on trace distance holds:
\begin{align}
    \frac{1}{2} \| \mathcal{N}_T(\rho_0) - \widetilde{\mathcal{N}}_T(\rho_0) \|_1 \le T \sum_{j=1}^r \| L_j^\dagger L_j \|.
\end{align}
\end{lemma}
\noindent For our local noise model, this upper bound is simply $\Gamma n_S T$. 

We now describe our protocol, assuming Lemma~\ref{lemma: effect_of_noise} to be true. By the lemma, as long as QSS terminates in time $\Tnoise = O((\Gamma n_S)^{-1})$, then the output will be correct, corresponding to the noiseless case, with high probability.~\footnote{There is a minor subtlety here, related to the observation that QSS never falsely concludes that the signal exists, which is not necessarily the case with noise. Therefore, we must modify our argument for amplifying the success probability of each Grover search. This is easily done, without changing the conclusions of this section: let $\Tnoise$ be a small enough (constant) fraction of $(\Gamma n_S)^{-1}$ that the gap between the acceptance probabilities on ``YES'' and ``NO'' instances is positive. By repeating each Grover search $\Omega(\log(|\Delta \omega| / |\Delta  \omega_{\scriptscriptstyle \mathrm{noise}} |))$ times for every short bandwidth $\Delta  \omega_{\scriptscriptstyle \mathrm{noise}}$ and outputting ``YES'' or ``NO'' based on the fraction of acceptances, a Chernoff bound shows that the error probability is at most $O((|\Delta \omega| / |\Delta  \omega_{\scriptscriptstyle \mathrm{noise}} |)^{-1})$ for each individual round (while modifying the sensing time by only a polylogarithmic factor). This is enough to conclude that ``YES'' instances are accepted with high probability by considering the particular round such that $\omega$ belongs to the bandwidth being probed. To conclude that ``NO'' instances are rejected with high probability, take a union bound over the number of rounds.} We can probe a bandwidth $\Delta \omega_{\scriptscriptstyle \mathrm{noise}}$ in time $\Tnoise$ as long as
\begin{align}
    |\Delta  \omega_{\scriptscriptstyle \mathrm{noise}} | = \widetilde{\Omega}((n_S B_{\min})^3 \Tnoise^2).
\end{align}
We repeat over $|\Delta \omega| /  |\Delta  \omega_{\scriptscriptstyle \mathrm{noise}} |$ intervals to cover the bandwidth. The protocol takes total sensing time
\begin{align}
    \tau = \frac{|\Delta \omega|}{|\Delta  \omega_{\scriptscriptstyle \mathrm{noise}}|} \Tnoise = \widetilde{O}\left(\frac{\Gamma}{B_{\min}} \frac{|\Delta \omega|}{(n_S B_{\min})^2} \right)
\end{align}
as desired.

We conclude by proving Lemma~\ref{lemma: effect_of_noise}:
\begin{proof}
    Define $U(t)$ as the unitary operator describing time evolution under $H(t')$ for $0 \le t' \le t$, such that $\mathcal{N}_t(\rho_0) = U(t) \rho_0 U(t)^\dagger$. Therefore
    \begin{align}\label{eq: unitary_deriv}
        \dv{t} U(t) = -i H(t) U(t) \quad \text{and} \quad  \dv{t} U(t)^\dagger = i U(t)^\dagger H(t).
    \end{align}
    We can rewrite the trace distance of interest as
    \begin{align}
        \frac{1}{2} \| \mathcal{N}_T(\rho_0) - \widetilde{\mathcal{N}}_T(\rho_0) \|_1 = \frac{1}{2} \| \rho_0 - U(T)^\dagger \widetilde{\mathcal{N}}_T(\rho_0) U(T) \|_1 
    \end{align}
    using unitary invariance of the Schatten one-norm. This difference can be rewritten as an integral as follows:
    \begin{align}\label{eq: td_as_integral}
        \frac{1}{2} \left\| \int_0^T \mathrm{d}t~\dv{t} \left( U(t)^\dagger \widetilde{\mathcal{N}}_t(\rho_0) U(t) \right) \right\|_1 .
    \end{align}
    Define $\tilde{\rho}(t) = \widetilde{\mathcal{N}}_t(\rho_0)$ and $\mathcal{L}(\cdot) = \sum_{j=1}^r L_j (\cdot) L_j^\dagger - \frac{1}{2} \{L_j^\dagger L_j, \cdot \}$. Expanding the derivative, we can use~\eqref{eq: unitary_deriv} and the Markovian master equation to write~\eqref{eq: td_as_integral} as
    \begin{align}
        \frac{1}{2} \left\| \int_0^T dt~i U(t)^\dagger H(t) \tilde{\rho}(t) U(t) + U(t)^\dagger \left(-i [H(t), \tilde{\rho}(t)] + \mathcal{L}(\tilde{\rho}(t)) \right) U(t) - i U(t)^\dagger \tilde{\rho}(t) H(t) U(t) \right\|_1.
    \end{align}
    Simplifying, we conclude that
    \begin{align}
        \frac{1}{2} \| \mathcal{N}_T(\rho_0) - \widetilde{\mathcal{N}}_T(\rho_0) \|_1 = \frac{1}{2} \big\| \int_0^T dt~U(t)^\dagger \mathcal{L}(\tilde{\rho}(t)) U(t) \big\|_1. 
    \end{align}
    Using triangle inequality (with respect to the integral over time as well as the sum over jump operator index) and unitary invariance, we obtain
    \begin{align}
        \frac{1}{2} \big\| \int_0^T dt~U(t)^\dagger \mathcal{L}(\tilde{\rho}(t)) U(t) \big\|_1 \le \frac{1}{2} \sum_{j=1}^r \int_0^T dt~\| L_j \tilde{\rho}(t) L_j^\dagger \|_1 + \frac{1}{2} \| L_j^\dagger L_j \tilde{\rho}(t) \|_1 + \frac{1}{2} \| \tilde{\rho}(t) L_j^\dagger L_j \|_1.
    \end{align} 
    We can upper bound the first term in the integrand using $ \| L_j \tilde{\rho}(t) L_j^\dagger \|_1 = \text{tr}(L_j \tilde{\rho}(t) L_j^\dagger ) = \text{tr}(L_j^\dagger L_j \tilde{\rho}(t)) \le \| L_j^\dagger L_j \tilde{\rho}(t) \|_1$, where the first equality holds since $\tilde{\rho}(t)$ is positive semi-definite. Therefore, using H{\"o}lder's inequality on each term in the master equation, we obtain the upper bound
    \begin{align}
        \frac{1}{2} \sum_{j=1}^r \int_0^T dt~2 \| L_j^\dagger L_j \| = T  \sum_{j=1}^r \| L_j^\dagger L_j \| 
    \end{align}
    as desired.
\end{proof}
\newpage 

\section{Experimental Platforms Compatible with the QSS Algorithm}
\label{SM:Exp_Plat}

In this section, we explain how the different ingredients of the QSS algorithm can be naturally implemented in many modern quantum platforms: solid state spin defects (e.g., nitrogen-vacancy centers), Rydberg atoms in a tweezer array, trapped ions, and qubits in a cavity (both neutral atoms in an optical cavity and superconducting qubits in a microwave cavity).
The necessary ingredients are: (i) arbitrary rotations of the sensing qubit, (ii) global rotations of the computational qubits, and (iii) Ising interactions between the sensor and computational qubits.

\vspace{0.1cm}
\ul{\emph{Solid state spin defects}}:
In the main text, we focused our discussion on nitrogen-vacancy (NV) centers in diamond.
However, there is a much broader landscape of quantum spin defects which can be used for computation and metrology~\cite{wolfowicz:2021}, including, but not limited to, defects in diamond (both substitutional and interstitial~\cite{bradac:2019}), defects in silicon, silicon carbide, and hexagonal boron-nitrate~\cite{son:2020,gottscholl:2020}, as well as rare-earth ions embedded within a crystal host~\cite{zhong:2019}.

Many defects of this kind can be optically polarized and coherently controlled via radio and microwave fields, enabling the first two ingredients~\cite{wolfowicz:2021}. The third ingredient arises when the defects, which naturally interact via dipolar or contact interactions, are placed in a large external magnetic field. The difference in gyromagnetic ratios between the different spin qubits induces an energy mismatch that suppresses all non-Ising terms, enabling the desired interactions.

In the limit of low electron spin density, the coherence time of such platforms becomes limited by the interactions between different nuclear spin qubits and the interactions with a nuclear spin bath~\cite{bradley:2019}.
The energy scale of these interactions can be three orders of magnitude smaller than the electronic-nuclear spin-spin interaction.

\vspace{0.1cm}
\ul{\emph{Rydberg atom arrays}}:
Rydberg atom arrays have emerged as a powerful and versatile platform for quantum science and technology, offering the ability to perform fast, high-fidelity gates through strong Rydberg-Rydberg interactions, as well as precise (static and dynamic) control over qubit positions~\cite{madjarov:2019,cao:2024a,evered:2023,bluvstein:2024}.
The hyperfine manifold of the ground state has been demonstrated as a long-lived computational qubit, and high-fidelity multi-qubit gates are possible using the excited Rydberg manifold~\cite{levine:2018}. The Rydberg manifold, by contrast,  serves as good sensing qubit.
The delocalized nature of the electronic wavefunction makes these states particularly suitable for sensing electric fields~\cite{yuan:2023}.
In both manifolds, arbitrary single-qubit control is available via a combination of microwave and optical fields~\cite{bornet:2023, bluvstein:2023}, enabling the first two ingredients.

There are different methods to achieve the third ingredient, all of which utilize the van der Waals Ising interactions arising from the highly delocalized Rydberg states~\cite{beguin:2013}.
These interactions can be accessed from the ground state hyperfine manifold in a gate-based manner, by coherently exciting to the Rydberg manifold, or in a continuous manner, by dressing the hyperfine states with laser light to couple the hyperfine and Rydberg manifolds~\cite{zeiher:2017}.
The form and strength of the resulting Hamiltonian can be controlled by the choice of array geometry and atomic species~\cite{anand:2024}.

\vspace{0.1cm}
\ul{\emph{Trapped ions}}: Trapped ions are one of the foundational platforms for quantum information processing~\cite{monroe:1995} and can be utilized for both sensing and computation, owing to their controllability and long coherence times~\cite{bruzewicz:2019,moses:2023}. In this platform, charged ions are spatially confined via a combination of static and oscillating electric fields.
The repulsion between the charged particles ensures they form a crystalline structure. The hyperfine levels of the ion can be used as a qubit and addressed by local optical beams, enabling the first two ingredients. Multi-qubit gates are mediated through the motional modes of the crystal~\cite{sorensen:1999}: 
by locally controlling the laser intensities and detunings, one can engineer tunable Ising interactions between ions and achieve the third ingredient~\cite{monroe:2020}.

\vspace{0.1cm}
\ul{\emph{Qubits in a cavity}}: Strong light-matter interactions, which are realized by both neutral atoms in an optical cavity~\cite{pedrozo-penafiel:2020} and superconducting qubits dispersively coupled to a quantized photonic mode~\cite{wallraff:2004}, are a powerful resource both for mediating interactions between qubits and interfacing with the outside world~\cite{reiserer:2015}.

Similarly to trapped ions,  where Ising interactions are mediated by a bosonic mode, a dispersive coupling between the qubits and a photonic mode can induce Ising or XY interactions between different qubits~\cite{viehmann:2013, periwal:2021}, realizing the third ingredient.
The strength and connectivity of these interactions is controlled by the location of the neutral atoms in the optical cavity or the geometry of the superconducting circuit. Individual qubit control can be achieved by addressing qubits from the side of the cavity in the neutral atom setting~\cite{ho:2024}, or with additional resonators in the superconducting setting, leading to the first two ingredients.

\newpage

\section{Detailed Analysis of a Proof-of-Principle Demonstration of Quantum Computing Enhanced Sensing}\label{sec: numsim}

In the main text, we introduced a simplified protocol that demonstrates the key ideas and speedup of the QSS algorithm, but forgoes many of the components which make our approach robust to continuous signal strengths and frequencies.
This simplified protocol succeeds if the signal has additional, known structure. Namely, the signal strength $B$ must be known, and the signal frequency $\omega$ must belong to a known, discrete set $\{\omega_k\}$ --- for this reason, in these supplementary materials, we call the simplified protocol \emph{discrete Quantum Search Sensing (dQSS)}. We emphasize that dQSS solves a more restricted sensing problem than QSS. Nevertheless, we will use it to highlight some key considerations for experimental implementations of quantum computing enhanced sensing and quantify near-term, achievable improvements in sensing time from the use of quantum algorithms. This analysis will take into account the role of decoherence, as well as the finite coupling strength between the sensor and computational qubits.

This section is organized as follows.
In Sec.~\ref{SM:dQSS_Description}, we describe dQSS in detail and illustrate how it can be straightforwardly implemented using the same ingredients as the QSS algorithm.
For comparison, we also introduce and discuss a conventional quantum sensing protocol for the same sensing task.
In Sec.~\ref{SM:Improvement}, we define and compute the \emph{improvement factor}, which quantifies the decrease in sensing time afforded by our quantum computing enhanced approach. 
In Sec.~\ref{SM:Exp_Num}, we return our attention to the experimental platform considered in the main text --- an NV center coupled to an ensemble of nuclear spins --- and provide additional details of our proposed experiment and numerical simulations.
We conclude in Sec.~\ref{SM:Extra} by presenting and discussing additional numerical results, including an optimization procedure which informed the choice of experimental parameters for the data in the main text.

\subsection{Proof-of-Principle Protocol (dQSS) for Quantum Computing Enhanced Sensing}
\label{SM:dQSS_Description}

We begin by discussing the dQSS protocol introduced in the main text.
For simplicity, we assume that the signal $H(t)$ couples in the transversal direction to the sensor quantization axis $Z_S$:
\begin{equation}
H(t) = B \cos(\omega t + \phi) X_S~.
\end{equation}

We assume that the signal strength $B$ is known and the signal frequency $\omega$ is restricted to a set of $N$ different values, $\omega = \omega_{k_*} \in \{ \omega_k\}$ for $k = 1, \hdots, N$. As noted, when the sensing task is restricted in this way, the components of the QSS algorithm which enabled robustness (namely, the ESU and QSP sequence) are no longer necessary. Instead, the discrete nature of the frequency search space means that the sensing oracle $\orac$ resembles the usual Grover oracle and should act as $\orac \ket{k}_Q = (-1)^{\delta_{k,k_*}}\ket{k}_Q$, where $\ket{k}_Q$ is the state associated to frequency $\omega_k$. $\orac$ is then simply a controlled phase gate conditional on state $\ket{k_*}_Q$ and can be implemented analogously to the gate $R_0 = 2 \ketbra{0 \cdots 0}_Q - \mathbbm{1}_Q$ as described in the main text. More specifically, if the sensor and computational qubits interact via $H_c = (Z_S / 2) \otimes \sum_k \omega_k \ketbra{k}_Q$, then by exposing the sensing qubit to the signal for time $2\pi/B$, the sensing qubit undergoes a $2\pi$ rotation and acquires a $-1$ phase, conditional on the computational qubits being in state $\ket{k_*}_Q$. As noted in the main text, this assumes we conclude by evolving under $-H_c$ for time $2\pi/B$, without the influence of the signal, to cancel the conditional phases arising solely from $H_c$.

Using this simplified sensing oracle, we implement the dQSS protocol as follows. First, prepare the system in state $\ket{+}_Q^{\otimes{n_Q}}\ket{0}_S$ by global rotation of the computational qubits. Then, apply $N_G$ Grover iterations, each composed of the sensing oracle $\orac$, global Hadamard gates on the computational qubits, the $R_0$ gate, and another set of Hadamard gates (see FIG.\,4 in the main text). Finally, measure the computational qubits in the computational basis --- let $k_o$ denote the measurement outcome. We conclude by checking for the signal using a short conventional sensing protocol targeting frequency $\omega_{k_o}$.
The restrictions on the signal also simplify this checking procedure (as compared to Sec.~\ref{sec: conventproof}): the sensing qubit is initialized in state $\ket{0}_S$ with bare frequency tuned to $\omega_{k_o}$, then let to evolve under the signal for time $\tau_{\scriptscriptstyle \mathrm{Check}} = \pi/B$. If $k_* = k_o$, the sensor undergoes a $\pi$-pulse and transitions to state $\ket{1}_S$ (in the absence of decoherence). Measuring the sensing qubit in state $\ket{1}_S$ confirms the presence of the signal, while measuring $\ket{0}_S$ means that either the signal does not exist or some error has occurred.

The probability that some step of the protocol fails depends on the relationship between the coherence time $T_2$ of the system and the duration of the Grover search (which in turn depends on $N_G, B,$ and $B_{R_0}$), as well as the relationship between the minimum frequency spacing $\min_{k \neq k_*} |\omega_k - \omega_{k_*}|$ and the drive strengths $B$ and $B_{R_0}$, due to the effects of power broadening. Depending on this failure rate, the Grover search and checking procedure should be repeated multiple times to ensure accuracy of the result, increasing the overall sensing time. We define $\tau^{ \scriptscriptstyle(k_*)}_{\scriptscriptstyle\text{ dQSS}}$ as the expected duration of the dQSS protocol given that there exists a signal at frequency $\omega=\omega_{k_*}$, assuming we repeat the Grover search and checking procedure until the signal is detected (i.e., the checking procedure measures $\ket{1}_S$). We carefully analyze the expected duration in Sec.~\ref{sec:SM_tauQSS}.

\subsubsection{Conventional Protocol}
\label{SM:Conventional_Algo}

In addition, we consider a conventional quantum sensing protocol for detecting an unknown signal in this restricted setting, against which we will compare the dQSS protocol. The protocol is simply to randomly sample (without replacement) a target frequency from the discrete set $\{ \omega_k \}$, then apply the checking procedure above at that frequency. 
Measuring $\ket{1}_S$ confirms the presence of the signal, while measuring $\ket{0}_S$ indicates either the absence of the signal or that some error has occurred. As in dQSS, these errors mean that if the signal exists, it might not be detected even after checking all $N$ target frequencies. We define $\tauconv^{\scriptscriptstyle (k_*)}$ as the expected duration of this conventional protocol given that there exists a signal at frequency $\omega = \omega_{k_*}$, assuming we resample the target frequencies and repeat the checking procedure until the signal is detected. We analyze the expected duration of the conventional protocol in Sec.~\ref{SM:Tau_conv}. 

\subsection{Defining and Computing the Improvement Factor \texorpdfstring{$\mathcal{I}$}{I}}
\label{SM:Improvement}

We define the improvement factor $\mathcal{I}$ as the ratio between the expected durations of the conventional protocol and dQSS, averaged over the $N$ possible signal frequencies:

\begin{equation}
\mathcal{I} = \frac{1}{N} \sum_{k_*=1}^N \frac{\tauconv^{\scriptscriptstyle (k_*)}}{\tau^{\scriptscriptstyle (k_*)}_{\scriptscriptstyle \text{ dQSS}}}~.
\end{equation}
$\mathcal{I}$ captures the improvement in sensing time achieved through quantum computing enhancement. In the following sections, we compute $\tauconv^{\scriptscriptstyle (k_*)}$ and $\tau_{\scriptscriptstyle \text{ dQSS}}^{\scriptscriptstyle (k_*)}$, carefully considering the various error sources and their impact on the performance of both protocols.

\subsubsection{Computing Conventional Sensing Time}
\label{SM:Tau_conv}

We now discuss how the expected conventional sensing time $\tau_{\scriptscriptstyle \text{conv}}^{\scriptscriptstyle (k_*)}$ is calculated in our analysis of the improvement factor. First, suppose that the protocol is error-free, so the checking procedure detects the signal if and only if the target frequency is $\omega_{k_*}$. Since the target frequencies are sampled without replacement, the expected sensing time is $\tauconv^{\scriptscriptstyle (k_*)} = \tau_{\scriptscriptstyle \text{Check}} (N+1)/2  = (N+1)\pi / (2B)$.

Next, we consider the impact of errors on the expected duration of the protocol. As noted in the main text and discussed further in Sec.~\ref{SM:Extra}, the dominant sources of error are finite coherence time and power broadening of the signal. If the coherence time $T_2$ of the sensing qubit is comparable to the duration of the checking procedure, $T_2/\tau_{\scriptscriptstyle \text{Check}} \lesssim 1$, then the signal does not implement a perfect $\pi$-pulse during the checking procedure. This decreases the probability of detecting the signal, thus increasing the expected protocol duration.  If the signal strength $B$ is comparable to the frequency difference between $\omega_{k_*}$ and some nearby frequency $\omega_k$, $B \gtrsim |\omega_{k_*} - \omega_k|$, then due to power broadening it is possible to detect the signal when probing at the target frequency $\omega_k$. This increases the probability of detecting the signal, thus decreasing the expected protocol duration. 

We can account for both of these effects, as well as any other potential error sources, by defining the probability $p_{\scriptscriptstyle k}^{\scriptscriptstyle (k_*)}$ of detecting a signal at frequency $\omega_{k_*}$ when checking frequency $\omega_k$. The distribution $p_{\scriptscriptstyle k}^{\scriptscriptstyle (k_*)}$ depends on the signal strength $B$, the set of discrete frequencies $\{\omega_k\}$, and the coherence time $T_2$, and it can be calculated analytically by solving the driven-dissipative dynamics of the two-level sensing qubit~\cite{scully:1997}. 

The probability of detecting the signal by checking a single randomly sampled target frequency is $\sum_{k=1}^N p_{\scriptscriptstyle k}^{\scriptscriptstyle (k_*)}/N$. Therefore, if the conventional sensing protocol were to randomly sample target frequencies \emph{with} replacement, it would have an expected sensing time of $N \tau_{\scriptscriptstyle \text{Check}} / (\sum_{k=1}^N p_{\scriptscriptstyle k}^{\scriptscriptstyle (k_*)})$. Numerically, we find that instead sampling \emph{without} replacement never improves the duration of the sensing protocol by more than a factor of 2. Therefore, we have the following lower bound:
\begin{align}
    \tauconv^{\scriptscriptstyle (k_*)} \gtrsim \frac{\pi}{2B} \frac{N}{\sum_{k=1}^N p^{\scriptscriptstyle (k_*)}_{\scriptscriptstyle k}}~.
\end{align}
This gives a lower bound on the improvement factor, which we use in our further analysis.

\subsubsection{Computing dQSS Sensing Time}
\label{sec:SM_tauQSS}

We now discuss how the expected sensing time $\tau_{\scriptscriptstyle \text{dQSS}}^{\scriptscriptstyle (k_*)}$ of the dQSS protocol is calculated. Recall that the protocol consists of repeatedly performing a Grover search over frequencies, then applying the checking procedure to probe the predicted frequency. We define $G_{\scriptscriptstyle k}^{\scriptscriptstyle (k_*)}$ as the probability distribution over measurement outcomes $k$ at the conclusion of the Grover search, when the signal is at frequency $\omega_{k_*}$. $G_{\scriptscriptstyle k}^{\scriptscriptstyle (k_*)}$ is defined to include all potential errors in the implementation of the Grover search, including those due to finite coherence time and power broadening. $G_{\scriptscriptstyle k}^{\scriptscriptstyle (k_*)}$ plays the role of a prior distribution for the checking procedure, which again succeeds with probability $p_{\scriptscriptstyle k}^{\scriptscriptstyle (k_*)}$ when checking target frequency $\omega_k$. Therefore, if the duration of the Grover search is $\tau_{\scriptscriptstyle \text{Grover}}$ and the duration of the checking procedure is $\tau_{\scriptscriptstyle \text{Check}}$, the expected dQSS sensing time is
\begin{align}
    \tau_{\scriptscriptstyle \text{dQSS}}^{\scriptscriptstyle (k_*)} = \frac{\tau_{\scriptscriptstyle \text{Grover}} + \tau_{\scriptscriptstyle \text{Check}}}{\sum_{k=1}^{N} G_{\scriptscriptstyle k}^{\scriptscriptstyle (k_*)}p_{\scriptscriptstyle k}^{\scriptscriptstyle (k_*)}}~.
\end{align}
\indent
It should be noted that $\tau_{\scriptscriptstyle \text{Grover}}$ depends on the tunable parameters $N_G$, the number of Grover iterations, and $B_{R_0}$, the strength of the $R_0$ gate driving field. Depending on the error sources, it may be possible to decrease the expected sensing time by optimizing over these parameters. As an extreme example, if $T_2$ is very small, then performing $N_G \sim \sqrt{N}$ Grover iterations is not an effective use of sensing time, since the system will decohere and the measurement outcome will be uniformly random. In this scenario, it is better to perform the conventional sensing protocol, which is equivalent to choosing $N_G = 0$. We discuss the optimization over $N_G$ and $B_{R_0}$ in greater detail in Sec.~\ref{Sec:SM_opt}, but first we provide further details of the experimental platform considered in our work.

\subsection{Details of the Experimental Implementation}
\label{SM:Exp_Num}

We now provide further details related to an NV center based implementation of the dQSS protocol, motivated by recent experiments which use an ensemble of solid state spin defects as a quantum processor~\cite{abobeih:2022,abobeih:2019,bradley:2019}. We also specify the choice of experimental parameters used in our numerical simulations. 

We consider using an NV electronic spin defect as a sensing qubit and $n_Q$ nearby nuclear spins as computational qubits. More precisely, since the NV is a spin-1 defect, we define a two-level sensing qubit using the $\ket{m_s=0}$ and $\ket{m_s=-1}$ states, where $m_s$ denotes the magnetic quantum number of the state along the NV quantization axis.~\footnote{As in the main text, we absorb all coupling constants (including the factor of $1/\sqrt{2}$ arising from the fact that the NV is spin-1) into our definition of the magnetic field $B$, so that $X_S$ is the usual Pauli-$X$ operator.} For our computational qubits, we consider using the host $^{14}$N nuclear spin, which is closest in proximity and most strongly coupled to the sensing qubit, as well as nearby $^{13}$C nuclear spins. In our numerical simulations, we use the experimentally measured hyperfine couplings $A_i$ in~\cite{abobeih:2019} to define our interaction Hamiltonian. We also assume that the system is exposed to a large external magnetic field with strength $B_\text{ext} \approx 3000~\mathrm{G}$.
In this regime, the perpendicular dipolar terms $\sim Z_S X_{Q,i}$ can be neglected (the overlap of the eigenstates of the dipolar Hamiltonian with product states in the $Z$-basis exceeds $0.999$ for $n_Q=7$). We comment that these terms can more generally be understood to ``tilt'' the quantization axis of the nuclear spins, and by carefully initializing and controlling the nuclear spins with respect to this tilted quantization axis, the strength of the applied external field can be dramatically reduced --- we leave such optimizations and analysis to future work.
We also neglect the weak interactions ($\sim 1-10 \text{ Hz}$) between different nuclear spins, which are three orders of magnitude smaller than the interactions between the NV and the nuclear spins ($\sim 20\text{ kHz}$). With these approximations, the interaction Hamiltonian is given by Ising interactions between the two types of spin defects, projected onto the spin-1 manifold:
\begin{equation}
H_{\text{NV-nuc}} = -\ketbra{m_s=-1}_S \otimes \sum_{i=1}^{n_Q} A_i \frac{Z_{Q,i}}{2}.
\end{equation}
We assume that the signal frequency $\omega$ belongs to the set $\{\omega_k\}$ of the $N = 2^{n_Q}$ conditional splittings of the NV. Each $\omega_k$ corresponds to a particular configuration $| k \rangle_Q = | k_1  k_2  \dots k_{n_Q} \rangle_Q$ of the nuclear spins, with $k_i \in \{0,1\}$:
\begin{equation}
\omega_k = \Delta_0 + \sum_{i=1}^{n_Q}  \frac{(-1)^{k_i}}{2} A_i,
\end{equation}
where $\Delta_0$ is the frequency splitting between the $\ket{m_s=0}$ and $\ket{m_s=-1}$ states in the absence of interactions with the nuclear spins.

We now briefly discuss how the gates in the dQSS protocol are realized in this system. As noted, the sensing oracle and the gate $R_0$ are applied by driving the NV at frequency $\omega = \omega_{k_*}$ for time $2\pi/B$ (with the external signal), and at frequency $\omega_{0}$ for time $2\pi/B_{R_0}$ (with an additional driving field), respectively.
Notably, because the NV is spin-1, the sensing qubit and computational qubits are decoupled when the NV is in state $\ket{m_s=0}_S$; this means that no additional, state-dependent phase is accumulated, and the final corrective evolution under $-H_{\text{NV-nuc}}$ is not necessary. This simplifies and speeds up our implementation of the sensing oracle. In our numerical simulations, we choose $B_{R_0} = (32 \times 2^{-n_Q} + 10\sqrt{B/1~\mathrm{kHz}})~\mathrm{kHz}$ (this choice is determined by a small-size optimization procedure, see Sec.~\ref{Sec:SM_opt}). Finally, we implement the Hadamard gates on the computational qubits by globally driving the nuclear spins near their bare frequency (set by the Larmor frequency under the external field). In our numerical simulations, we use a fast drive (with Rabi frequency $\Omega_{H} = 10~\text{kHz}$).
For this choice of parameters, the duration of the protocol is limited by the sensing time, not the speed of the Hadamard gates or $R_0$. 

We conclude by discussing the role of decoherence in our proposed experiment and the choice of noise model in our numerical simulations. For simplicity, we suppose that all qubits (both the NV and nuclear spins) undergo dephasing at the same rate $\Gamma$, determined by the coherence time of the electronic spin as $\Gamma=1/T_2$. Because the coherence time of the nuclear spin is in general much longer than that of the electronic spin, our calculation is conservative. That is, it will \emph{overestimate} the impact of decoherence and thus \emph{underestimate} the performance of the dQSS protocol. The choice to include only dephasing noise is experimentally well-motivated; the depolarization time $T_1$ of the NV is approximately an hour, much longer than any other time scale in our analysis~\cite{abobeih:2018}. Note that our sensing protocol will still never falsely conclude that the signal exists, since dephasing cannot induce a spin-flip of the NV during the checking step.

Our numerical simulations will use coherence times ranging from $T_2=10~\mathrm{ms}$ to $10~\mathrm{s}$.
Electronic spin coherence times of $T_2=10~\mathrm{ms}$~\cite{bradley:2019} and $0.5~\mathrm{s}$~\cite{abobeih:2018} have been demonstrated in experiment, by performing dynamical decoupling. For simplicity, our numerical simulations do not take into account details of the dynamical decoupling sequence. However, it is important to note that these details can enter non-trivially into the design of the sensing sequence. For example, the electronic coherence time $T_2$ is limited by interactions with other nuclear defects, which induce additional, uncontrollable shifts to the NV energy level.
Fortunately, these defects are quasi-static, so a simple echoing procedure can greatly enhance the coherence time of the NV.
However, this echoing procedure will also decouple the NV from the nuclear spins used to implement entangling gates.
To avoid this repercussion, one must rotate both the NV and the nuclear spins; while this capability has been experimentally demonstrated, it adds additional complexity to our proposed experiment~\cite{bradley:2019}.
As an alternative approach, one could polarize a greater number of nuclear spins than is necessary for the dQSS protocol, thereby suppressing thermal fluctuations of the NV energy levels and increasing $T_2$.
We leave an analysis of these details for future work.

\subsubsection{Details of the Numerical Simulation}

We numerically simulate the above protocol for different values of $n_Q$ and $T_2$.
In general, it is numerically challenging to solve for the evolution of a quantum system subject to a fast oscillating drive and decoherence, due to the large separation in time scales (e.g., MHz driving frequencies vs.~Hz decoherence rates) and the size of the full density matrix ($2^{n_Q+1}\times 2^{n_Q+1}$ in our case).
To study systems of size up to $n_Q=10$, we leverage some simplifications which are valid within the parameter regimes of interest for our problem.

First, the effect of the transverse drive (arising from the external signal or $R_0$ driving field) can be accounted for by computing the evolution in the frame rotating at the drive frequency. In this frame, the fast oscillating drive transforms into an effective static term, up to small, far off-resonant terms that can be safely ignored by the rotating wave approximation, Sec.~\ref{sec: esu_error_proof}.

Second, because the state of the computational qubits is unchanged by the NV driving operations (the sensing oracle and the $R_0$ gate), we can further simplify the computation of these dynamics. 
In particular, the dynamics of the full system can be decomposed into the dynamics of $\ketbra{k_1}{k_2}_Q \otimes \rho_S$ for every pair of states $\ket{k_1}_Q$ and $\ket{k_2}_Q$.
We then solve for the dynamics of each $2\times 2$ density matrix $\rho_S$ by exact diagonalization, which is dramatically faster than numerically integrating the full $2^{n_Q+1}\times 2^{n_Q+1}$ density matrix. 

Finally, because the nuclear spin Hadamard gates are much faster than the decoherence rate of the system, the effect of dissipation during these gates is negligible. Therefore, we assume that the Hadamard gates are implemented perfectly and infinitely fast. This simplification results in only negligible deviations from the full many-body calculation, due to a small occupation of the $\ket{m_s=-1}$ state after the sensing oracle and $R_0$ gates.
When the NV is in $\ket{m_s=-1}$, the Ising coupling shifts the nuclear spin resonance frequency, decreasing the fidelity of the global drives.
Crucially, this simplification does not modify our estimation of the achievable improvement or other conclusions.

\subsection{Numerical Simulation Results}
\label{SM:Extra}

In this section, we provide further numerical results, complementing the data and discussions in the main text. 
In Sec.~\ref{Sec:SM_nq}, we compute the improvement factor for different numbers $n_Q$ of computational qubits and discuss the broader landscape of attainable improvements and limitations in different parameter regimes.
In Sec.~\ref{Sec:SM_opt}, we optimize the improvement factor over the parameters $N_G$ and $B_{R_0}$ of the dQSS protocol for small systems of size $n_Q \le 7$. These results inform our parameter choices when simulating larger systems (of size $n_Q \ge 8$), for which optimization is excessively cumbersome. 

\subsubsection{Improvement Factor for Different \texorpdfstring{$n_Q$}{nQ}}
\label{Sec:SM_nq}

\begin{figure}
    \centering
\includegraphics[width=0.9\linewidth]{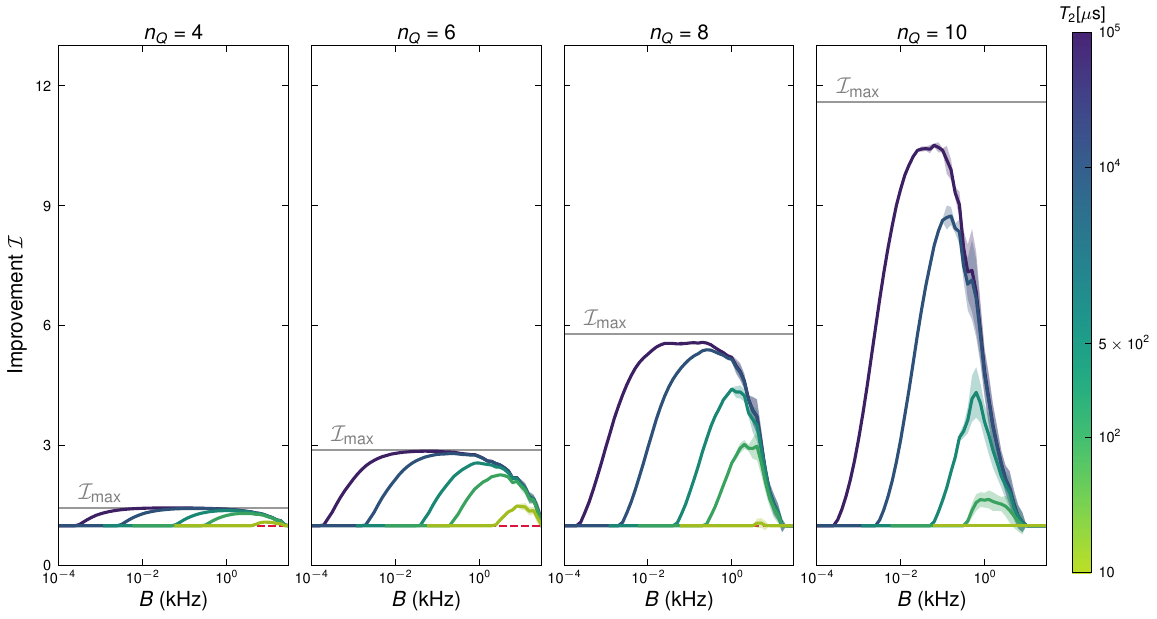}
    \caption{Improvement factor $\mathcal{I}$ as a function of $B$ and $T_2$, for different numbers of computational qubits $n_Q$.
    The improvement factor is averaged over $8$ random signal frequencies, and the shaded areas indicate the variance over frequencies. 
    The variance is only non-negligible in the large $B$ regime, where power broadening limits the protocol performance.
    }
    \label{fig:SMDiff_NQ}
\end{figure}

In the main text, we reported the optimal improvement factor for a single value of $n_Q=8$ (see FIG.~4c in the main text); we extend these results to different $n_Q$ in FIG.~\ref{fig:SMDiff_NQ}. The results further illustrate how the improvement factor can be limited by both details of the sensing problem (e.g., the relationship between signal strength and frequency spacing) and the experimental capabilities (e.g., the available $n_Q$ and $T_2$).

For small signal strengths $B$, the protocol is limited mainly by the system coherence time; for larger $T_2$, we can perform a greater number of Grover iterations, increasing the improvement factor. The improvement increases up to an $n_Q$-dependent maximum $\mathcal{I}_{\max}$, which we compute as $\mathcal{I}_{\max}\approx 0.362 \sqrt{2^{n_Q}}$ in the large $n_Q$ limit. 
This maximum value takes into account the fact that the duration of each Grover iteration adds to the duration of the protocol; as such, the optimal $N_G$ does not maximize the success probability of the Grover search (corresponding to $N_G = \lfloor \pi \sqrt{2^{n_Q}}/4\rfloor$). Rather, it strikes a balance between the duration of each Grover iteration and the marginal gain in success probability attained by increasing the number of iterations, minimizing the overall duration, see Sec.~\ref{Sec:SM_opt}.

For large signal strengths $B$, the protocol is limited mainly by power broadening; the external signal can no longer spectroscopically resolve the different frequencies $\{\omega_k\}$, so the sensing oracle imparts coherent phase errors. This effect is independent of the coherence time, so $\mathcal{I}$ converges to the same reduced value at large $B$, regardless of $T_2$. 
Because the different frequencies $\{\omega_k\}$ are not equally spaced, power broadening affects dQSS differently for different signal frequencies; this results in an increase in the variance of the individual improvement factors for the different possible signal frequencies (shaded regions in FIG.~\ref{fig:SMDiff_NQ}).
As $n_Q$ increases, the onset of power broadening occurs at lower $B$, due to the decrease in the minimum spacing among the frequencies $\{\omega_k\}$.

\subsubsection{Parameter Optimization}
\label{Sec:SM_opt}

The duration of the dQSS protocol is determined by $N_G$, the number of Grover iterations, and $B_{R_0}$, the strength of the $R_0$ gate driving field. In realistic settings, there are important error trade-offs to consider when choosing these parameters. As $N_G$ increases, the success probability of the Grover search increases, but the protocol becomes longer and more susceptible to decoherence. As $B_{R_0}$ increases, each $R_0$ gate is implemented faster, but has lower fidelity due to power broadening. Consequently, to determine the maximal improvement factor, one must optimize over these parameters.

\begin{figure}[h]
    \centering
\includegraphics[width=0.97\linewidth]{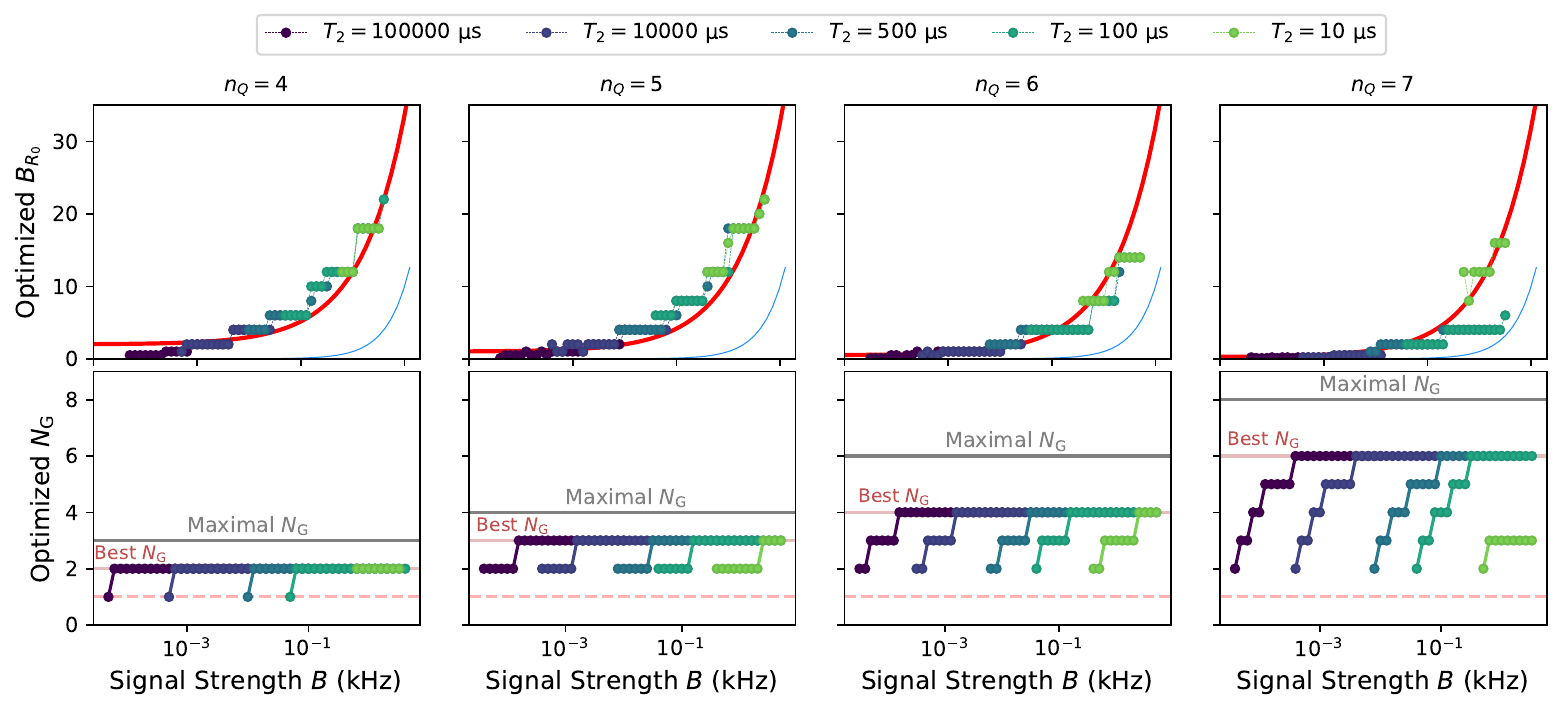}
    \caption{The results of optimizing $B_{R_0}$ and $N_G$ for small system sizes $n_Q \le 7$ (different columns) over a wide range of signal strengths $B$ (horizontal axis) and coherence times $T_2$ (different colors). The optimal $B_{R_0}$ is well-captured by the simple ansatz  $\widehat{B}_{R_0}$ (red line), which exceeds $B$ (blue line) over the full range of parameters. The optimal $N_G$ depends non-trivially on $n_Q$ and $T_2$.}
    \label{fig:SM_Opt}
\end{figure}

We numerically simulate the dQSS protocol for a range of discrete values of $B_{R_0}$ (from $0.01$ to $50.0$ kHz) and $N_G$ (from $0$ to $\lfloor \pi \sqrt{2^{n_Q}}/4\rfloor$). For each choice of parameters, we estimate the improvement factor by averaging over 8 random signal frequencies; the optimal parameters are taken as those which maximize this estimated improvement factor. In FIG.~\ref{fig:SM_Opt}, we present the results of this optimization for different small system sizes $n_Q$, signal strengths $B$, and coherence times $T_2$. There are two immediate conclusions from the data. First, though there is some variation in the optimal $B_{R_0}$, it is well-captured by the following simple ansatz (red line):
\begin{equation}
\widehat{B}_{R_0} = (32 \times 2^{-n_Q} + 10\sqrt{B/1~\mathrm{kHz}})~\mathrm{kHz}~.
\end{equation}
$\widehat{B}_{R_0}$ is significantly larger than the signal strength $B$ (blue line) over the parameter range of interest, so the dQSS protocol is indeed limited by the sensing time rather than the speed of the $R_0$ gate. We fix $B_{R_0}=\widehat{B}_{R_0}$ in our numerical simulations for larger system sizes.

Second, as the signal strength increases (but is not so large that power broadening is the dominant error source), the optimal $N_G$ increases, provided the coherence time is sufficiently large. At the same time, the optimal $N_G$ never saturates the maximum number of Grover iterations, $\lfloor \pi \sqrt{2^{n_Q}}/4\rfloor$, but instead plateaus at a smaller value (which agrees with the previously reported value of $\mathcal{I}_{\max}$).
This further emphasizes that it is not optimal to maximize the success probability of the Grover search;
as $N_G$ approaches the maximal number of Grover iterations, the improvement in the Grover effectiveness is not sufficient to compensate for the increase in the duration of the Grover search. 
Because this trade-off depends on all the details of the protocol, we optimize over $N_G$ in all our numerical simulations.

\newpage

\section{Grover-Heisenberg Limit}\label{sec: gh_limit}

In this section, we prove Theorem 2 in the main text by deriving a lower bound on the duration $\tau$ of any quantum sensing protocol which solves $\ACProb$. We call this fundamental lower bound the Grover-Heisenberg limit. We first comment briefly on the relationship between the Grover-Heisenberg limit and the standard Heisenberg limit. Then, we consider a general quantum sensing protocol, as defined in Sec.~\ref{sec: sensing_protocol}, and present a simple, necessary condition for such a protocol to solve $\ACProb$. Finally, we show how this condition can be used to derive a lower bound on the duration $\tau$ of the sensing protocol.

We begin by observing that, to detect a signal of strength $B_{\min}$ using $n_S$ sensors at a single, known frequency, the Heisenberg limit implies that the total sensing duration must satisfy $\tau = \Omega((n_S B_{\min})^{-1})$~\cite{degen:2017}. Since solving $\ACProb$ is at least as difficult as solving this task, this lower bound also applies to any quantum sensing protocol which solves $\ACProb$. This proves the Grover-Heisenberg limit in the edge case that $|\Delta \omega|/(n_S B_{\min})$ is at most a constant (narrowband sensing), so in what follows we assume $|\Delta \omega|/(n_S B_{\min})$ is unbounded (broadband sensing).

Consider a general quantum sensing protocol which uses $n_S$ sensors, $n_Q$ ancilla qubits, and total sensing time $\tau$ to solve $\ACProb$. As in Sec.~\ref{sec: sensing_protocol}, suppose the protocol applies unitaries $V_0, V_1, \hdots, V_P$ to all $n_S + n_Q$ qubits at times $0 = t_0 < t_1 < \cdots < t_P = \tau$. We allow $n_Q$ and $P$ to be arbitrarily large. For simplicity, define the initial state $\mathbf{\ket{0}} \equiv \ket{0}^{\otimes n_S + n_Q}$. If the signal does not exist, the state immediately prior to measurement is simply
\begin{align}
    |\psi_0 \rangle = \left( \prod_{j=0}^{P} V_j \right) \mathbf{\ket{0}}.
\end{align}
Conversely, if the signal exists with Hamiltonian $H_\omega(t) = B_{\min} \cos(\omega t + \phi) \sum_{i = 1}^{n_S} Z_i$, the state is
\begin{align}
    |\psi_\omega \rangle = V_P \left( \prod_{j=0}^{P-1} \text{exp}\left(-i \int_{t_j}^{t_{j+1}} dt~H_\omega(t) \otimes \mathbbm{1}_Q \right) V_j \right) \mathbf{\ket{0}}.
\end{align}
To simplify matters slightly, we can re-express $|\psi_\omega \rangle$ as the outcome of time evolution under a single time-dependent Hamiltonian, by writing $H_\omega(t)$ in the so-called toggling frame~\cite{mansfield:1971}. We define $\widetilde{H}_\omega(t) = B_{\min} \cos(\omega t + \phi) \widetilde{Z}(t)$ for $0 \le t \le \tau$, where
\begin{align}
        \widetilde{Z}(t) = V^\dagger(t) \left(\sum_{i=1}^{n_S} Z_i \otimes \mathbbm{1}_Q \right) V(t) \quad \text{with} \quad V(t) = \prod_{j:~t_j \le t} V_j.
\end{align}  
Let $\widetilde{U}_\omega$ be the unitary corresponding to time evolution under $\widetilde{H}_\omega(t)$ for $0 \le t \le \tau$. Then $\ket{\psi_\omega}$ can be written as follows:
\begin{align}
    \ket{\psi_\omega} = \left( \prod_{j=0}^{P} V_j \right) \widetilde{U}_\omega \mathbf{\ket{0}}.
\end{align}

If the sensing protocol solves $\ACProb$, there must be some two outcome measurement which distinguishes between $\ket{\psi_0}$ and $\ket{\psi_\omega}$ with probability at least $2/3$. This implies that $\| \ket{\psi_0} - \ket{\psi_\omega} \|_2^2$ exceeds some fixed constant (by combining the Holevo-Helstrom theorem with $\frac{1}{2} \| \ketbra{\psi_0} - \ketbra{\psi_\omega} \|_1 \le \| \ket{\psi_0} - \ket{\psi_\omega} \|_2$). This is true for all $\omega \in \Delta \omega$, so in particular we must have
\begin{align}\label{eq: avg_dist}
    D_{\Delta \omega}(\tau) = \frac{1}{|\Delta \omega|} \int_{\Delta \omega} d\omega~\| \ket{\psi_0} - \ket{\psi_\omega} \|_2^2 = \Omega(1).
\end{align}
We call $D_{\Delta \omega}(\tau)$ the \emph{average distinguishability}. Therefore, our strategy for proving the Grover-Heisenberg limit is to upper bound $D_{\Delta \omega}(\tau)$ in terms of $\tau$, which will imply a lower bound on $\tau$. Note that this strategy is essentially identical to a well-known proof of optimality for Grover's algorithm~\cite{bennett:1997, zalka:1999}; the only difference is that discrete variables, such as the query complexity and the index enumerating the search space, are replaced by continuous variables, such as the total sensing time and frequency of the signal.

We upper bound $D_{\Delta \omega}(\tau)$ in two steps. First, we bound $D_{\Delta \omega}(T)$ for short times $T = O((n_S B_{\min})^{-1})$. Second, we extend the short time bound to arbitrarily long times $\tau$. The key insight is that the average distinguishability for long times can grow at most quadratically in the number of short time steps, as with Grover's algorithm.

We begin with the short time bound:
\begin{lemma}[$D_{\Delta \omega}(T)$ for Short Times]\label{lemma: short_times} If $T \le (n_S B_{\min})^{-1}$, then $D_{\Delta \omega}(T) \le 4\pi (n_S B_{\min}) / |\Delta \omega|$. 
\end{lemma}
\begin{proof}
First, we upper bound the average distinguishability by extending the integral over $\omega$ to include all frequencies:
\begin{align}
    D_{\Delta \omega}(T) \le \frac{1}{|\Delta \omega|} \int_{-\infty}^{\infty} d\omega~ \left\| \mathbf{\ket{0}} - \widetilde{U}_\omega \mathbf{\ket{0}} \right\|_2^2 .
\end{align}
Noting that $\| \ket{\psi} - \ket{\phi} \|_2^2 = 2(1 - \text{Re}[\langle \psi | \phi \rangle])$ for any normalized $\ket{\psi}$ and $\ket{\phi}$, we can rewrite this upper bound as:
\begin{align}
    D_{\Delta \omega}(T) \le \frac{2}{|\Delta \omega|} \text{Re}\left[ \int_{-\infty}^{\infty} d\omega~\mathbf{\bra{0}} \mathbbm{1} - \widetilde{U}_\omega \mathbf{\ket{0}} \right] \le \frac{2}{|\Delta \omega|} \left| \int_{-\infty}^{\infty} d\omega~\mathbf{\bra{0}} \mathbbm{1} - \widetilde{U}_\omega \mathbf{\ket{0}} \right| .
\end{align}
Next, we express the time evolution operator $\widetilde{U}_\omega$ as a series expansion $\widetilde{U}_\omega = \sum_{m = 0}^\infty W_m$, where $W_0 = \mathbbm{1}$ and
\begin{align}
    W_m = (-i B_{\min})^m \int_0^T dt'_1 \int_0^{t'_1} dt'_2 \cdots \int_0^{t'_{m-1}} dt'_m \cos(\omega t'_1 + \phi) \cdots \cos(\omega t'_m + \phi) \widetilde{Z}(t'_1) \cdots \widetilde{Z}(t'_m)
\end{align}
for $m \ge 1$. In terms of this expansion, the upper bound on average distinguishability becomes
\begin{align}\label{eq: dist_bound_with_dyson}
    D_{\Delta \omega}(T) \le \frac{2}{|\Delta \omega|} \left| \sum_{m=1}^\infty \int_{-\infty}^{\infty} d\omega~\mathbf{\bra{0}} W_m \mathbf{\ket{0}} \right|.
\end{align}
Note that the sum begins at $m = 1$. We will upper bound each term in the sum individually, then show that the overall sum converges to a value at most linear in $n_S B_{\min}$, as long as $T \le (n_S B_{\min})^{-1}$.

Consider the $m$-th term in the sum. Change variables to $\theta = \omega T$ and $s_l = t'_l / T$ for $l = 1, \hdots, m$. Then, we have
\begin{align}\label{eq: Dyson_term_upper}
    \int_{-\infty}^{\infty} d\omega~\mathbf{\bra{0}} W_m \mathbf{\ket{0}} = \frac{(-i B_{\min} T)^m}{T} \int_0^1 ds_1 \cdots \int_0^{s_{m-1}} ds_m  \mathbf{\bra{0}} \widetilde{Z}(T s_1) \cdots \widetilde{Z}(T s_m) \mathbf{\ket{0}} \int_{-\infty}^{\infty} d\theta \prod_{l = 1}^m \cos(\theta s_l + \phi).
\end{align}
Consider the integral over $\theta$, which can be written in terms of Dirac delta functions as follows:~\footnote{For the reader averse to Dirac delta functions: all results in the section can be equivalently derived by appealing to the Fourier inversion theorem applied to Fourier transforms in the variables $\sum_{l=1}^m s_l b_l$, but we find the current presentation more clear.}
\begin{align}
    \int_{-\infty}^{\infty} d\theta \prod_{l = 1}^m \cos(\theta s_l + \phi) = \frac{1}{2^m} \sum_{\vec{b} \in \{\pm 1\}^m} \int_{-\infty}^{\infty} d\theta~e^{i \theta \sum_{l=1}^m b_l s_l + i \phi \sum_{l = 1}^m b_l} = \frac{2 \pi}{2^m} \sum_{\vec{b} \in \{\pm 1\}^m} e^{i \phi \sum_{l=1}^m b_l} ~\delta\left(\sum_{l = 1}^m s_l b_l \right).
\end{align}
Using $\delta(x) = \delta(-x)$, we may pair the terms corresponding to any $\vec{b}$ and $-\vec{b}$ in this sum to rewrite this expression as
\begin{align}
    \frac{2 \pi}{2^m} \sum_{\vec{b} \in \{\pm 1\}^m} \cos(\phi \sum_{l=1}^m b_l)~\delta\left(\sum_{l = 1}^m s_l b_l \right) 
\end{align}
Next, observe the following upper bound on $\cos(\phi')$, where $\phi'$ is an arbitrary angle, times the expectation value of the product of the $\widetilde{Z}(Ts_l)$ for $l = 1, \hdots, m$:
\begin{align}
    \cos(\phi') \mathbf{\bra{0}} \widetilde{Z}(T s_1) \cdots \widetilde{Z}(T s_m) \mathbf{\ket{0}} \le \| \widetilde{Z}(T s_1) \cdots \widetilde{Z}(T s_m) \| \le \left( \left\| \sum_{i=1}^{n_S} Z_i \otimes \mathbbm{1}_Q \right\| \right)^m \le n_S^m.
\end{align}
Therefore, we have $\int_0^1 ds_1 \cdots \int_0^{s_{m-1}} ds_m \left(n_s^m - \cos(\phi \sum_{l=1}^m b_l) \mathbf{\bra{0}} \widetilde{Z}(T s_1) \cdots \widetilde{Z}(T s_m) \mathbf{\ket{0}} \right) \delta(\sum_{l = 1}^m s_l b_l ) \ge 0$, since this is the integral of a non-negative function. We conclude that~\eqref{eq: Dyson_term_upper} is upper bounded as
\begin{align}
    \int_{-\infty}^{\infty} d\omega~\mathbf{\bra{0}} W_m \mathbf{\ket{0}} &\le \frac{(n_S B_{\min} T)^m}{T}  \left| \int_0^1 ds_1 \cdots \int_0^{s_{m-1}} ds_m  \int_{-\infty}^{\infty} d\theta \prod_{l = 1}^m \cos(\theta s_l)  \right|
    \\
    &= \frac{(n_S B_{\min} T)^m}{T} \frac{1}{m!}  \left|  \int_{-\infty}^{\infty} d\theta \left(\int_0^1 ds \cos(\theta s) \right)^m \right|
    \\
    &= \frac{(n_S B_{\min} T)^m}{m!} \frac{1}{T}  \left|  \int_{-\infty}^{\infty} d\theta \left(\frac{\sin \theta}{\theta} \right)^m \right|.
\end{align}
Since $\sin(\theta) / \theta \le 1$ for any $\theta$, the integral over $\theta$ is largest when $m = 1$, in which case it equals $\pi$. We conclude that
\begin{align}
     \int_{-\infty}^{\infty} d\omega~\mathbf{\bra{0}} W_m \mathbf{\ket{0}} \le \frac{\pi}{T} \frac{(n_S B_{\min} T)^m}{m!}.
\end{align}
Returning to~\eqref{eq: dist_bound_with_dyson} and using triangle inequality, the average distinguishability is bounded by
\begin{align}
    D_{\Delta \omega}(T) \le \frac{2}{|\Delta \omega|} \sum_{m=1}^\infty \frac{\pi}{T} \frac{(n_S B_{\min} T)^m}{m!} = \frac{2}{|\Delta \omega|} \frac{\pi}{T} \left(e^{n_S B_{\min} T} - 1 \right) \le \frac{4 \pi}{|\Delta \omega|} n_S B_{\min}
\end{align}
where we have used $e^x - 1 \le 2x$ for $0 \le x \le 1$, concluding the proof.
\end{proof}

We now use this short time bound to upper bound the average distinguishability for an arbitrary duration $\tau$. First, suppose that the protocol solves $\ACProb$ in a short time $\tau \le (n_S B_{\min})^{-1}$. Then, Lemma~\ref{lemma: short_times} implies $D_{\Delta \omega}(\tau) \le 4 \pi (n_S B_{\min})/|\Delta \omega| 
 \le 4 \pi / (|\Delta \omega| \tau)$, so we must have $\tau \le c/|\Delta\omega|$ for some constant $c$. By the Heisenberg limit, we also have $\tau \ge c'/(n_S B_{\min})$ for some constant $c'$. Combining the two inequalities, we conclude that $\Delta \omega/(n_S B_{\min})$ is at most a constant. As already noted, the Grover-Heisenberg limit reduces to the Heisenberg limit in this case. Therefore, we may assume $\tau > (n_S B_{\min})^{-1}$.

As such, consider dividing $\tau$ into $\tau / T$ intervals, each of duration  $T = (n_S B_{\min})^{-1}$. (We assume $\tau/T$ is an integer for convenience; our results change by at most a constant factor if not.) Note that the time step $T$ has nothing to do with the structure of the protocol -- it is introduced only for the purpose of the proof. Let $U_k$ denote the unitary corresponding to time evolution during the $k$-th interval, and let $\ket{\psi_k}$ denote the state at time $kT$ :
\begin{align}
    U_k = \mathcal{T} \text{exp}\left(-i \int_{(k-1) T}^{kT} dt~\widetilde{H}_\omega(t) \right) \quad \text{and} \quad | \psi_k \rangle = U_k \cdots U_1 \mathbf{\ket{0}}.
\end{align}
Note that $\widetilde{U}_\omega = U_{\tau/T} \cdots U_2 U_1$. Also, define the intermediate average distinguishability $D_k = D_{\Delta \omega}(kT)$. We therefore have
\begin{align}
    D_k &= \frac{1}{|\Delta \omega|} \int_{\Delta \omega} d\omega~ \| | \psi_k \rangle - \mathbf{\ket{0}} \|_2^2 = \frac{1}{|\Delta \omega|} \int_{\Delta \omega} d\omega~ \| U_k | \psi_{k-1} \rangle - \mathbf{\ket{0}} + U_k \mathbf{\ket{0}} - U_k \mathbf{\ket{0}} \|_2^2.
\end{align}
Applying triangle inequality, we have
\begin{align}\label{eq: grover_eq_one}
    D_k  \le \frac{1}{|\Delta \omega|} \int_{\Delta \omega} d\omega~ \| U_k | \psi_{k-1} \rangle - U_k \mathbf{\ket{0}} \|_2^2 + 
    \| U_k \mathbf{\ket{0}}  -\mathbf{\ket{0}} \|_2^2 + 
    2 \| U_k | \psi_{k-1} \rangle - U_k \mathbf{\ket{0}} \|_2 \cdot \| U_k \mathbf{\ket{0}}  - \mathbf{\ket{0}} \|_2.
\end{align}
Applying unitary invariance of Euclidean distance, we note that the first term is simply $D_{k-1}$. The final term can be bound using integral Cauchy-Schwarz:
\begin{align}\label{eq: grover_eq_two}
    \frac{2}{|\Delta \omega|} \int_{\Delta \omega} d\omega \| | \psi_{k-1} \rangle - \mathbf{\ket{0}} \|_2 \cdot \| U_k \mathbf{\ket{0}}  - \mathbf{\ket{0}} \|_2 \le 2 \sqrt{ \frac{1}{|\Delta \omega|} \int_{\Delta \omega} d\omega \| | \psi_{k-1} \rangle - \mathbf{\ket{0}} \|_2^2 \cdot \frac{1}{|\Delta \omega|} \int_{\Delta \omega} d\omega \| U_k \mathbf{\ket{0}} - \mathbf{\ket{0}} \|_2^2 }.
\end{align}
Finally, define $d_{k-1} = \frac{1}{|\Delta \omega|} \int_{\Delta \omega} d\omega \| U_k \mathbf{\ket{0}}  - \mathbf{\ket{0}} \|_2^2$. For each $k$, $d_{k-1}$ gives the average distinguishability over a duration $T$. Combining~\eqref{eq: grover_eq_one} and~\eqref{eq: grover_eq_two}, we conclude that
\begin{align}
    D_k \le D_{k-1} + d_{k-1} + 2 \sqrt{D_{k-1} \cdot d_{k-1}} = \left( \sqrt{D_{k-1}} + \sqrt{d_{k-1}} \right)^2
\end{align}
or, taking the square root of each side, $\sqrt{D_k} \le \sqrt{D_{k-1}} + \sqrt{d_{k-1}}$. This implies that $D_{\Delta \omega}(\tau) \le \left(\sum_{k=1}^{\tau/T} \sqrt{d_{k-1}} \right)^2$. Applying Lemma~\ref{lemma: short_times}, we have that $d_{k-1} \le 4\pi (n_S B_{\min}) / |\Delta \omega|$ for each $k$, and thus
\begin{align}
    D_{\Delta \omega}(\tau) \le \left( \frac{\tau}{T} \sqrt{\frac{4 \pi}{|\Delta \omega|}} \sqrt{n_S B_{\min}} \right)^2 = \frac{4 \pi}{|\Delta \omega|} (n_S B_{\min})^3 \tau^2.
\end{align}
The necessity $ D_{\Delta \omega}(\tau) = \Omega(1)$ therefore implies $\tau = \Omega((n_S B_{\min})^{-1} \sqrt{|\Delta \omega|/(n_S B_{\min})})$. This concludes our proof of the Grover-Heisenberg limit.

\newpage

\section{Quantum Fisher Information Based Sensing Protocols}

In this section, we prove Theorem 3 in the main text. We begin by recalling the definition of the quantum Fisher information (QFI) and its interpretation in quantum sensing, which motivates our definition of a QFI-based sensing protocol. We then lower bound the duration $\tau$ of any such protocol, using similar techniques as in the proof of the Grover-Heisenberg limit in Sec.~\ref{sec: gh_limit}. 
This result means that the QFI has limited utility as a measure of success in AC sensing; any protocol performing at or near the Grover-Heisenberg limit necessarily has low QFI.

We note that our lower bound matches results which were previously and independently derived in~\cite{polloreno:2023}. However, we explicitly consider protocols which use an unbounded number of ancilla qubits, unlike that work. Consequently, our lower bound holds even against protocols which perform intermediate measurements and adapatively chosen gates.

The QFI is a well-established measure of performance in quantum sensing and quantifies the susceptibility of a sensing protocol to infinitesimal changes in some parameter to be estimated. Let $B$ be a continuous, unknown parameter which is encoded in a family of pure quantum states $\ket{\psi(B)}$ (which we imagine are output by a quantum sensing protocol). Assuming that the family of states satisfies certain regularity conditions (which are necessary for the QFI to be well-defined), the QFI (at $B = 0$) is~\cite{pang:2017}
\begin{align}\label{eq: qfi_def}
    \mathcal{F}_Q = 4 \left( \langle \partial_B \psi(B) | \partial_B \psi(B) \rangle - |\langle \psi(B) | \partial_B \psi(B)\rangle|^2\right)\big|_{B = 0}.
\end{align}
The quantum Cramér-Rao bound relates the QFI to the precision of any unbiased estimator $\widehat{B}$ of $B$ constructed from a measurement of $\ket{\psi(B)}$. Specifically, the variance of such an estimator satisfies $\text{Var}(\widehat{B}) \ge 1/\mathcal{F}_Q$. Therefore, to differentiate $B = 0$ from $B = B_{\min}$ with constant probability, we must have $\mathcal{F}_Q \ge C/B_{\min}^2$ for some constant $C > 0$.

In our setting, the states $\ket{\psi_\omega(B)}$ produced by the quantum sensing protocol depend on $\omega$, so the QFI $\mathcal{F}_Q(\omega)$ does too. Motivated by the discussion above, we define a QFI-based sensing protocol for $\ACProb$ as a protocol with well-defined QFI satisfying
\begin{align}\label{eq: qfi_based}
    \mathcal{F}_Q(\omega) \ge C/B_{\min}^2 \quad \text{for all }~\omega \in \Delta\omega
\end{align}
where $C$ is a constant. As in Sec.~\ref{sec: gh_limit}, we allow the protocol to use an arbitrary number of ancilla qubits and unitary control gates, so this definition includes protocols which perform intermediate measurements and adaptively chosen unitaries. We place no restrictions on the QFI with respect to $\omega$; in other words, our focus is on protocols which solve $\ACProb$ by estimating the single parameter $B$. We will prove that an arbitrary quantum sensing protocol satisfies
\begin{align}\label{eq: qfi_bound}
    \frac{1}{|\Delta \omega|} \int_{\Delta \omega} d\omega~\mathcal{F}_Q(\omega) \le 4 \pi \frac{n_S^2}{|\Delta \omega|} \tau.
\end{align}
Together,~\eqref{eq: qfi_based} and~\eqref{eq: qfi_bound} imply that any QFI-based quantum sensing protocol for $\ACProb$ must take time
\begin{align}
    \tau \ge \frac{C}{4 \pi} \frac{1}{n_S B_{\min}} \left( \frac{|\Delta \omega|}{n_S B_{\min}} \right).
\end{align}
Combining this lower bound with the Heisenberg limit in the edge case that $|\Delta \omega| < n_S B_{\min}$, this proves Theorem 3 in the main text. An immediate and important corollary of this result is that the QSS algorithm, as well as any other quantum sensing protocol which asymptotically outperforms this lower bound, has QFI which vanishes on average over $\Delta \omega$ in the large bandwidth limit ($n_S B_{\min} / |\Delta \omega| \to 0$).

We now prove the inequality~\eqref{eq: qfi_bound}. The proof is almost immediate using the techniques for proving Lemma~\ref{lemma: short_times}. We first observe that the second term in~\eqref{eq: qfi_def} is non-positive, so we have
\begin{align}
    \mathcal{F}_Q(\omega) \le 4 \langle \partial_B \psi_{\omega}(B) | \partial_B \psi_{\omega}(B) \rangle\big|_{B = 0} = 4 \int_0^\tau dt \int_0^\tau dt' \cos(\omega t + \phi) \cos(\omega t' + \phi) \mathbf{\bra{0}} \widetilde{Z}(t) \widetilde{Z}(t') \mathbf{\ket{0}}.
\end{align}
The second step follows by explicit computation, where $\widetilde{Z}(t)$ is as defined in Sec.~\ref{sec: gh_limit}. We take the average of both sides over $\Delta \omega$ and extend the range of integration to all of frequency space on the right hand side to obtain
\begin{align}
    \frac{1}{|\Delta \omega|} \int_{\Delta \omega} d\omega~\mathcal{F}_Q(\omega) \le \frac{4}{|\Delta \omega|} \int_{-\infty}^{\infty} d\omega \int_0^\tau dt \int_0^\tau dt' \cos(\omega t + \phi) \cos(\omega t' + \phi) \mathbf{\bra{0}} \widetilde{Z}(t) \widetilde{Z}(t') \mathbf{\ket{0}}.
\end{align}
By the analysis used to prove Lemma~\ref{lemma: short_times}, the right hand side is at most
\begin{align}
    \frac{4 n_S^2}{|\Delta \omega|} \int_{-\infty}^{\infty} d\omega \left(\int_0^\tau dt \cos(\omega t) \right)^2 = 4 \pi \frac{n_S^2}{|\Delta \omega|} \tau
\end{align}
as desired.
\vspace{-8mm}
\newpage

\section{Restricted Quantum Sensing Models}

We now explore the necessity of quantum computation for optimal AC sensing by proving Theorems 4 and 5 in the main text. We begin by reviewing our two restricted models of quantum sensing.

A sensing protocol with constant memory lifetime can fully control the sensors with a quantum computer, but only for a limited sensing duration. Specifically, we can utilize an unbounded number of ancilla qubits, prepare arbitrary initial states, and perform general unitary gates interleaved with signal evolution. However, we must measure all sensors and ancillae after a time less than or equal to $\Tmem$, before repeating with a new initial state.

A sensing protocol with classical signal processing can sense coherently for arbitrarily long times, but has limited control over the sensor. Specifically, we prepare arbitrary initial states and perform general measurements on the sensors, but our control is limited to applying a time-dependent modulation function $\chi_i(t) \in [-1,1]$ to each sensor $i = 1, \hdots, n_S$ to classically vary the interaction with the external field. We note that this model encompasses many existing quantum sensing protocols, in which all control gates either commute or anticommute with each $Z_i$. 

We refer to a single round of state preparation, unitary evolution, and measurement as an experiment. In both models, we allow adaptivity, meaning that later experiments can be chosen conditional on all prior measurement results. To describe adaptive quantum sensing algorithms, we will utilize the learning tree formalism of~\cite{chen:2021}, which we describe in Sec.~\ref{sec: learning_tree} and use in Sec.~\ref{sec: ctqs} and~\ref{sec: cpqs} to prove Theorems 4 and 5, respectively.

\subsection{Learning Tree Formalism}\label{sec: learning_tree}

We present the learning tree formalism, describing algorithms that conduct quantum experiments adaptively.

\begin{definition}[Learning Tree Representation] \label{def:tree_sensing}
    Suppose the external world has an unknown state indexed by $\omega \in \Omega$, for $\Omega$ the set of possible states. A classical algorithm $\mathcal{A}$ that tries to learn some property of the external world $\omega$ can be represented by a \emph{learning tree} $\mathcal{T}$. $\mathcal{T}$ is a rooted tree, where each node in the tree encodes the transcript for all experiments the algorithm has conducted so far and the corresponding experimental outcomes. The tree satisfies the following properties:
    \begin{itemize}[leftmargin=*,itemsep=0pt]
        \item Each node $u$ is associated with a value $p_\omega(u)$ corresponding to the probability that the transcript observed so far is given by the path from the root $r$ to $u$.
        $\mathcal{T}$ naturally induces a distribution over its leaves. At root $r$, $p_\omega(r) = 1$.
        \item At each non-leaf node $u$, an experiment is conducted. An experiment creates the final state $\ket{\phi_{u, \omega}}$ that depends on both $u$ and $\omega$, and performs a POVM measurement $\mathcal{F}_u = \{ F_{u, s} \}_s$ that depends on $u$. The children $v$ of $u$ are indexed by the possible measurements outcomes $s$. The distribution $q_{\omega, u}(s)$ over measurement outcomes $s$ depends on the transcript $u$ of all experiments and the external world $\omega$,
        \begin{equation}
            q_{\omega, u}(s) = \Tr( F_{u, s} \ketbra{\phi_{u, \omega}} ).
        \end{equation}
        We refer to the edge between $u$ and $v$ as $(u, s)$. The probability distribution over $v$ is
        \begin{equation}
            p_\omega(v) = p_\omega(u) \cdot q_{\omega, u}(s) \, .
        \end{equation}
        The cost of the experiment conducted at node $u$ is given by $c_u$.
        \item If along any root-to-leaf path, the sum of the costs $c_u$ is at most $C$, we say the complexity of $\mathcal{A}$ is at most $C$. In applications to query complexity, $c(u) = 1, \forall~u$. In applications to quantum sensing, $c(u)$ is evolution time under the signal Hamiltonian during the experiment which prepares $\ket{\phi_{u, \omega}}, \forall~u$ and independent of $\omega$.
    \end{itemize}
    After all experiments, the classical algorithm $\mathcal{A}$ will reach a leaf node $\ell$. From the leaf node $\ell$, the classical algorithm makes a prediction about the desired property about the unknown external world $\omega$.
\end{definition}

In \cite{chen:2023}, the authors give a proof for the NISQ complexity of unstructured search. In particular, the authors prove a lower bound for unstructured search against bounded-depth computation.
While the proof is focused on unstructured search, the result can be stated more abstractly.
By following the same steps as the proof given in \cite{chen:2023}, one can establish the following two general theorems.

\begin{theorem}[Learning tree lower bound from two-norm indistinguishability; adapted from \cite{chen:2023}] \label{thm:learning-tree-lowerbound-alt}
    Consider the task of distinguishing between two cases: $\omega = \bot$ and $\omega \in \Omega \setminus \{ \bot \}$. Suppose that in the $\omega \neq \bot$ case, the unknown $\omega$ is sampled from a distribution $\mathcal{D}$ over $\Omega \setminus \{ \bot \}$.
    Assume that there exists $L > 0$ and $c_{\max} > 0$ such that for any node $u$, we have $c(u) \leq c_{\max}$ and the following holds,
    \begin{equation} \label{eq:L2-norm-bound-alt}
        \E_{\omega \sim \mathcal{D}} \norm{\ket{\phi_{u, \omega}} - \ket{\phi_{u, \omega = \bot }}}_2^2 \leq L \cdot c(u)^2,
    \end{equation}
    where $c(u)$ is the cost of the experiment conducted at node $u$.
    Then the complexity $C$ to distinguish between $\omega = \bot$ and $\omega \in \Omega \setminus \{ \bot \}$ with a constant probability must satisfy
    $C = \Omega\left( 1 / (c_{\max} L) \right).$
\end{theorem}

\begin{theorem}[Alternative learning tree lower bound from two-norm indistinguishability; adapted from \cite{chen:2023}] \label{thm:learning-tree-lowerbound}
    Consider the task of distinguishing between two cases: $\omega = \bot$ and $\omega \in \Omega \setminus \{ \bot \}$. Suppose that in the $\omega \neq \bot$ case, the unknown $\omega$ is sampled from a distribution $\mathcal{D}$ over $\Omega \setminus \{ \bot \}$.
    Assume that there exists $L > 0$ such that for any node $u$ the following holds,
    \begin{equation} \label{eq:L2-norm-bound}
        \E_{\omega \sim \mathcal{D}} \norm{\ket{\phi_{u, \omega}} - \ket{\phi_{u, \omega = \bot}}}_2^2 \leq L \cdot c(u),
    \end{equation}
    where $c(u)$ is the cost of the experiment conducted at node $u$.
    Then the complexity $C$ to distinguish between $\omega = \bot$ and $\omega \in \Omega \setminus \{ \bot \}$ with a constant probability must satisfy
    $C = \Omega\left( 1 / L \right).$
\end{theorem}

Both theorems follow straightforwardly from the proof given in \cite{chen:2023} after noting the following basic changes:
\begin{itemize}
\item While the proof of \cite{chen:2023} is written as if $\Omega$ is a finite discrete set, the proof extends to any set $\Omega$ after replacing averages over the discrete set $\Omega$ by expectations over $\omega \sim \Omega$, collections of $\omega$ satisfying some desired property by subsets of $\Omega$ with the same property (namely, the $\omega$ such that a given path in the decision tree is $\omega$-good, $\omega$-balanced, or $\omega$-concentrated, see Definitions D.7, D.8, and D.13 in the arXiv version of~\cite{chen:2023}), etc. This applies to both theorems.
\item When~\eqref{eq:L2-norm-bound} is assumed, the potential function $\tau(\mathbf{z})$ for a given path $\mathbf{z} = (u_1, \ldots, u_n)$ considered in \cite{chen:2023} is replaced by $\sum_{i=1}^n c(u_i)$ without changing the proof. In this case, one does not need to apply an additional step of H\"older's inequality (Lemma D.6 in the arXiv version of \cite{chen:2023}). When~\eqref{eq:L2-norm-bound-alt} is assumed, the potential function $\tau(\mathbf{z})$ is replaced by $\sum_{i=1}^n c(u_i)^2$ without changing the proof. Similar to \cite{chen:2023}, one needs to apply an additional step of H\"older's inequality.
\end{itemize}
Hence, proving a lower bound for the distinguishing problem reduces to bounding $\E_{\omega \sim \mathcal{D}} \norm{\ket{\phi_{u, \omega}} - \ket{\phi_{u, \omega = \bot}}}_2^2$.
Depending on whether the norm bound is quadratic or linear in $c(u)$, we can utilize Theorem~\ref{thm:learning-tree-lowerbound-alt} 
or Theorem~\ref{thm:learning-tree-lowerbound} accordingly to obtain a complexity lower bound. 

In our case, $\bot$ indicates that the Hamiltonian is zero, while $\Omega \setminus \{ \bot \}$ gives the bandwidth $\Delta \omega$ of possible signal frequencies. It will suffice to let $\mathcal{D}$ be the uniform distribution over $\Delta \omega$, so that $\E_{\omega \sim \mathcal{D}} \norm{\ket{\phi_{u, \omega}} - \ket{\phi_{u, \omega = \bot}}}_2^2$ equals the average distinguishability $D_{\Delta \omega}(c(u))$ defined in~\eqref{eq: avg_dist} for a single experiment of duration $c(u)$.

\subsection{Sensing with Constant Memory Lifetime}\label{sec: ctqs}

Recall that a sensing protocol with constant memory lifetime can conduct multiple experiments adaptively, i.e., each experimental setup can depend on all prior experimental outcomes. Each experiment consists of the preparation of an arbitrary initial state, an evolution under arbitrary unitary gates interleaved with the action of the external field, and an arbitrary measurement. However, the evolution time within a single experiment is bounded by $\Tmem$. To establish Theorem 4, we need to bound the average distinguishability for an arbitrary quantum experiment and utilize the learning tree formalism to handle many adaptive experiments.

From the proof of the Grover-Heisenberg limit in Sec.~\ref{sec: gh_limit}, we know that over duration $T$ the average distinguishability for an arbitrary quantum experiment satisfies
\begin{align}
    D_{\Delta \omega}(T) \le \frac{4 \pi}{|\Delta \omega|} (n_S B_{\min})^3 T^2.
\end{align}
For a sensing protocol with constant memory lifetime, $T \le \Tmem$, so by applying Theorem~\ref{thm:learning-tree-lowerbound-alt} with $L = 4 \pi (n_S B_{\min})^3/|\Delta \omega|$, we conclude that the total duration $\tau$ is lower bounded by
\begin{align}
    \tau = \Omega\left(\frac{1}{n_S B_{\min} \Tmem} \cdot \frac{1}{n_S B_{\min}} \left( \frac{|\Delta \omega|}{n_S B_{\min}} \right)\right).
\end{align}
In the edge case that $|\Delta \omega| < n_S B_{\min}$, we use the Heisenberg limit $D_{\Delta \omega}(T) \le (n_S B_{\min})^2 T^2$ and apply Theorem~\ref{thm:learning-tree-lowerbound-alt} with $L = (n_S B_{\min})^2$ to conclude
\begin{align}
    \tau = \Omega\left(\frac{1}{n_S B_{\min} \Tmem} \cdot \frac{1}{n_S B_{\min}} \bigg\lceil \frac{|\Delta \omega|}{n_S B_{\min}} \bigg\rceil \right).
\end{align}
Together, these two lower bounds prove Theorem 4.

\subsection{Sensing with Classical Signal Processing}\label{sec: cpqs}

We now upper bound the average distinguishability for protocols which sense with classical signal processing. A sensing protocol with classical signal processing can also conduct multiple experiments adaptively. Each experiment senses the time-varying field coherently for arbitrarily long times, but it is limited to modulating the interaction between the sensor and the external field classically.

More precisely, an experiment can prepare arbitrary initial states and perform general measurements on the sensors, but the control is limited to applying a time-dependent modulation function $\chi_i(t) \in [-1,1]$ to each sensor $i = 1, \hdots, n_S$ to classically vary the interaction with the external field. Depending on the choice of modulation function, the phase response of the sensor can realize any linear system~\cite{oppenheim:1996}; this is why we refer to the model as sensing with classical signal processing. We do not rule out the possibility that quantum sensing protocols inspired by more advanced methods in classical signal processing might outperform the class of protocols considered here. 

For a given experiment (with signal strength $B_{\min}$), the Hamiltonian and the corresponding unitary evolution can be written in the following form:
\begin{align}
    \widetilde{H}_{\omega}(t) = B_{\min} \cos(\omega t + \phi) \cdot \sum_{i = 1}^{n_S} \chi_i(t) Z_i \quad \text{and} \quad \widetilde{U}_\omega = \text{exp}\left(-i \int_0^T dt~ \widetilde{H}_{\omega}(t)  \right)~.
\end{align}
Note that defining $\widetilde{U}_\omega$ does not require time-ordering, since $\widetilde{H}_{\omega}(t)$ commutes with itself at all times. Letting $\ket{\psi}$ indicate the arbitrary initial state of the $n_S$ sensors, we have
\begin{align}
    D_{\Delta \omega}(T) = \frac{1}{|\Delta \omega|} \int_{\Delta \omega} d\omega~ \left\| \ket{\psi} - \widetilde{U}_{\omega} \ket{\psi} \right\|_2^2 \le \frac{1}{|\Delta \omega|} \int_{-\infty}^{\infty} d\omega~ \left\| \ket{\psi} - \widetilde{U}_{\omega} \ket{\psi} \right\|_2^2 \le \frac{1}{|\Delta \omega|} \int_{-\infty}^{\infty} d\omega~ \left\| I - \widetilde{U}_{\omega} \right\|^2 
\end{align}
where the first step is to extend the range of integration to all of frequency space, and the second follows by the definition of operator norm. For any Hermitian operator $A$, $\|I - e^{-iA}\| \le \|A\|$. Therefore we have
\begin{align}
   \hspace*{-3ex} D_{\Delta \omega}(T) \le \frac{B_{\min}^2}{|\Delta \omega|} \int_{-\infty}^{\infty} d\omega\left\| \sum_{i = 1}^{n_S} \int_0^T dt~\chi_i(t) \cos(\omega t + \phi) Z_i \right\|^2 = \frac{B_{\min}^2}{|\Delta \omega|} \int_{-\infty}^{\infty} d\omega\left\| \sum_{i = 1}^{n_S} \text{Re}\left(e^{i \phi} \int_0^T dt~\chi_i(t) e^{i \omega t} \right) Z_i \right\|^2.
\end{align}
Define $\zeta_i(t)$ as $\chi_i(t)$ for $0 \le t \le T$ and zero otherwise, so $\int_0^T \chi_i(t) e^{i \omega t} = \hat{\zeta}_i(\omega)$, the Fourier transform of $\zeta_i(t)$. Hence
\begin{align}
    D_{\Delta \omega}(T) \le \frac{B_{\min}^2}{|\Delta \omega|} \int_{-\infty}^{\infty} d\omega~\left\| \sum_{i = 1}^{n_S} \text{Re} \big( e^{i \phi} \hat{\zeta}_i(\omega) \big) Z_i \right\|^2 =  \frac{B_{\min}^2}{|\Delta \omega|} \int_{-\infty}^{\infty} d\omega \left(\max_{\vec{b} \in \{\pm 1\}^{n_S}} \sum_{i=1}^{n_S}b_i~\text{Re} \big( e^{i \phi} \hat{\zeta}_i(\omega) \big) \right)^2
\end{align}
where the equality follows by computing the operator norm. The argmax may depend on $\omega$; call it $\vec{b}(\omega)$. Therefore
\begin{align}
     D_{\Delta \omega}(T) &\le \frac{B_{\min}^2}{|\Delta \omega|} \sum_{i,j = 1}^{n_S} \int_{-\infty}^{\infty} d\omega~b_i(\omega) \text{Re} \big(e^{i \phi} \hat{\zeta}_i(\omega) \big)  b_j(\omega) \text{Re} \big(e^{i \phi} \hat{\zeta}_j(\omega) \big) 
     \\
     &\le \frac{B_{\min}^2}{|\Delta \omega|} \sum_{i,j = 1}^{n_S} \sqrt{\int_{-\infty}^{\infty} d\omega~\text{Re} \big(e^{i \phi} \hat{\zeta}_i(\omega)\big)^2 \int_{-\infty}^{\infty} d\omega~\text{Re} \big(e^{i \phi} \hat{\zeta}_j(\omega)\big)^2}
     \\
     &\le \frac{B_{\min}^2}{|\Delta \omega|} \sum_{i,j = 1}^{n_S} \sqrt{\int_{-\infty}^{\infty} d\omega~ | \hat{\zeta}_i(\omega) |^2 \int_{-\infty}^{\infty} d\omega~| \hat{\zeta}_j(\omega) |^2}
\end{align}
where in the second line we have used Cauchy-Schwarz on the integral, along with the fact that $b_i(\omega)^2 = 1$ for all $i$ and $\omega$, and in the final line we upper bounded the real part of $e^{i \phi} \hat{\zeta}_i(\omega)$ by its absolute value. Next, we relate the integral of $|\hat{\zeta}_i(\omega)|^2$ over frequency space to the integral of $|\zeta_i(t)|^2$ over time:
\begin{align}
    \int_{-\infty}^{\infty} d\omega~|\hat{\zeta}_i(\omega)|^2 = 2 \pi \int_{-\infty}^{\infty} d\omega~|\zeta_i(t)|^2 = 2 \pi \int_{0}^{T} d\omega~\chi_i(t)^2 \le 2 \pi T.
\end{align}
The first equality is Plancherel's theorem, the second follows from the definition of $\zeta_i(t)$, and the last holds since $\chi_i(t) \in [-1,1]$. We conclude that the average distinguishability is bound by
\begin{align}
    D_{\Delta \omega}(T) \le 2 \pi \frac{(n_S B_{\min})^2}{|\Delta \omega|} T
\end{align}
Utilizing Theorem~\ref{thm:learning-tree-lowerbound} with $L = 2 \pi (n_S B_{\min})^2/|\Delta \omega|$, we conclude that $\tau$ satisfies 
\begin{align}
    \tau = \Omega\left(\frac{1}{n_S B_{\min}} \cdot  \frac{|\Delta \omega|}{n_S B_{\min}} \right).
\end{align}
Combining this result with the Heisenberg limit in the edge case that $|\Delta \omega| < n_S B_{\min}$, this proves Theorem 5. 
\newpage

\section{Beyond Quadratic Metrological Gain in AC Sensing}

In this section, we demonstrate how, in some regimes, there is a beyond-quadratic separation in sensing time between quantum computing enhanced sensing and conventional sensing approaches, including classical signal processing and quantum Fisher information based protocols.

\subsection{Strong Field, Large Bandwidth}

The difficulty of solving $\ACProb$ depends on both the sensitivity $B_{\min}$ and bandwidth $|\Delta \omega|$; therefore it is not in general possible to characterize an instance of the problem by a single ``size'' parameter. Here, we consider a simple setting where the bandwidth and sensitivity are both determined by a size parameter $N$. We suppose the bandwidth scales linearly with problem size, $|\Delta \omega| \propto N$, and the field is strong with minimum strength scaling as $B_{\min} \propto N^\mu$ for some $0 \le \mu < 1/2$. In this strong field, large bandwidth regime, the sensing time for quantum computing enhanced sensing is
\begin{equation}
\widetilde{O} \left( \frac{1}{n_S B_{\min}} \sqrt{\bigg\lceil \frac{|\Delta \omega|}{n_S B_{\min}}} \bigg\rceil \right) = \Theta(N^{1/2- 3\mu/2}),
\end{equation}
while the sensing time for conventional approaches (including both classical signal processing and quantum Fisher information based protocols) is
\begin{equation}
\Theta \left( \frac{1}{n_S B_{\min}} \bigg\lceil \frac{|\Delta \omega|}{n_S B_{\min}} \bigg\rceil \right) = \Theta(N^{1 - 2 \mu}).
\end{equation}
This comparison reveals three distinct regimes:
\begin{itemize}
    \item For $1/3 < \mu < 1/2$, quantum computing enhanced sensing achieves constant time in $N$, while conventional sensing requires polynomial time.
    \item For $0 < \mu < 1/3$, quantum computing enhanced sensing provides a beyond-quadratic speedup.
    \item At $\mu = 0$, the advantage reduces to the standard quadratic speedup of Grover's search.
\end{itemize}
This analysis illuminates a fundamental distinction between query complexity advantage and metrological gain. While the metrological gain stems from the quadratic quantum computational advantage of Grover's search, its characteristics can differ significantly from those of Grover's algorithm.

\subsection{Power Law Signal Strength}

We now consider an alternative scenario where similar, beyond-quadratic metrological gain is made possible by quantum computing enhanced sensing. Consider a setting where a signal of frequency $\omega$, if present, has strength satisfying $B \ge B_{\min}(\omega) = B_0^{1 - \mu} \omega^{\mu}$ for some constant $B_0 > 0$ and power law exponent $\mu > 0$. This scenario can be interpreted as arising from a frequency-dependent coupling constant $g(\omega) \sim \omega^\mu$ with fixed signal strength. We demonstrate that for certain power law exponents, quantum computing enhanced sensing can solve $\mathsf{AC}[B_{\min}(\omega), \Delta \omega]$ in constant time, while conventional protocols require time that scales polynomially with $|\Delta \omega|$. For our asymptotic analysis, we assume $|\Delta \omega| \gg n_S B_0$.

Our quantum computing enhanced protocol partitions the bandwidth $\Delta \omega$ into subintervals $\Delta \omega_j$ ($j = 1, \hdots, j_{\max}$), with interval sizes growing as $|\Delta \omega_j| = |\Delta \omega_0| j^{\alpha}$. Here, $|\Delta \omega_0| \ge B_0$ is constant and $\alpha \ge 1$. Within interval $\Delta \omega_j$, any present signal must have strength $B \ge B_j \sim B_0 j^{\beta}$, where $\beta = \mu(1+\alpha)$. We then employ QSS to solve $\mathsf{AC}[B_j, \Delta \omega_j]$ for each interval. The total sensing time $\tau_Q$ for this protocol is:
\begin{align}
    \tau_Q = \widetilde{O} \left( \sum_{j = 1}^{j_{\max}} \frac{1}{n_S B_j} \sqrt{\frac{|\Delta \omega_j|}{n_S B_j}} \right) =  \widetilde{O} \left( \frac{1}{n_S B_0} \sqrt{\frac{|\Delta \omega_0|}{n_S B_0}} \sum_{j = 1}^{j_{\max}} j^{\frac{\alpha - 3\beta}{2}} \right)
\end{align}
Crucially, when $(\alpha - 3 \beta)/2 < -1$, or equivalently $(1-3\mu) \alpha < 3 \mu + 2$, this sum converges to yield an $O(1)$-time algorithm\footnote{The result is $O(1)$ rather than merely $\widetilde{O}(1)$ because $\sum_{j = 1}^\infty (j \log(j)^p)^{-1 - \epsilon}$ converges for any $p$ when $\epsilon > 0$.}.

In contrast, for protocols using classical signal processing or quantum Fisher information based sensing, the runtime $\tau_C$ is lower bounded by:
\begin{align}
    \tau_C = \Omega\left( \frac{|\Delta \omega|}{(n_S (B_0^{1-\mu} \omega_{\max}^{\mu}))^2} \right) = \Omega\left(\frac{|\Delta \omega|^{1-2\mu}}{(n_S B_0^{1-\mu})^2}\right)
\end{align}
where we use that $B_0^{1 - \mu} \omega_{\max}^\mu = B_{\min}(\omega_{\max})$ is a possible signal strength for all $\omega \in \Delta \omega$, and $\omega_{\max} \sim |\Delta \omega|$. This scaling is nontrivial when $\mu < 1/2$.

Comparing $\tau_Q$ and $\tau_C$, we find a constant-versus-polynomial separation between quantum computing enhanced sensing and conventional protocols when $(1-3\mu) \alpha < 3 \mu + 2$ and $\mu < 1/2$. Taking $\alpha \to \infty$ in the large bandwidth limit, these inequalities reduce to $1/3 < \mu < 1/2$. This gives an alternative demonstration for how Grover's quadratic query complexity advantage translates into a beyond-quadratic metrological gain.

While both approaches require polynomial gate counts in $|\Delta \omega|$, the quantum computing enhanced protocol's performance in the regime $1/3 < \mu < 1/2$ is limited only by gate speed, whereas conventional approaches are constrained by both gate speed and sensing time. This advantage becomes particularly relevant when sensing time significantly exceeds quantum gate implementation time.

\newpage
\section{Synthesizing Arbitrary Quantum Oracles}

In this section, we establish a broader connection between oracular problems in quantum complexity theory and the detection of time-dependent signals. Specifically, we demonstrate that any quantum oracle can be realized by time evolution under a time-dependent AC signal with multiple frequencies. This construction establishes a foundation for greater metrological gains in quantum computing enhanced sensing, based on oracular problems with larger quantum-classical query separations.
However, it remains an open question to establish sensing time lower bounds for restricted sensing models, such as sensing with classical signal processing, from the classical query complexity lower bounds for the corresponding oracular problem. Developing a general lifting framework \cite{chattopadhyay:2021} is an important future direction.

Let $f: \{0,1\}^n \to \{0,1\}^m$ be an arbitrary vector-valued Boolean function and let $O_f$ be the associated quantum oracle, defined by
\begin{align}
    O_f \ket{x}_X \ket{y}_Y = \ket{x}_X \ket{y \oplus f(x)}_Y.
\end{align}
Here, $X$ and $Y$ are systems of $n$ and $m$ qubits, respectively. We will show that there exists a Hamiltonian $H_f(t)$ acting on $m$ qubits (denoted $S$) such that time evolution under $H_f(t)$, interleaved with a sequence of $f$-independent control unitaries, implements $O_f$. 

We begin with the case $m = 1$, so $f$ is an arbitrary Boolean function. Consider the following Hamiltonian $H_f(t)$:
\begin{align}
    H_f(t) = B \sum_{x = 1}^{2^n} f(x) \cos(\omega_x t) Z_S
\end{align}
and let $\omega_x = p_x \omega_0$, where $p_x$ is the $x$-th prime number. For our control unitaries, we will apply a CPMG sequence to the sensor (see Sec.~\ref{sec: conventproof}), but we let the interpulse spacing between $\pi$ pulses be conditional on the state of $X$. Specifically, prepare the sensor in state $\ket{+}_S$, then conditioned on state $\ket{k}_X$, apply a CPMG sequence with target frequency $\omega_k$ and duration $T = 2 \pi/\omega_0$.
The conditional state of the sensor is $\frac{1}{\sqrt{2}}(e^{- i \theta_k} \ket{0}_S + e^{i \theta_k} \ket{1}_S)$, where $\theta_k$ is
\begin{align}
    \theta_k = B T \sum_{x=1}^{2^n} f(x) \sinc\left(\frac{\omega_x T}{2} \right) \left(1 - \sec(\frac{\pi}{2} \frac{\omega_x}{\omega_k}) \right) \cos(\frac{\omega_x T}{2}).
\end{align}
The filter function of the CPMG sequence vanishes at any frequency which is not an odd integer multiple of the target frequency~\cite{degen:2017}. Therefore, the $x$-th term in this sum is zero as long as $\omega_x/\omega_k = p_x/p_k$ is not an odd integer, which is true for all $x \neq k$. Explicitly computing the $x = k$ term, we obtain
\begin{align}
    \theta_k = \frac{2 B T}{\pi} f(k).
\end{align}
By letting $B = \pi^2 / (4 T)$, the conditional phase angle is $\theta_k = \frac{\pi}{2} f(k)$. The conditional state of the sensor is therefore $\frac{1}{\sqrt{2}}((-i)^{f(k)} \ket{0}_S + (i)^{f(k)} \ket{1}_S)$. Let $W$ be the single qubit unitary defined by
\begin{align}
    W = \frac{1}{\sqrt{2}}
    \begin{pmatrix}
        1 & 1 \\
        i & -i
    \end{pmatrix}.
\end{align}
Apply $W$ to $S$, then a CNOT gate between $S$ and $Y$, then $W^T$ on $S$. This flips qubit $Y$ if and only if $f(k) = 1$ and maps the sensor to state $\frac{1}{\sqrt{2}}((i)^{f(k)} \ket{0} + (-i)^{f(k)} \ket{1})$. Finally, evolve under $H_f(t)$ for duration $T$ while applying the same conditional CPMG sequence. This uncomputes the sensor state,
completing the oracle construction for $m = 1$.

We can easily generalize to arbitrary $m$ by using a different sensor for each output bit of $f$. Define
\begin{align}
    H_f(t) = B \sum_{x = 1}^{2^n} \sum_{i=1}^m f_i(x) \cos(\omega_{x} t) Z_{S, i}
\end{align}
where $f_i(x)$ denotes the $i$-th bit of $f(x)$ and $Z_{S,i}$ acts on the $i$-th sensor. By evolving under $H_f(t)$ while applying the gate sequence above to all sensors (matching each sensor to the corresponding qubit in $Y$ for the CNOT), we implement the oracle $O_f$.

\vspace{-4mm}
\bibliography{QuantumSensing}